\documentclass[structabstract]{aa} 
\usepackage{graphicx}
\usepackage{lscape}
\usepackage{natbib}
\usepackage[]{hyperref} 
\usepackage{txfonts}
\usepackage{longtable}
\begin{document}
\newcommand{\kms}{km~s$^{-1}$}
\newcommand{\Msun}{M$_{\odot}$}
\newcommand{\Teff}{$T_{\rm eff}$}
\newcommand{\FeH}{[Fe/H]}
\newcommand{\doe}{{\scriptsize DOE}}
\newcommand{\ha}{H$\alpha$}

\title {The \emph{Gaia}-ESO Survey\thanks{Based on data products from observations made with ESO Telescopes at the La Silla Paranal Observatory under programme ID 188.B-3002. These data products have been processed by the Cambridge Astronomy Survey Unit (CASU) at the Institute of Astronomy, University of Cambridge, and by the FLAMES/UVES reduction team at INAF/Osservatorio Astrofisico di Arcetri. These data have been obtained from the \emph{Gaia}-ESO Survey Data Archive, prepared and hosted by the Wide Field Astronomy Unit, Institute for Astronomy, University of Edinburgh, which is funded by the UK Science and Technology Facilities Council.}: double, triple and quadruple-line spectroscopic binary candidates}
   \author{T. Merle\inst{1}
          \and S. Van Eck\inst{1}          
          \and A. Jorissen\inst{1}
          \and M. Van der Swaelmen\inst{1}
          \and T. Masseron\inst{2}
          \and T. Zwitter\inst{3}
          \and D. Hatzidimitriou\inst{4,5}
          \and A. Klutsch\inst{6} 
          \and D. Pourbaix\inst{1}
          \and R. Blomme\inst{7}         
          \and C.~C. Worley\inst{2}          
          \and G. Sacco\inst{8}
          \and J. Lewis\inst{2}
          \and C. Abia\inst{9}
          \and G. Traven\inst{3}
          \and R. Sordo\inst{10}
          \and A. Bragaglia\inst{11}
          \and R. Smiljanic\inst{12}
          \and E. Pancino\inst{8,21}
          \and F. Damiani\inst{13}
          \and A. Hourihane\inst{2}
          \and G. Gilmore\inst{2}
          \and S. Randich\inst{8}
          \and S. Koposov\inst{2}
          \and A. Casey\inst{2}
          \and L. Morbidelli\inst{8}
          \and E. Franciosini\inst{8}
          \and L. Magrini\inst{8}
          \and P. Jofre\inst{2,22}
          \and M.~T. Costado\inst{14}
          \and R.~D. Jeffries\inst{15}
          \and M. Bergemann\inst{16}
          \and A.~C. Lanzafame\inst{6,17}
          \and A. Bayo\inst{18}    
          \and G. Carraro\inst{19}
          \and E. Flaccomio\inst{13}
          \and L. Monaco\inst{20}
          \and S. Zaggia\inst{10}}

 \institute{$^1$ Institut d'Astronomie et d'Astrophysique, Universit\'e Libre de Bruxelles, CP. 226, Boulevard du Triomphe, 1050 Brussels, Belgium \\
                 \email{tmerle@ulb.ac.be} \\
            $^2$  Institute of Astronomy, University of Cambridge,  Madingley Road, Cambridge CB3 0HA, UK\\
            $^3$  Faculty of Mathematics and Physics, University of Ljubljana, Jadranska 19, 1000, Ljubljana, Slovenia\\
            $^4$  Department of Astrophysics, Faculty of Physics, National and Kapodistrian University of Athens, Panepistimiopolis, GR15784 Zografos, Athens, Greece\\
            $^5$  IAASARS, National Observatory of Athens, 15236 Penteli, Greece \\
            $^6$  INAF – Osservatorio Astrofisico di Catania, via S. Sofia 78, 95123 Catania, Italy\\
            $^7$  Royal Observatory of Belgium, Ringlaan 3, 1180, Brussels, Belgium\\
            $^8$  INAF – Osservatorio Astrofisico di Arcetri, Largo E. Fermi 5, 50125 Firenze, Italy\\
            $^9$  Dpto. Física Teórica y del Cosmos, Universidad de Granada, 18071, Granada, Spain\\
            $^{10}$  INAF – Osservatorio Astronomico di Padova, vicolo dell'Osservatorio 5, 35122 Padova, Italy\\
            $^{11}$  INAF – Osservatorio Astronomico di Bologna, via Ranzani 1, 40127, Bologna, Italy\\
            $^{12}$  Nicolaus Copernicus Astronomical Center, Polish Academy of Sciences, ul. Bartycka 18, 00-716, Warsaw, Poland\\
            $^{13}$  INAF – Osservatorio Astronomico di Palermo, Piazza del Parlamento 1, 90134, Palermo, Italy  \\ 
            $^{14}$  Instituto de Astrof\'{i}sica de Andaluc\'{i}a-CSIC, Apdo. 3004, 18080 Granada, Spain\\
            $^{15}$  Astrophysics Group, Keele University, Keele, Staffordshire ST5 5BG, United Kingdom\\
            $^{16}$ Max-Planck Institute for Astronomy, Koenigstuhl 17, Heidelberg, D-69117, Germany \\
            $^{17}$ Universit\`a di Catania, Dipartimento di Fisica e Astronomia, Sezione Astrofisica, Via S. Sofia 78, I-95123 Catania, Italy\\
            $^{18}$  Instituto de F\'isica y Astronomi\'ia, Universidad de Valpara\'iso, Chile\\
            $^{19}$  European Southern Observatory, Alonso de Cordova 3107 Vitacura, Santiago de Chile, Chile\\
            $^{20}$  Departamento de Ciencias Fisicas, Universidad Andres Bello, Republica 220, Santiago, Chile\\
            $^{21}$  ASI Science Data Center, Via del Politecnico SNC, 00133 Roma, Italy
            $^{22}$ N\'ucleo de Astronom\'ia, Facultad de Ingenier\'ia, Universidad Diego Portales,  Av. Ej\'ercito 441, Santiago, Chile}

\date{Received ...; accepted ...}

\abstract 
{The \emph{Gaia}-ESO Survey (GES) is a large spectroscopic survey that provides a unique opportunity to study the distribution of spectroscopic multiple systems among different populations of the Galaxy.} 
{We aim at detecting binarity/multiplicity for stars targeted by the GES from the analysis of the cross-correlation functions (CCFs) of the GES spectra with spectral templates.} 
{We develop a method based on the computation of the CCF successive derivatives to detect multiple peaks and determine their radial velocities, even when the peaks are strongly blended. The parameters of the detection of extrema (\doe) code have been optimized for each GES GIRAFFE and UVES setup to maximize detection. The \doe\ code therefore allows to automatically detect multiple line spectroscopic binaries (SB$n$, $n \ge 2$).}
{We apply this method on the fourth GES internal data release and detect 354 SB$n$ candidates (342~SB2, 11~SB3 and even one~SB4), including only nine SBs known in the literature. This implies that about $98$\,\% of these SB$n$ candidates are new (because of their faint visual magnitude that can reach $V=19$). Visual inspection of the SB$n$ candidate spectra reveals that the most probable candidates have indeed a composite spectrum. 
Among SB2 candidates, an orbital solution could be computed for two previously unknown binaries: CNAME 06404608+0949173 (known as V642~Mon) in NGC~2264  and CNAME 19013257-0027338 in Berkeley~81 (Be~81). A detailed analysis of the unique SB4 (four peaks in the CCF) reveals that CNAME 08414659-5303449 (HD~74438) in the open cluster IC~2391 is a physically bound stellar quadruple system. The SB candidates belonging to stellar clusters are reviewed in detail to discard false detections. We warn against the use of atmospheric parameters for these system components rather than by by SB-specific pipelines.} 
{Our implementation of an automatic detection of spectroscopic binaries within the GES has allowed an efficient discovery of many new multiple systems. With the detection of the SB1 candidates that will be the subject of a forthcoming paper, the study of the statistical and physical properties of the spectroscopic multiple systems will soon be possible for the entire GES sample.} 


\titlerunning{GES: Double, triple and quadruple-line SBs}

\maketitle

\vspace{-1cm}

\section{Introduction}

Binary stars play a fundamental role in astrophysics since they allow  direct measurements of masses, radii, and luminosities  that put constraints on stellar physics, Galactic archaeology, high-energy physics, etc. Binary systems are found at all evolutionary stages, and after strong interaction, some may end up as double degenerate systems or merged compact objects.

Spectroscopic binaries (SBs) exist in different flavours. On the one hand, SB1 (SB with one observable spectrum) can only be detected from the Doppler shift of the stellar spectral lines. On the other hand, SB$n$ ($n \ge 2$) are characterized by a composite spectrum made out of $n$ stellar components, and are detected either from the composite nature of the spectrum or from the Doppler shift of the spectral lines.   
SBs are certainly the binaries that cover the widest range of masses (from brown dwarfs to massive twins) and all ranges of periods (from hours to hundreds of years as observed so far, \emph{e.g.} \citealt{pourbaix2000}). To date, more than 3500 SBs with orbital elements have been catalogued and, among them, about 1126 are SB2 \citep[][and the latest online version of the SB9 catalogue]{pourbaix2004}. The Geneva-Copenhagen Survey catalogue \citep{nordstrom2004, holmberg2009} contains approximately 4000 SB1, 2100 SB2, and 60 SB3 out of 16700 F and G dwarf stars in the solar neighborhood, most without orbits. In the vast majority of cases, these binaries are not yet confirmed  but correspond to an overall binary fraction {in the Milky Way of almost 40~\%. A census of binary fraction is also available from the Hipparcos catalogue \citep{frankowski2007} though the binary fraction per spectral type is probably biased due to selection biases in the Hipparcos entry catalogue. 
New recent Galactic surveys like APOGEE \citep{majewski2015} or LAMOST \citep{luo2015} allow new investigations of binarity over large sample of stars \citep[see, \emph{e.g.},][]{gao2014,troup2016, fernandez2017}. For instance, the RAVE survey has led to the detection of 123 SB2 candidates out of 26\,000 objects \citep{matijevic2010, matijevic2011}. We refer the reader to \cite{duchene2013} for a recent review of the physical properties of multiplicity among stars and more specifically to \cite{raghavan2010} for a complete volume-limited sample of solar-type stars in the solar neighborhood (distances closer than 25 pc).

The \emph{Gaia}-ESO Survey (GES) is an on-going ground-based high-resolution spectroscopic survey of 10$^5$ stellar sources \citep{gilmore2012, randich2013} covering the main stellar populations (bulge, halo, thin and thick disks) of the Galaxy as well as a large number of open clusters spanning large metallicity and age ranges. All evolutionary stages are encountered within the GES, from pre-main sequence objects to red giants. It aims at complementing the spectroscopy of the \emph{Gaia} ESA space mission \citep{wilkinson2005}. The GES uses the FLAMES multi-fibre back end at the high resolution UVES ($R \sim 50\,000$) and moderate resolution GIRAFFE ($R \sim 20\,000$) spectrographs. The visual magnitude of the faintest targets reaches $V\sim20$. The spectral coverage spans the optical wavelengths (from 4030 to 6950~\AA) and the near infrared around the \ion{Ca}{ii} triplet and the Paschen lines (from 8490 to 8900~\AA\ including the wavelength range of the Radial Velocity Spectrometer of the \emph{Gaia} mission). The median signal-to-noise ($S/N$) ratio per pixel is similar for UVES and GIRAFFE single exposures ($\sim 30$) whereas the most frequent values are around 20 and 5 respectively.

The motivation of the present work is to take the advantage of a very large sample to detect automatically SBs with more than one visible component\footnote{Since SB1 systems require a special treatment by analyzing temporal series, their analysis should await the completion of the observations.} that are not always detected by the GES single-star main analysis pipelines. SBs may be a potential source of error when deriving atmospheric parameters and detailed abundances. This project presents (i) a new method to identify automatically the number of velocity components in each cross-correlation functions (CCFs) using their successive derivatives and (ii) the analysis of about 51\,000 stars available within the GES internal data release 4 (iDR4).  

In Sect.~\ref{sect:data_selection}, we describe the iDR4 stellar observations, their associated CCFs and the selection criteria applied to them. The method on which the detection of the velocity components in a CCF relies, its parameters and the formal uncertainty are presented in Sect.~\ref{sect:methods}. In Sect.~\ref{sect:results}, the set of SB$n$ ($n\ge 2$) detected in iDR4 using this method is discussed, organized according to the stellar populations they belong to.

\begin{table*}
\center
\caption{Setups used in GES and the associated estimated best parameters of the \doe\ code.}
 \begin{tabular}{lcclccc}
 \hline\hline
 Instrumental & Spectral & $\lambda$ range & Main spectral features & THRES0  & THRES2 & SIGMA \\
 setup & resolution & [nm] & & [\%] & [\%] & [\kms] \\
 \hline
 \\
 UVES\\
U520 low & 47\,000 & $420 - 520$ & G band, H$\gamma$, H$\beta$ & 35 & 8 & 5.0 \\
U520 up  & 47\,000 & $525 - 620$ & Fe~I~E, Na~I~D               & 35 & 8 & 5.0 \\
U580 low & 47\,000 & $480 - 575$ & H$\beta$, Mg~I~b             & 35 & 5 & 5.0 \\
U580 up  & 47\,000 & $585 - 680$ & Na~I~D, \ha                  & 35 & 5 & 5.0 \\
\\
GIRAFFE\\
HR3      & 24\,800 & $403 - 420$ & H$\delta$ & 55 & 8 & 3.0 \\
HR5A     & 18\,470 & $434 - 457$ & H$\gamma$ & 55 & 8 & 3.0 \\
HR6      & 20\,350 & $454 - 475$ & He~I \& II, Si III \& IV, C III, N~II, O~II   & 55 & 8 & 3.0 \\
HR9B     & 25\,900 & $514 - 535$ & Mg~I~b, Fe~I~E & 55 & 8 & 3.0 \\
HR10     & 19\,800 & $534 - 561$ & many weak lines & 55 & 8 & 2.1 \\
HR14A    & 17\,740 & $631 - 670$ & \ha & 55 & 8 & 3.0 \\
HR15N    & 17\,000 & $645 - 681$ & \ha, Li I & 55 & 8 & 3.0 \\
HR15     & 19\,300 & $660 - 695$ & O$_2$ A, Li I & 55 & 8 & 3.0 \\
HR21     & 16\,200 & $849 - 900$ & \ion{Ca}{ii} triplet, Paschen lines & 55 & 8 & 5.0 \\
\\
\hline
\end{tabular}
\label{tab:doe_param}
\end{table*}

\begin{figure*}
 \includegraphics[width=0.5\linewidth]{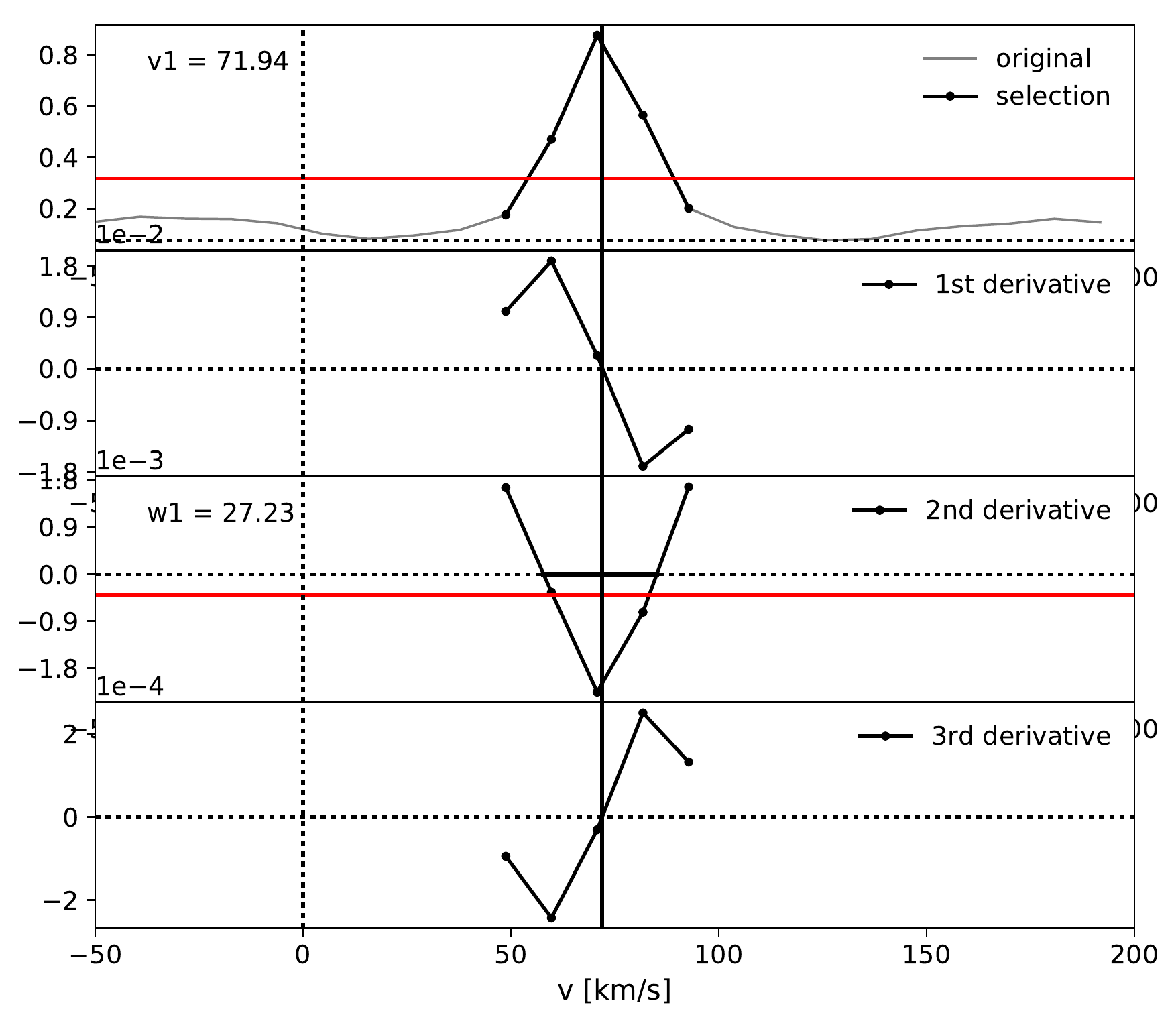}
 \includegraphics[width=0.47\linewidth]{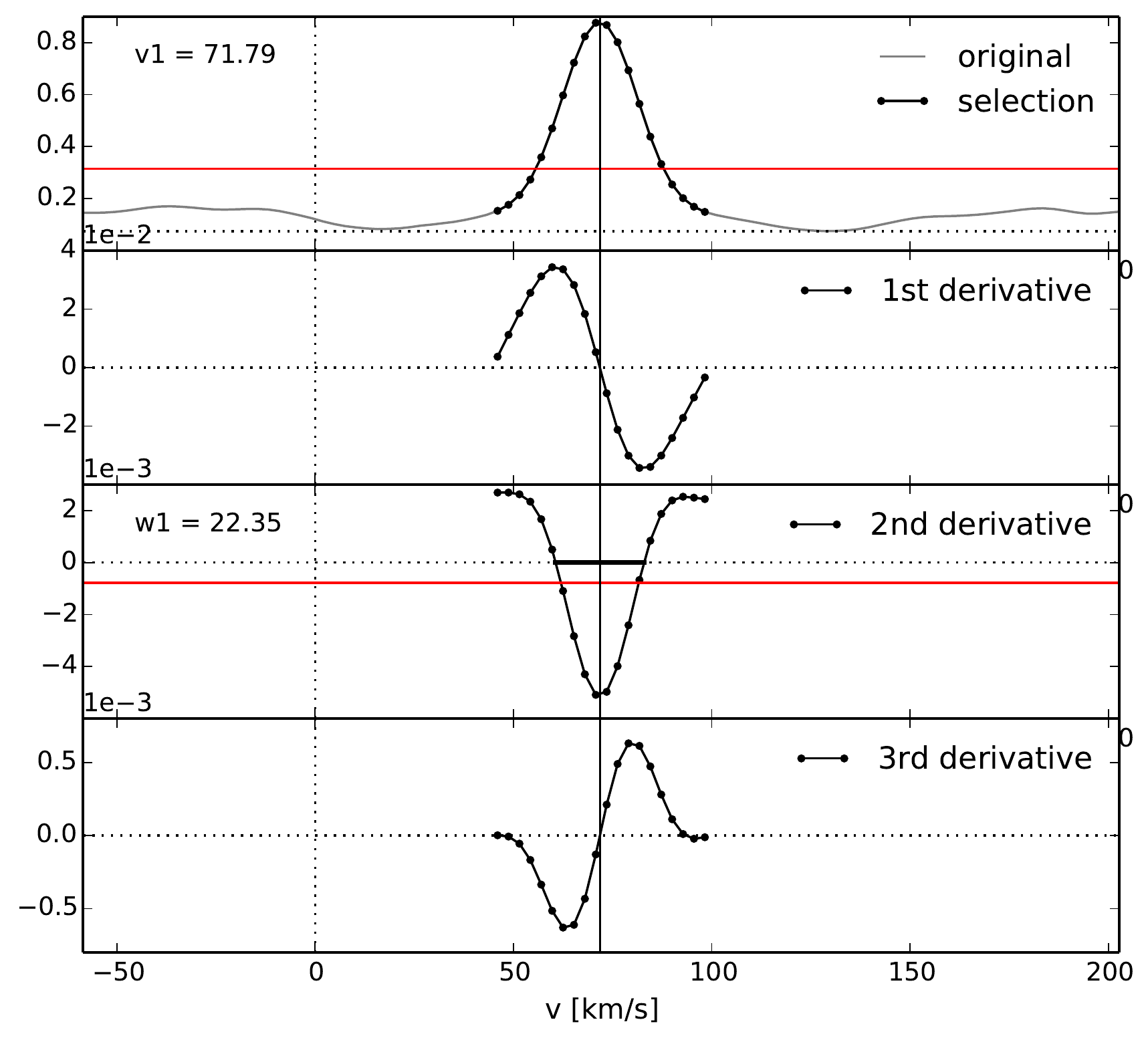}
 \caption{Simulated CCF at limiting numerical resolution to test the computation of successive derivatives and the detection of the peak (left), and with a more realistic sampling (right). The spectrum used to simulate these CCFs has a radial velocity of 72.0 \kms and $S/N=5$.}
 \label{fig:ccf_test}
\end{figure*}

\begin{figure*}
 \centering
 \includegraphics[width=0.70\linewidth]{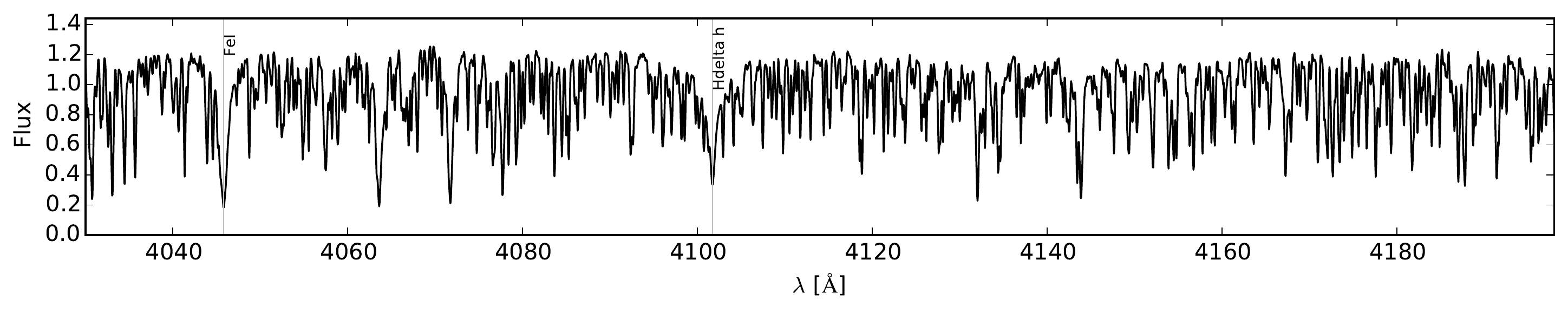}
 \includegraphics[width=0.21\linewidth]{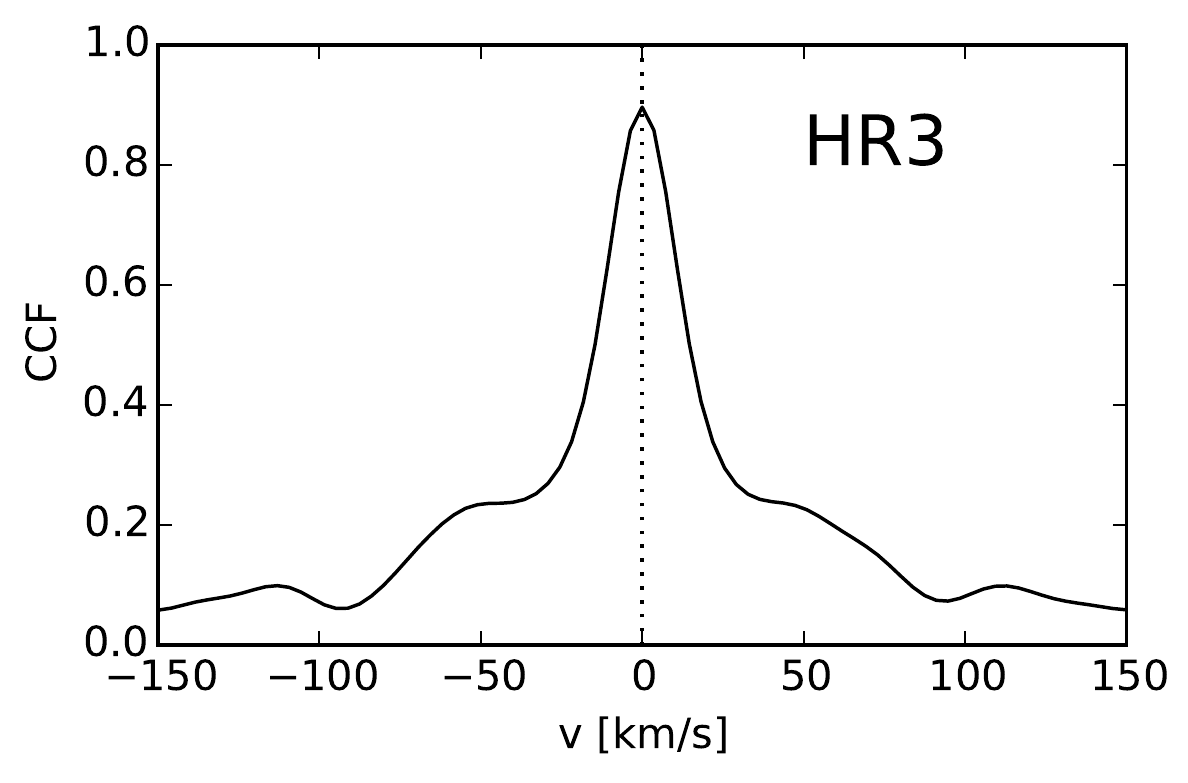}

 \includegraphics[width=0.70\linewidth]{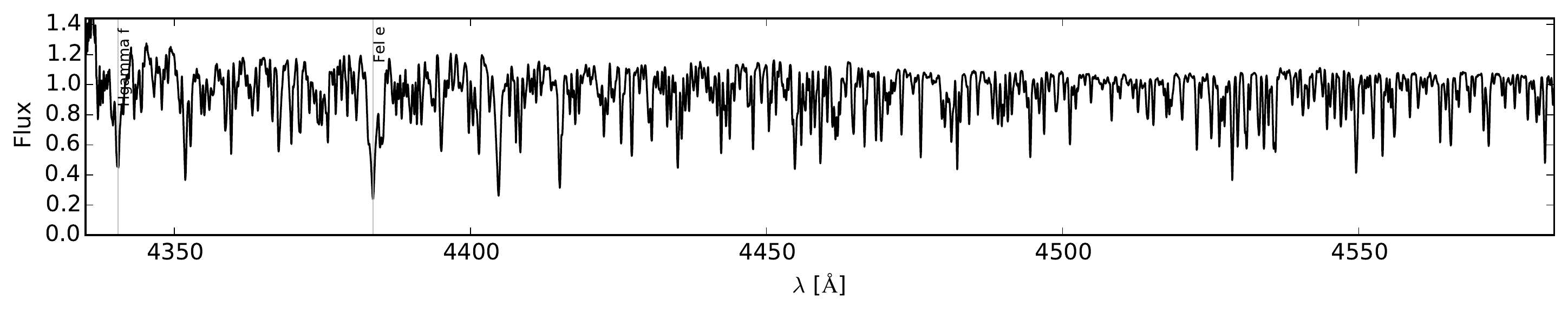}
 \includegraphics[width=0.21\linewidth]{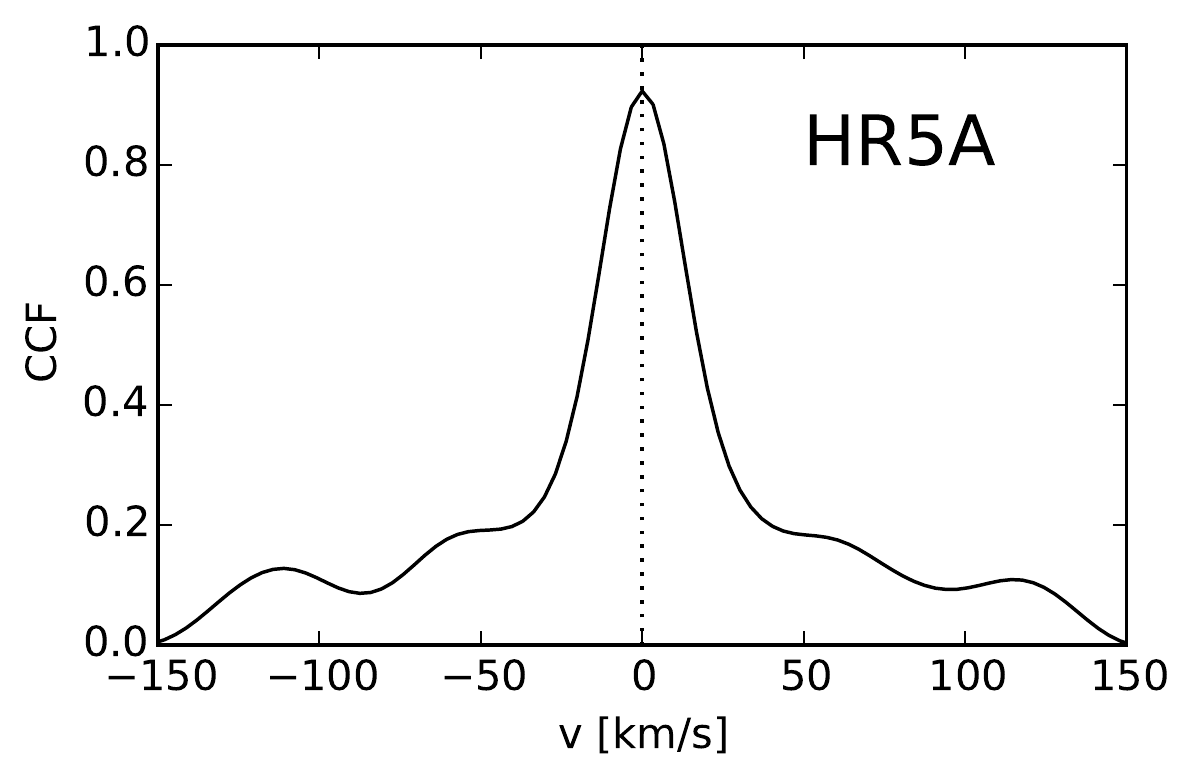}

 \includegraphics[width=0.70\linewidth]{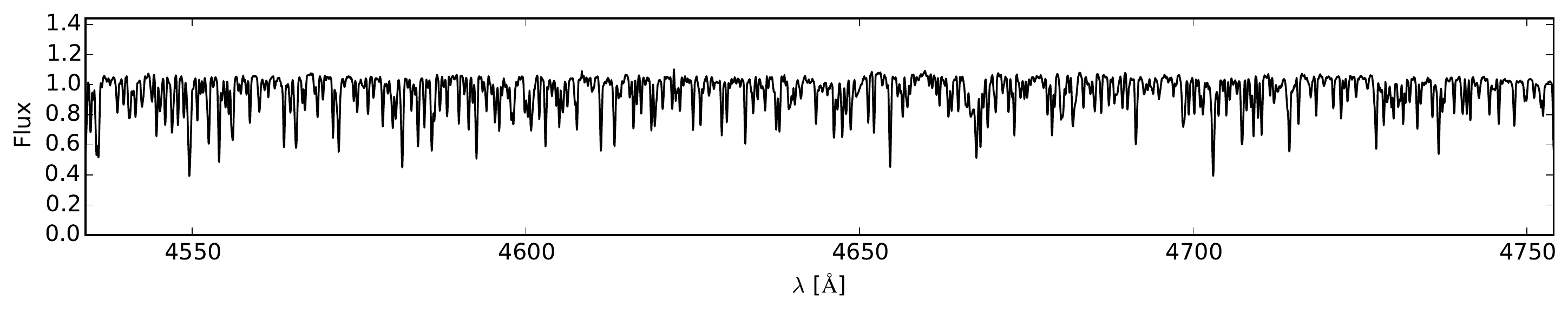}
 \includegraphics[width=0.21\linewidth]{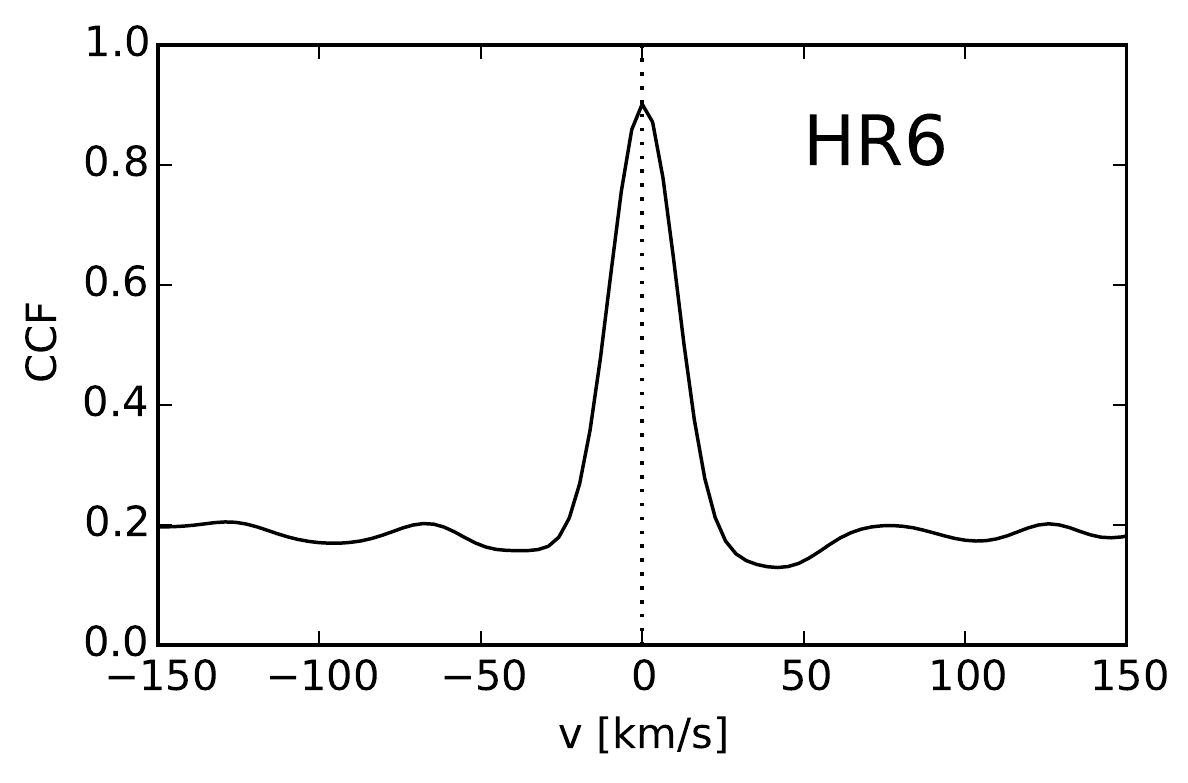}

 \includegraphics[width=0.70\linewidth]{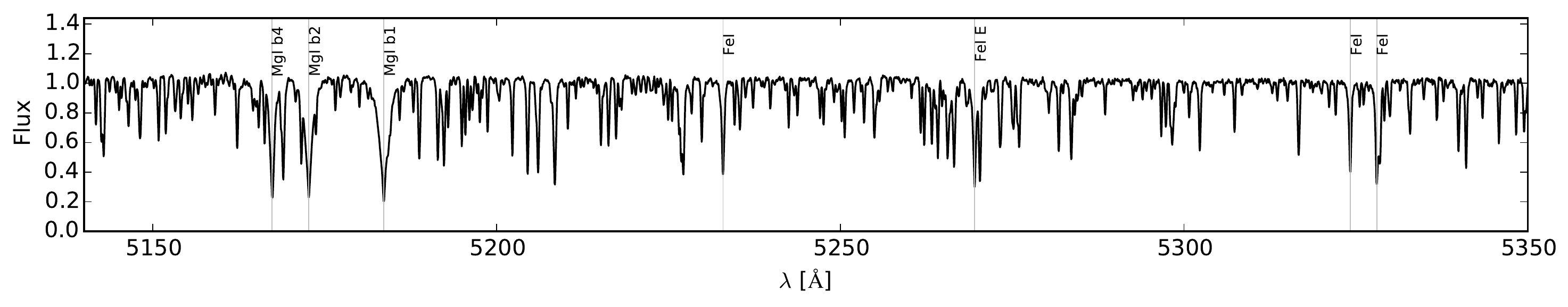}
 \includegraphics[width=0.21\linewidth]{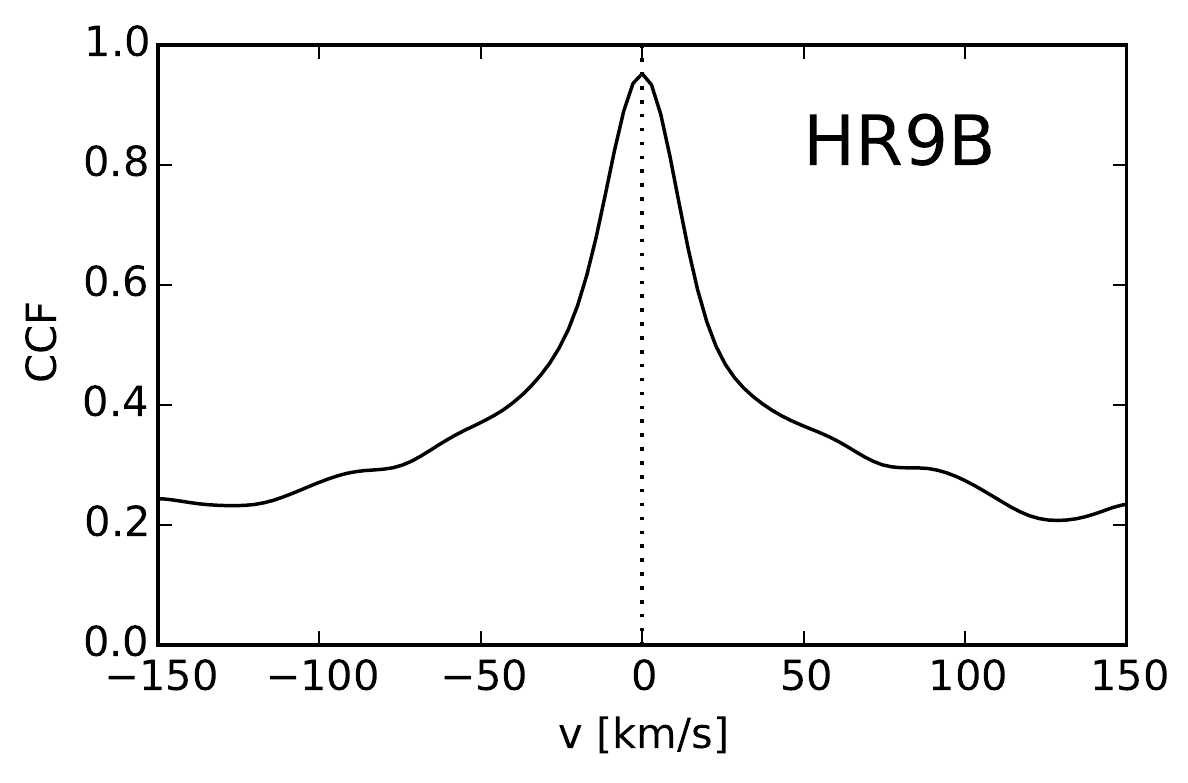}

 \includegraphics[width=0.70\linewidth]{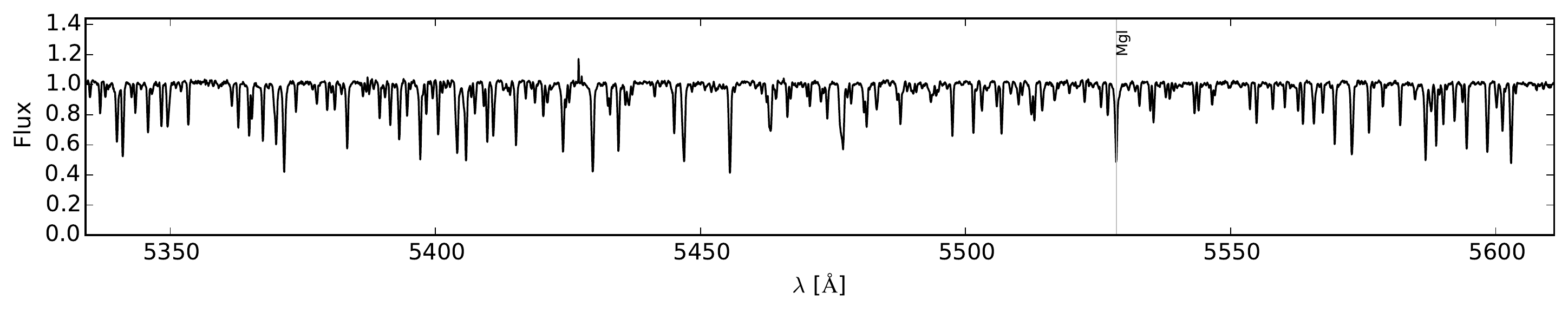}
 \includegraphics[width=0.21\linewidth]{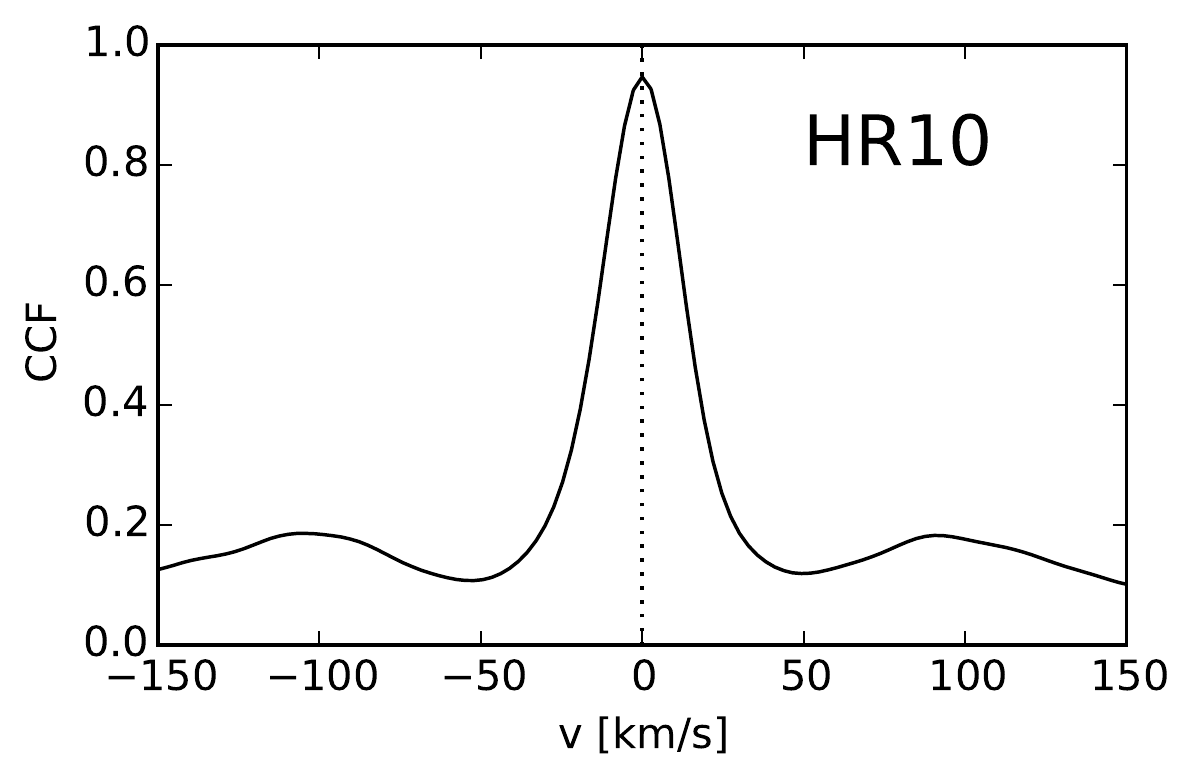}

 \includegraphics[width=0.70\linewidth]{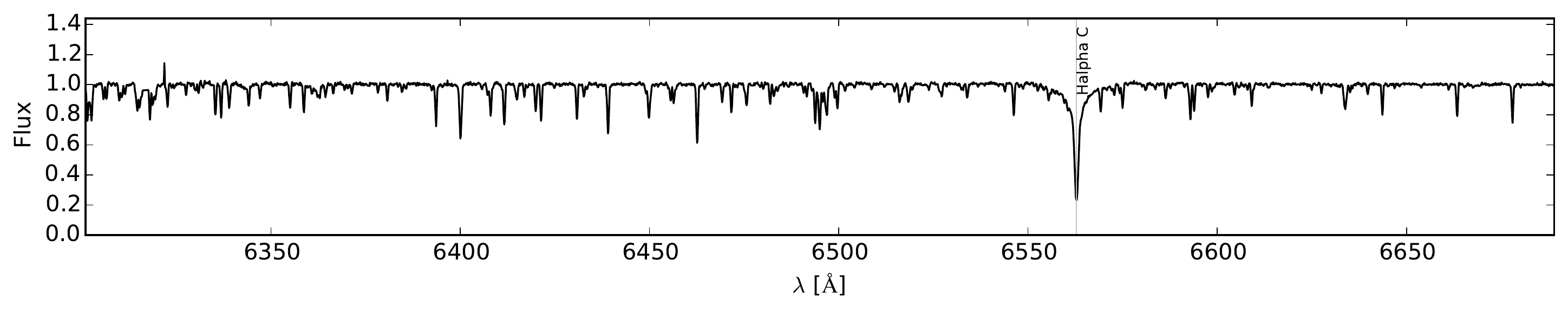}
 \includegraphics[width=0.21\linewidth]{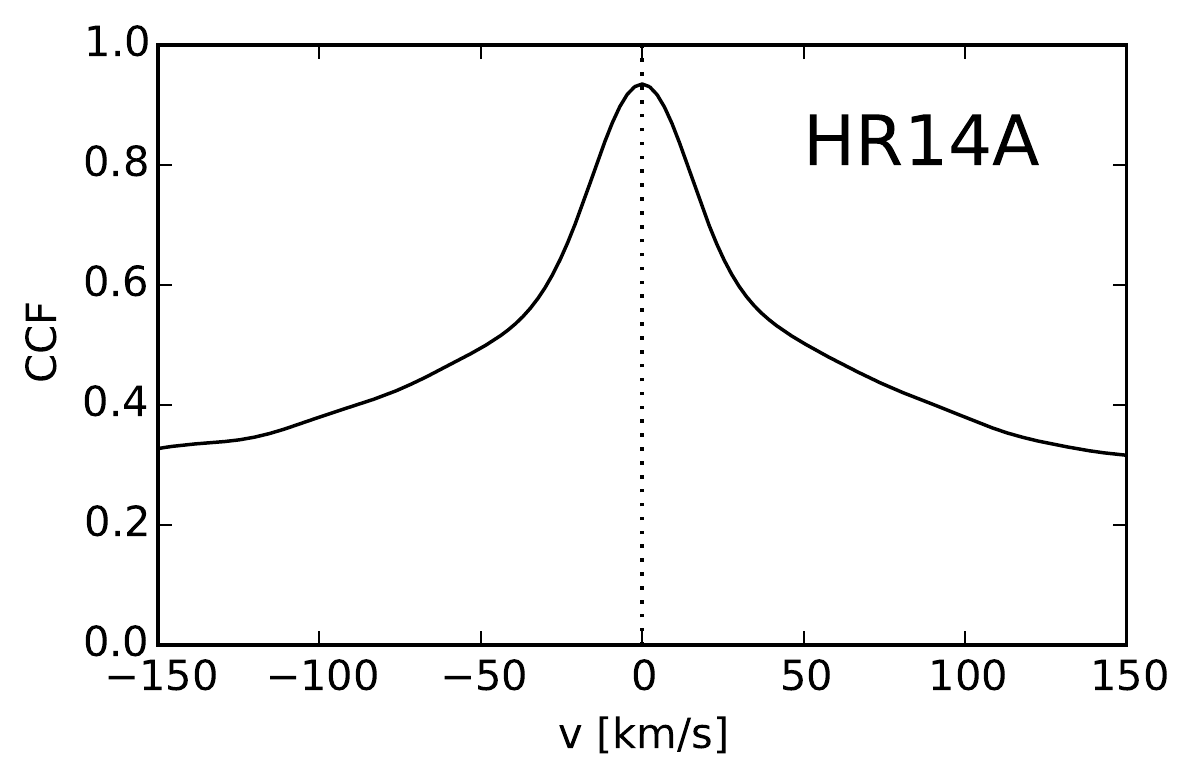}

 \includegraphics[width=0.70\linewidth]{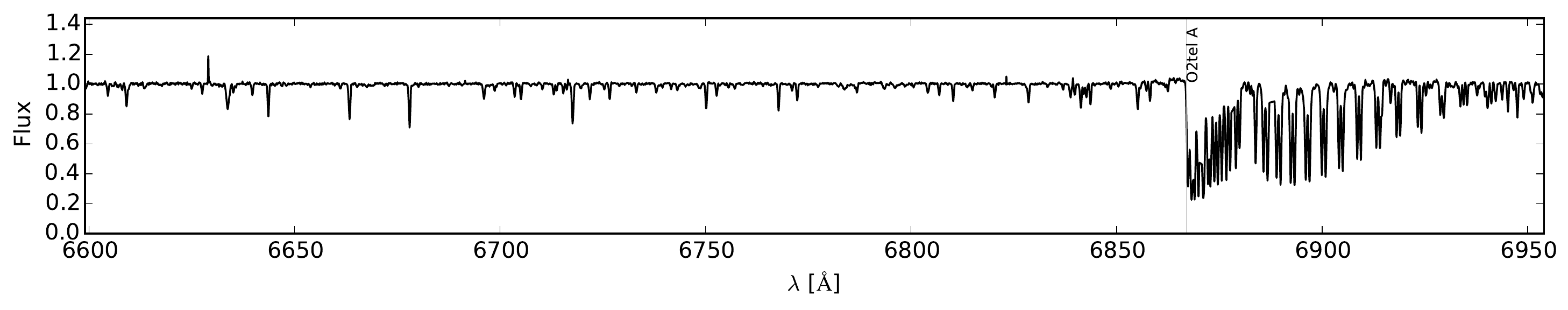}
 \includegraphics[width=0.21\linewidth]{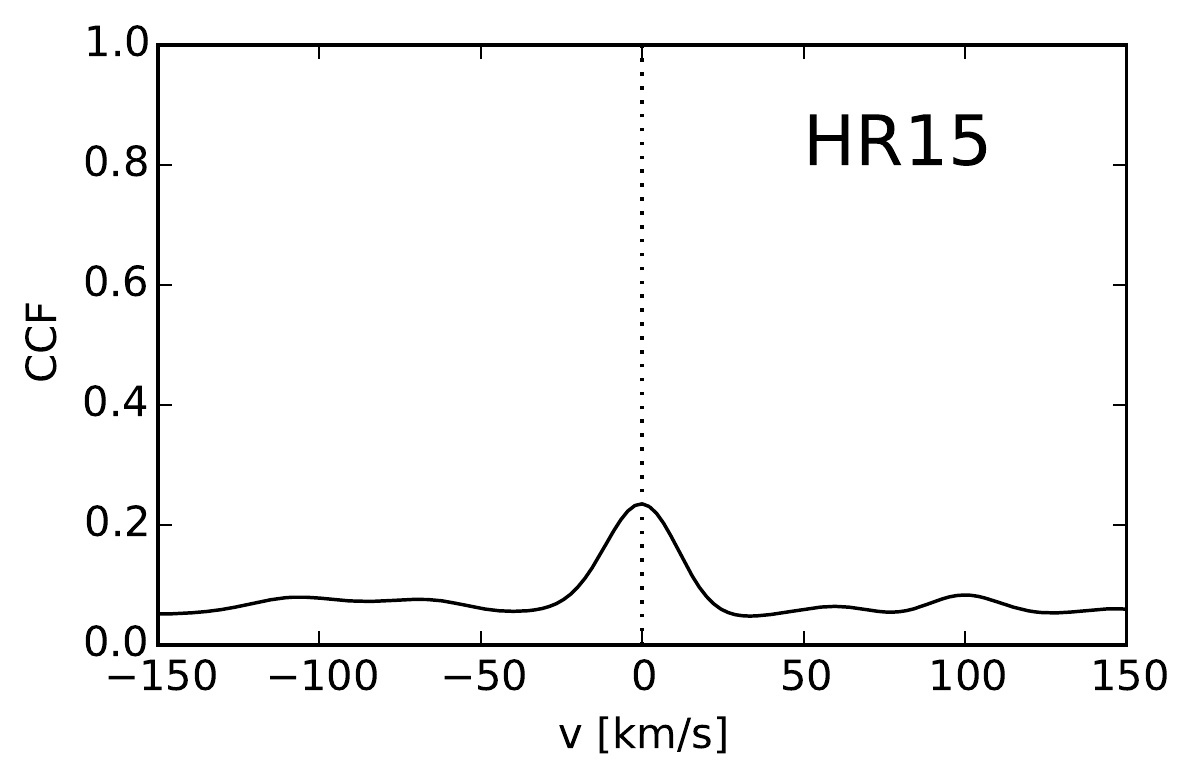}

 \includegraphics[width=0.70\linewidth]{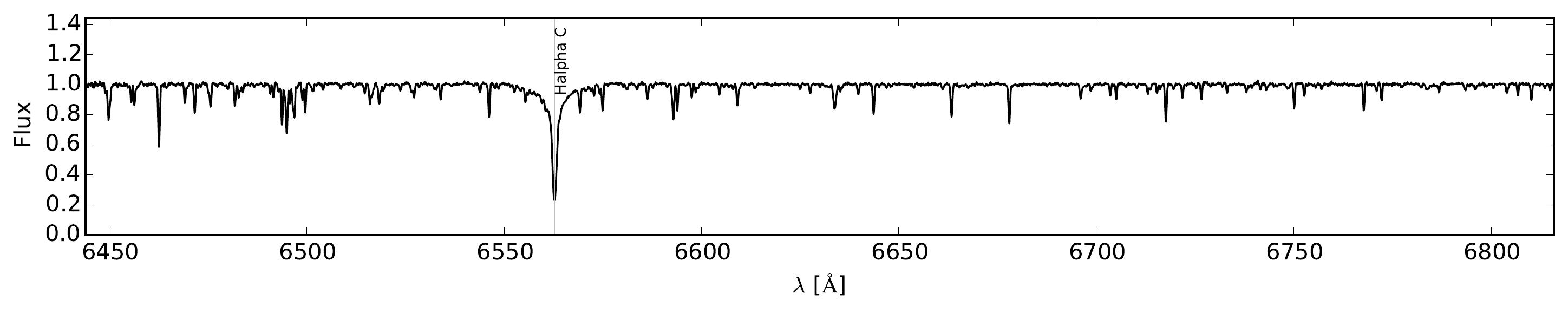}
 \includegraphics[width=0.21\linewidth]{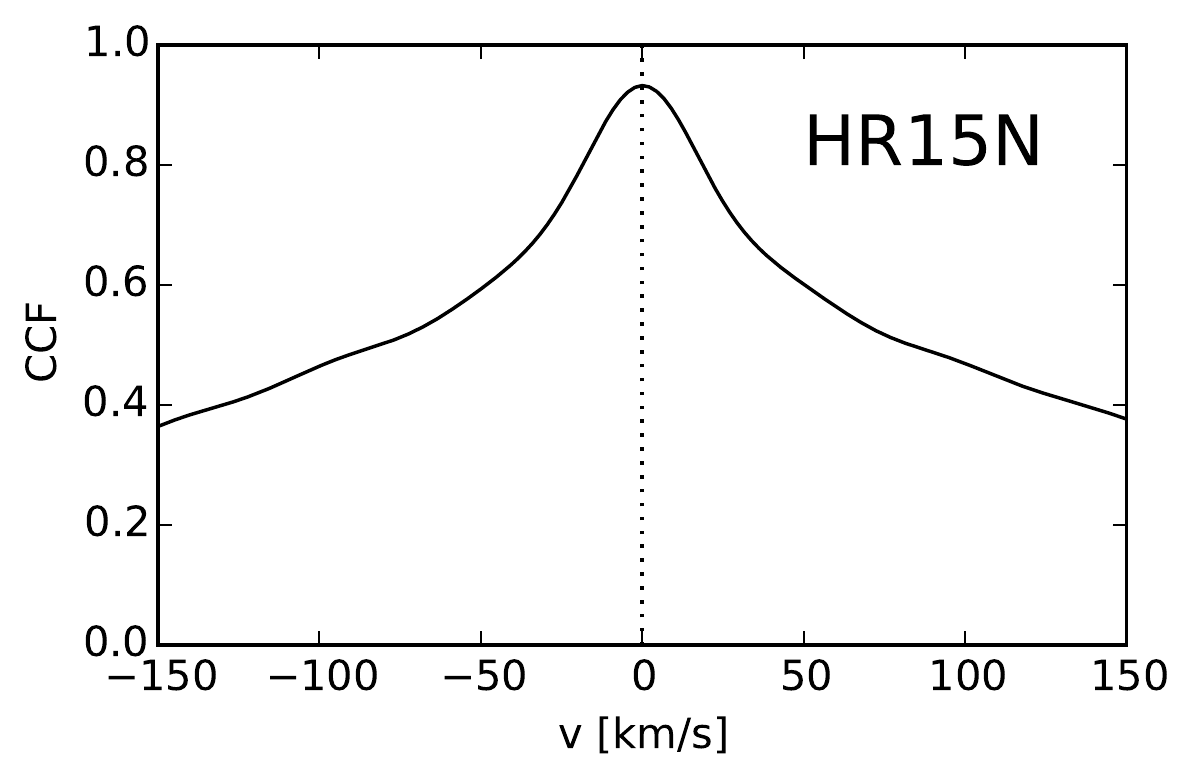}

 \includegraphics[width=0.70\linewidth]{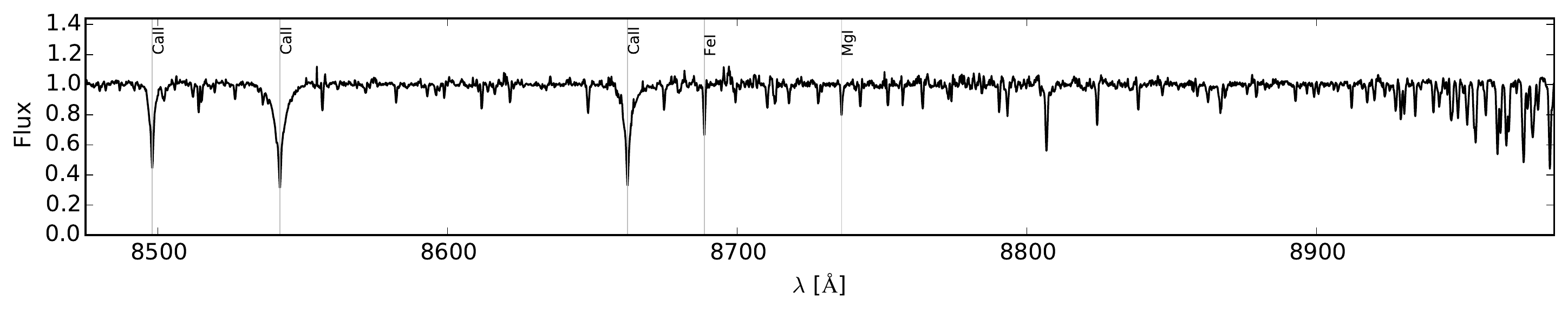}
 \includegraphics[width=0.21\linewidth]{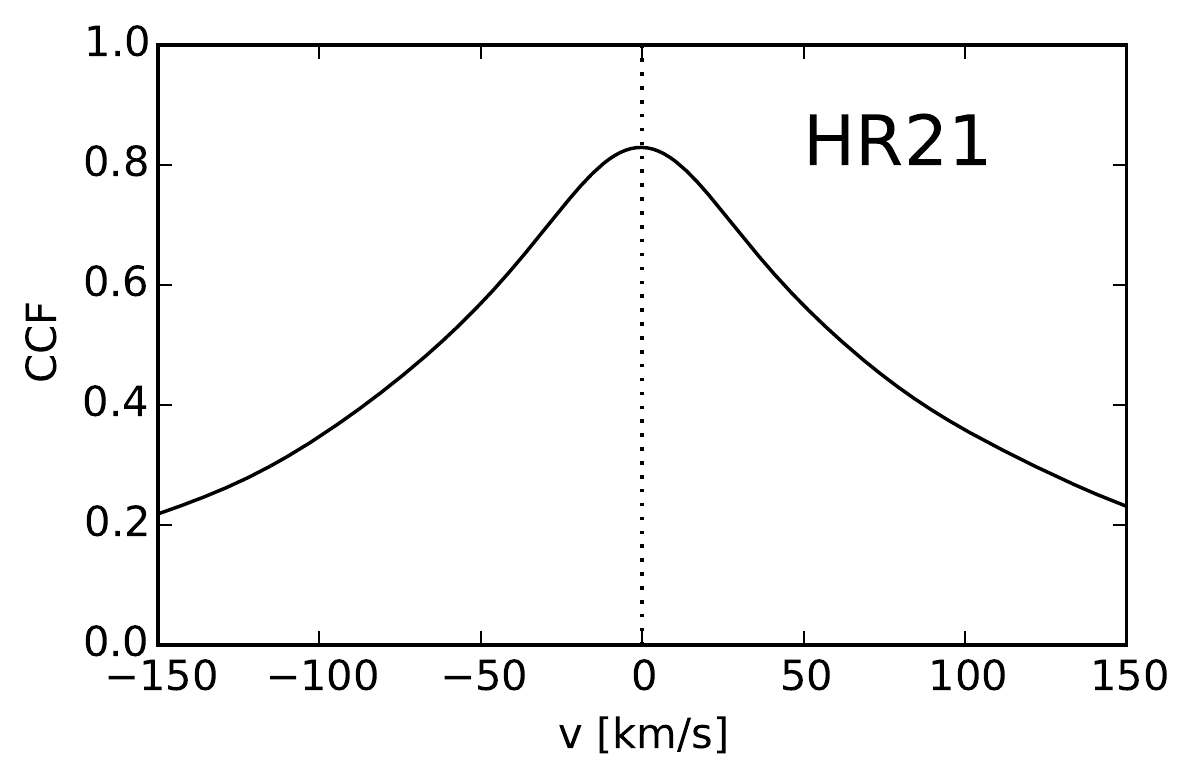}

 \caption{Solar spectra acquired by GES in GIRAFFE setups with high $S/N$ ($>1000$) except for setup HR9B where $S/N$~$\approx 700$. The normalised spectra are shown together with the identification of the main spectral features (left); the associated CCFs are shown on the right panels.}
 \label{fig:sun_spec_ccf}
\end{figure*}

\begin{figure*}
 \centering
\includegraphics[width=0.70\linewidth]{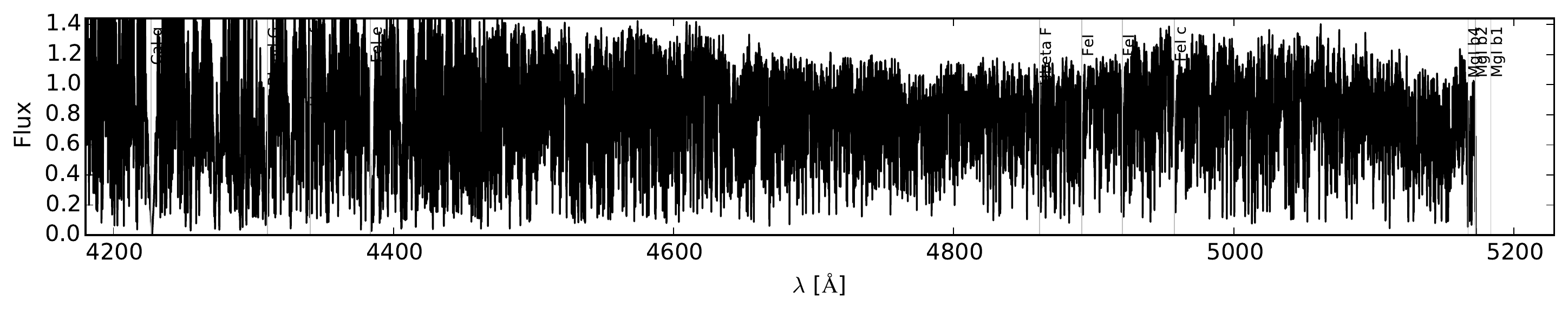}
 \includegraphics[width=0.21\linewidth]{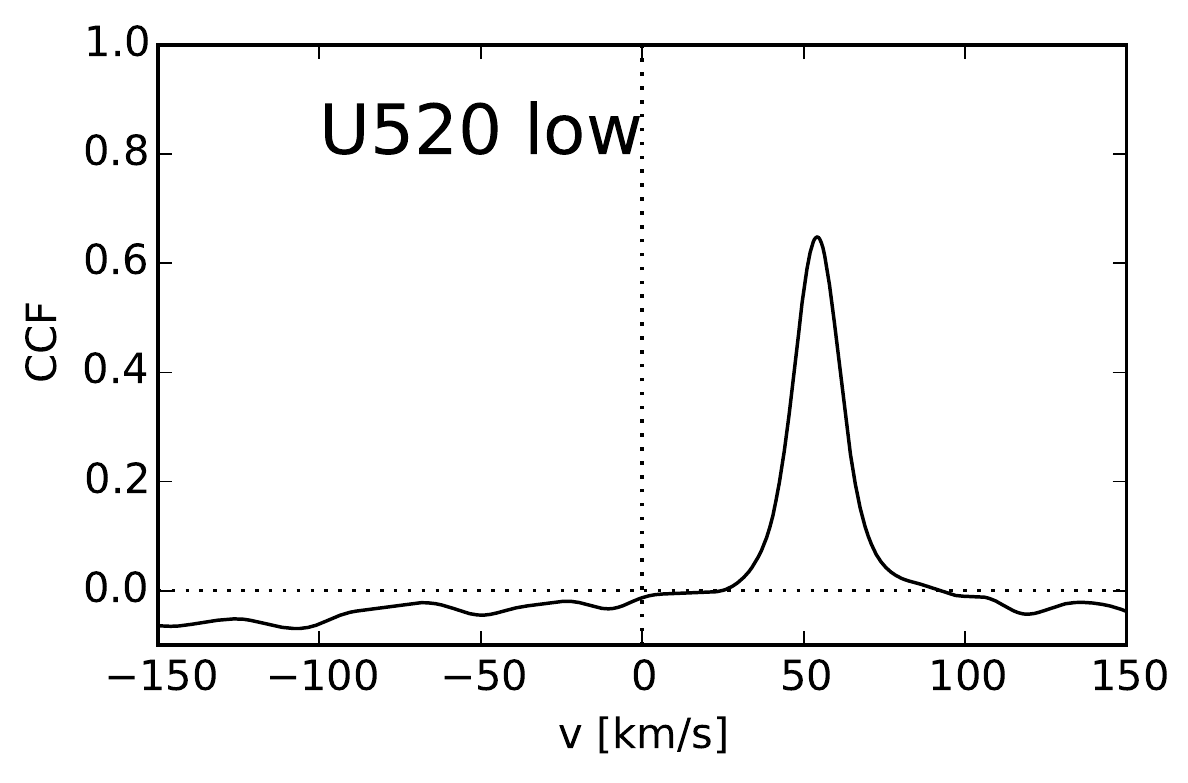}

 \includegraphics[width=0.70\linewidth]{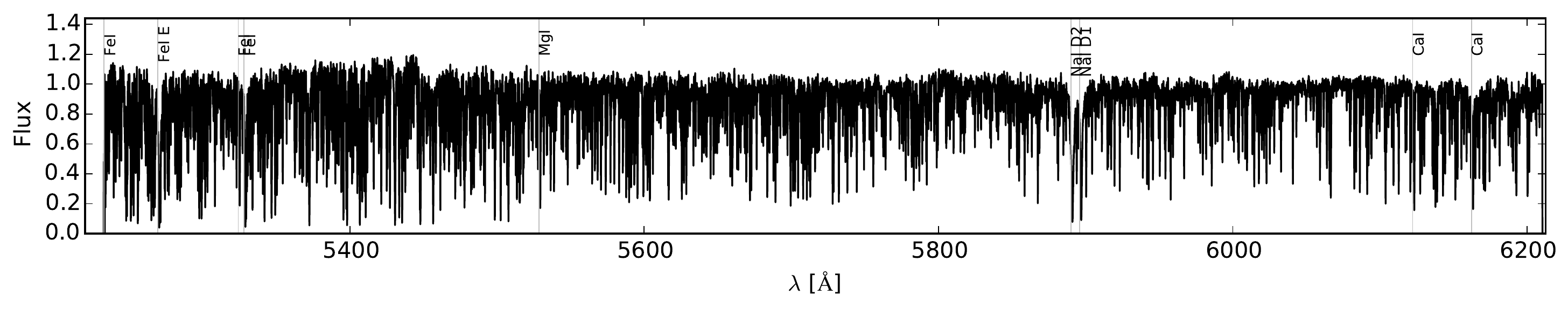}
 \includegraphics[width=0.21\linewidth]{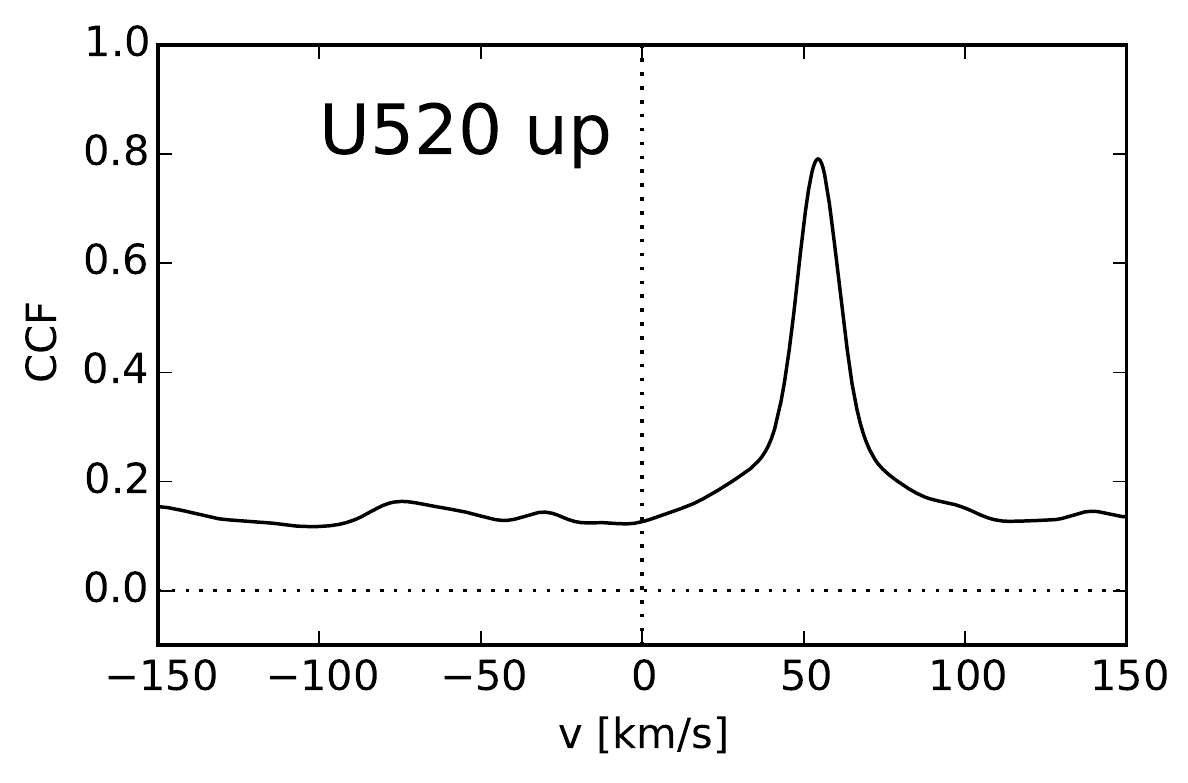}

 \includegraphics[width=0.70\linewidth]{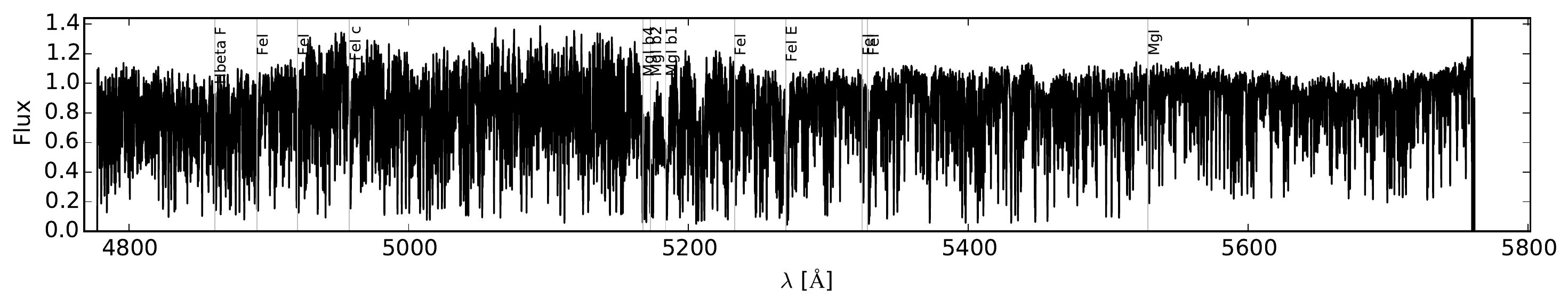}
 \includegraphics[width=0.21\linewidth]{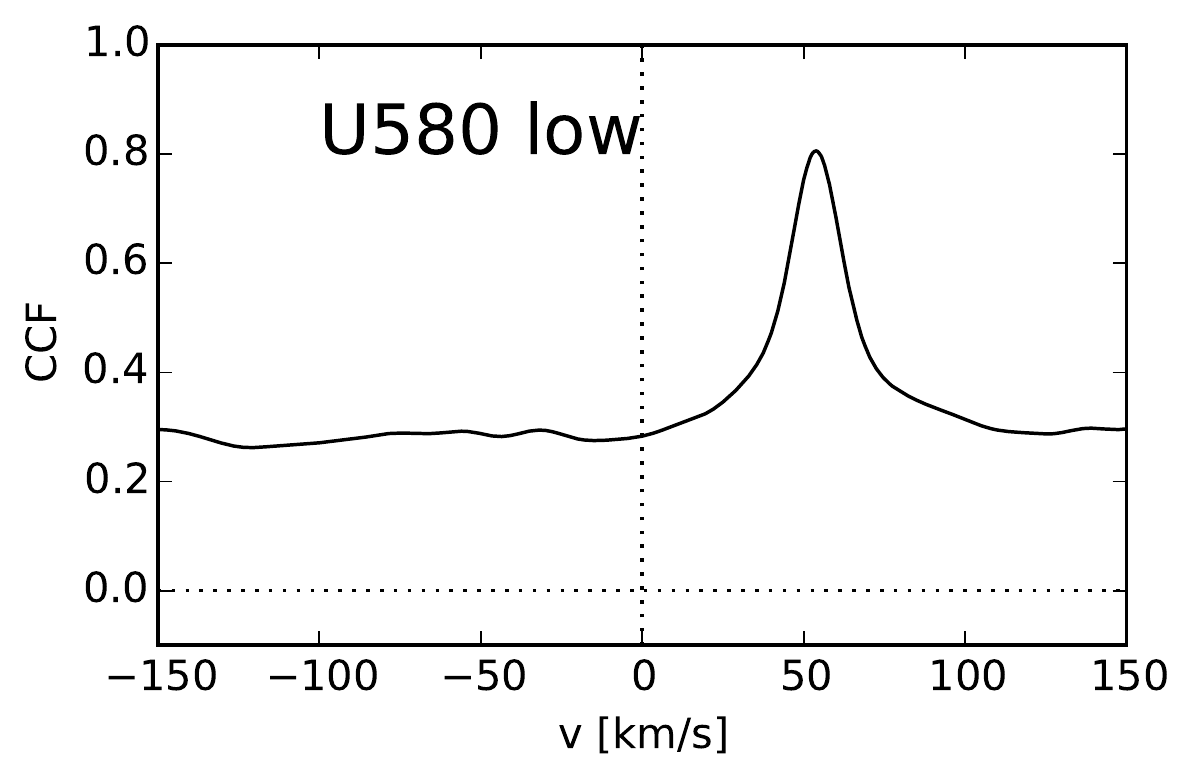}

 \includegraphics[width=0.70\linewidth]{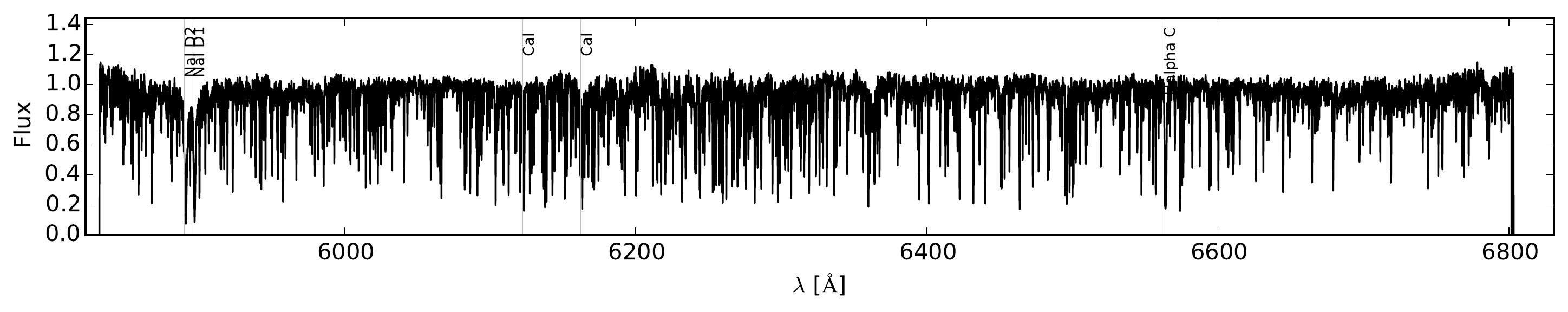}
 \includegraphics[width=0.21\linewidth]{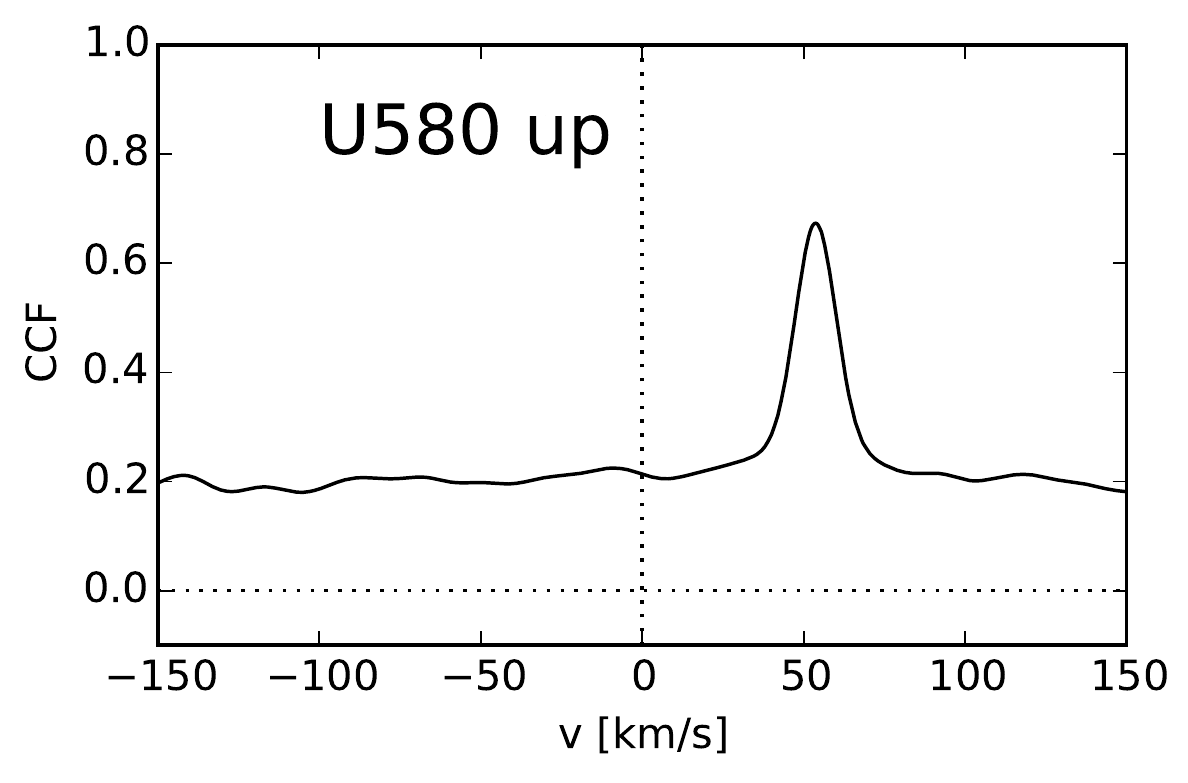}
 \caption{Spectra of Aldebaran ($\alpha$~Tau) by GES in UVES setups with $S/N$~$\approx 70$ for the low spectral chunks and $S/N$~$>100$ for the upper chunks.}
 \label{fig:ald_spec_ccf}
\end{figure*}

\section{Data selection}
\label{sect:data_selection}

\subsection{Observations and CCF computation}
\label{sect:obs_ccf}
Our analysis was performed on the iDR4 consisting of $\sim 260\, 000$ single exposures (corresponding to $\sim 100\, 000$ stacked spectra) of about $51\,000$ distinct stars observed with the FLAMES instrument feeding the optical spectrographs GIRAFFE (with setups HR3, HR5A, HR6, HR9B, HR10, HR14A, HR15N, HR15, HR21) and UVES (with setups U520 and U580) covering the optical and near-IR wavelength ranges given in Table~\ref{tab:doe_param}.

The classical definition of a CCF function applied to the stellar spectra is:
\begin{equation}
\textrm{CCF}(h) = \int_{-\infty}^{+\infty} f(x)g(x+h)~\textrm{d}x
\end{equation}
where $f$ is a normalised spectrum, $g$ a normalised template spectrum and $h$ is the lag expressed in \kms. The computation of the CCFs is performed by pipelines at CASU (Cambridge Astronomy Survey Unit\footnote{\url{http://casu.ast.cam.ac.uk/gaiaeso}}) for GIRAFFE spectra (Lewis et al., in prep.) and at INAF-Arcetri for UVES spectra \citep{sacco2014}. For UVES CCFs, spectral templates from the library produced by \citet{delaverny2012}, and based on MARCS models \citep{gustafsson2008}, are used. 
For GIRAFFE CCFs, spectral templates from the library produced by \citet{munari2005}, and based on Kurucz's models \citep{kurucz1993, castelli2003}, are used.
We stress that for a given spectrum, CCFs are calculated for all the templates and the CCF with the highest peak is selected. For UVES spectra, \ha\ and H$\beta$ are masked in the observations.
As illustrated \emph{e.g.} in Fig.~\ref{fig:ccf_test}, CCFs are characterized by a maximum value (CCF ``peak''), a minimum value (lowest point of the CCF ``tail'') and a full amplitude (maximum - minimum). The constant velocity steps of GIRAFFE and UVES CCFs are 2.75 (mainly) and 0.50~\kms\ (for a sampling of 401 and 4000 velocity points), respectively.

Examples of spectra and CCFs in the setups mentionned above are displayed in Figs.~\ref{fig:sun_spec_ccf} and~\ref{fig:ald_spec_ccf}. These figures are built from the solar and Aldebaran spectra.  The CCFs are represented over the same velocity range to allow an easy comparison between the various setups. When a lot of weak absorption lines are present (as in setups HR6 and HR10), the CCF peak is narrow and well defined with a width smaller than for setups with strong features like H$\delta$ (HR3), H$\gamma$ (HR5A), the Mg b triplet (HR9B), \ha\ (HR14A and HR15N) and the Ca II triplet (HR21). For HR15, the presence of telluric lines from 685~nm onwards reduces the maximum amplitude of the CCF to a value as low as 0.25, even with a $S/N$ larger than 1000.  

For the UVES setups, Aldebaran ($\alpha$ Tau, spectral type K5III) spectra and corresponding CCFs are presented in Fig.~\ref{fig:ald_spec_ccf}. Each setup is composed of two spectral chunks. In the present case, the lower chunk comes with $S/N \approx 70$ and the upper one with $S/N > 100$. For the setup U520 low, the leftward CCF tail is negative, probably as a result of poor spectrum normalisation due to the co-existence of lots of weak and strong lines. Since the wavelength range of the UVES setups is 2 or 3 times wider than those of GIRAFFE, the UVES setups are well suited to the detection of SB$n$ candidates.

The final GES spectrum of a given object is a stack of all individual exposures, wavelength calibrated, sky substracted and heliocentric radial velocity corrected. This could be a source of confusion in the case of composite spectra where the radial velocity of the different components changes between exposures. Moreover, a double-lined CCF coming from stacked spectra (and mimicking an SB2) can be the result of the SB1 combination  taken at different epochs and stacked. To avoid this problem, we performed the binarity detection on the individual exposures (rather than on the stacked ones). This choice avoids spurious spectroscopic binary detection, at the expense of  using spectra with lower $S/N$ ratios which will be shown not to be detrimental as long as $S/N > 5$ (see Sect.~\ref{sec:error}).

The number of individual observations per target is plotted in Fig.~\ref{fig:hist_nb}. The majority of stars observed with GIRAFFE has 2 or 4 observations because generally observed with HR10 and HR21 setups, whereas there are 4 or 8 observations in the case of UVES due to the presence of two spectral chunks per setup. Moreover, the time span between consecutive observations is very often less than three days, as shown on Fig.~\ref{fig:time-span}. Benchmark stars \citep[\emph{i.e.}, a sample of stars with well-determined parameters, to be used as reference; see][]{2015A&A...582A..49H} are the most observed objects, some having more than 100 observations. 

\begin{figure}
 \includegraphics[width=\linewidth]{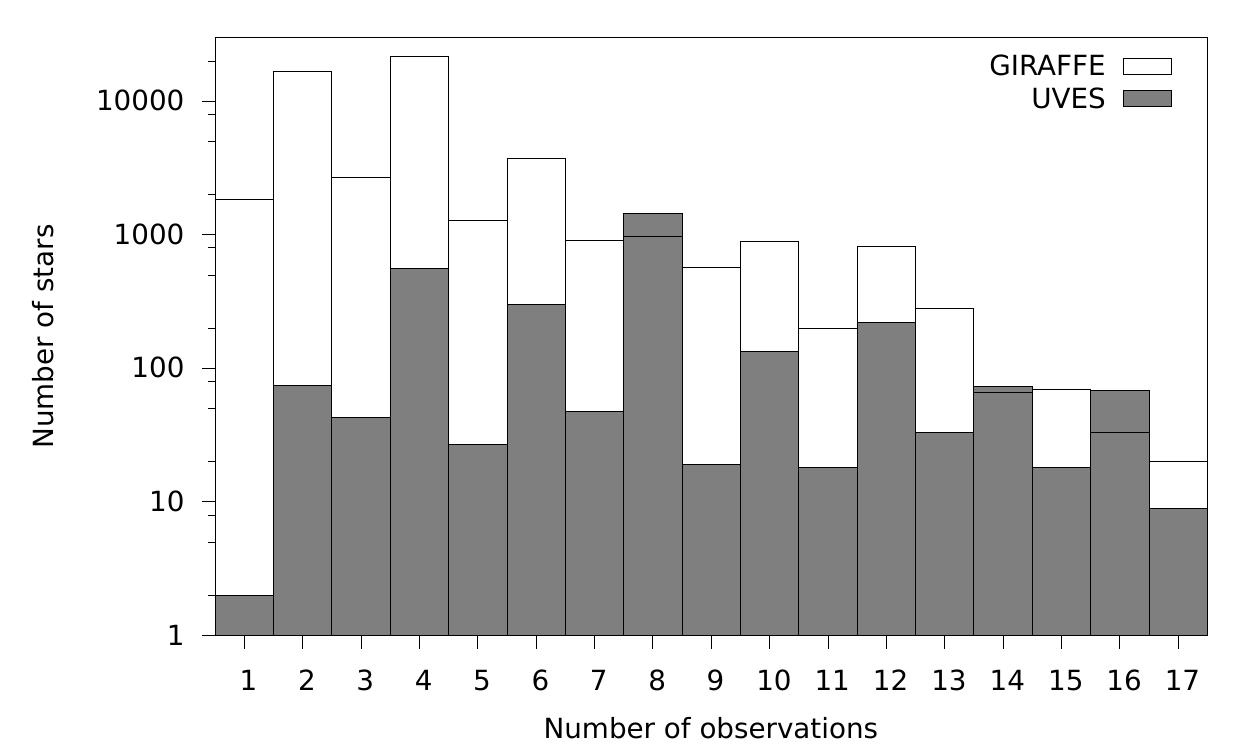}
 \caption{Number of stars observed as a function of the number of observations per star. A tiny fraction, including benchmark stars, have a number of observations that can reach $\sim 100$.}
 \label{fig:hist_nb}
\end{figure}

\begin{figure}
\includegraphics[width=9cm]{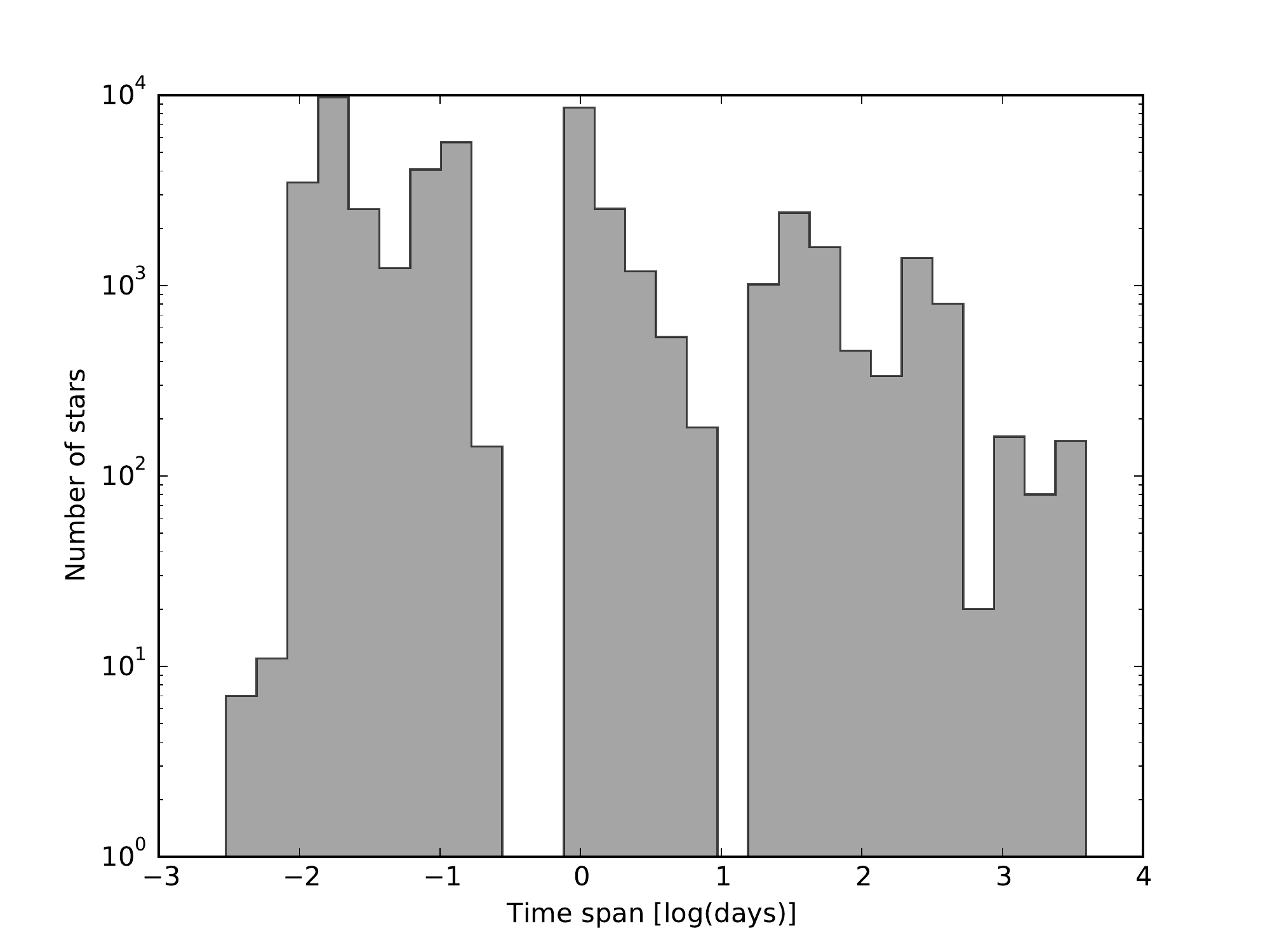}
\caption[]{\label{fig:time-span} Histogram of the full time span between observations if more than one is available for a given target.}
\end{figure}

\begin{figure*}
\includegraphics[width=0.49\linewidth]{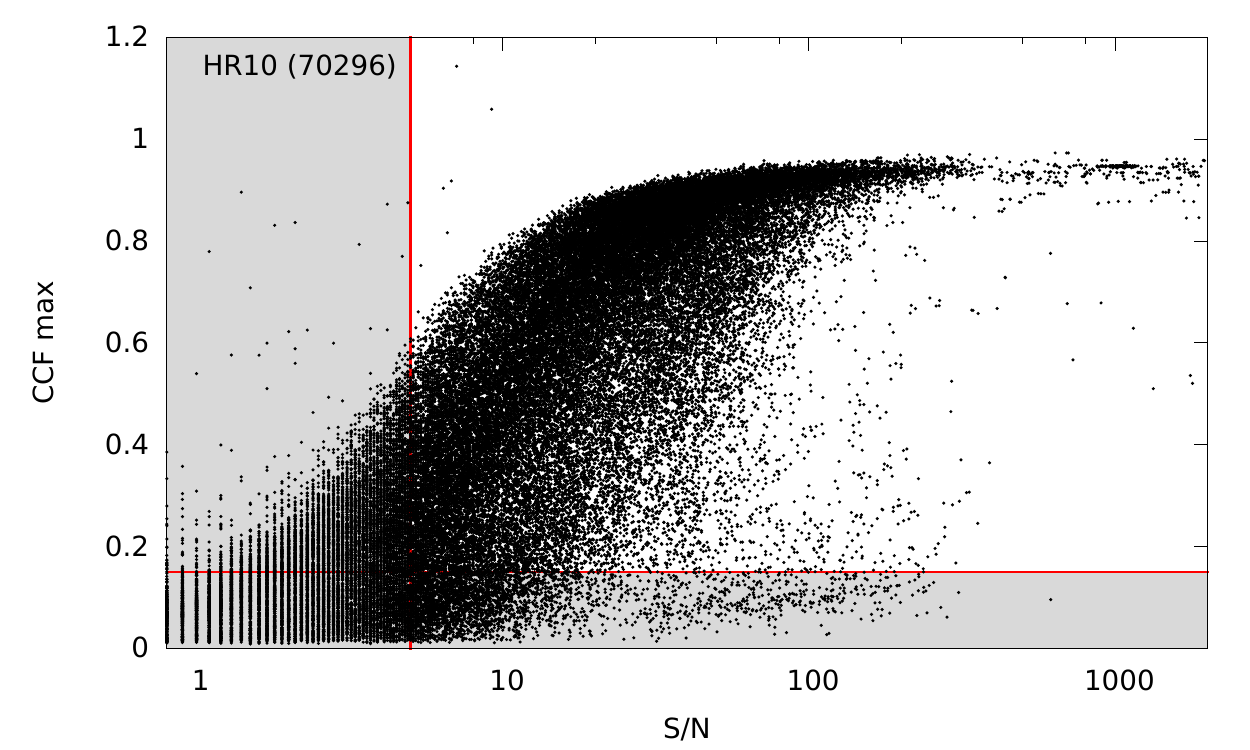}
\includegraphics[width=0.49\linewidth]{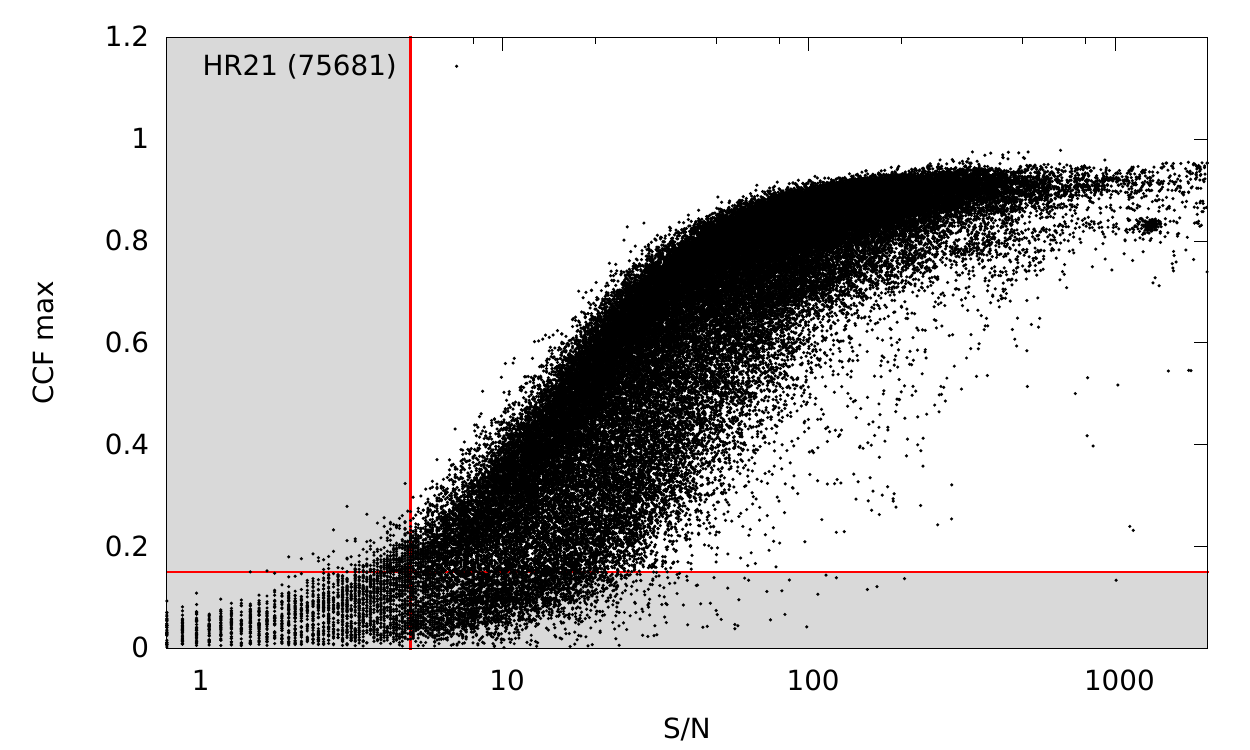}
\caption{\label{fig:ccfmax_snr} CCF maximum amplitude versus $S/N$ for HR10 (left panel) and for HR21 (right panel). Solid red lines are the criteria on the $S/N$ (vertical, $S/N=5$) and on the lowest value of the CCF maximum (horizontal, $0.15$). The shadow area shows the observations excluded from the analysis. In parenthesis is the number of single exposures in each setup.}
\end{figure*}

\subsection{Data selection in iDR4}
\label{sect:sel}

Our sample has been drawn from the individual spectra database of the GES iDR4\footnote{\label{note:GES}GES public data releases may be found at https://www.\emph{Gaia}-eso.eu/data-products/public-data-releases}, covering observations until June 2014, to which the following selection criteria were applied:
\begin{itemize}
 \item $S/N$ larger than 5;
 \item CCF maximum larger than $0.15$; 
 \item CCF minimum larger than $-1$;
 \item CCF full amplitude larger than $0.10$;
 \item left CCF continuum  $-$ right CCF continuum lower than $0.15$.
\end{itemize}
These criteria were empirically determined thanks to a visual inspection of a representative sample of CCFs.
We allow negative values for the CCF minimum to keep CCFs computed on unperfectly normalised spectra (without allowing spectra with a completely wrong normalisation). Criteria on the $S/N$ and on the CCF maximum are presented in Fig.~\ref{fig:ccfmax_snr} for setups HR10 and HR21 which contain the most numerous observations. This figure clearly shows the impact of the $S/N$ of a spectrum on its associated CCF: the higher the $S/N$, the higher the CCF maximum. For a given $S/N$, the interval spanned by the CCF maximum is mainly due to spectrum -- template mismatch. For HR10, the over-density located at $30<S/N<200$ and CCF max $<0.15$ is mainly due to NGC~6705 members. In HR21, the clump located at $1000<S/N<2000$ and $0.80<\mathrm{CCF~max}< 0.85$ is due to repeated observations of the solar spectrum.

These criteria allow us to avoid detecting spurious (noise-induced) CCF peaks. Over the 260\,000 individual science spectra (corresponding to the  100\,000 stacked spectra) within the iDR4, $9.3$~\% have a $S/N$ lower than 5, $1.0$~\% have a null CCF (data processing issues), $7.8$~\% have a CCF maximum lower than $0.15$, $0.2$~\% have a CCF minimum lower than $-1.0$, and $0.02$~\% have a CCF full amplitude lower than $0.10$. We ended up with about 205\,000 CCFs ($77.7$~\%), corresponding to $\sim 51\,000$ different stars.

\begin{figure*}
 \includegraphics[width=0.33\linewidth]{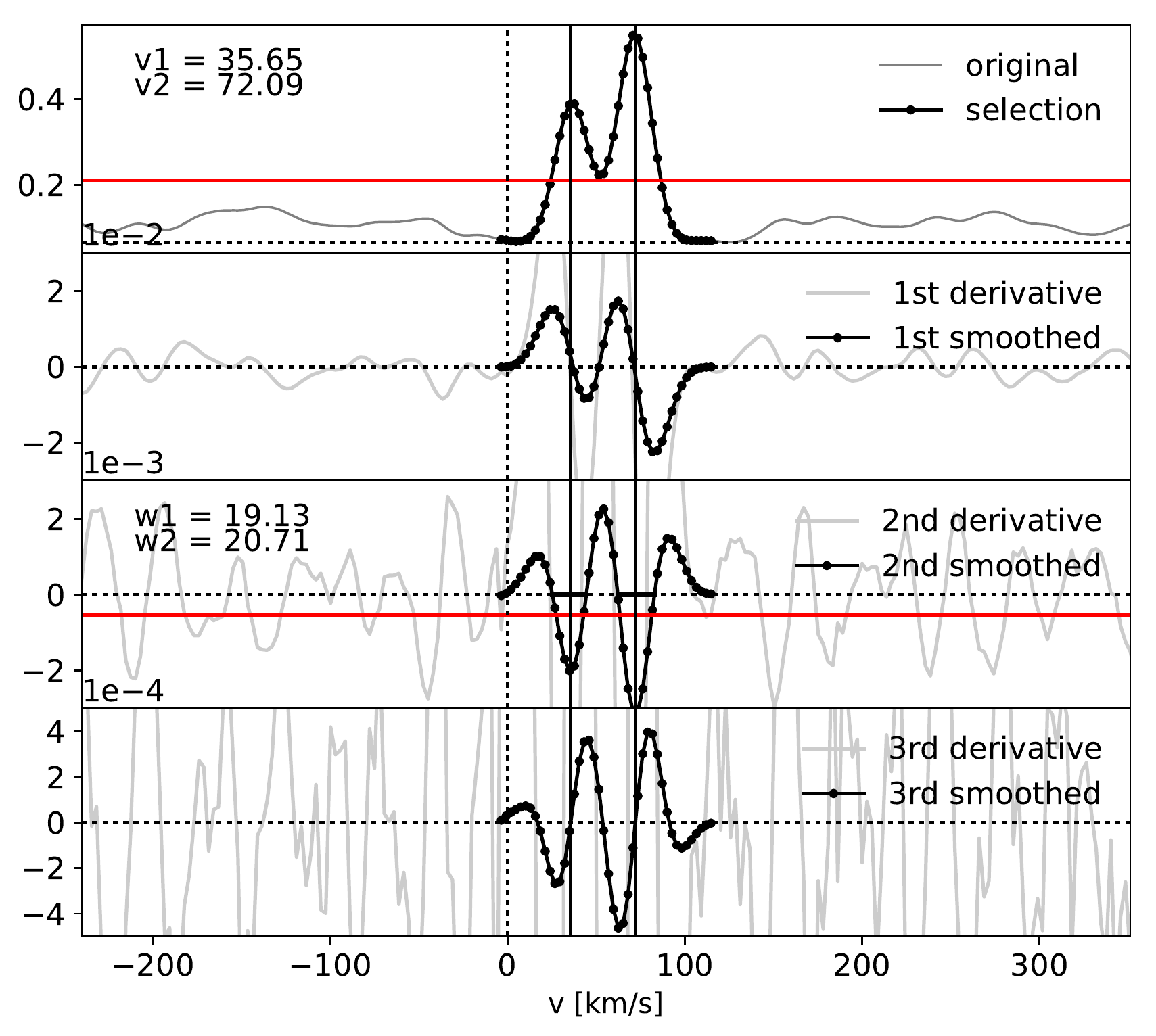}
 \includegraphics[width=0.33\linewidth]{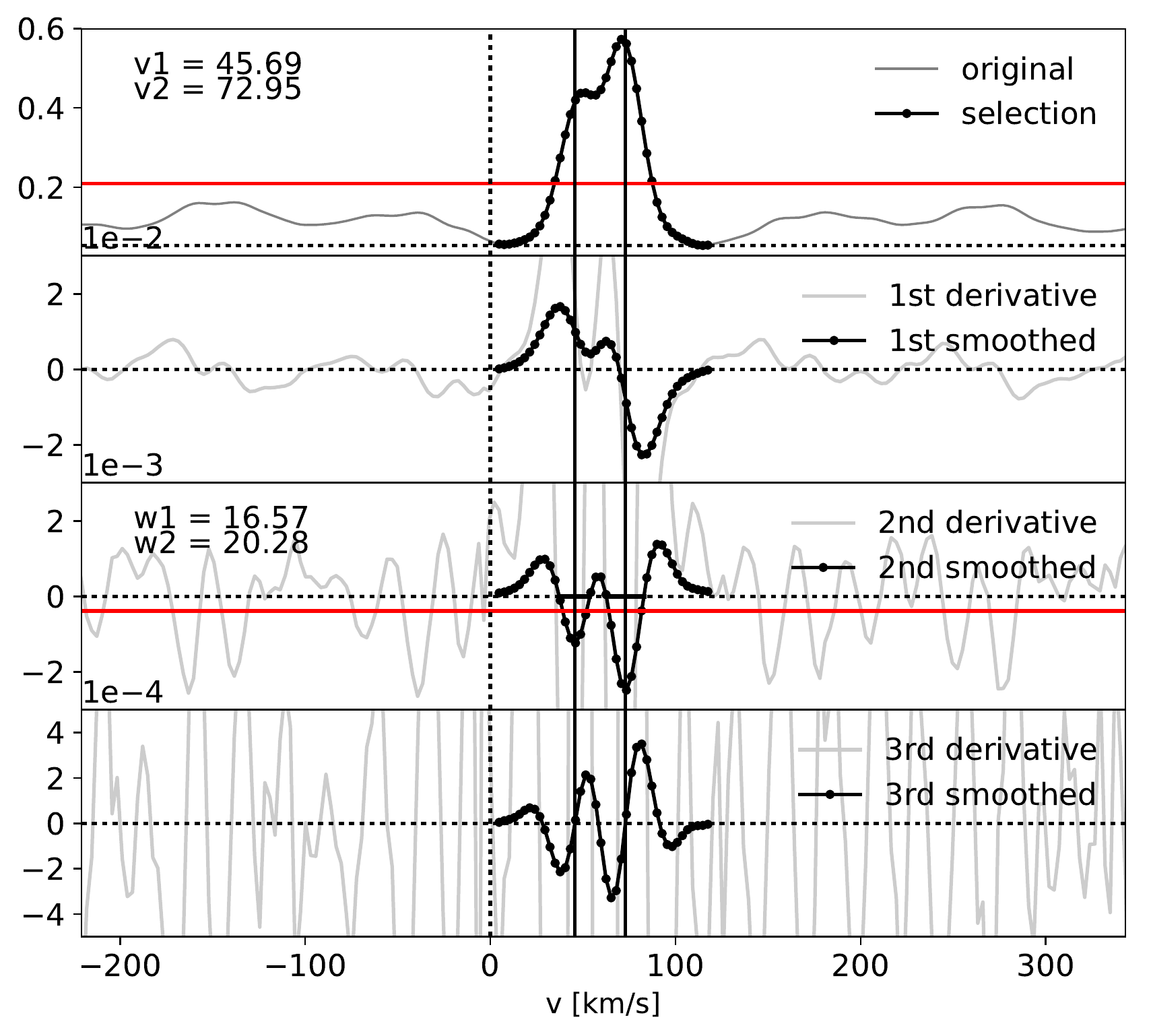}
 \includegraphics[width=0.33\linewidth]{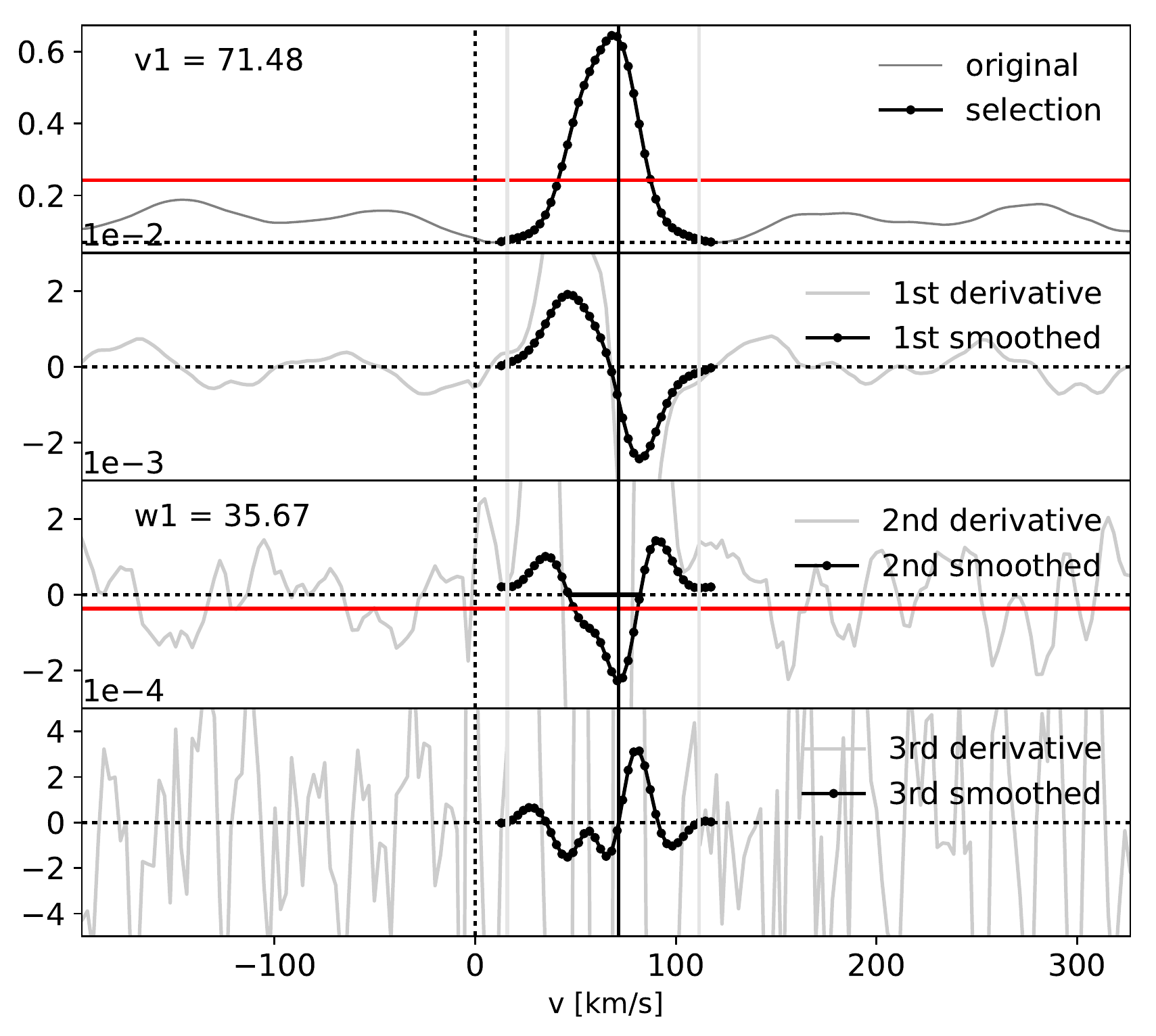}
 \caption{Simulated noisy double-peak CCF with peaks located at 36.0~\kms\ and 72.0~\kms\ (left),  48.0~\kms\ and 72.0~\kms\ (centre), and 54.0~\kms\ and 72.0~\kms\ (right). 
 Grey lines show derivatives from a simple finite differences method which have the drawback to be very noisy. Instead, black line with dots (in panels below the top one) show the smoothed derivatives computed with Eq.~\ref{eq:conv}. 
 Red lines in top panels show the threshold parameter on the CCF (THRES0) and in the middle-low panels the threshold parameter on the second derivative (THRES2).}
 \label{fig:ccf2_test}
\end{figure*}

\section{Methods}
\label{sect:methods}

 \subsection{Detection of extrema (\doe) code}
\label{sect:doe}
The Detection of extrema (\doe) code has been designed to identify the (local and global) extrema in a given signal even in case these extrema are strongly blended. By using successive derivatives of a function, it is possible to characterize it in a powerful way. Applied on spectral-line profiles for instance, the method makes it possible to identify all contributing blends \citep{sousa2007}. 
Here we apply it to the CCFs. The method is inspired from signal-processing techniques \citep{Foster2013} which convolve the signal (here the CCF) with the derivatives of a Gaussian kernel to smooth and calculate the derivative of the CCF in a single operation. In other words, the first, second and third derivatives of the Gaussian kernel are used to obtain the smoothed derivatives of the CCFs. Indeed, one of the interesting properties of the convolution of two generalized functions is defined as follows: 
\begin{equation}
 (f' * g)(x) = (f * g')(x)
 \label{eq:conv}
\end{equation}
where $f'$ and $g'$ are the first derivatives of the generalized functions $f$ and $g$. Convolving the CCF with the derivative of a Gaussian kernel is equivalent to compute the derivative of the CCF and to convolve (\emph{i.e.}, smooth) it by a Gaussian kernel. We use the routine \textit{gaussian\_filter1d} of the sub-module \textit{ndimage} of the \textit{scipy} module \citep{scipy2001} in Python. The routine calculates first the derivative of the Gaussian kernel before correlating it with the CCF function. The width of the Gaussian kernel controls the amount of smoothing. 

A zero in the descending part of the first derivative obviously provides the position of the maximum of the CCF. However, in the case of a CCF composed of two or more peaks, the zeros of the first derivative will only provide the positions of well-separated  peaks, \emph{i.e.}, peaks with a local minimum in between them. Blended peaks might thus be missed. However, this difficulty may be circumvented by using the third derivative, whose zeros occurring in an \emph{ascending} part provide the positions of all the peaks including the blended ones. Fig.~\ref{fig:ccf2_test} shows that the use of the first derivative only does not allow a satisfactory detection of the CCF components. Indeed, although the CCF in the middle panel clearly exhibits two peaks, the first derivative has only one descending zero-crossing, thus resulting in the detection of one component only. However, the second derivative shows two local minima corresponding to the two CCF velocity components. The position of these two minima can be found by detecting the ascending zero-crossing of the third derivative.
By using the third derivative, the different CCF components may thus be identified as regions where the CCF curvature is sufficiently negative (minima of the second derivative, or ascending zeros of the third derivative), separated by a region of larger curvature. To get the velocities of the various components, the CCF third derivative is simply interpolated to find its intersection with the $x$-axis. Some detection thresholds had to be set to automate the process in order to match the results obtained from an eye inspection of multiple-component CCFs.

The procedure is illustrated on simulated CCFs with or and two peaks (Figs.~\ref{fig:ccf_test} and \ref{fig:ccf2_test} respectively). We first test the operation of the \doe\ code on single peaks at the lowest numerical resolution, \emph{i.e.}, peaks defined with only six velocity points (left panel of Fig.~\ref{fig:ccf_test}). The  \doe\ code applied on a more realistic (more noisy) simulated single-peaked CCF (as shown on the right panel of Fig.~\ref{fig:ccf_test}) also provides satisfactory results, with an accuracy on the radial velocity of the order of 0.20~\kms. We will show in Sect.~\ref{sec:error} that the \doe\ code has a small internal error of 0.25 \kms.

The first threshold (THRES0), expressed as a fraction of the full CCF amplitude, defines the considered velocity range: the \doe\ code is applied only in the region where the CCF is larger than THRES0. The THRES0 threshold is represented by the horizontal red line in the top panels of Fig.~\ref{fig:ccf_test} and subsequent figures. However, if several well-defined peaks are identified in the CCF, the THRES0 criterion is overridden, and all data points between the CCF peaks are included in the analysis of the derivatives, even though the CCF may be lower than THRES0. 

A second threshold, THRES2, is set on the second CCF derivative. The THRES2 parameter is expressed as a fraction of the full amplitude of the CCF second derivative. This negative threshold is represented by the horizontal red line in the ``2nd derivative'' panel in Fig.~\ref{fig:ccf_test} (and subsequent figures) such that only minima lower than this threshold are selected for the final peak detection (vertical black lines) whereas second-derivative minima larger than this threshold are not considered to be related to real components (vertical light grey lines in \emph{e.g.} Fig.~\ref{fig:cal_u580}).

The width of the Gaussian kernel for the convolution of the CCF, SIGMA, is the third parameter. It is a smoothing parameter and aims at making the successive derivatives of the CCF less sensitive to the data noise. 

The three parameters of \doe\ (THRES0, THRES2 and SIGMA) have to be set by the user. Their value may have an impact on the number of detected peaks and the radial velocities associated to them. These three parameters need to be adjusted in order to give meaningful results (\emph{i.e.}, matching the efficiency of an eye-detection) on all CCFs, but once fixed for each instrumental setup (see Table~\ref{tab:doe_param} and Sect.~\ref{Sect:parameters}), they are kept constant to ensure homogeneous detection efficiency over the whole GES sample.

The parameter values result from a compromise between antagonistic requirements:
\begin{itemize}
 \item the THRES0 parameter must not be too low to avoid an unrealistically large velocity range, neither too high in order to be able to detect real albeit low secondary peaks;
 \item the THRES2 parameter must be calibrated on extreme cases (two very close or very separated peaks). The choice of this parameter is important: it ensures that the second derivative (\emph{i.e.} the curvature) of the CCF is negative enough, therefore corresponding to real components;
 \item the SIGMA parameter must not be too large, resulting in a too strong smoothing which would endanger the detection of close peaks, and not too small to reduce the impact of the numerical noise induced by the successive derivatives.
\end{itemize}
The empirical method used to set these parameters is described in Sect.~\ref{Sect:parameters}.

\begin{figure*}
\includegraphics[width=0.33\linewidth]{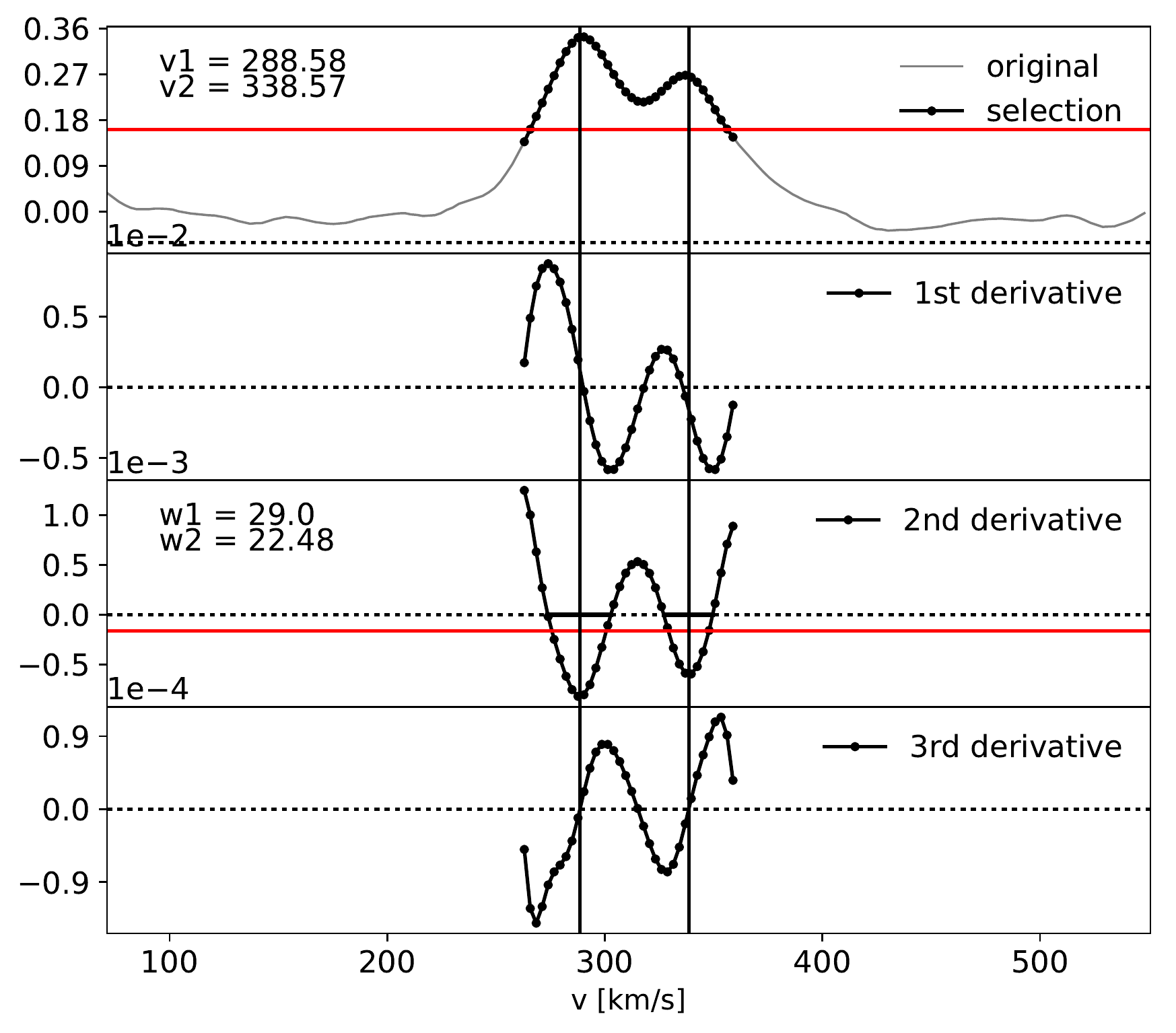}
\includegraphics[width=0.33\linewidth]{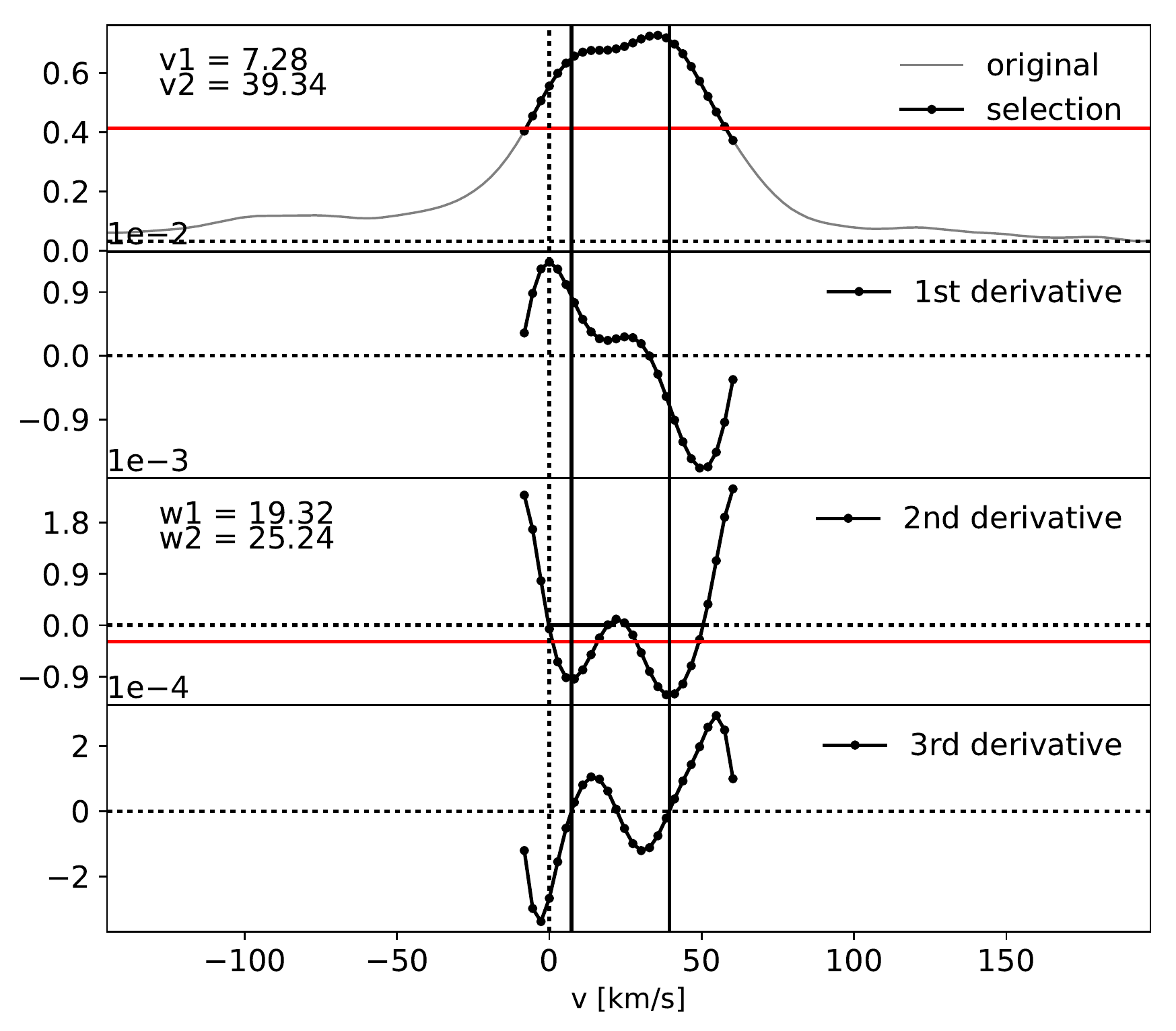}
\includegraphics[width=0.33\linewidth]{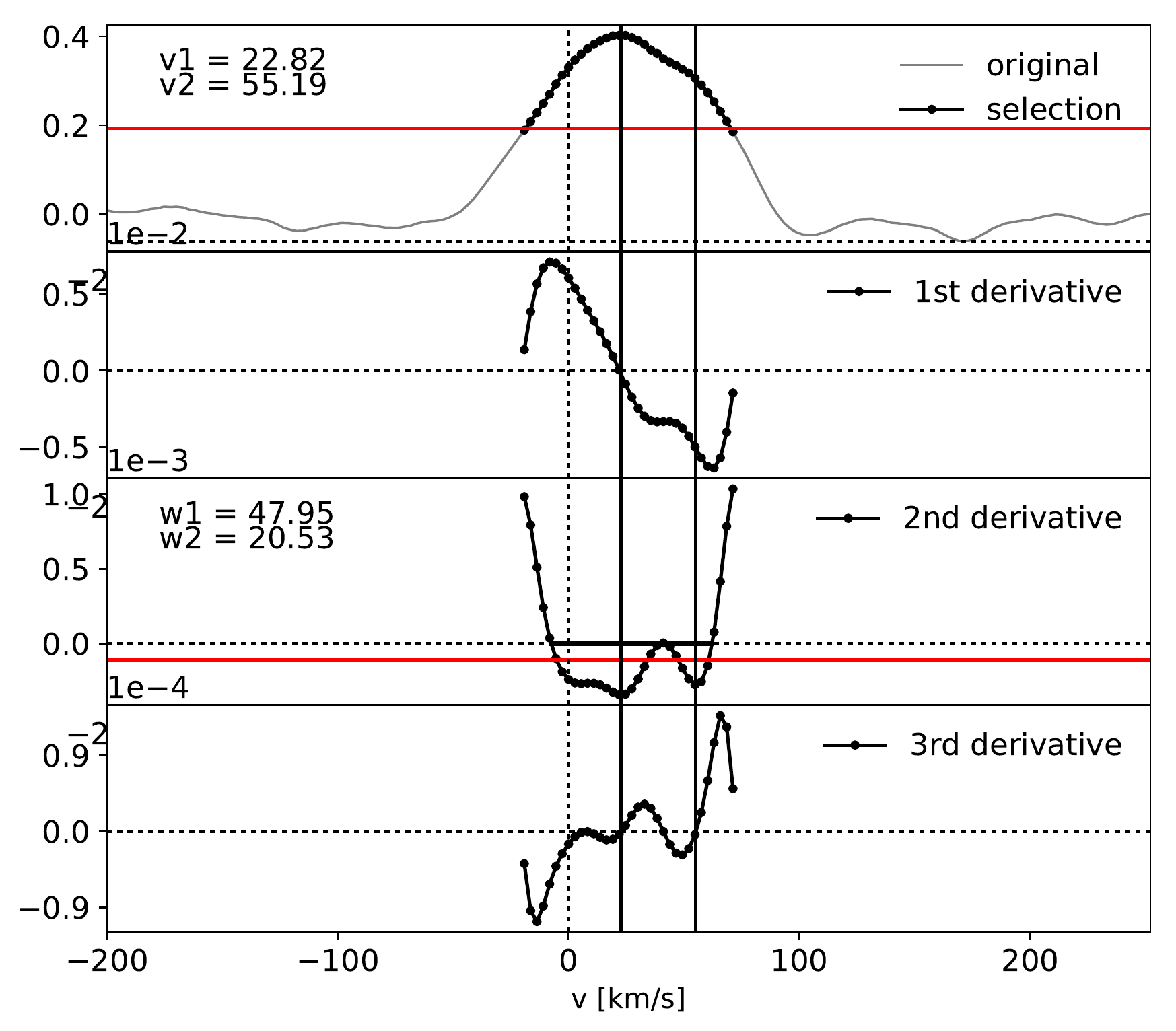}
 \caption{Examples of iDR4 HR10 double-peak CCFs used to calibrate the parameters of  the \doe\ code. These parameters (THRES0, THRES2 and SIGMA) have been fine-tuned in order to detect multiple components even when they are severely blended as in the case of the rightmost panel.}
 \label{fig:cal_hr10}
 \end{figure*}
\begin{figure*}
 \includegraphics[width=0.33\linewidth]{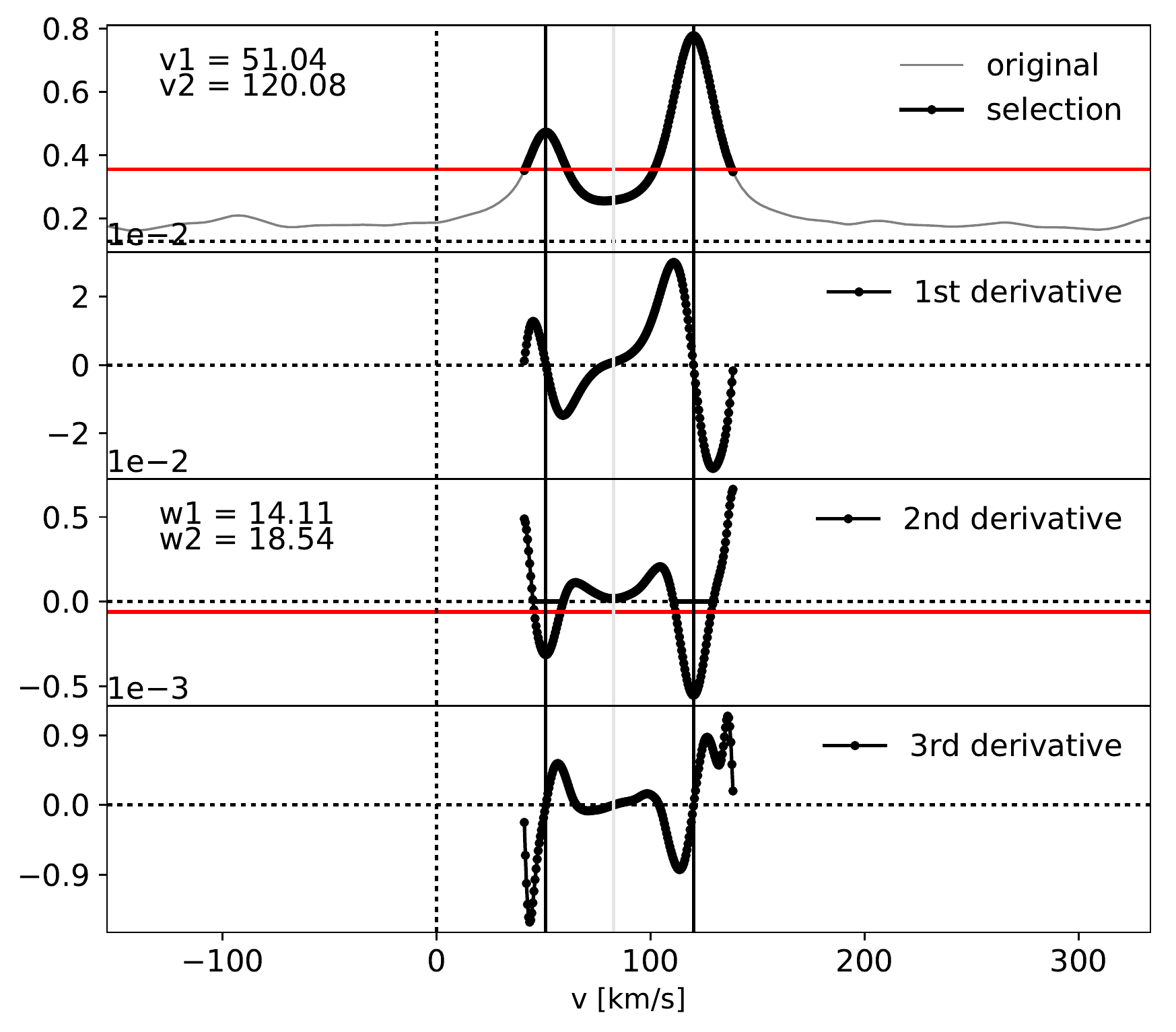}
 \includegraphics[width=0.33\linewidth]{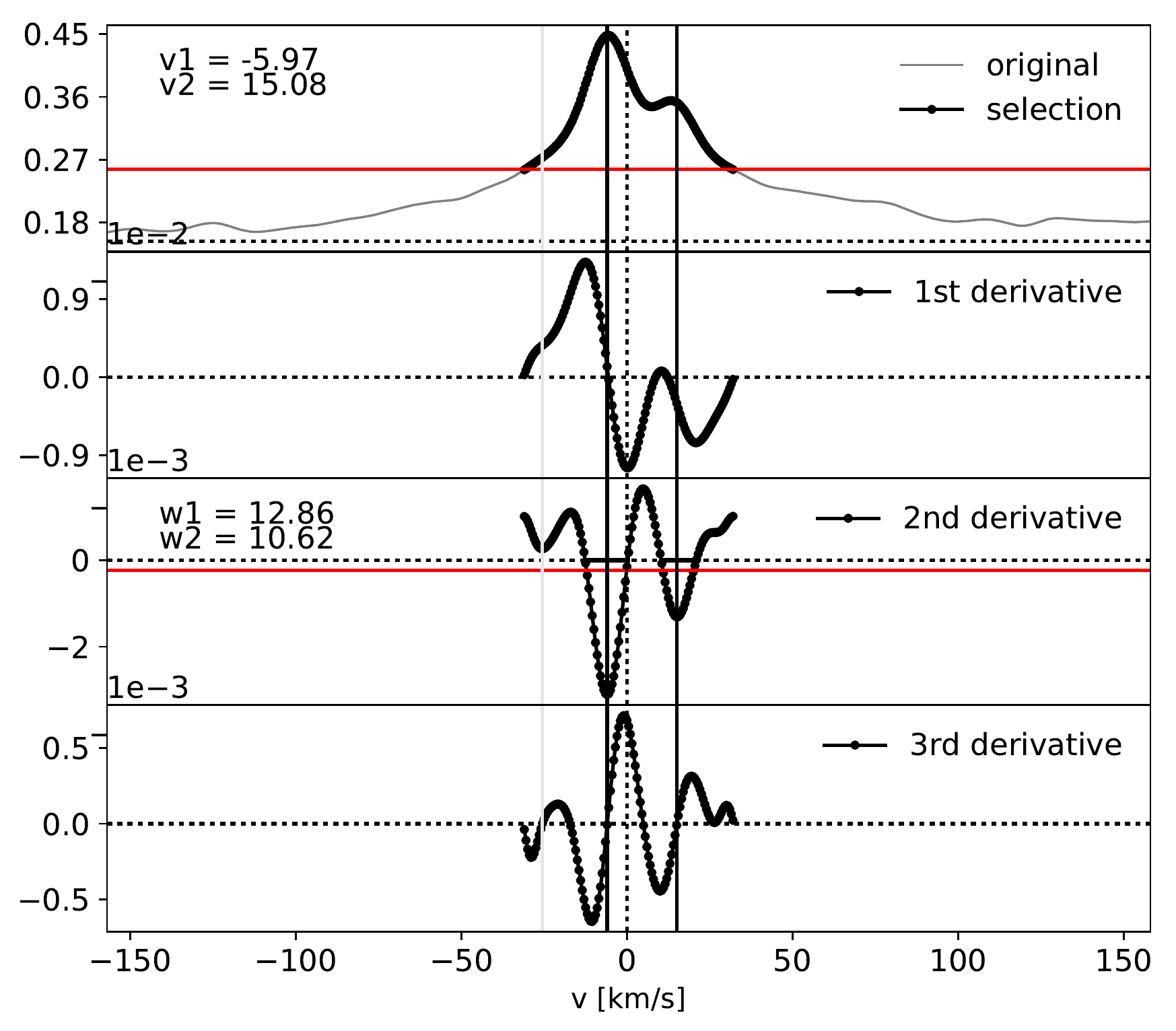}
 \includegraphics[width=0.33\linewidth]{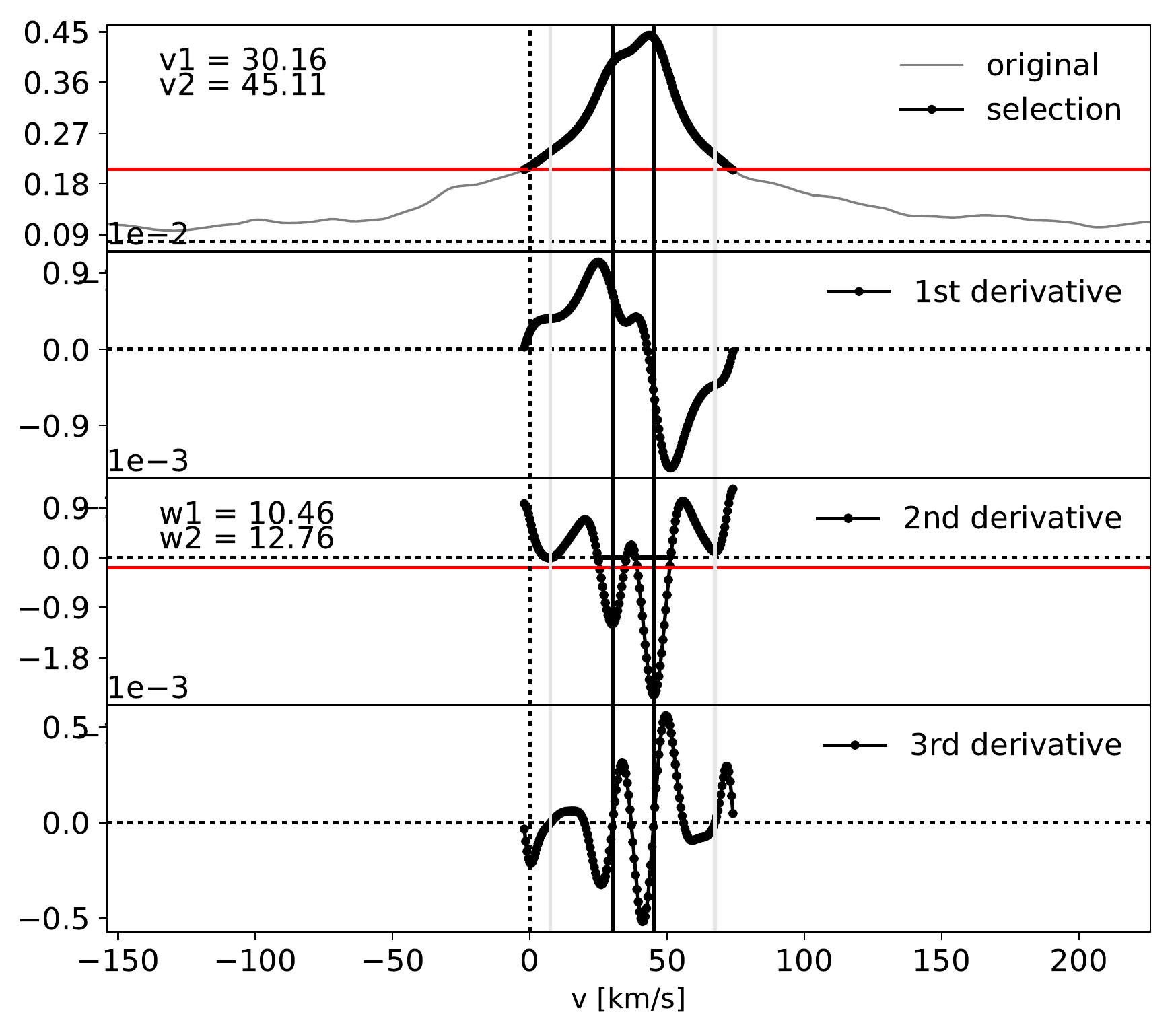}
 \caption{As Fig.~\ref{fig:cal_hr10} but for the U580 setup.}
 \label{fig:cal_u580}
 \end{figure*}

\subsection{Detection of peaks on simulated CCFs}
\label{sec:detection_of_peaks}
We tested the efficiency of the \doe\ code on simulated double-peak CCFs.
Using the radiative transfer code \emph{turbospectrum} \citep{plez2012, delaverny2012}, the MARCS library of model atmospheres with spherical geometry \citep{gustafsson2008} and the GES atomic linelist \citep{heiter2015}, we computed the synthetic spectrum of a star with the following stellar parameters: $T_{\mathrm{eff}} = 5000\,\mathrm{K}$, $\log g = 1.5$, $[\mathrm{Fe}/\mathrm{H}] = 0.$ and $\xi_{\mathrm{t}} = 1.5\,\mathrm{km\,s}^{-1}$, between 5330~\AA\ and 5610~\AA\ for a resolution of $R\sim 20\,000$, \emph{i.e.} to reproduce an HR10 spectrum (see Sect.~\ref{sect:obs_ccf}). Then, we shifted this spectrum so that the radial velocity of this simulated star is $v_{\mathrm{rad,0}} = 72$~\kms.

We also add a Gaussian noise to reproduce spectra with $S/N=20$. Then we combine the spectra shifted at different radial velocities to simulate a composite spectrum. Assuming a flux ratio between the two components of $2/3$, we set the main peak at a fixed velocity of 72.0 \kms whereas the position of the second peak is set at either 36.0, 48.0 or 54.0 \kms.The cross-correlation function between the composite and the initial spectrum is calculated and the CCF is normalised by the maximum value of the mask auto-correlation (auto-correlation of the initial spectrum).

The three simulated CCFs and their derivatives are shown on Fig.~\ref{fig:ccf2_test}, the value of SIGMA being $2.1$~\kms. From the first derivative, only one crossing of the $x$-axis leads to the detection of one single peak. From the second derivative, we see clearly two minima in the left and middle panel whereas we see only one minimum in the right panel.
This leads to the conclusion that the detection limit between two components is 18~\kms. 
This detection limit depends on the typical width of absorption lines in the tested spectrum but also on the SIGMA parameter.
However reducing the SIGMA parameter too much could increase false peak detections for bumpy CCFs. A compromise had to be adopted, as described in Sect.~\ref{Sect:parameters}. 

\subsection{Choice of the \doe\ parameters for the different setups}
\label{Sect:parameters}

The three parameters of the \doe\ code described in Sect.~\ref{sect:doe} have to be adjusted to optimize the CCF components detection. These parameters were adjusted by performing individual calibrations for the different setups (GIRAFFE HR10, HR15N, HR21, and UVES U520 and U580) using examples of single-, double-, and triple-peak CCFs with different separations between the components, and different component widths (\emph{i.e.} different degree of blending). For the remaining GIRAFFE setups, a standard value of the SIGMA parameter (3~\kms) was adopted. The adopted values are listed in Table~\ref{tab:doe_param}. The parameter adjustment aims at obtaining the same detection efficiency on the test CCFs as through eye inspection, especially in the extreme cases (blended CCFs). Figures~\ref{fig:cal_hr10} and \ref{fig:cal_u580} illustrate favourable and extreme cases. The value of THRES0 is larger for the GIRAFFE CCFs than for the UVES ones because the correlation noise (\emph{i.e.}, the signal level in the CCF continuum) was observed to be larger in GIRAFFE CCFs.

Depending on the setup resolution along with the number and strength of lines, the minimum separation for peak detection was empirically found to be in the range [20-60] \kms\ for GIRAFFE setups (15~\kms\ for UVES ones). As an example in Sect.~\ref{sec:detection_of_peaks} and Fig.~\ref{fig:ccf2_test}, we showed with simulated CCFs that the detection limit is reached for a minimum separation of 18~\kms\ at $R\sim 20\,000$ for slowly rotating stars.
The spectrograph resolution and the CCF sampling are not the only relevant parameters here, since the intrinsic line broadening (macroturbulence and stellar rotation) also impacts the CCF width.

\doe\ includes a procedure to compare the number of valleys in the second derivative with the number of detected peaks. When these numbers are not identical, iteration on the detection occurs after increasing the SIGMA parameter. This procedure prevents false detections since in these situations, the wide CCF often exhibits inflexion points which cause zeros in the third derivative (see left panels of Fig.~\ref{fig:doe_it}). The number of valleys, defined as regions where the second derivative is continuously negative, is assessed first. For example, in the left ``2nd derivative'' panel of Fig.~\ref{fig:doe_it}, one valley is detected. For low values of the SIGMA parameter, the number of detected velocity components is systematically larger than the number of valleys (left panels of Fig.~\ref{fig:doe_it}). As long as the number of valleys is lower than the number of velocity components detected from the 3rd derivative, the SIGMA parameter is increased by 2~\kms, until the number of detected velocity components equals the number of valleys. The iterative process is then stopped and the radial velocities of the detected velocity components are identified. 

\begin{figure*}[t]
  \includegraphics[width=0.49\linewidth]{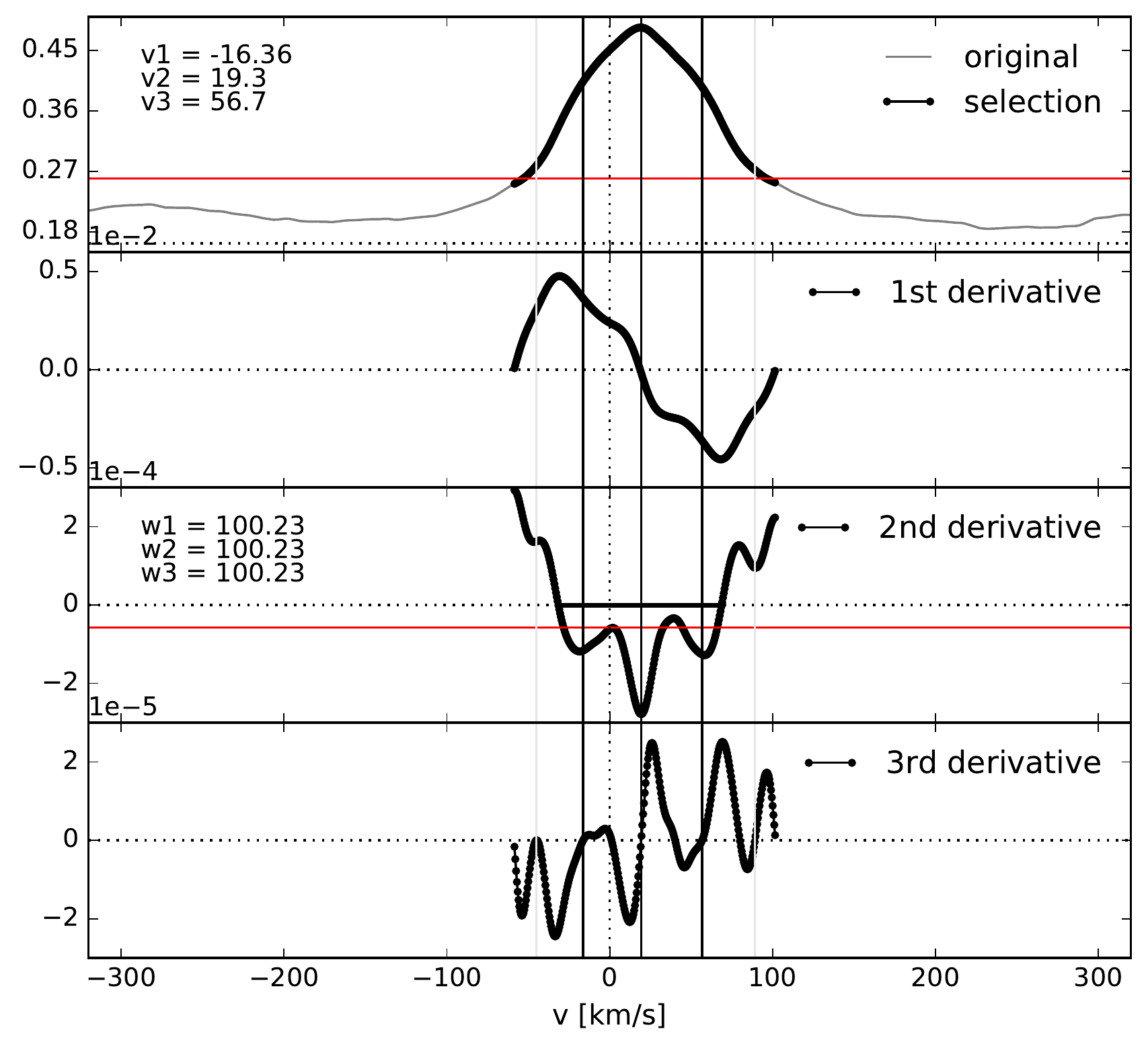}
 \includegraphics[width=0.49\linewidth]{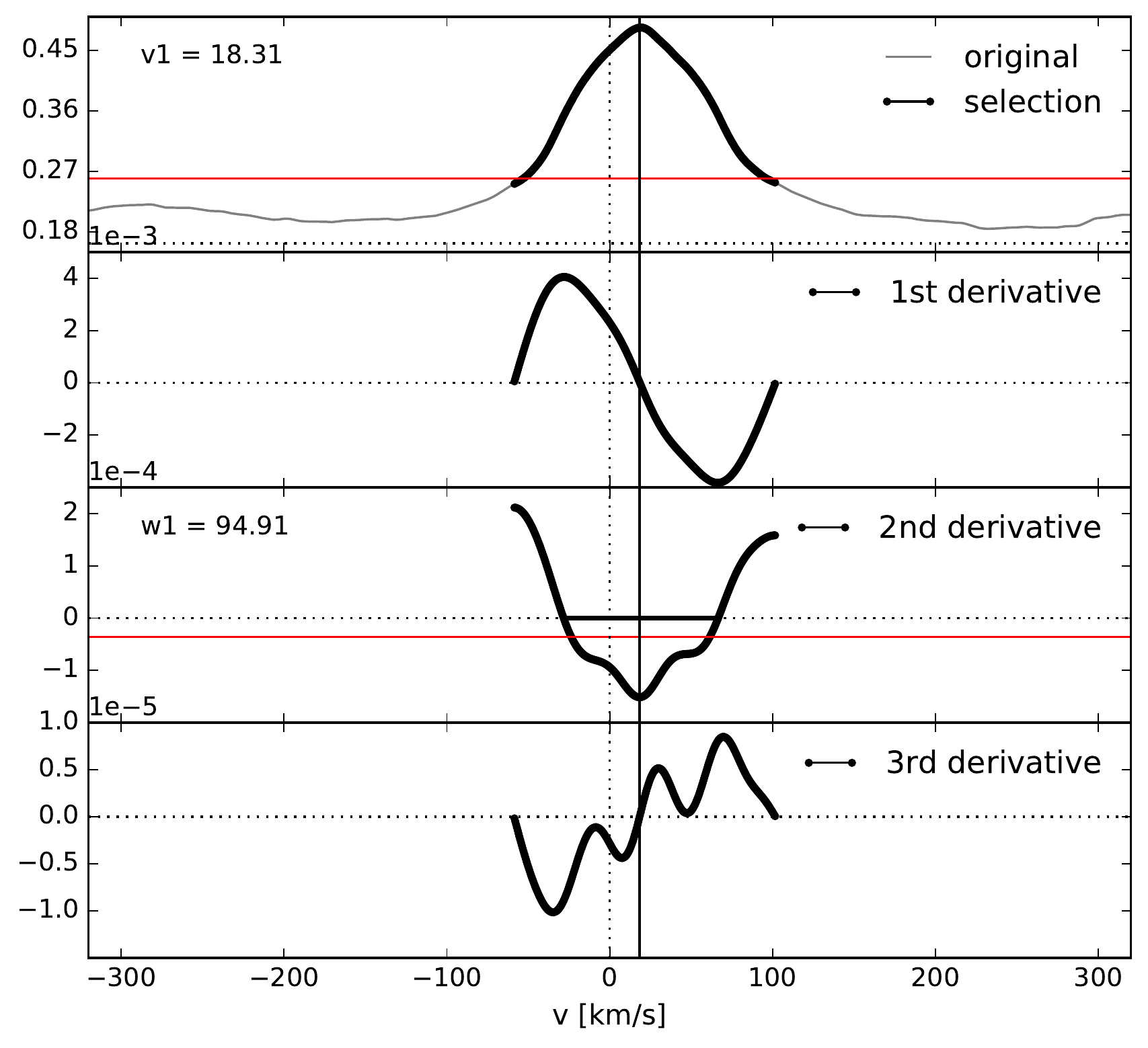}
 \caption{Special procedure for fast rotators. Left panel: after few iterations three velocity components and one valley are detected. Right panel: after 11 iterations, one velocity component associated to one valley is identified. The associated spectrum has $S/N=65$.}
 \label{fig:doe_it}
\end{figure*}

Figure~\ref{fig:doe_it} shows an example of this procedure applied on the K1 pre-main sequence object 2MASS J06411542+0946396 (CNAME\footnote{By convention within the GES, the sources are referred to by a `CNAME' identifier formed from the ICRS (J2000) equatorial coordinates of the sources. For instance, the J2000 coordinates of the source CNAME 08414659-5303449 are $\alpha=8$~h 41~min 46.59~s and $\delta=-53~^\circ$ 3' 44.9''.} 06411542+0946396) member of the cluster NGC~2264 \mbox{\citep{furesz2006}}. The \doe\ run starts with the standard SIGMA value of  5~\kms. Initially, the \doe\ code detects three valleys in the second derivative and six velocity components from the third derivative, which are clearly spurious detections. After three iterations, one valley and three velocity components are identified (left panel of Fig.~\ref{fig:doe_it}). After 11 iterations, SIGMA increases from 5 to 27~\kms\ and the process ends up with one velocity component located at 18.31~\kms\ (right panel of Fig.~\ref{fig:doe_it}, to be compared with the velocity of 19.86~\kms\ found by \citealt{furesz2006}). The case of CCF multiplicity that can be due to physical processes different from binarity (like pulsating stars, nebular lines in spectra, etc.) is discussed in Sect.~\ref{sect:ccf_degeneracy}.

\subsection{Estimation of the formal uncertainty of the method}
\label{sec:error}
In this section, we assess the choice of the SIGMA parameter and its effect on the derived radial velocities and their uncertainty.
The uncertainty on the derived radial velocity for single-peak CCF depends mainly on the $S/N$ of the spectrum used to compute the CCF, the normalisation of this spectrum and the mismatch between the spectrum and the mask (spectral type, atomic and molecular profiles, rotational velocity, etc.).

We performed Monte-Carlo simulations to compute single-peak CCFs from spectra of different $S/N$ ratios but using the same atmospheric parameters defined in Sect.~\ref{sec:detection_of_peaks}. We sliced this synthetic spectrum and degraded its resolution in order to match the following settings: GIRAFFE HR10 and HR21, UVES U520 and U580 (up and low). For each $S/N$ level, we computed $251$ realisations of our simulated GIRAFFE and UVES spectra by adding a Gaussian noise and computed the corresponding CCFs using a mask made of a noise-free spectrum with a null radial velocity. We finally ran \doe, with different values of SIGMA (from $1$ to $15$ by step of 1~\kms). Figures~\ref{fig:erv_sigma_giraffe} and \ref{fig:erv_sigma_uves} show the difference $\Delta v_{\mathrm{rad}} = v_{\mathrm{rad,doe}} - v_{\mathrm{rad,0}}$, where $v_{\mathrm{rad,0}}=72.0$~\kms, as a function of the \doe\ parameter SIGMA (right panel) and the $251$ CCFs (left panel) along with the noise-free CCF (labeled ``$+\infty$''). We show the results for the lowest $S/N$ (\emph{i.e.}, the most unfavorable cases) for the setups GIRAFFE HR10 and HR21 and UVES U580 (low and up). The mean and standard deviation of $\Delta v_{\mathrm{rad}}$ are also superimposed with dark dots and error bars in the right panels.

Comparing the noise-free CCF (blue curve) in the left panels of Figs.~\ref{fig:erv_sigma_giraffe} and \ref{fig:erv_sigma_uves} shows striking differences from one setup to the other. This is directly related to the spectral information contained by the spectrum used in the CCF computation. For our simulated star, the HR10 and U580 (low) spectra are more crowded than the HR21 and the U580 (up) spectra. This results in a higher level of the CCF continuum. In addition, in HR21, the large wings of the CCFs are due to the strong \ion{Ca}{ii} IR triplet that completely dominates this spectral range (see Fig.~\ref{fig:sun_spec_ccf}). Figures~\ref{fig:erv_sigma_giraffe} and \ref{fig:erv_sigma_uves} also show that the spectral noise tends to shift downward the CCF in comparison to the noise-free CCF because the noisy spectra are less similar to the mask than the noise-free ones. In U580 (Fig.~\ref{fig:erv_sigma_uves}), we see that the distance between the noisy CCFs and the reference one is not similar in upper and lower left panels, despite the same $S/N$. This larger distance in U580 low compared to U580 up could be due to the fact that, for our simulated star, there are more weak lines in the low setup, and therefore, they quickly vanish in the noise when the $S/N$ drops.

\begin{figure*}
\center
 \includegraphics[width=0.45\linewidth]{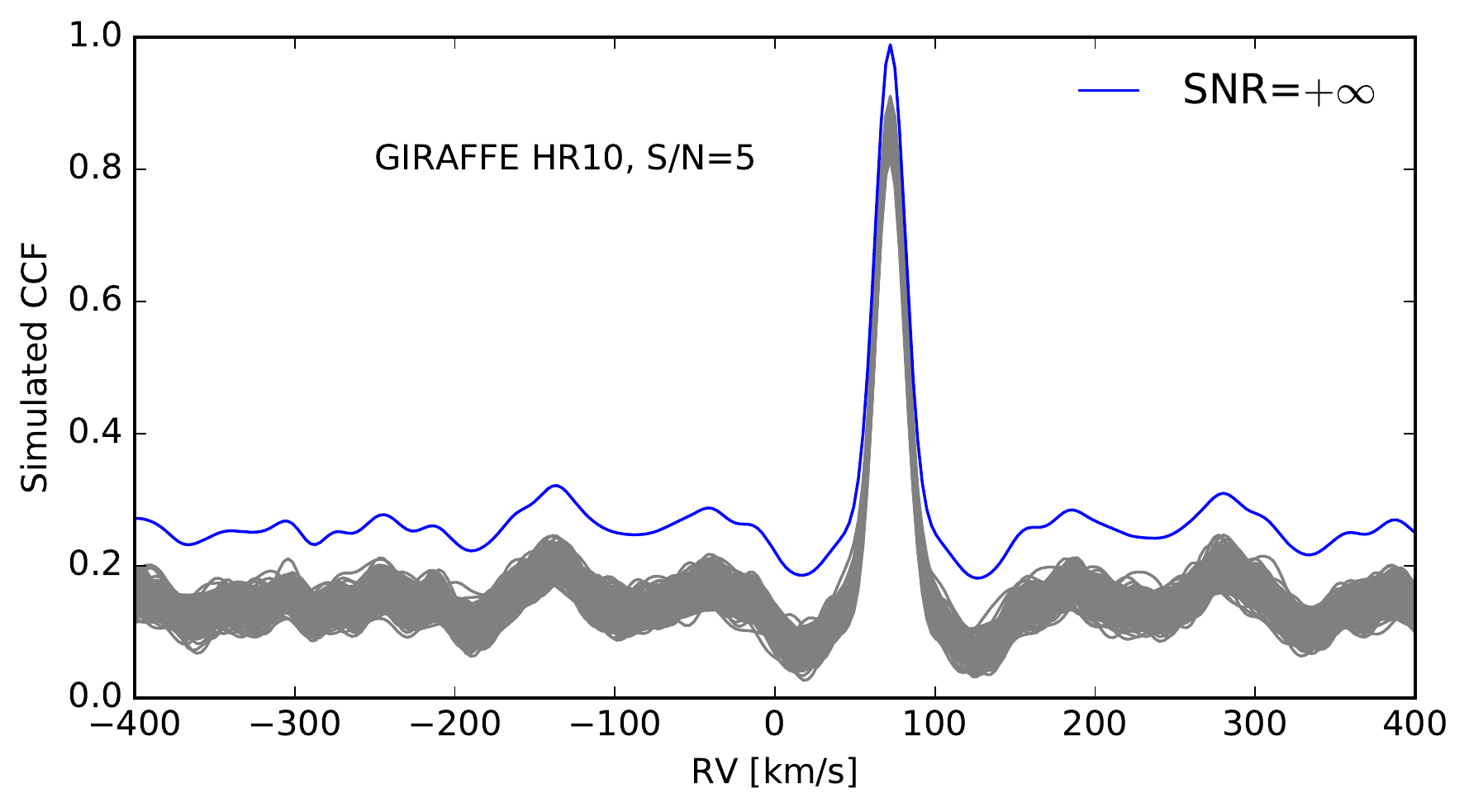} 
 \includegraphics[width=0.45\linewidth]{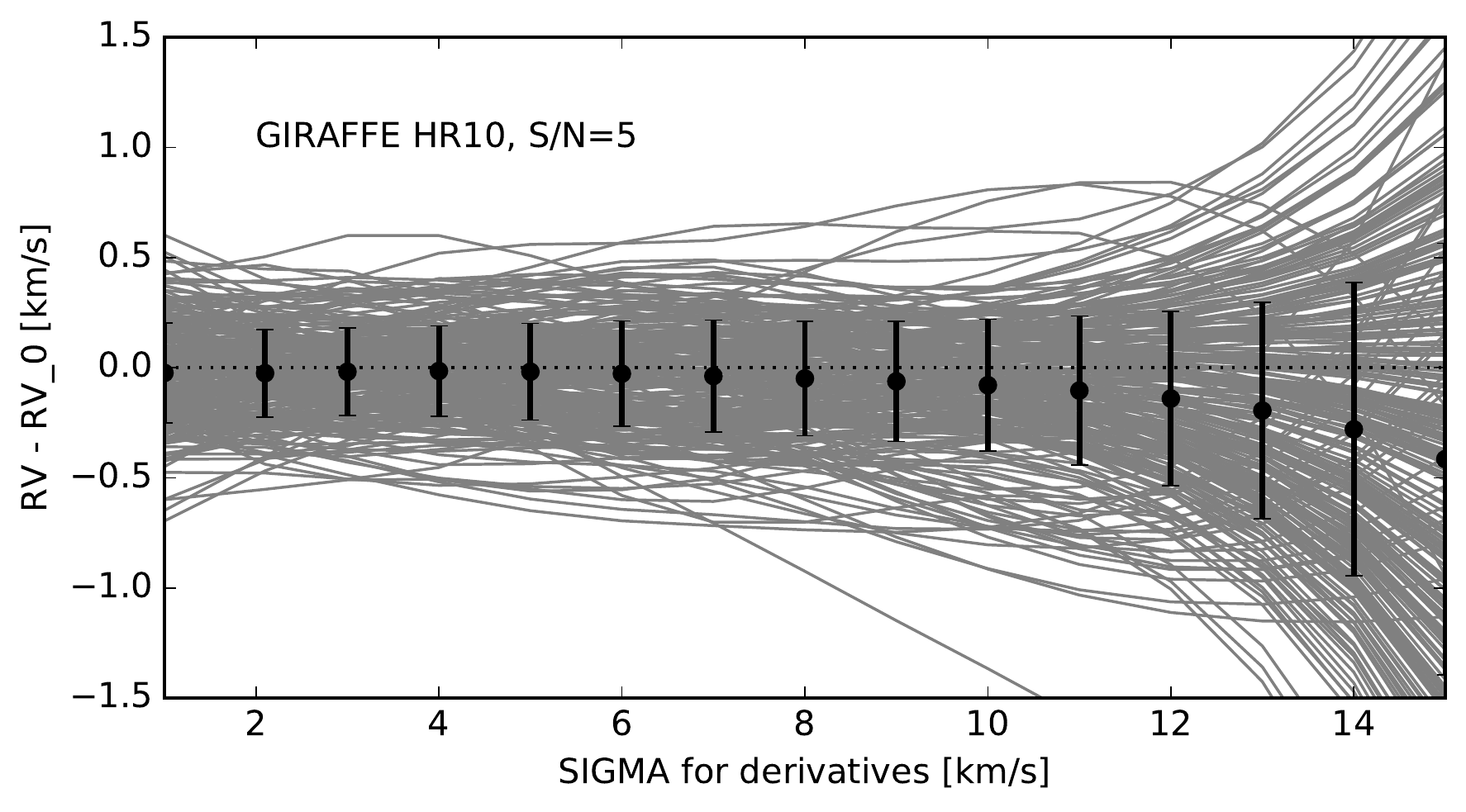}

 \includegraphics[width=0.45\linewidth]{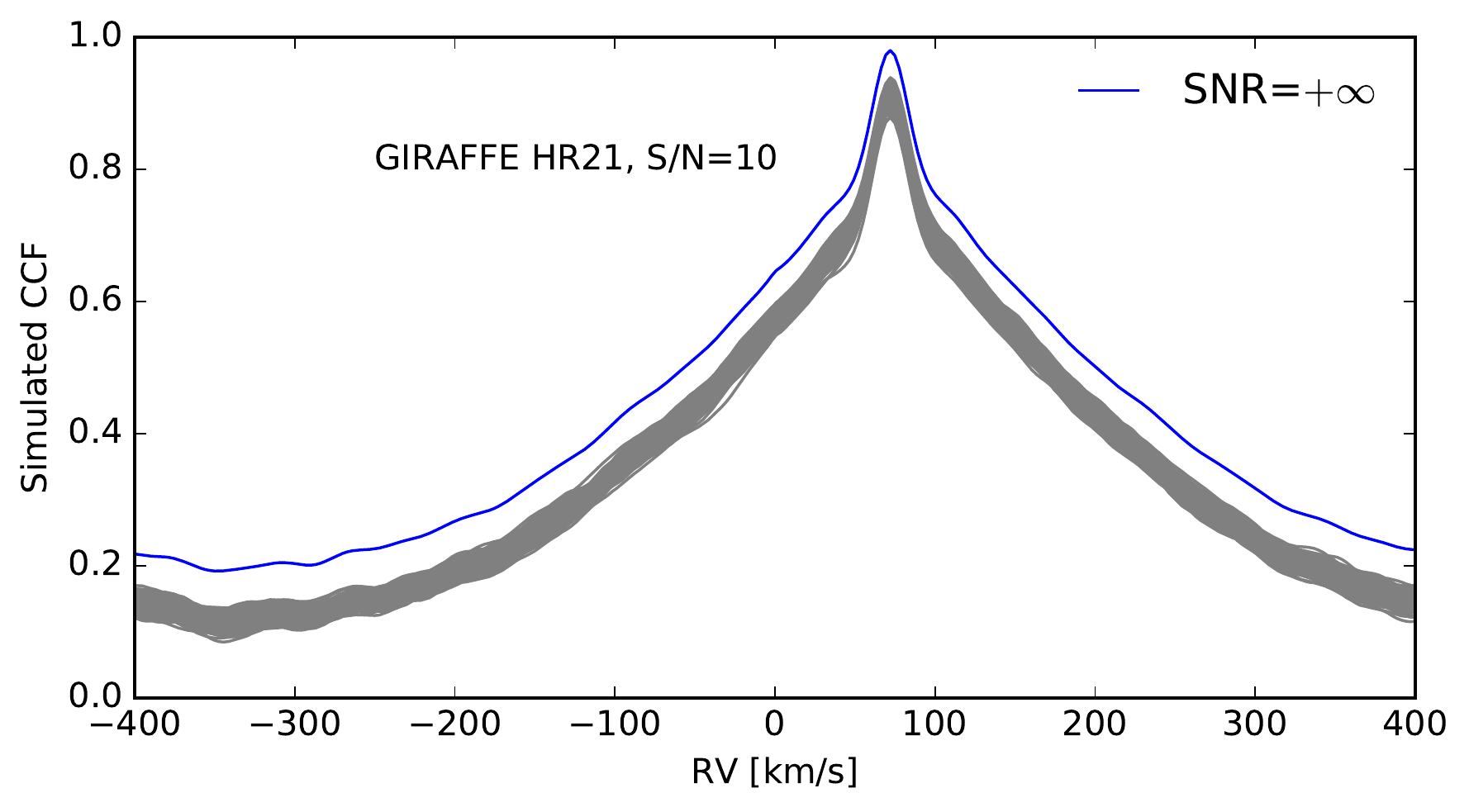}
 \includegraphics[width=0.45\linewidth]{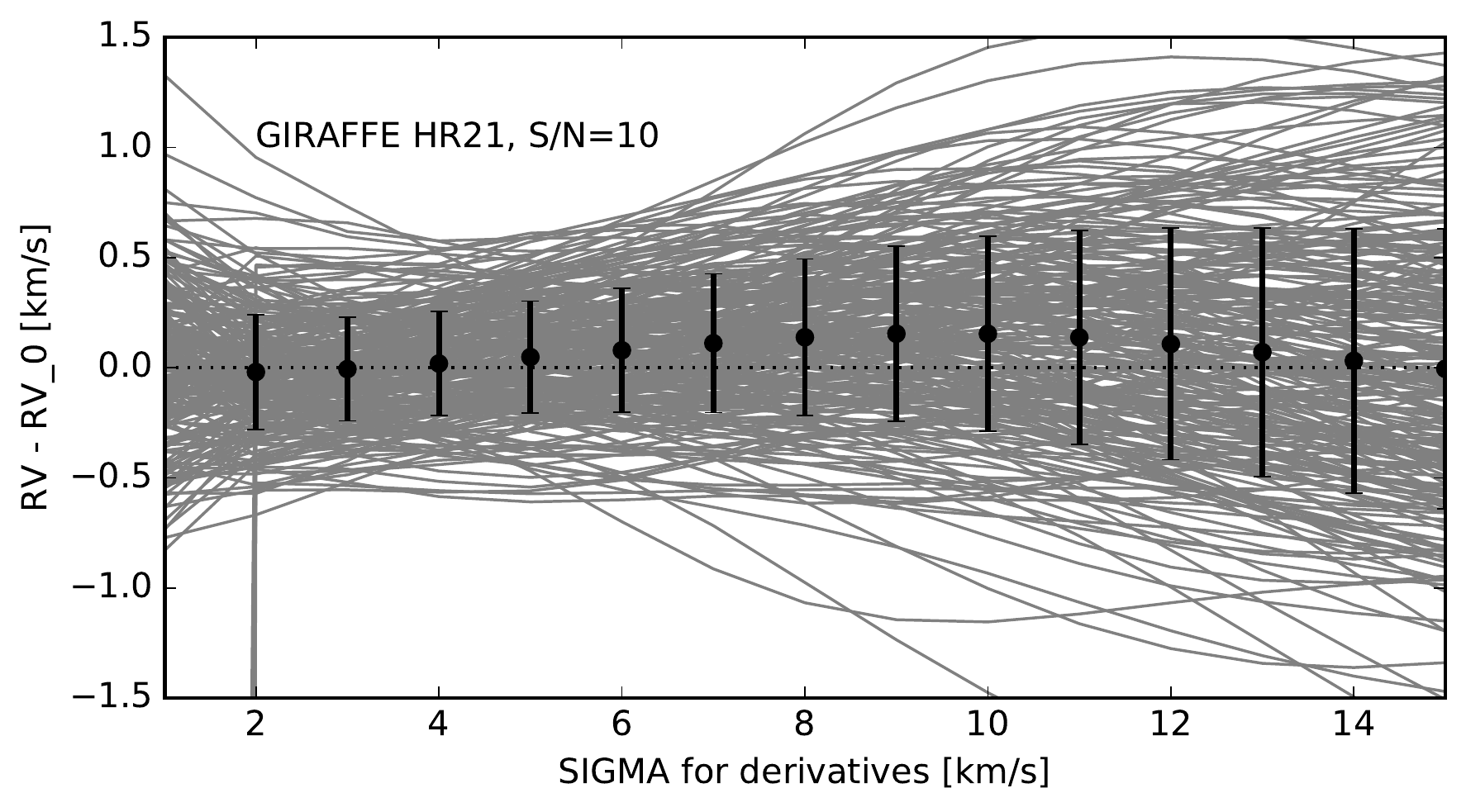}
 \caption{Estimation of the accuracy of the radial velocities determined by the \doe\ code on GIRAFFE setups HR10 and HR21 (\ion{Ca}{ii} triplet region). In each case, 251 simulated CCFs with a S/N ratio as labelled and the blue line representing a noise-free CCF (left panels) were analyzed with \doe\ varying the value of SIGMA for the calculation of the smoothed successive derivatives and of the radial velocity (right panels).}
 \label{fig:erv_sigma_giraffe}
\end{figure*}
\begin{figure*}[h!]
\center
 \includegraphics[width=0.45\linewidth]{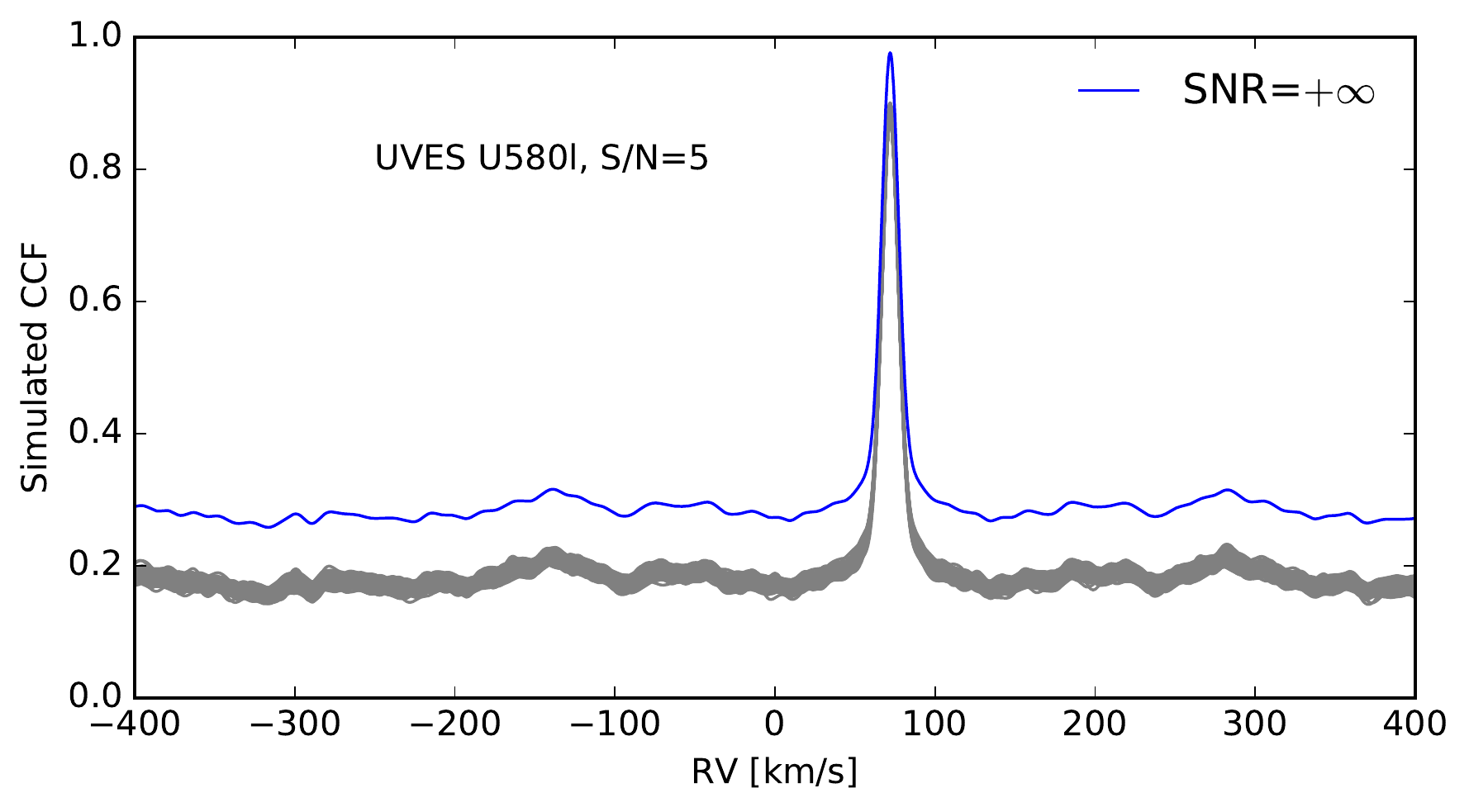}
 \includegraphics[width=0.45\linewidth]{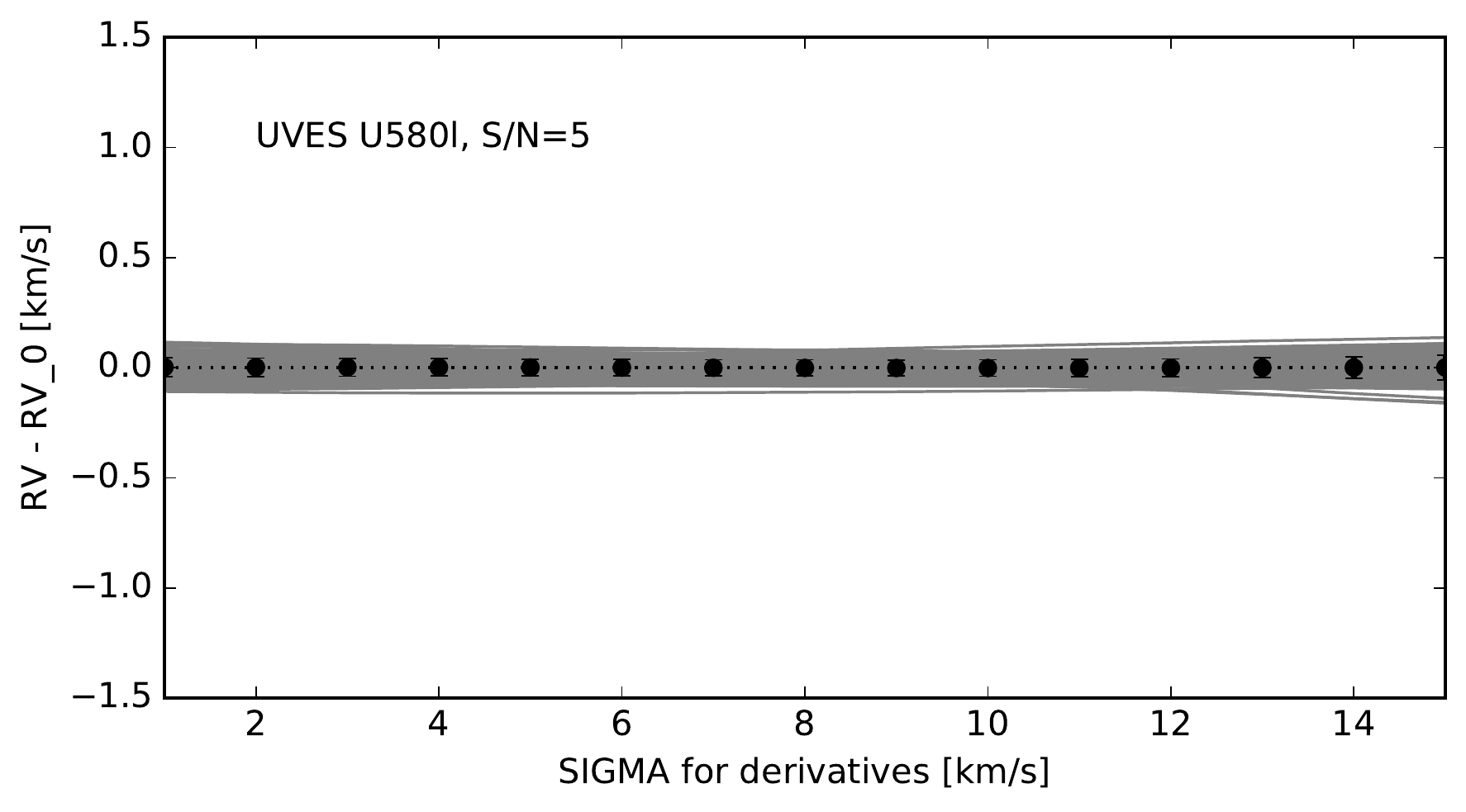}

 \includegraphics[width=0.45\linewidth]{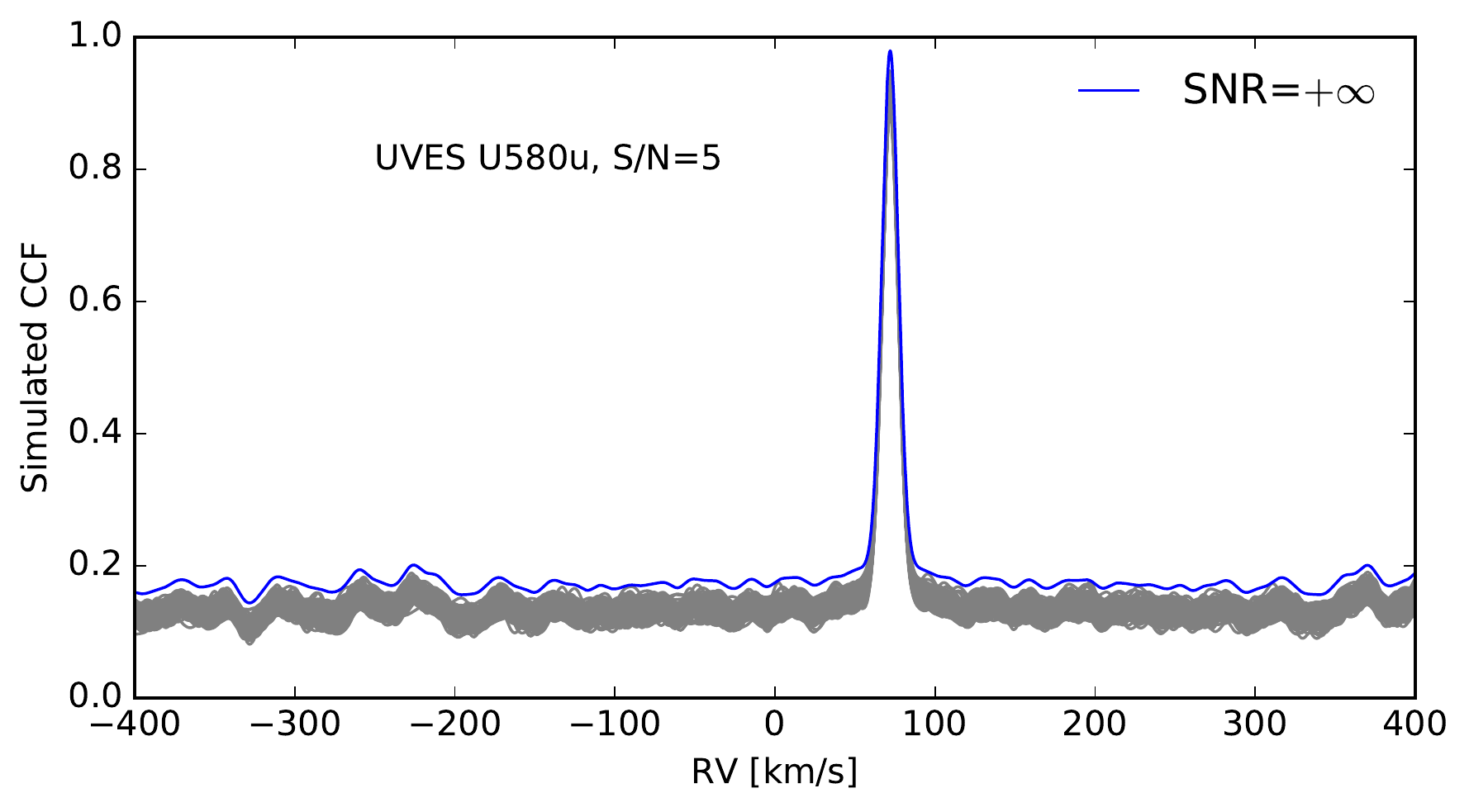}
 \includegraphics[width=0.45\linewidth]{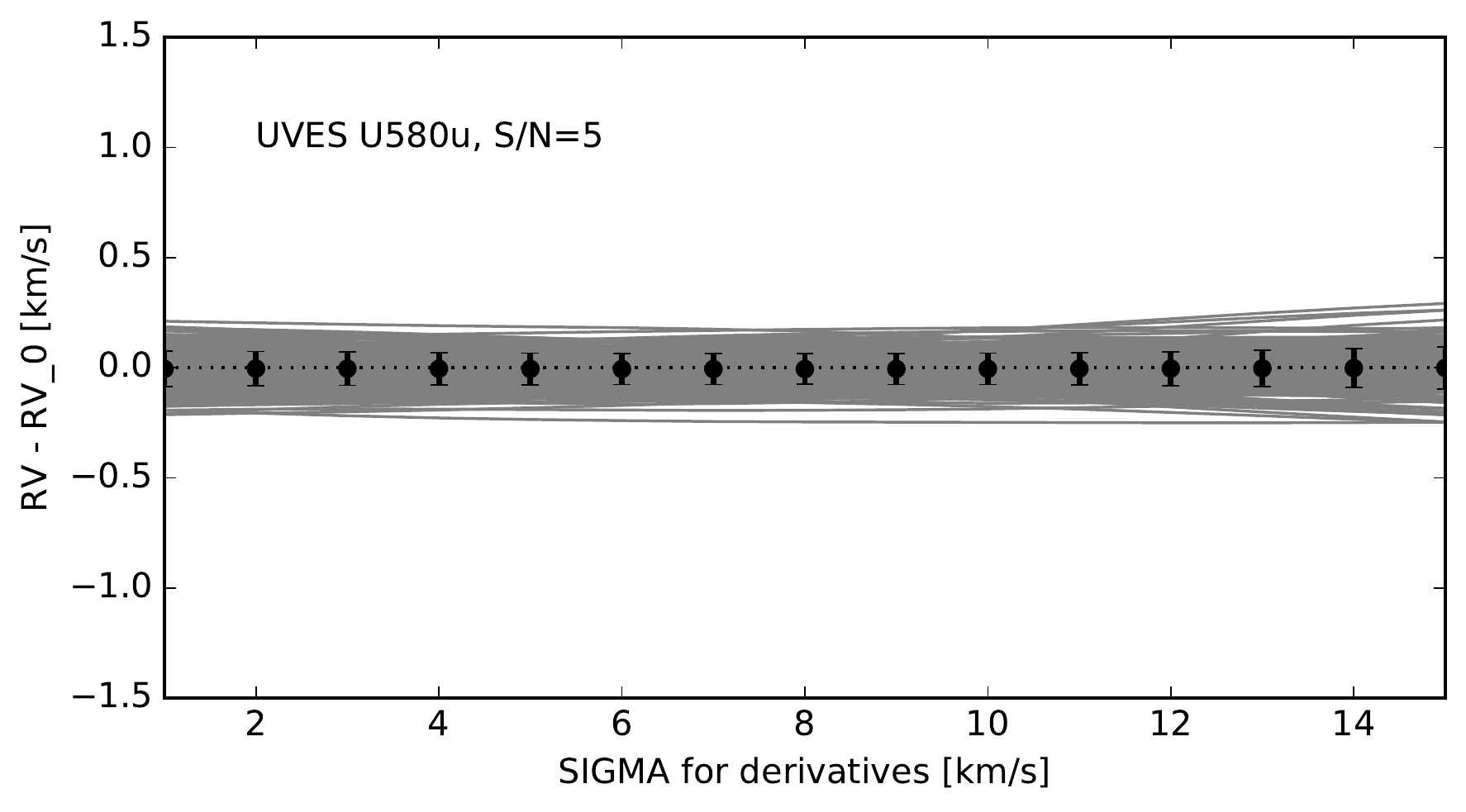}
 \caption{
Same as Fig.~\ref{fig:erv_sigma_giraffe} for the UVES setups U580 low (H$\beta$ + Mg~I b triplet region) and U580 up (\ha\ + Na~I~D doublet region).}
 \label{fig:erv_sigma_uves}
\end{figure*}

\begin{figure*}
 \includegraphics[width=0.49\linewidth]{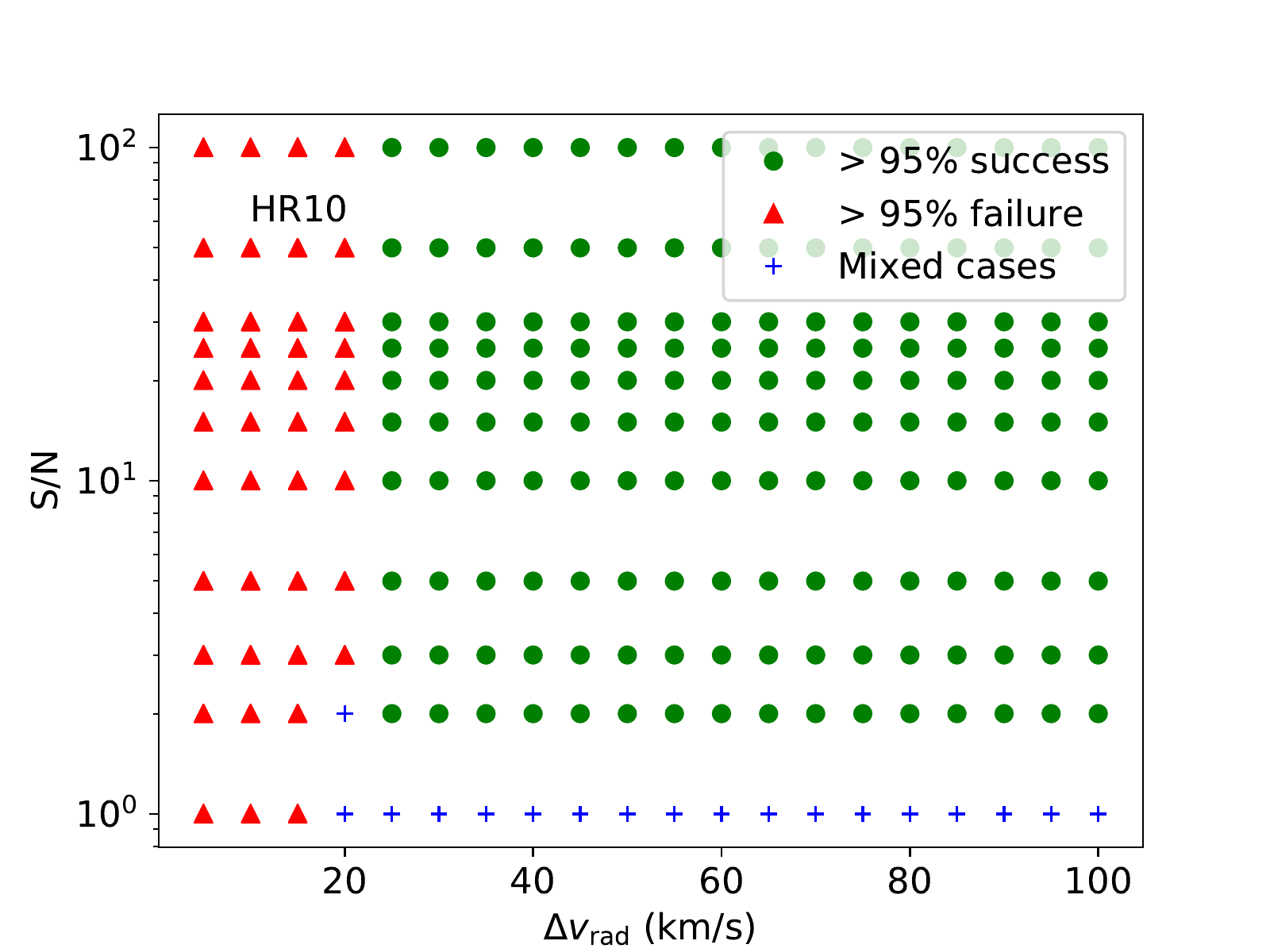}
 \includegraphics[width=0.49\linewidth]{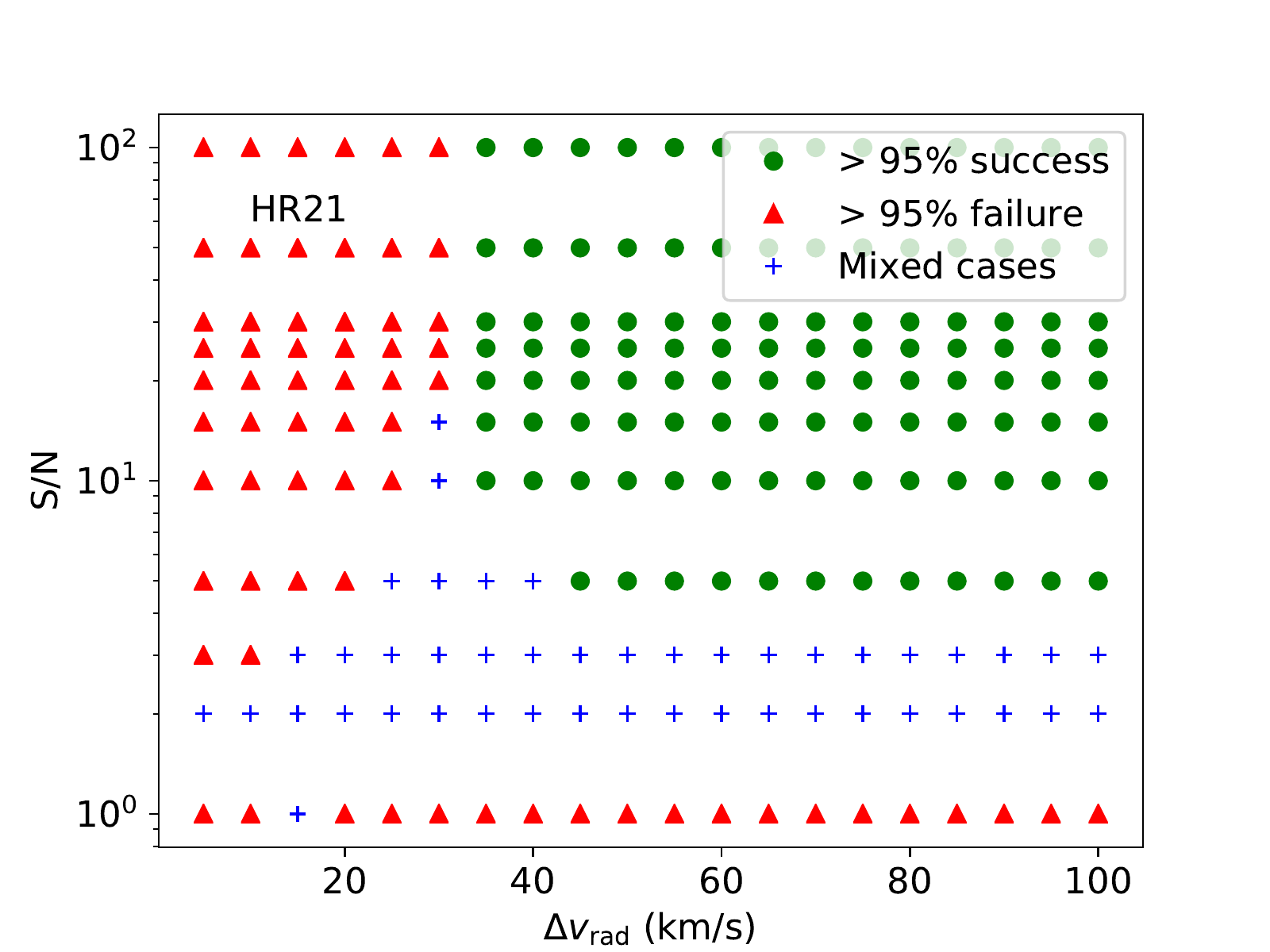}
 \caption{Assessement of the \doe\ detection efficiency of the two radial velocity components of simulated SB2 CCFs as a function of the   
 $S/N$ ratio and the radial velocity differences for GIRAFFE HR10 (left panel) and HR21 (right panel) setups.}
 \label{fig:doe_eff}
\end{figure*}

The right panels of Figs.~\ref{fig:erv_sigma_giraffe} and \ref{fig:erv_sigma_uves} show the effect of SIGMA on the derived radial velocity (uncertainty and/or bias). Our simulations clearly demonstrate that SIGMA has to be chosen in a specific range to ensure reliable results. While our simulated UVES CCFs show that the \doe\ performance is very stable for any value of SIGMA, our simulated GIRAFFE CCFs show that only a limited range of SIGMA values can ensure reliable velocity measurements. Figure~\ref{fig:erv_sigma_giraffe} suggests to keep SIGMA between $\sim 2$ and $\sim 8$~\kms\ for HR10 and $\sim 2$ and $\sim 7$ \kms\ for HR21, in agreement with our empirical calibration on a sub-sample of real GES CCFs (see Table~\ref{tab:doe_param}). The behavior of \doe, while varying SIGMA, is different for GIRAFFE and UVES CCFs (Figs.~\ref{fig:erv_sigma_giraffe} and \ref{fig:erv_sigma_uves}). This is not due to the $S/N$ ratio but rather to the sampling of the velocity grid onto which the GIRAFFE and UVES CCFs are computed, \emph{i.e.} SIGMA is related to the velocity step of the CCFs. Indeed, in Sect.~\ref{sect:obs_ccf}, we recalled that the sampling frequency of the CCF is lower for GIRAFFE CCFs than for UVES CCFs: as SIGMA increases, a pronounced asymmetry on the second derivative appears for GIRAFFE CCFs, resulting in the high scatter displayed by Fig.~\ref{fig:erv_sigma_giraffe}.

Our simulations allow us to quantify the effect of the $S/N$ of the spectra on the method. For U520 and U580, the standard deviation on the radial velocity at the recommended SIGMA goes from 0.05~\kms\ at $S/N=5$ to lower than $0.01$~\kms\ at $S/N=50$. For GIRAFFE HR10, it goes from 0.20~\kms at $S/N=5$ to 0.02~\kms\ at $S/N=50$. For GIRAFFE HR21, the situation is the worst of all the setups with a standard deviation going from 0.25~\kms\ at $S/N=10$ to 0.06~\kms at $S/N=50$. The obvious conclusion is that the UVES setups tend to give more precise results for a given $S/N$ compared to GIRAFFE setups. This is understandable since a single UVES spectrum has a higher resolution and awavelength coverage larger than any GIRAFFE spectrum. For our simulated star, the precision on the radial velocity derived by \doe\ is up to five times higher for UVES setups than for GIRAFFE HR10 (this is even worse when compared with HR21).

This first approach of simulated CCFs shows that the method is quite robust with respect to the noise level in the GES spectra. Obviously, the presence of multiple components in the CCF may shift the detected radial velocities especially when the peaks blend one another. In such a case, the inaccuracy on the radial velocity can reach several \kms\ (increasing as the blending degree increases). No quantitative calculations have been performed so far but the middle panel of Fig.~\ref{fig:ccf2_test} shows a good illustration: the main peak is detected at 0.95~\kms\ of its expected position and the second peak at 2.3~\kms, with a simulated distance of 24~\kms\ between the two peaks. We conclude that the (conservative) random uncertainty on the radial velocity derived by \doe\ is of the order of $\pm0.25$\,\kms\ while the systematic uncertainty is lower than $0.05$\,\kms\ for single-peak CCF and may reach a few \kms\ for multi-peak CCF. 
Other effects, like template mismatch or imperfect normalisation, may have an effect on the uncertainty on the derived radial velocity.
We also refer the reader to \cite{jackson2015} where a discussion on the radial velocity uncertainties may be found, along with their empirical calibrations as a function of $S/N$, $v\sin{i}$ and the effective temperature of the source for GIRAFFE HR10, HR15N and HR21 setups. As shown by \cite{sacco2014} and \mbox{\cite{jackson2015}}, the errors on the GES radial velocities for most of the stars are dominated by the zero-point systematic errors of the wavelength calibration that are not discussed here.

\subsection{Detection efficiency as a function of the $S/N$ ratio}

Using Monte-Carlo simulations, we assessed the impact of the $S/N$ ratio of GIRAFFE HR10 and HR21 spectra on the detection efficiency of the double-peaked CCF of an SB2. For that purpose we simulated synthetic SB2 spectra (pair of twin stars) varying the $S/N$ (from $1$ to $100$) and varying the difference in radial velocity of the two components $\Delta v_\mathrm{rad}$ (from $5$ to $100$~\kms). For each pair ($\Delta v_\mathrm{rad}$, $S/N$), we computed as above 251 realisations of the spectra and their corresponding CCFs. We then applied \doe\ with the parameters adapted to each setup (see Table~\ref{tab:doe_param}).

The maps in Fig.~\ref{fig:doe_eff} show the detection efficiency in HR10 and HR21. The green dots (respectively the red triangles) indicate ($\Delta v_\mathrm{rad}$, $S/N$) conditions when \doe\ is able to detect the two expected peaks in more than 95\% of cases (respectively, conditions when \doe\ failed at detecting the two expected peaks in more than 95\% of cases). Blue plusses represent intermediate cases making detection efficiency dependent on the noise: due to the noise, spurious peaks may appear (\emph{i.e.} detection failed) or the two peaks have different height (despite being twin stars) and become discernible to \doe\ for small $\Delta v_\mathrm{rad}$ (\emph{i.e.} detection succeeded; \emph{e.g.}, for HR21, at $S/N = 10$ and $\Delta v_\mathrm{rad} = 25$~\kms).

These simulations demonstrate that even spectra with very low $S/N$ carry sufficient information to reveal the binary nature of the targets. Specifically, in the HR10 setup, double peaks are detected in 95\% of cases when $S/N \ge 2$ and $\Delta v_\mathrm{rad} \ge 25$~\kms while in HR21 setup, they are detected at the same rate when $S/N \ge 5$ and $\Delta v_\mathrm{rad} \ge 45$~\kms. Thus, the $S/N$ threshold that we adopted (\emph{i.e} analysis of CCFs for all spectra with $S/N \ge 5$) protect us from mixed cases, which tend to happen for the lowest levels of $S/N$ ratios. This shows also that the HR10 setup is more suitable to detect SB2 than HR21 because HR21 is located around the IR \ion{Ca}{ii} triplet whose lines have strong wings that decrease the detection efficiency. In Sect.~\ref{sec:sb2_candidates}, the histogram of the radial velocity separation of the effectively detected SB2 candidates is presented. Observationally, HR10 spectra (respectively, HR21) allow us to detect SB2 with $\Delta v_\mathrm{rad}$ as low as $\sim 25$~\kms (respectively, $\sim 60$~\kms): thus for both setups we are dealing with cases falling in the green dotted area of the maps. Thus, we expect in all cases an SB2 detection efficiency better than 95\%.

\section{iDR4 results and discussion}
\label{sect:results}

The \doe\ code is included in a specifically designed workflow to handle all the GES single-exposure spectra for all setups.
The automated workflow includes three steps: first, the CCFs are selected using the set of criteria described in Sect.~\ref{sect:sel}; second, the \doe\ code is applied to the CCFs to identify the number of peaks and a confidence flag is assigned; 
third, the CCFs in a given setup are combined per star and a last criterion is applied: for a given star, if more than 75\% of the CCFs in at least one setup show 2 peaks (respectively 3 and 4), then the star is classified as SB2 candidate (SB3 and SB4 respectively). This rather restrictive criterion (see Sect.~\ref{sect:ccf_degeneracy}) was adopted to prevent false positive SB detections (due to spectra normalisation, cosmics, nebular lines, etc.). 

After this automatic procedure, a visual inspection is performed to ensure that  (i) no false positive detection remains (ii) the confidence flag is relevant. We investigate the CCFs and the spectra of all the SB$n$ candidates one by one. When a clear false detection is encountered, the SB candidate is removed from the list. When an SB was flagged by the automatic process as probable (A) or possible (B), but the visual inspection of the CCF series (all setups considered) casts doubts on this classification, the corresponding spectra for that object are inspected. The choice of the final flag for an object can be downgraded in case other CCFs provide discrepant results.
Such a procedure ensures that processes other than binarity moderately contaminate SB candidates flagged C, marginally contaminate SB candidates flagged B and exceptionally contaminate those flagged A. 
Despite these difficulties, adopting clear classification criteria ensures the best possible consistency throughout the survey. 

\begin{table}
\center
\caption{Number of SB2, SB3 and SB4 candidates per confidence flag.}
 \begin{tabular}{crrrr}
 \hline\hline
 & \multicolumn{3}{c}{Confidence flag}& \\
 \cline{2-4}\
 Peculiarity index &  A  &  B  &  C  & Total \\
\hline \\
  SB2 (2020) & 127  & 107 & 108 & 342  \\
  SB3 (2030) &   7  &   1 &   3 &  11  \\
  SB4 (2040) &   1  &   0 &   0 &   1  \\
  \hline
 \end{tabular}
 \tablefoot{A: probable, B: possible, C: tentative}
 \label{tab:total_res}
\end{table}

The SB$n$ candidates reported in the present paper are much fainter on average than those already collected in the {\it Ninth Catalogue of Spectroscopic Binary Orbits} \citep[SB9; ][]{pourbaix2004} (Fig.~\ref{fig:histoV}). The average visual magnitude of SB2 within the SB9 catalogue is around $V\sim 8$. For the GES SB2 candidates, the average is $V\sim 15$. The \emph{Gaia}-ESO program targets both Milky Way field stars and stars in open and globular clusters. We refer the reader to \citet{stonkute2016} for the selection function of Milky Way field stars (excluding the bulge stars), to \citet{bragaglia2012} and Bragaglia et al. (in prep.) for the selection criteria in open clusters, and to \citet{pancino2017} in globular clusters and calibration open clusters. We emphasize that the targets observed in regions like the bulge, Cha~I \citep{sacco2017} and $\gamma^2$~Vel \citep{prisinzano2016} associations, as well as $\rho$ ~Oph \citep{rigliaco2016} molecular cloud are selected on the basis of coordinates and photometry (VISTA and 2MASS), thus providing a rough membership criterion.

The list of the SB2 and SB3 candidates in the Milky Way field is given in Tables~\ref{tab:mw_sb2} and \ref{tab:mw_sb3}. The list of SB2 in the bulge, the Cha~I, $\gamma^2$~Vel and $\rho$ Oph associations and the CoRoT field is given in Table~\ref{tab:other_sb}. Finally, the list of SB$n$ in stellar clusters is given in Table~\ref{tab:SB2cluster}.
The results (classification and confidence flag) are included in the GES public releases (see footnote~\ref{note:GES}) using the nomenclature as described in the GES outlier dictionary developed by the GES Working Group 14 (WG~14)\footnote{The aim of WG~14 is to identify non-standard objects which, if not properly recognised, could lead to erroneous stellar parameters and/or abundances. A dictionary of encountered peculiarities has been created, allowing each node to flag peculiarities in a homogeneous way.}. 

\begin{figure}[t]
 \includegraphics[width=9cm]{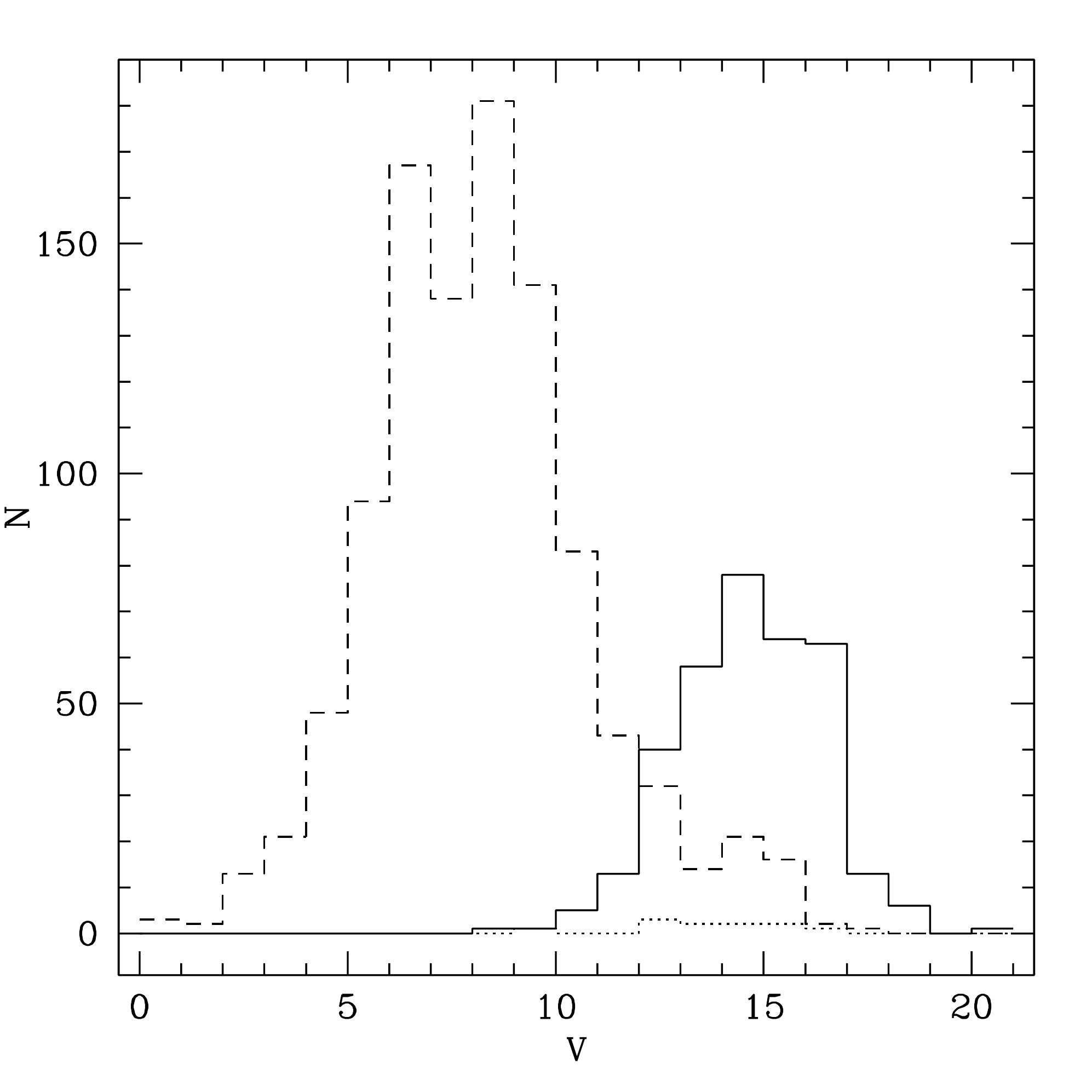}
\caption[]{\label{fig:histoV}
Magnitude distributions of SB2 systems in the \emph{Ninth Catalogue of Spectroscopic Binary Orbits} (SB9 Orbits; \protect\citealt[dashed line; ][]{pourbaix2004}, data downloaded in September 2016 from \url{http://sb9.astro.ulb.ac.be}, and in the GES (solid line). SB3 systems are shown as the dotted-line histogram.}
\end{figure}

\begin{figure*}[t]
 \includegraphics[width=0.33\linewidth]{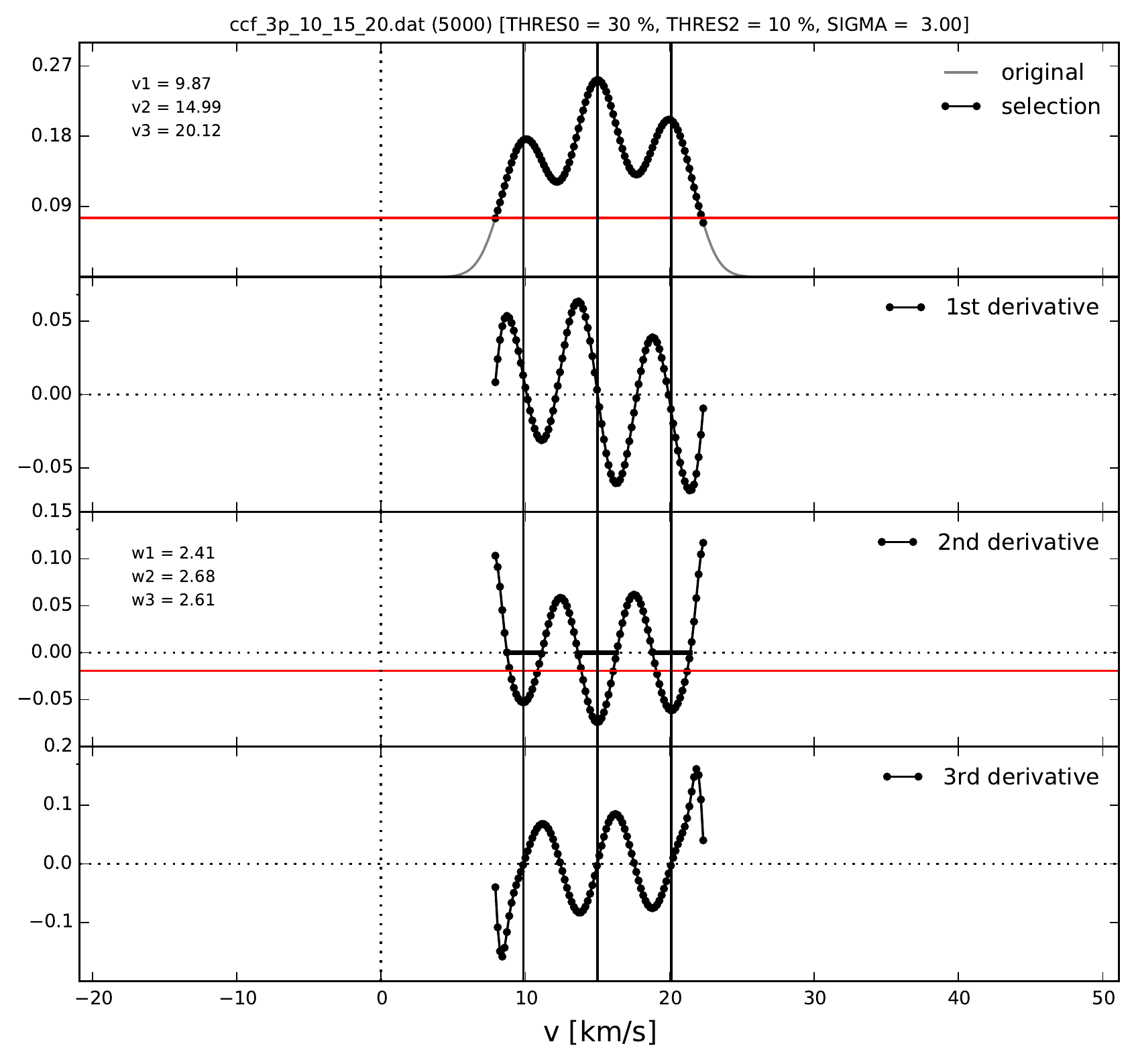}
 \includegraphics[width=0.33\linewidth]{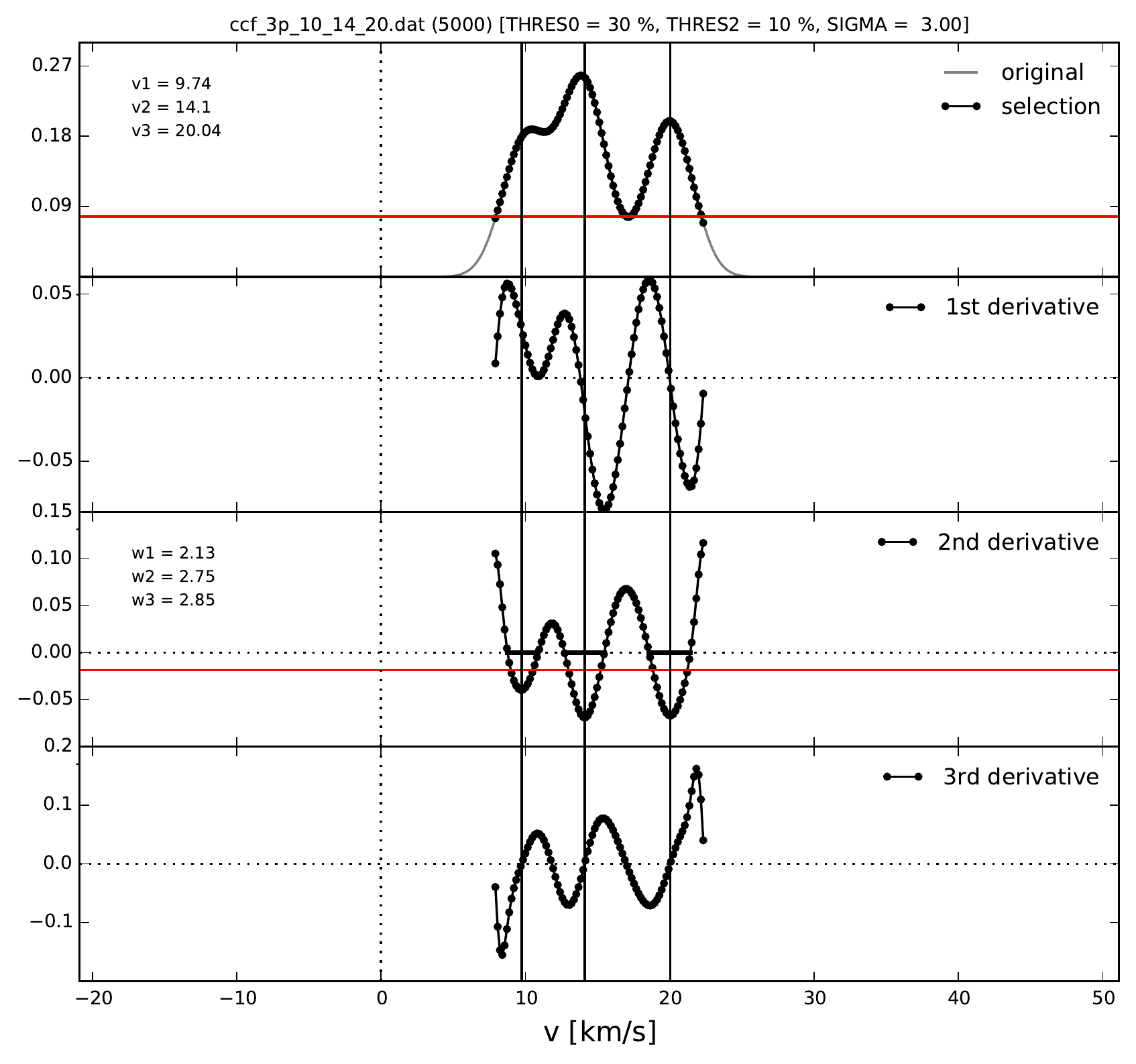}
 \includegraphics[width=0.33\linewidth]{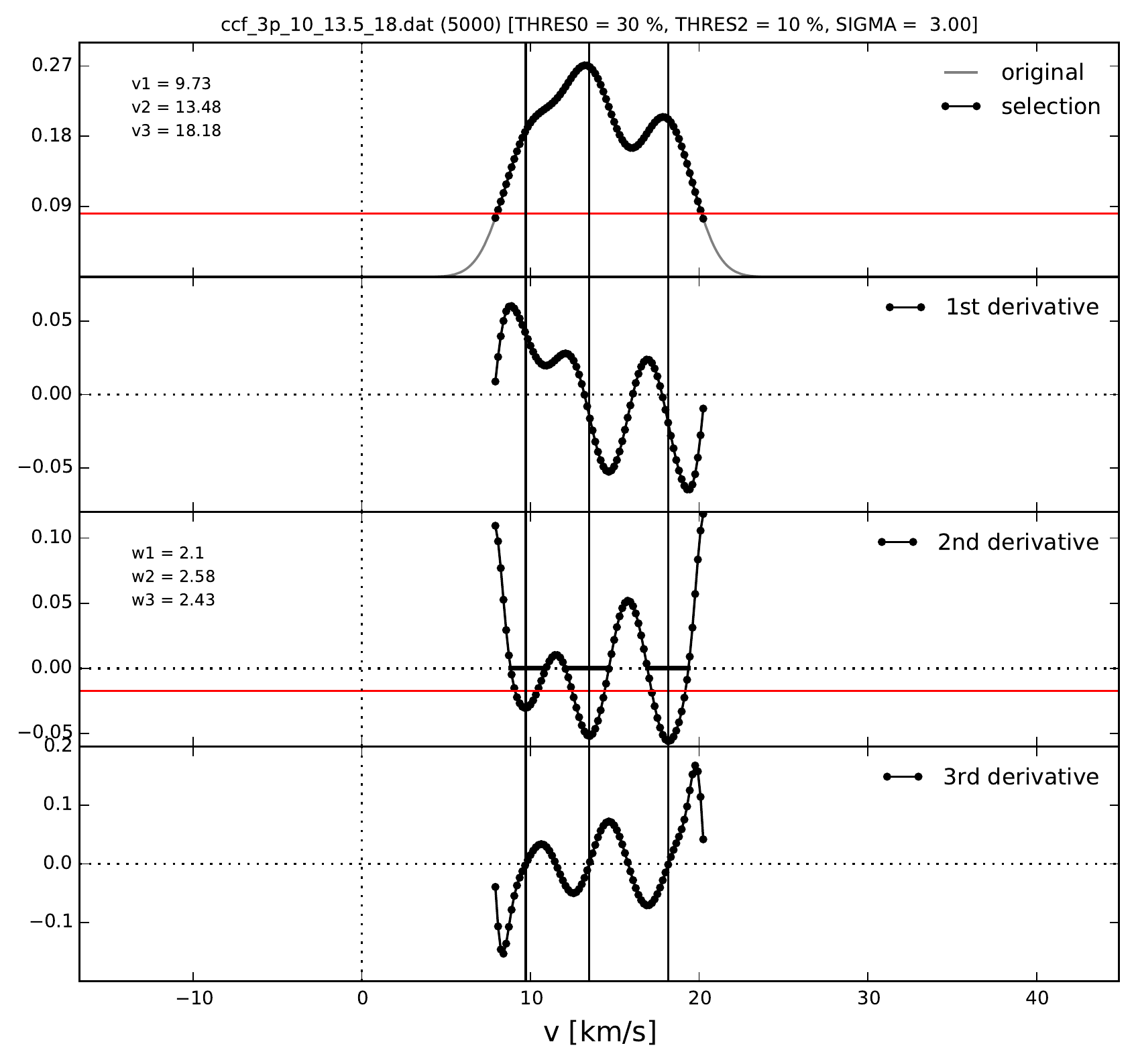}
 \caption{Triple-peak simulated CCFs with a main peak fixed at 10~\kms\ detected with confidence flags A (left; second and third peaks at 15 and 20~\kms), B (middle; second and third peaks at 14 and 20~\kms) and C (right; second and third peaks at 13.5 and 18~\kms).}
 \label{fig:flag_test3}
 \end{figure*}

\subsection{The binary classification$^{\ref{note:ib}}$}
The binary classification\footnote{\label{note:ib}See note \ref{note:GES}} has been developed for the GES within WG~14. The following scheme is adopted: the peculiarity flag is built from the juxtaposition of a peculiarity index, and a confidence flag letter. The peculiarity index is defined as $20n0$,  with $n\ge2$, where $n$ is the number of distinct velocity components in the CCF. With this peculiarity index, an SB2 is classified as 2020, an SB3 2030, etc. Of course, even though a star is flagged 2020 (\emph{i.e.} SB2), a third component may be present but not visible during the observation or undetectable with the resolution and $S/N$ of the considered exposure. 

Moreover, the WG14 dictionary recommends the use of confidence flags (A: probable, B: possible,  and C: tentative).
Clearly, the closer the CCF peaks are, the less certain the detection is. The criteria to allocate these flags were 
defined as follows:
\begin{itemize}
 \item A: the local minimum between peaks is deeper than 50\% of the full amplitude of the largest peak;
 \item B: the local minimum between peaks is higher than 50\% of the full amplitude of the largest peak;
 \item C: no local minimum is detected between peaks but the CCF slope changes.
\end{itemize}

With these definitions, the SB2 whose CCF is plotted on the left, middle and right panels of Figs.~\ref{fig:cal_hr10} and \ref{fig:cal_u580} would be flagged as A, B and C respectively.

For triple-peak CCFs, the same type of criterion is applied to the second local minimum. If this second local minimum is lower than 70\% of the full amplitude of the largest peak, then the confidence flag is set to A, else, B. Examples of these two cases are shown on simulated CCFs in Fig.~\ref{fig:flag_test3}. The CCF on the middle panel is classified as 2030B (due to the fact that the leftmost local minimum is higher than 0.5 times the largest amplitude) but also as 2020A because the middle and leftmost peaks, taken as a whole, are  well separated from the rightmost peak.

\subsection{iDR4 SB2 candidates}
\label{sec:sb2_candidates}
Table~\ref{tab:total_res} presents the breakdown of the detected SB$n$ candidates in terms of confidence flags, whereas Table~\ref{tab:sb_field} provides the detailed results of the analysis per field, in terms of automated detection (`\doe' column) and after visual checking (`confirmed' column). A total of 1092 sources were identified as SB2 candidates  by the automated procedure described in the previous section, out of which 342 are confirmed after visual inspection, giving a success rate of about 30\% similar to that of \citet{matijevic2010} for the RAVE survey. Typical rejected cases include distorted CCFs caused by negatives fluxes or pulsating stars. Some confidence flags were also changed during the visual inspection phase (see Sect. \ref{sect:ccf_degeneracy}).
The largest number of stars has been observed with the GIRAFFE setup HR21 because it corresponds to the \emph{Gaia} wavelength range of the radial velocity spectrometer. However, the rate of SB$n$ detection in this setup is very low because it is dominated by the presence of the \ion{Ca}{ii} triplet, which is a very strong feature in late-type stars, thus resulting in a broad CCF that can mask possible multiple peaks (Fig.~\ref{fig:sun_spec_ccf}, bottom panel). Moreover, emission in the line cores of this triplet induces fake double-peak CCFs because in the templates the lines are always in absorption. Consequently, it is very difficult to identify double peaks due to binarity based on HR21 CCFs (see Sect.~\ref{sect:ccf_degeneracy} for more details). This explains why we have only two firm detections among the 31\,970 stars observed with this setup only. Hence, this setup is not well-suited to detect stellar multiplicity at least in our situation 
(see \citealt{matijevic2010}: though they could discover 123 SB2 out of 26\,000 RAVE targets, they also had to deal with very broadened CCFs and could not detect binaries with $\Delta v_\textrm{rad} \le 50$~\kms).

The setup with the second largest number of observed objects is HR10. This setup covers the range [535-560]~nm with lots of small absorption lines that result in a narrow CCF, suitable for the detection of stellar multiplicity (see Fig.~\ref{fig:cal_hr10}). The largest number of probable SB2 candidates is indeed  detected with this setup.

\begin{table*}
\center
\caption{Distribution of SB2 and SB3 candidates among the different observed fields.}
\setlength{\tabcolsep}{3pt}
{\footnotesize
\begin{tabular}{lrcrrrcccccrcccrr}
\hline\hline
Field/cluster  & log age & $v_r$  &  \# stars &\multicolumn{5}{c}{\# SB2} && \multicolumn{5}{c}{\# SB3}  & SB2/total &  SB3/SB2\\
\cline{5-9}\cline{11-15}
               & [yr] &  [\kms]   &       &\doe & confirmed& A & B & C  &&\doe&confirmed& A & B & C&   [\%]  & [\%]\\
\hline
\\
Field          &      &                     & 27786 & 263 & 185 & 82 & 48 & 55 && 24 & 5 & 5 &   &   &  0.67 & 3 \\
Bulge          &      &                     &  2633 &   6 &   6 &  1 &  3 &  2 &&  0 & 0 &   &   &   &  0.23 &   \\
Cha I          &      &                     &   616 &   5 &   2 &    &  2 &    &&  1 & 0 &   &   &   &  0.49 &   \\
Corot          &      &                     &  1966 &  13 &   7 &  5 &  2 &    &&  0 & 0 &   &   &   &  0.36 &   \\
$\gamma^2$ Vel &      &                     &  1116 &  28 &  16 &  2 &  7 &  7 &&  2 & 0 &   &   &   &  1.43 &   \\
$\rho$ Oph     &      &                     &   278 &   2 &   1 &    &    &  1 &&  1 & 0 &   &   &   &  0.72 &   \\
\\
\hline
\\
 IC 2391         & 7.74 & $14.49\pm0.14$      &   398 &   4 &   3 &  2 &  1 &    &&  4 & 0 &   &   &   &  0.75 &  \\
 IC 2602         & 7.48 & $18.12\pm0.30$      &  1784 &   6 &   3 &  1 &  1 &  1 &&  3 & 0 &   &   &   &  0.17 &  \\
 IC 4665         & 7.60 & $-15.95\pm1.13$     &   559 &   6 &   5 &  2 &  2 &  1 &&  1 & 0 &   &   &   &  0.89 &  \\
 M 67            & 9.60 & $33.8\pm0.5$        &    25 &   4 &   4 &  4 &    &    &&  0 & 0 &   &   &   & 16.00 &  \\
 NGC 2243        & 9.60 & $59.5\pm0.8$        &   715 &  38 &   1 &    &  1 &    && 14 & 0 &   &   &   &  0.14 &  \\
 NGC 2264        & 6.48 & $24.69\pm0.98$      &  1565 &  78 &   4 &  2 &  2 &    && 18 & 0 &   &   &   &  0.26 &  \\
 NGC 2451        & 7.8 (A) & 22.70 (A) &  1599 &  18 &  11 &  3 &  5 &  3 &&  7 & 1 & 1 &   &   &  0.69 & 9 \\
                 & 8.9 (B) & 14.00 (B) \\
 NGC 2516        & 8.20 & $23.6\pm1.0$        &   726 &  19 &   8 &  1 &  4 &  3 && 10 & 1 &   &   & 1 &  1.10 & 13 \\
 NGC 2547        & 7.54 & $15.65\pm1.26$      &   367 &   7 &   1 &  1 &    &    &&  3 & 0 &   &   &   &  0.27 &  \\
 NGC 3293        & 7.00 & $-12.00\pm4.00$     &   517 & 158 &   9 &  1 &  5 &  3 && 55 & 0 &   &   &   &  1.74 &  \\
 NGC 3532        & 8.48 & $4.8\pm1.4$         &    94 &   1 &   1 &    &  1 &    &&  0 & 0 &   &   &   &  1.06 &  \\
 NGC 4815        & 8.75 & $-29.4\pm4$         &   174 &  11 &   2 &    &  1 &  1 &&  0 & 0 &   &   &   &  1.15 &  \\
 NGC 6005        & 9.08 & $-24.1\pm1.3$       &   531 &  12 &   4 &  2 &  1 &  1 &&  8 & 1 &   & 1 &   &  0.75 & 25 \\
 NGC 6530        & 6.30 & $-4.21\pm6.35$      &  1252 &  95 &   5 &  2 &    &  3 &&  1 & 0 &   &   &   &  0.40 &  \\
 NGC 6633        & 8.78 & $-28.8\pm1.5$       &  1643 &  17 &  15 &  3 &  7 &  5 &&  0 & 0 &   &   &   &  0.91 &  \\
 NGC 6705        & 8.47 & $34.9\pm1.6$        &   994 & 108 &  19 &  5 &  3 & 11 && 52 & 1 & 1 &   &   &  1.91 &  5 \\
 NGC 6752        & 10.13& $-24.5\pm1.9$       &   728 &   8 &   1 &    &  1 &    &&  0 & 0 &   &   &   &  0.14 &  \\
 NGC 6802        & 8.95 & $11.9\pm0.9$        &   156 &   7 &   2 &  2 &    &    &&  7 & 1 &   &   & 1 &  1.28 &  50\\
Tr 14            & 6.67 & $-15.0$             &   858 &  82 &   3 &  2 &  1 &    && 19 & 0 &   &   &   &  0.35 &  \\
 Tr       20     & 9.20 & $-40.2\pm1.3$       &  1316 &  84 &  19 &  3 &  7 &  9 && 24 & 1 &   &   & 1 &  1.44 &  5\\
 Tr       23     & 8.90 &  $-61.3\pm0.9$      &   164 &   5 &   1 &  1 &    &    &&  5 & 0 &   &   &   &  0.61 &  \\
Be        25     & 9.70 & $+134.3\pm0.2$      &    38 &   2 &   2 &    &  2 &    &&  1 & 0 &   &   &   &  5.26 &  \\
 Be       81     & 8.93 & $ 48.3\pm0.6  $     &   265 &   5 &   2 &  1 &    &  1 &&  6 & 0 &   &   &   &  0.75 &  \\
\\
\hline
\\
Total          &      &                     & 50863 &1092 & 342 &128 &107 &107 &&266 &11 & 7 & 1 & 3 & 0.68 &  3 \\
\\
\hline\\
 \end{tabular}
 }
\tablefoot{The column `log age' lists the logarithm of the cluster age (in years) from \citet{2014A&A...569A..17C} (NGC6705), Spina et al. (in prep.) (IC 2391, IC~2602, IC4665, NGC~2243, NGC~2264, NGC~2541, NGC~2547, NGC~3293, NGC~3532, NGC~6530), \citet{bellini2010} (M~67), \citet{bragaglia2006} (NGC~2243), \citet{sung2002} (NGC~2516), \citet{2014A&A...563A.117F} (NGC~4815), \citet{jacobson2016} (NGC~6005, NGC~6633), \citet{vendenberg2013} (NGC~6705), Tang et al. (submitted) (NGC~6802), \citet{2014A&A...561A..94D} (Tr~20 and Berkeley~81, written Be~81), \citet{overbeek2016} (Tr~23), and  \citet{2007A&A...476..217C} (Be~25). 
The column $v_r$ lists the radial velocity; for the clusters with ages larger than 100~Myr see \citet[][only UVES targets]{jacobson2016} excepted for M~67 \citep{casamiquela2016}, NGC~2243 \citep{smiljanic2016}; \citet{2014A&A...563A.117F} (NGC~4815),  \citet{1996AJ....112.1487H} (NGC 6752). For the young clusters, see \citet{dias2002} (IC~2391, IC~2602, IC~4665, NGC~2264, NGC2451, NGC~2547, NGC~3293, NGC~6530), and  \citet{2007A&A...476..217C} (Be~25).
The column `\# stars' lists the number of stars in that particular field/cluster observed by the GES. The columns `\doe'\, give the number of SB detected automatically, whereas the column `confirmed' represents the number of SB kept after eye inspection of CCFs and associated spectra. The columns labeled `A', `B' and `C' list the number of confirmed systems by confidence flag (probable, possible and tentative respectively).
 No SB2 or SB3 candidates have been found yet with the \doe\ code for the following clusters within the GES: Be~44 (93), M~15 (109), M~2 (110), NGC~104 (1138), NGC~1851 (127), NGC~1904 (113), NGC~2808 (112), NGC~362 (304), NGC~4372 (120), NGC~4833 (102) and NGC~5927 (124), where the numbers in parenthesis give the number of stars observed in each cluster.}
 \label{tab:sb_field}
\end{table*}

To illustrate the fact that some setups are more adapted than others to detect SB$n$, we show spectra and CCFs in these setups for single stars (the Sun and Arcturus in Figs.~\ref{fig:sun_spec_ccf} and \ref{fig:ald_spec_ccf}) and for an SB2 candidate 
(NGC~6705 1936 observed in most of the GIRAFFE setups where the composite nature of the spectrum is clearly visible in Fig.~\ref{fig:sb2_setups}).

Contrary to field stars which are observed in HR10 and HR21 only, cluster stars were observed with many different setups.  The number of SB2 candidates in the field is 185 out of 27786 stars (0.67\%) whereas in the clusters,  it amounts to  127 out of 16468 (0.77\%, see Table~\ref{tab:sb_field}). 

There are about 30 SB2 candidates detected with a double-peaked CCF in both GIRAFFE HR10 and HR21. For instance, the field star  02394731-0057248 (magnitude $V=13.8$) is identified as an SB2 candidate with HR10 and HR21 (see Fig.~\ref{fig:sb2_hr10_21}). This new candidate has no entry in the Simbad database.

\begin{figure*}
 \centering
 \includegraphics[width=0.73\linewidth]{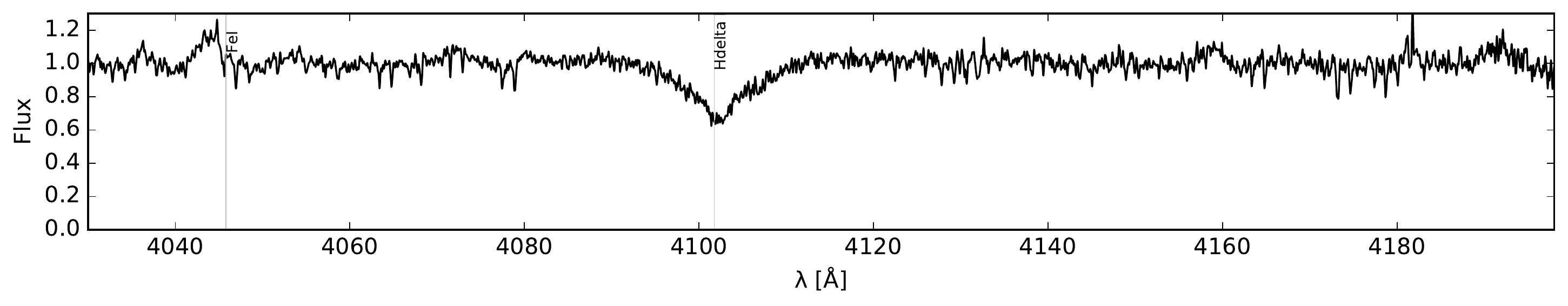}
 \includegraphics[width=0.24\linewidth]{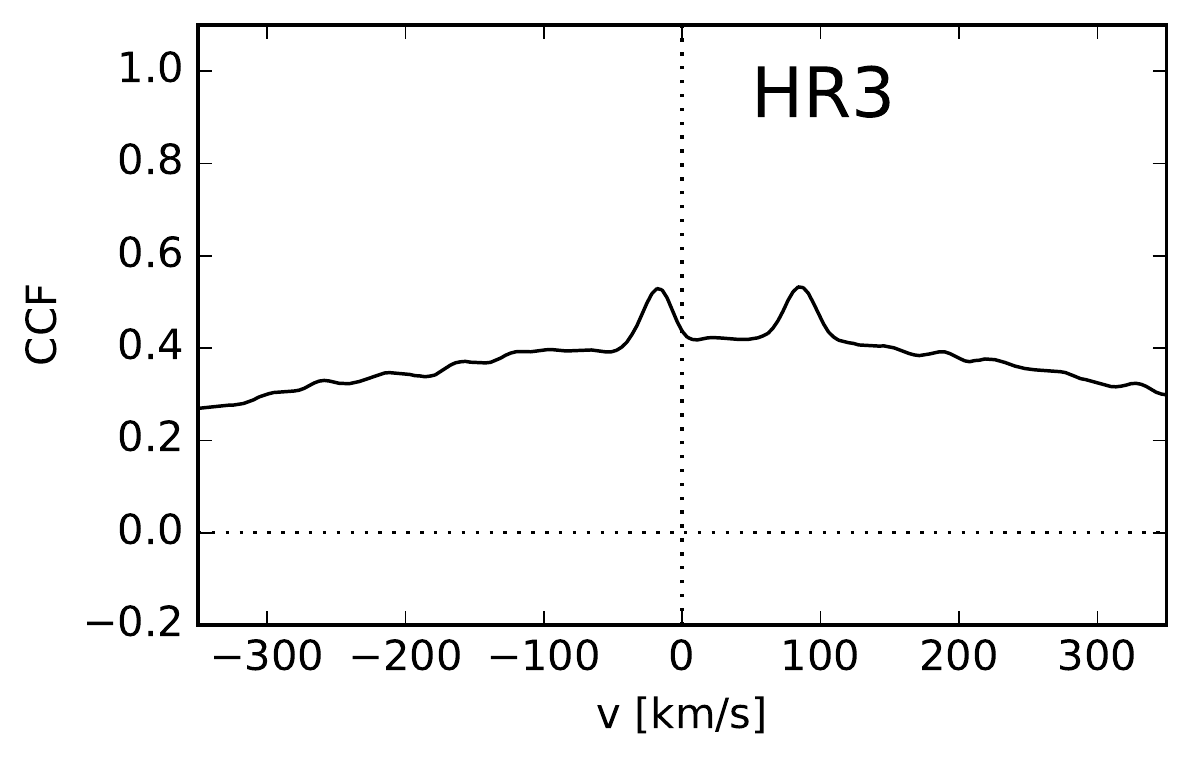}

 \includegraphics[width=0.73\linewidth]{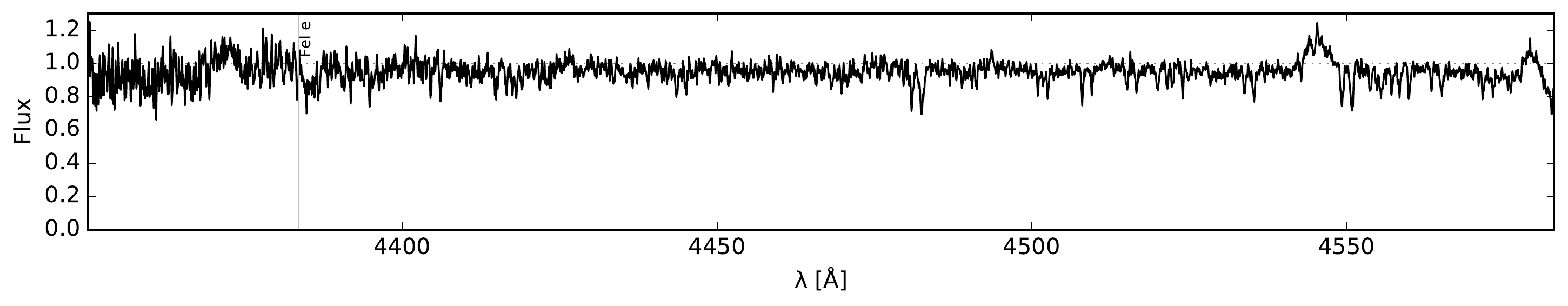}
 \includegraphics[width=0.24\linewidth]{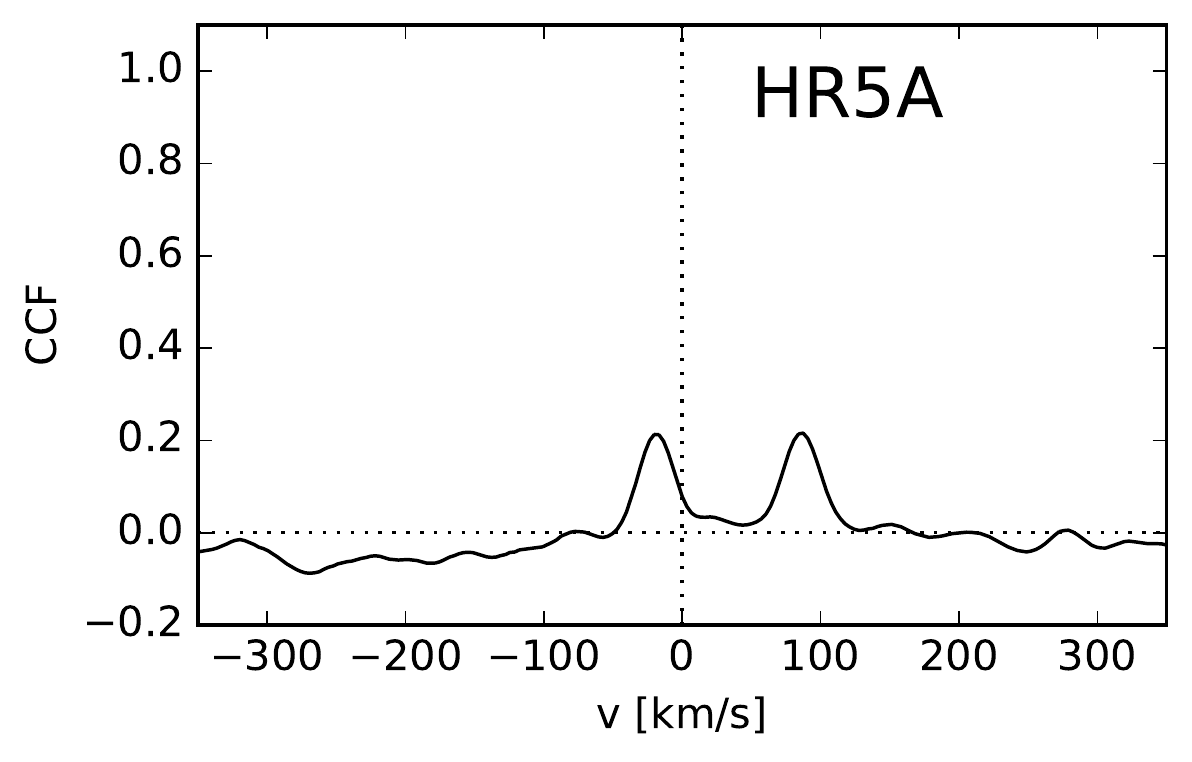}

 \includegraphics[width=0.73\linewidth]{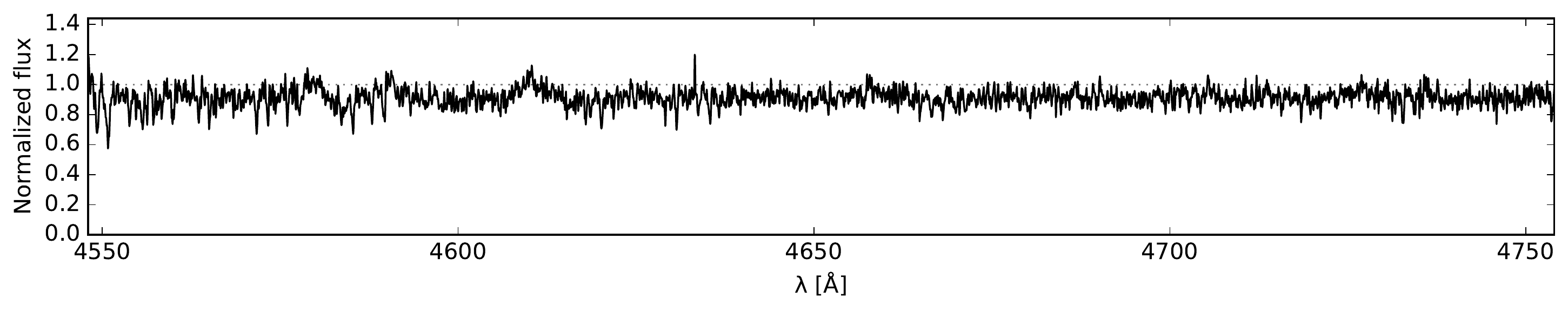}
 \includegraphics[width=0.24\linewidth]{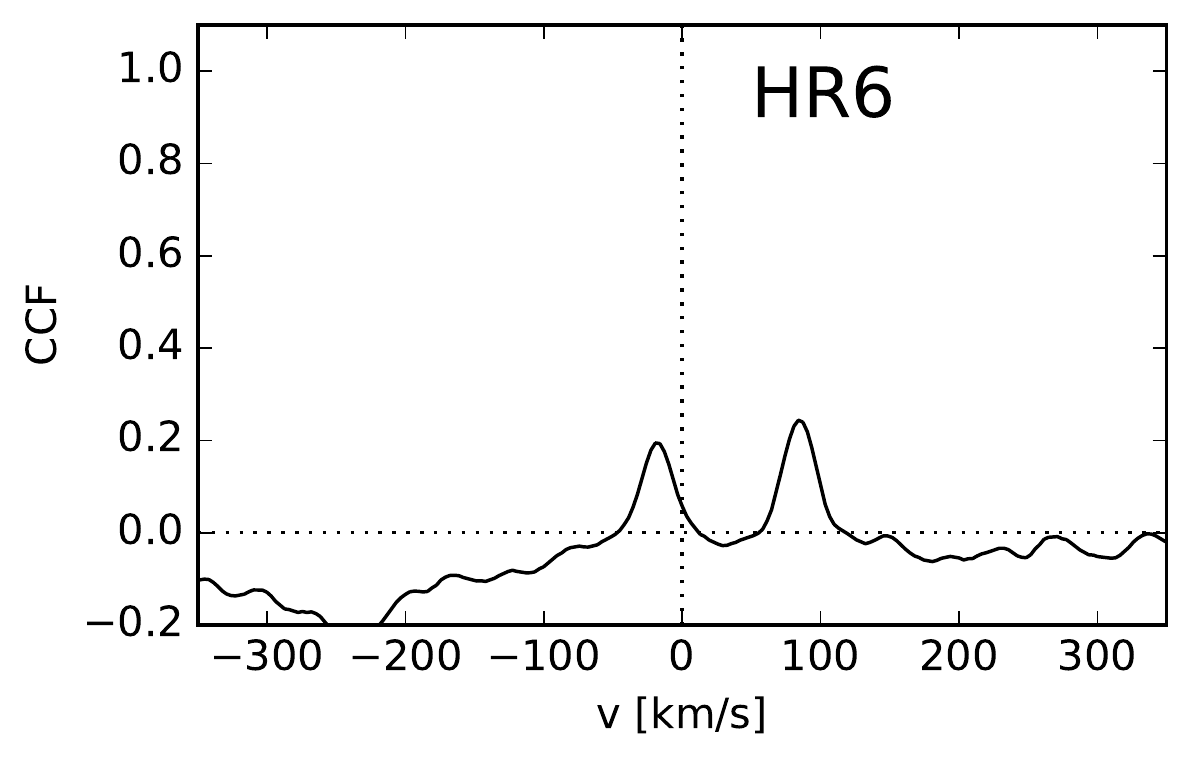}

 \includegraphics[width=0.73\linewidth]{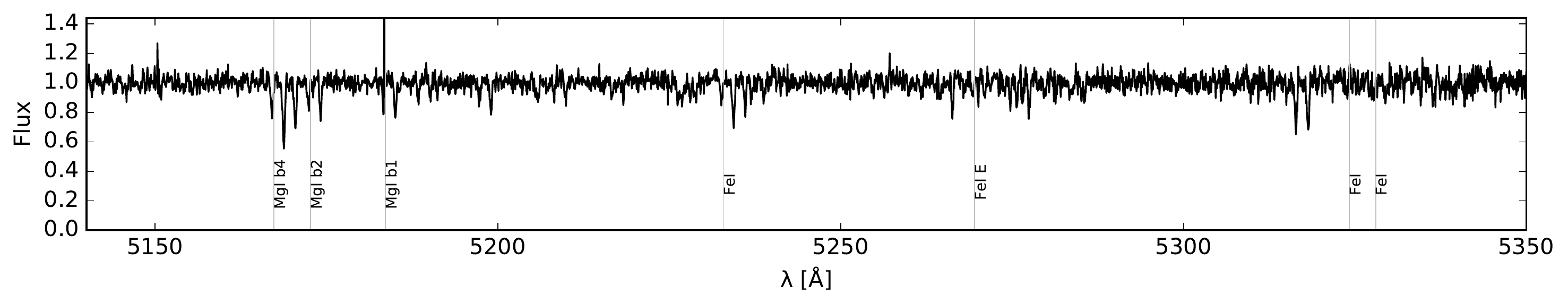}
 \includegraphics[width=0.24\linewidth]{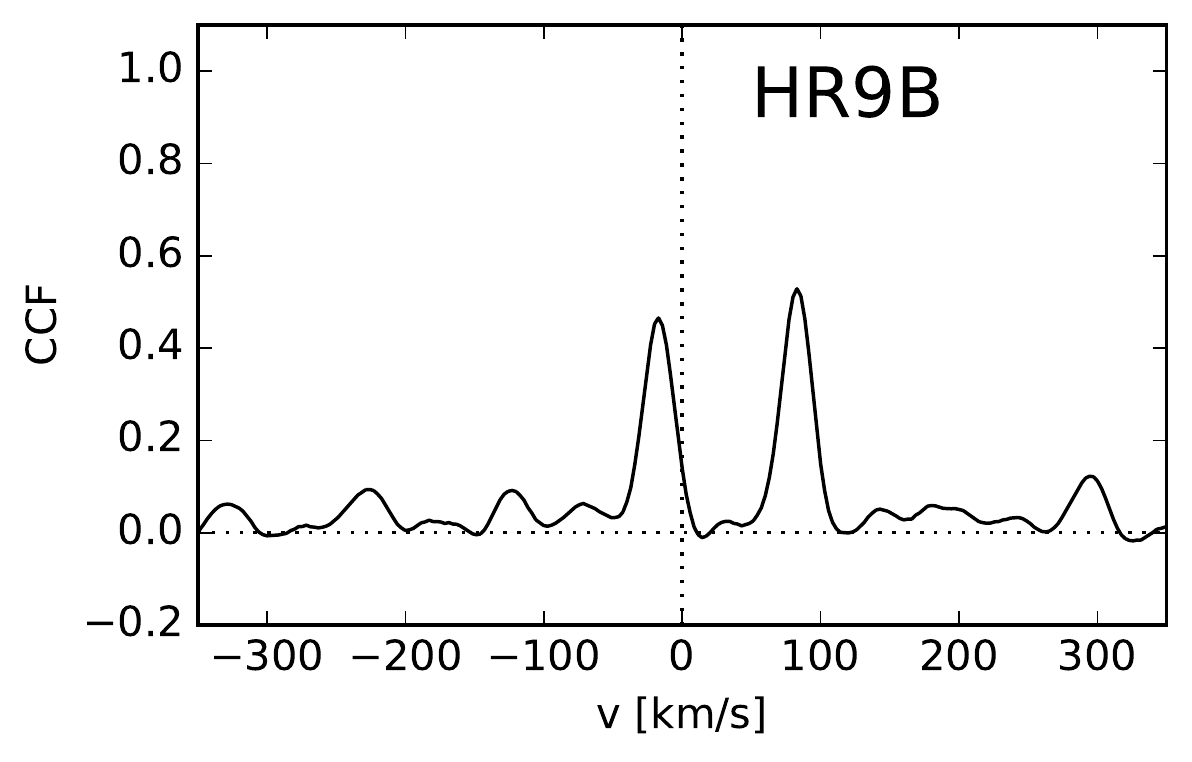}

 \includegraphics[width=0.73\linewidth]{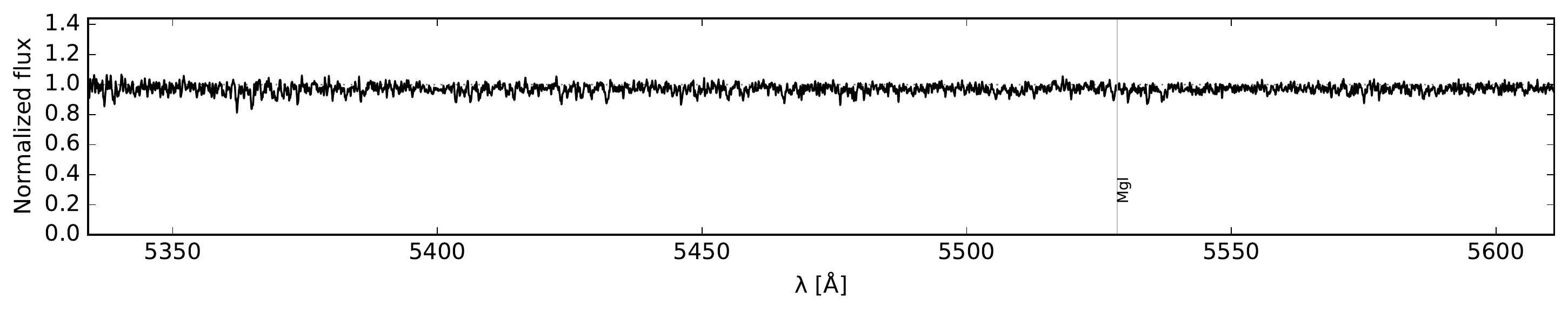}
 \includegraphics[width=0.24\linewidth]{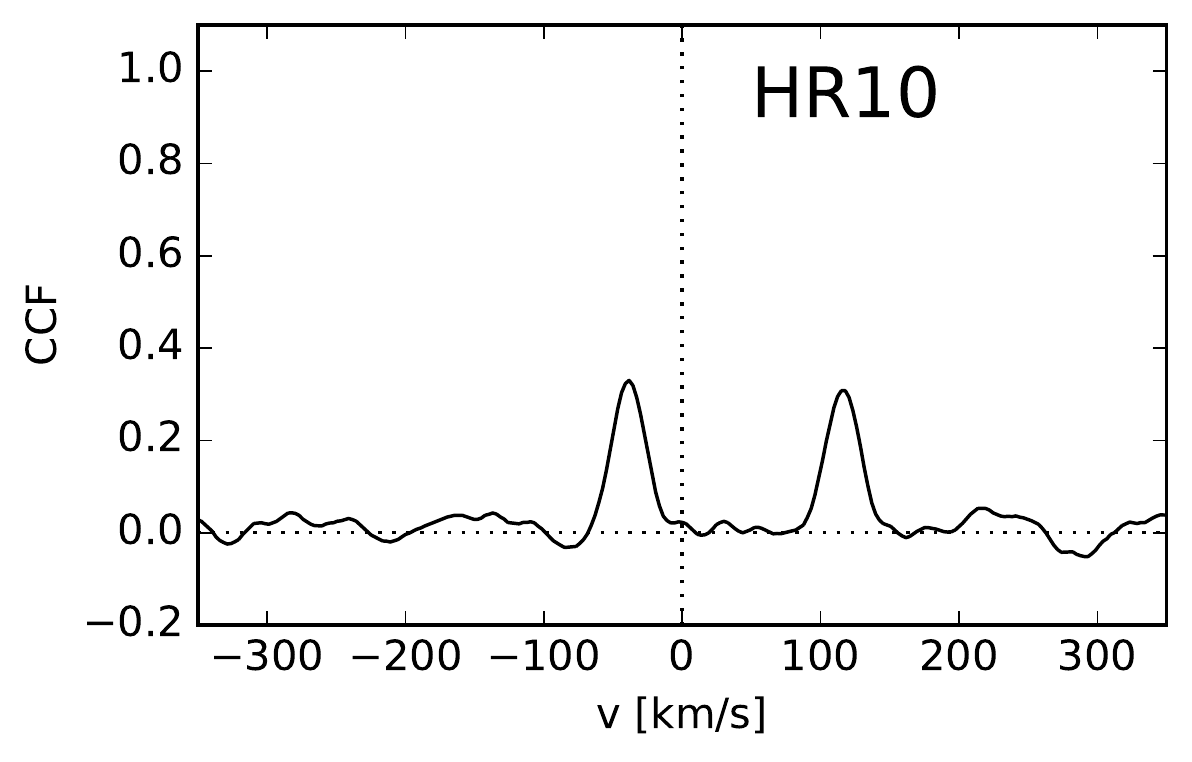}

 \includegraphics[width=0.73\linewidth]{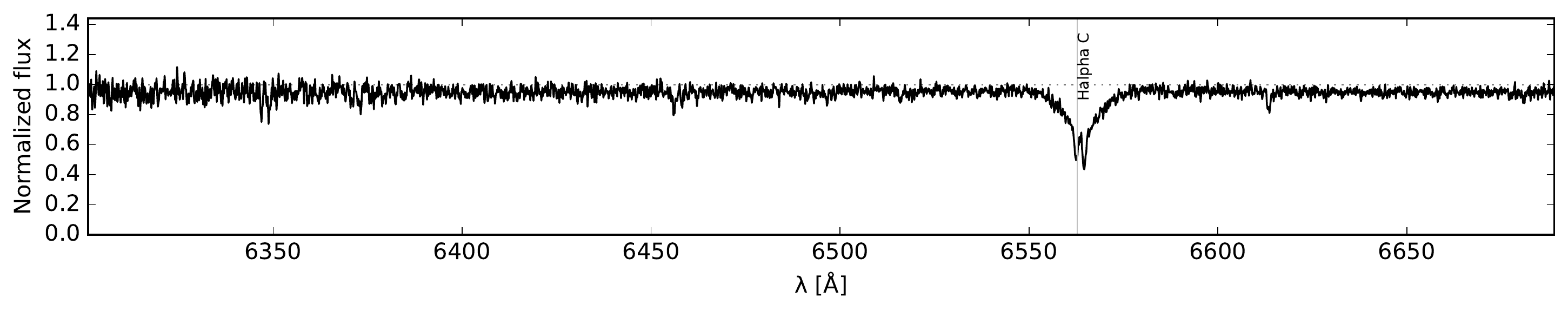}
 \includegraphics[width=0.24\linewidth]{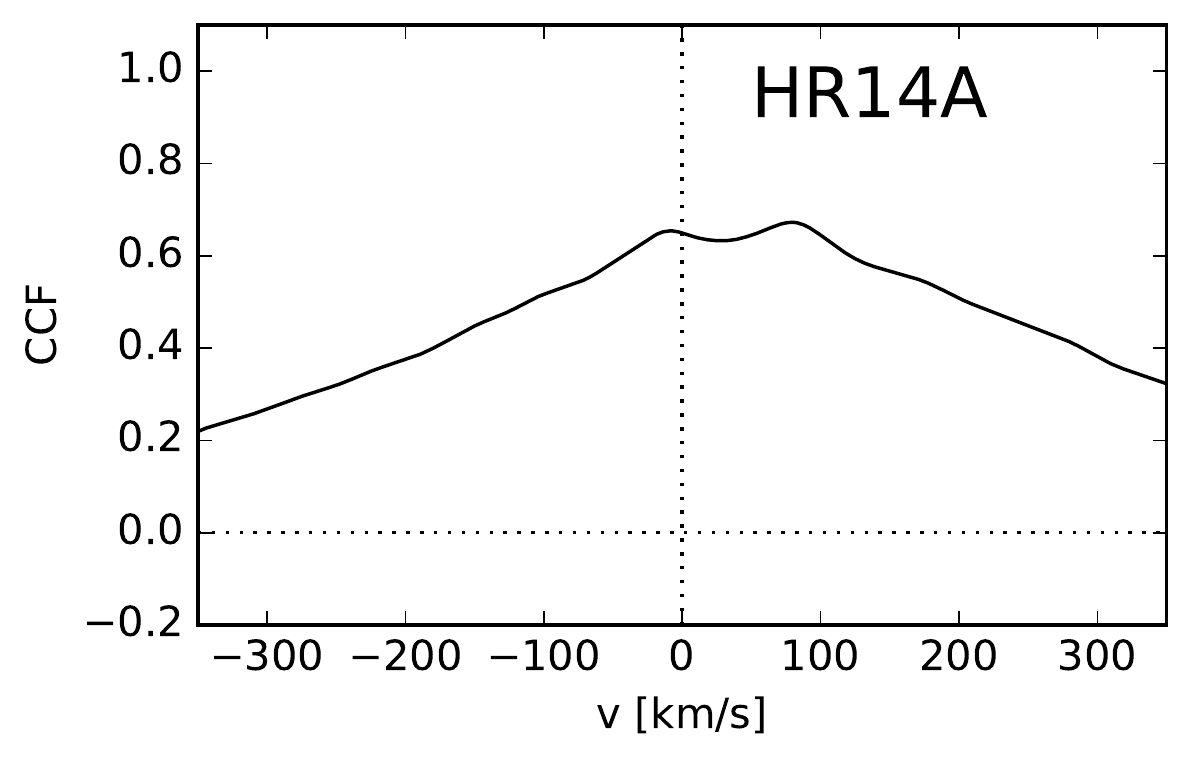}

 \includegraphics[width=0.73\linewidth]{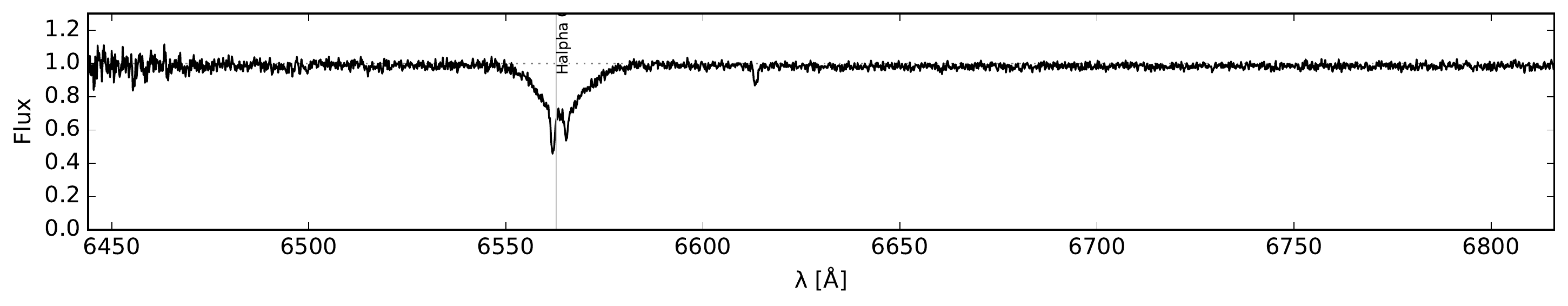}
 \includegraphics[width=0.24\linewidth]{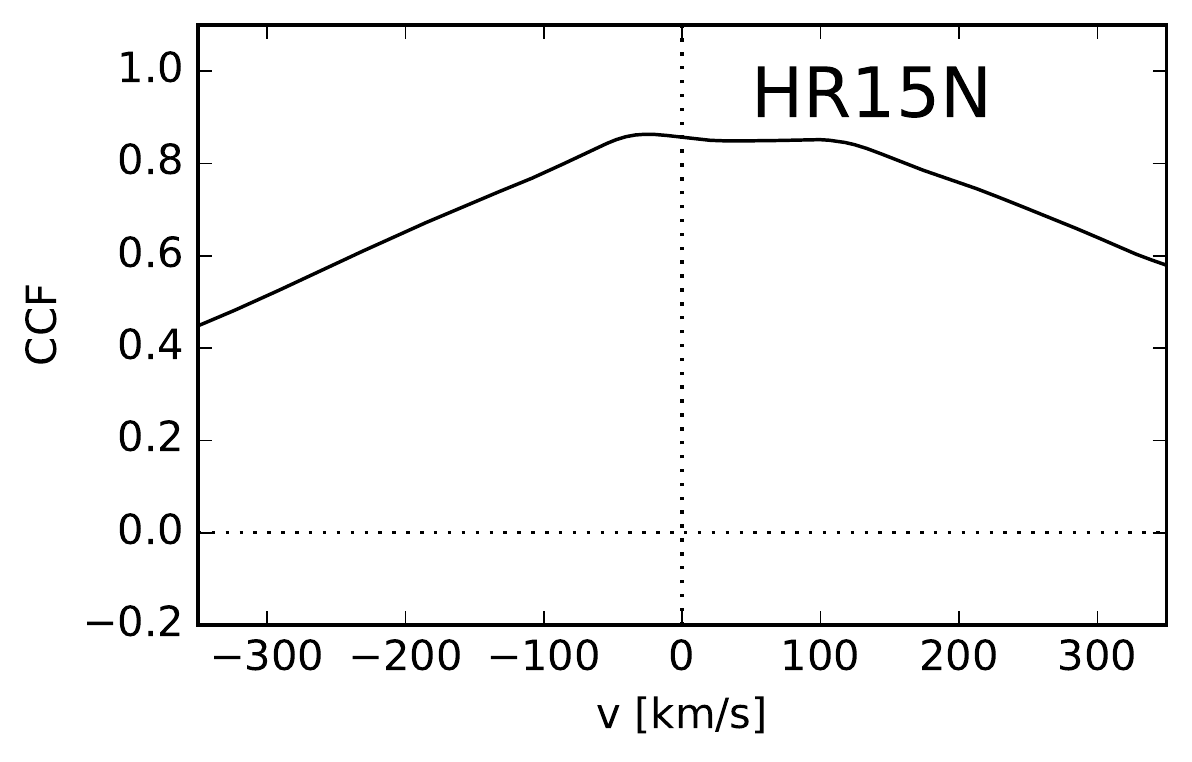}

 \includegraphics[width=0.73\linewidth]{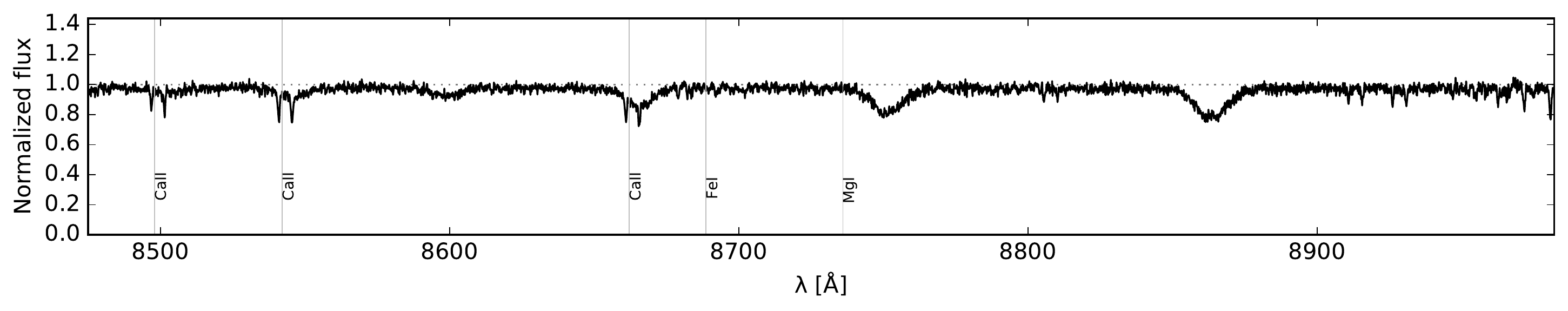}
 \includegraphics[width=0.24\linewidth]{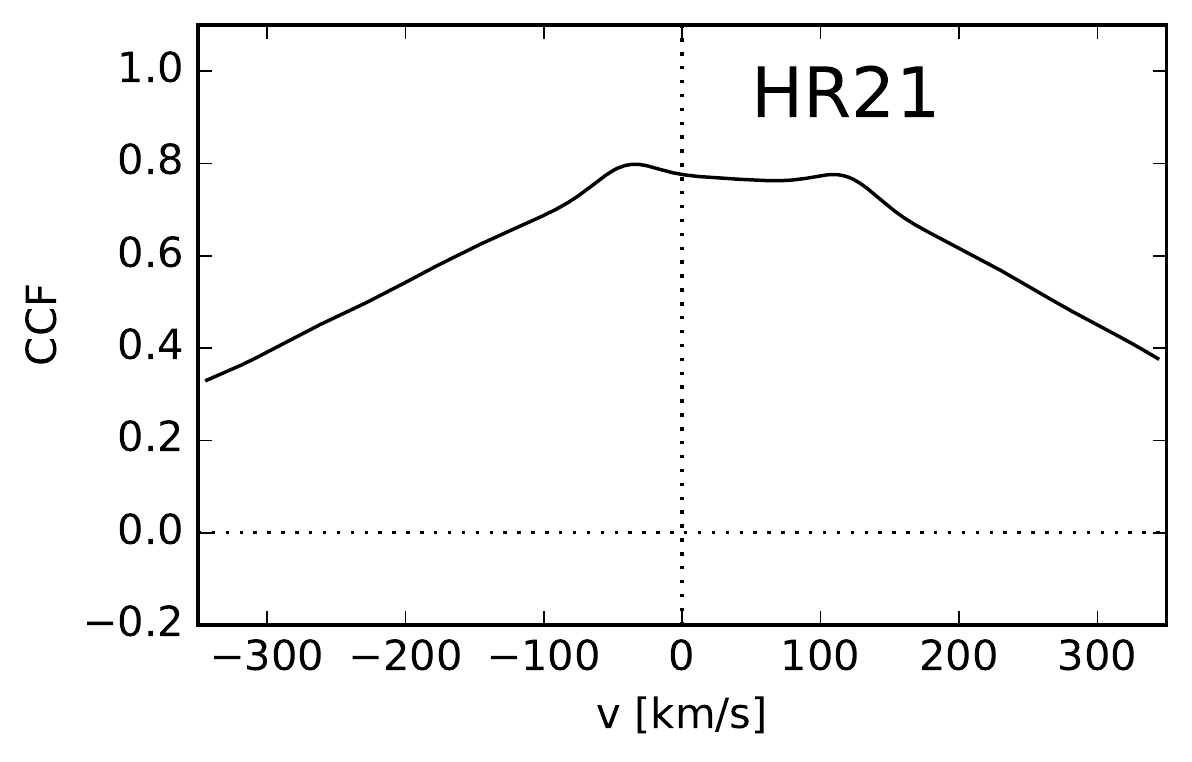}
\caption{Examples of composite spectra and CCFs associated with the new SB2 candidate 18503230-0617112 classified 2020A (NGC~6705 1936) with a visual magnitude of $V = 13.4$ ($B-V\sim0$). Broad emission lines in HR3, HR5A and HR6 are spill-over from strong Ar lines from a Th-Ar calibration lamp observed along with the target.}
\label{fig:sb2_setups}
\end{figure*}

\begin{figure*}
\includegraphics[width=0.49\linewidth]{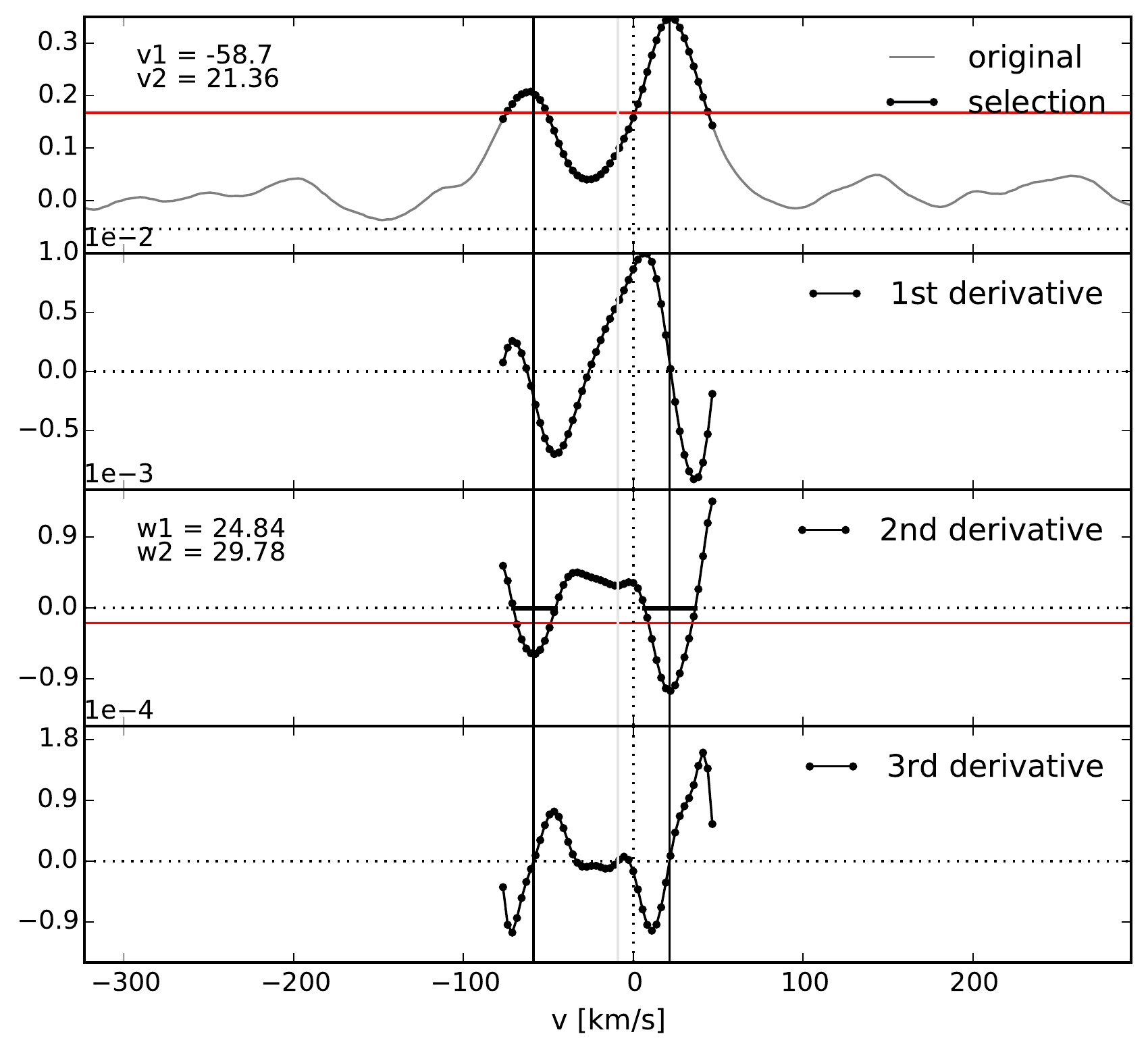}
\includegraphics[width=0.49\linewidth]{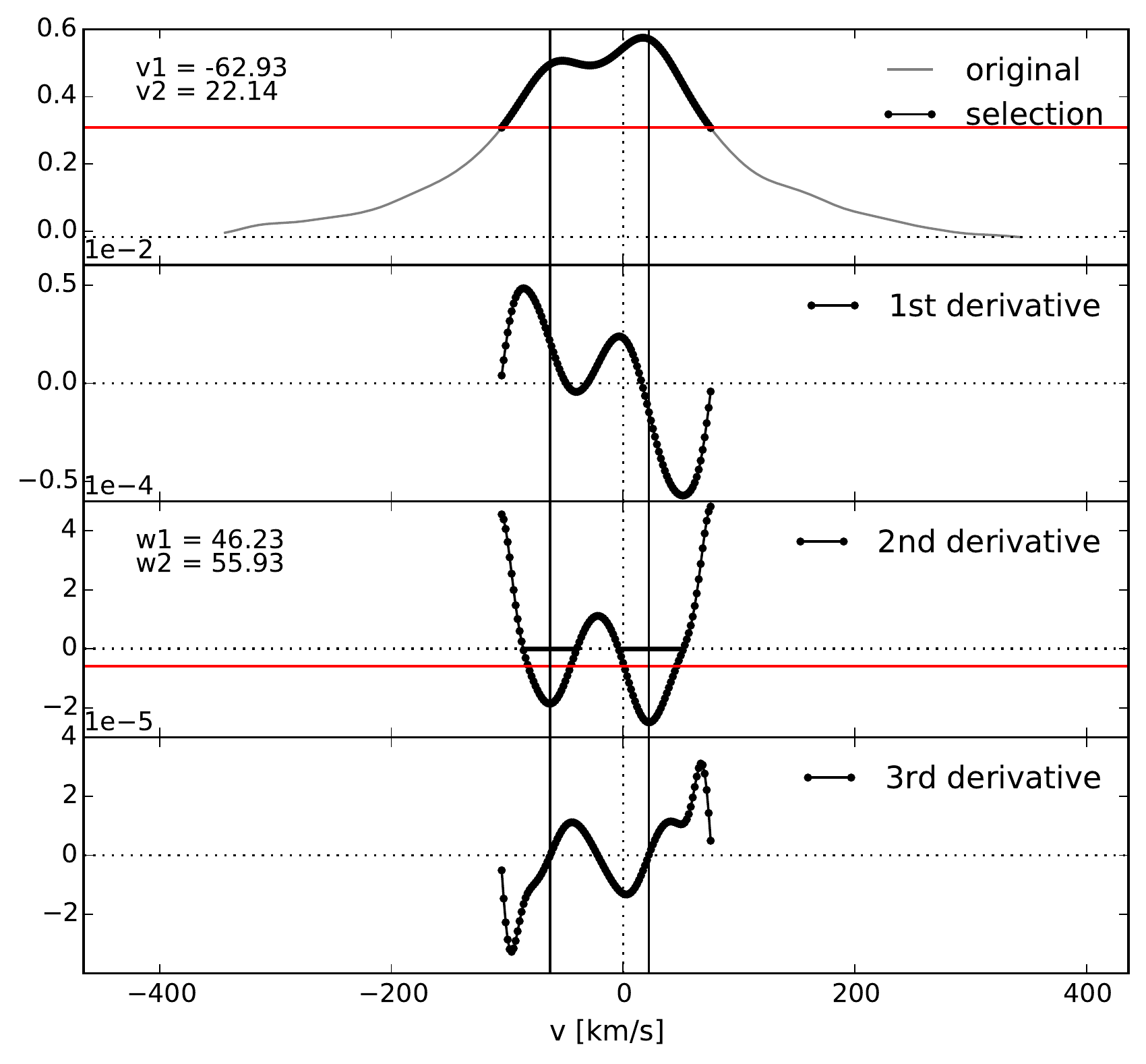}
\caption{Example of identification of a new SB2 candidate 02394731-0057248 not reported in Simbad. Left panel: GIRAFFE HR10 setup ($S/N\sim 10$). Right panel: GIRAFFE HR21 setup ($S/N\sim 140$).}
\label{fig:sb2_hr10_21}
\end{figure*}

The histograms of the radial velocity separation of SB2 candidates for GIRAFFE HR10 and HR21 as well as for UVES U580  are shown on Fig.~\ref{fig:hist_dvr} (U520 is not represented due to the small statistics). The smallest measured radial velocity separations are 23.3, 60.9 and 15.2~\kms\ for HR10, HR21 and U580, respectively. This is well in line with the detection capabilities of the \doe\ code as mentionned in Sect.~\ref{Sect:parameters} ($\sim30$~\kms\ for GIRAFFE and $\sim15$~\kms\ for UVES setups). In U580, the high bin value around 72~\kms\ is mainly due to the repeated observations of a specific object, the SB4 candidate 08414659-5303449 in IC~2391 (see Sect.~\ref{Sect:SB4}).

\begin{figure}
 \includegraphics[width=\linewidth]{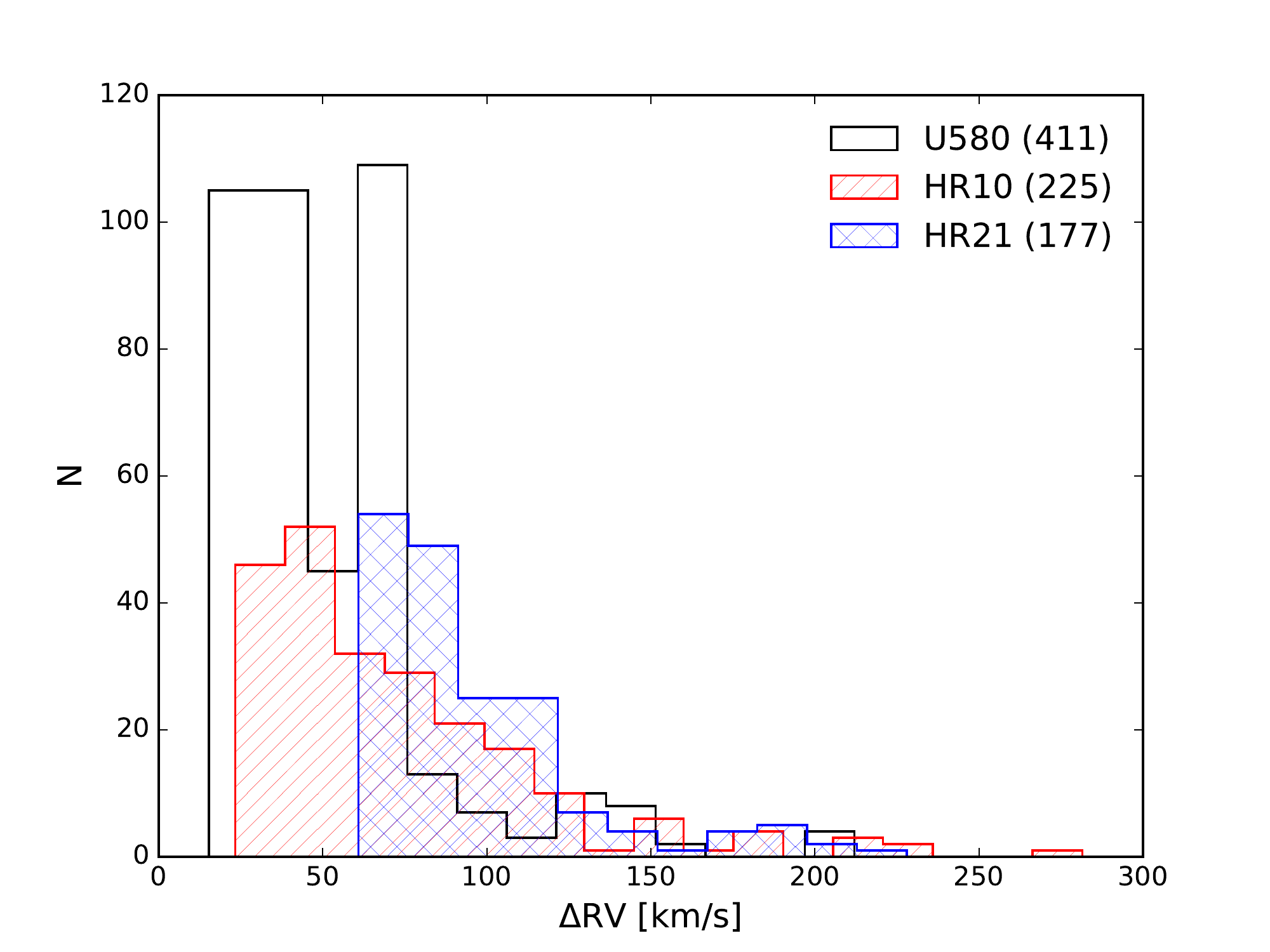}
 \caption{Histograms of the radial velocity separation of SB2 candidates for GIRAFFE HR10, HR21 and for UVES U580 single exposures. The numbers in parenthesis are the numbers of single exposures where two peaks were identified.}
 \label{fig:hist_dvr}
\end{figure}

Concerning the SB2 candidates in open clusters, not only did we check the cleanliness of the SB2 CCF profile, but we also compared the velocities of the two peaks with the cluster velocity. Assuming that most of the SB2 systems discovered by GES generally have components of about equal masses, then, an SB2 being member of the cluster should have the cluster velocity about midway between the two component velocities. This simple test allows us  to assess the likelihood that the SB2 system is a cluster member. This method is applied in full details for the SB2, SB3 and SB4 candidates analyzed in the present section and Sects.~\ref{Sect:SB3} and \ref{Sect:SB4}. The results are shown in Table~\ref{tab:SB2cluster}. The numbers of \emph{bona fide} SB2 candidates retained per cluster after this check are listed in the corresponding column of Table~\ref{tab:sb_field}.  The column labeled `Member' in Table~\ref{tab:SB2cluster} evaluates the likelihood of cluster membership based on the component velocities: if the cluster velocity falls in the range encompassed by the component  velocities, we assume that the centre of mass of the system moves at the cluster velocity, so that membership is likely. In that case, we put `y' in the column `Member'. On the contrary, if the CCF exhibits two well-defined peaks not encompassing the cluster velocity, the star is labeled as SB2 not member of the cluster (`n' in column `Member'). Another possibility is that one component has a velocity close to that of the cluster, and the second velocity is offset. In that  case, the SB2 nature is questionable and the star  is more probably a pulsating star (responsible for the secondary peak or bump) belonging to the cluster ('y' in column `Member'). The list of individual radial velocities will be given in  a forthcoming paper, based on iDR5 data. More extended remarks for each cluster are provided in Appendix~\ref{ap:sb_clusters}.

\subsection{Orbital elements of two confirmed SB2 in clusters}
With the data collected so far, we were able to confirm the binary nature of two SB2 candidates in clusters by deriving reliable orbital solutions for the systems 06404608+0949173 (NGC~2264 92) and 19013257-0027338 (Berkeley~81, written Be~81). 

The first system 06404608+0949173 (magnitude $V\sim~12$) is a \emph{bona fide} SB2 for which 24 spectra are available (20 GIRAFFE HR15N and 4 UVES U580), and an orbit can be computed, as shown on Fig.~\ref{fig:NGC2264_orbit}. 
Observations where only one velocity component is detected are not used to calculate the orbital solution because these velocities are not accurate (Fig.~\ref{fig:NGC2264_orbit}), since the two velocity components are blended.
The orbital elements are listed in Table~\ref{tab:orbit}. The short period of $2.9637\pm0.0002$~d implies that neither of the components can be a giant which is consistent with the classification of the system as K0~IV \citep{walker1956}. The centre-of-mass velocity of the system (14.6~\kms) is close to  the cluster velocity (17.7~\kms), as it should. The mass ratio is $M_B/M_A = 1.10$.
Classified as FK Com in the GCVS (=V642 Mon), this source is chromospherically active with X-ray emission (ROSAT and XMM). This system thus adds to the two SB2 systems with available orbits (VSB 111 and VSB 126) already known in NGC~2264 \citep{2013AJ....146..149K}.

The second system 19013257-0027338 (magnitude $V\sim~17$) is a confirmed SB2 (2020~A) for which 18 spectra are available (8 GIRAFFE HR15N and 10 GIRAFFE HR9B). This source is not listed in the Simbad database. The orbital elements  are given in Table~\ref{tab:orbit} and the orbit is displayed in Fig.~\ref{fig:Br81_orbit}. 
Strangely enough, a good SB2 solution for this system could only be obtained by adding an extra parameter to the orbital elements, namely an offset, between the systemic velocities derived from component A and from component B (see the $\Delta V_B$ term in Eq.~(2) of \citet{pourbaix2016}). In most cases, this offset is null but there could be situations where it is not, like in the presence of gravitational redshifts or convective blueshifts that are different for components A and B \citep{pourbaix2016}. Alternatively, if the spectrum of one of the components forms in an expanding wind (as in a Wolf-Rayet star), it would lead as well to such an offset. However, what is puzzling in the considered case is the large value of the offset ($24.8 \pm 1.2$) for which we could not find any convincing explanation. Indeed, no Wolf-Rayet stars are known in the Be~81 cluster according to the Simbad database. This very diffuse cluster of intermediate age lies towards the Galactic centre \citep{2014AJ....147...69H, 2014A&A...561A..94D}.

\subsection{SB3 candidates}
\label{Sect:SB3}

Tables~\ref{tab:total_res} and \ref{tab:sb_field} show that, in total, 11 SB3 candidates (7 probable: flag A, 1 possible: flag B, and 3 tentative: flag C) have been detected. 
Among those, five SB3 are found in the field (Fig.~\ref{fig:sb3_mw_ccf} and Table~\ref{tab:mw_sb3}) and six in clusters (Fig.~\ref{fig:sb3_oc_ccf} and within the Table~\ref{tab:SB2cluster}).
A total of 266 targets were initially labeled as SB3 candidates by the \doe\ code while only 11 were kept after visual inspection, giving a success rate of about 4~\% (compared to 30~\% for SB2 detection).
The SB3 candidates are essentially detected in UVES setups and in GIRAFFE setups HR9B and HR10. SB3 candidates in the stellar clusters have been examined on a case by case basis, and the results are reported below.
\medskip\\
\begin{table*}
\caption[]{\label{tab:orbit}
Orbital elements for 06404608+0949173 in NGC~2264, and 19013257-0027338 in Be~81.}
\center
\begin{tabular}{lllllllllll}
\hline \hline
CNAME & 06404608+0949173 & 19013257-0027338\\
\hline\\
$P$ (d) & $2.9637\pm0.0002$ & $15.528\pm0.002$\\
$e$       & $0.092\pm0.006$ & $0.170\pm0.006$\\
$\omega$ ($^\circ$) & $56.8\pm3.9$ & $265.7\pm3.9$\\
$T_0$-2\,400\,000 (d)& $56072.4085\pm0.0351$ & $56470.531\pm0.140$\\
$V_0$ (\kms) & $14.32\pm0.55$  & $34.51\pm0.66$\\
$\Delta V_B$ & $0.00$ (adopted) & $24.8\pm1.2$\\
$K_A$ (\kms) & $106.3\pm0.7$ & $86.0\pm0.9$\\
$K_B$ (\kms) & $117.0\pm0.6$  & $97.0\pm0.9$\\
$\sigma_A(\mathrm{O}-\mathrm{C})$ (\kms) & 20.2   & 6.1 \\
$\sigma_B(\mathrm{O}-\mathrm{C})$ (\kms) & 9.3    & 6.8 \\
$a_A \sin i$ (Gm) & $4.315\pm0.030$  & $18.1\pm0.2$ \\
$M_A/M_B$ & $1.10$& $1.13$\\
$N$       & 16    & 18 \\
\hline
\end{tabular}
\tablefoot{The orbital elements are the orbital period $P$, the eccentricity $e$, the argument of the periastron $\omega$ from the ascending node, the time of passage at periastron $T_0$, the velocity of the centre-of-mass $V_0$, the primary and secondary velocity amplitudes $K_A$ and $K_B$, the projected primary semi-major axis on the plane of the sky $a_A \sin i$ and the primary to the secondary mass ratio $M_A/M_B$. $\sigma_A(\mathrm{O}-\mathrm{C})$ and $\sigma_B(\mathrm{O}-\mathrm{C})$ are the standard deviation of the residuals (observed $-$ calculated) of components A and B. $N$ is the number of avalaible CCFs on which two velocity components are identified. For the meaning of $\Delta V_B$ see Eq.~(2) of \citep{pourbaix2016}.}
\end{table*} 
\begin{figure}[t]
\includegraphics[width=9cm]{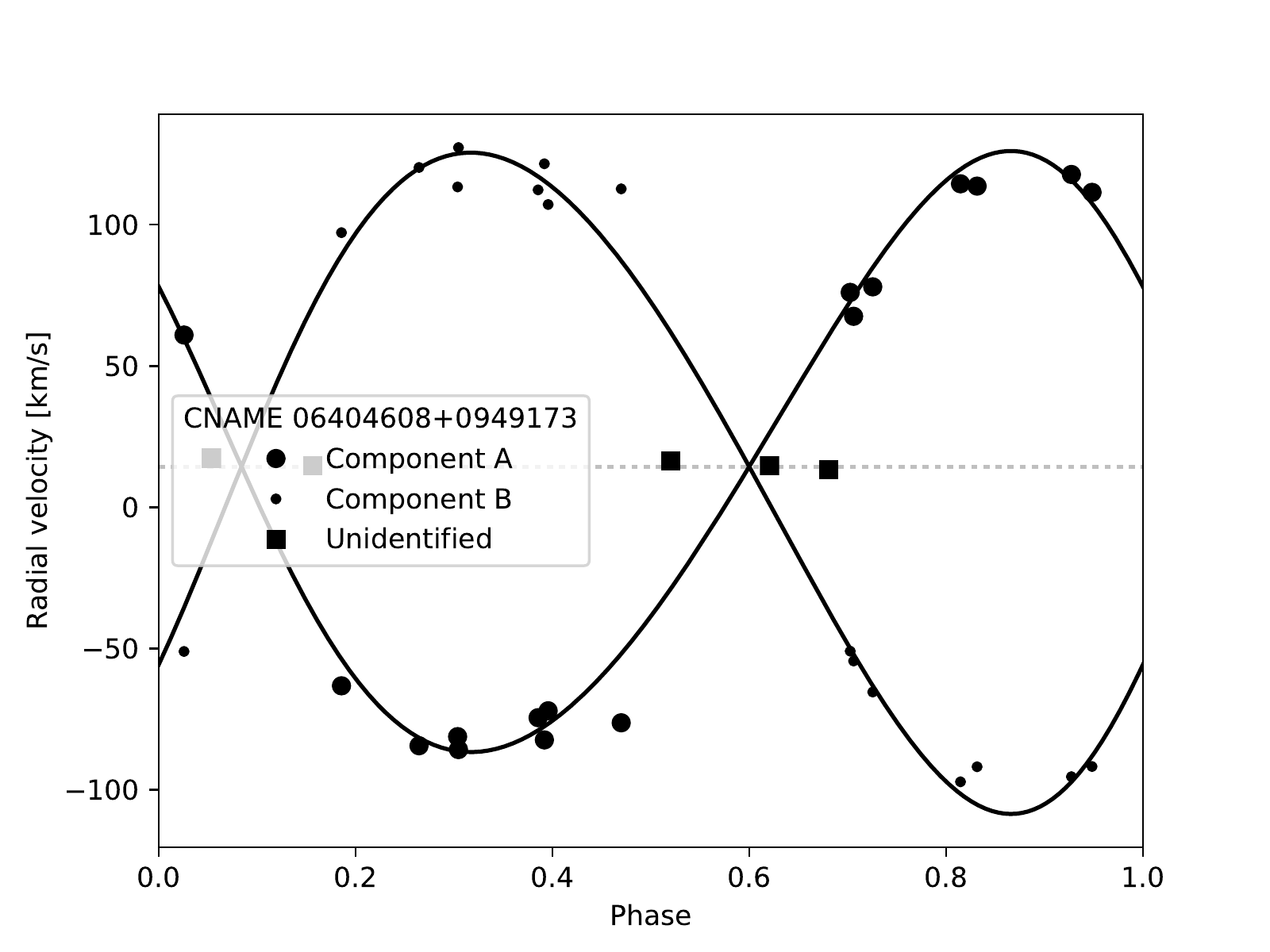}
\caption[]{\label{fig:NGC2264_orbit}
The SB2 orbit of 06404608+0949173 in NGC~2264. Component A is represented by large circles and component B by small circles. Squares represent the single radial velocity obtained when only one peak is visible in the CCF; these are not used to calculate the orbital solution, due to their larger uncertainties. The error on radial velocities amounts to $\pm0.25$~\kms. Horizontal dotted line is $V_0$.}
\end{figure}
\begin{figure}[t]
\includegraphics[width=9cm]{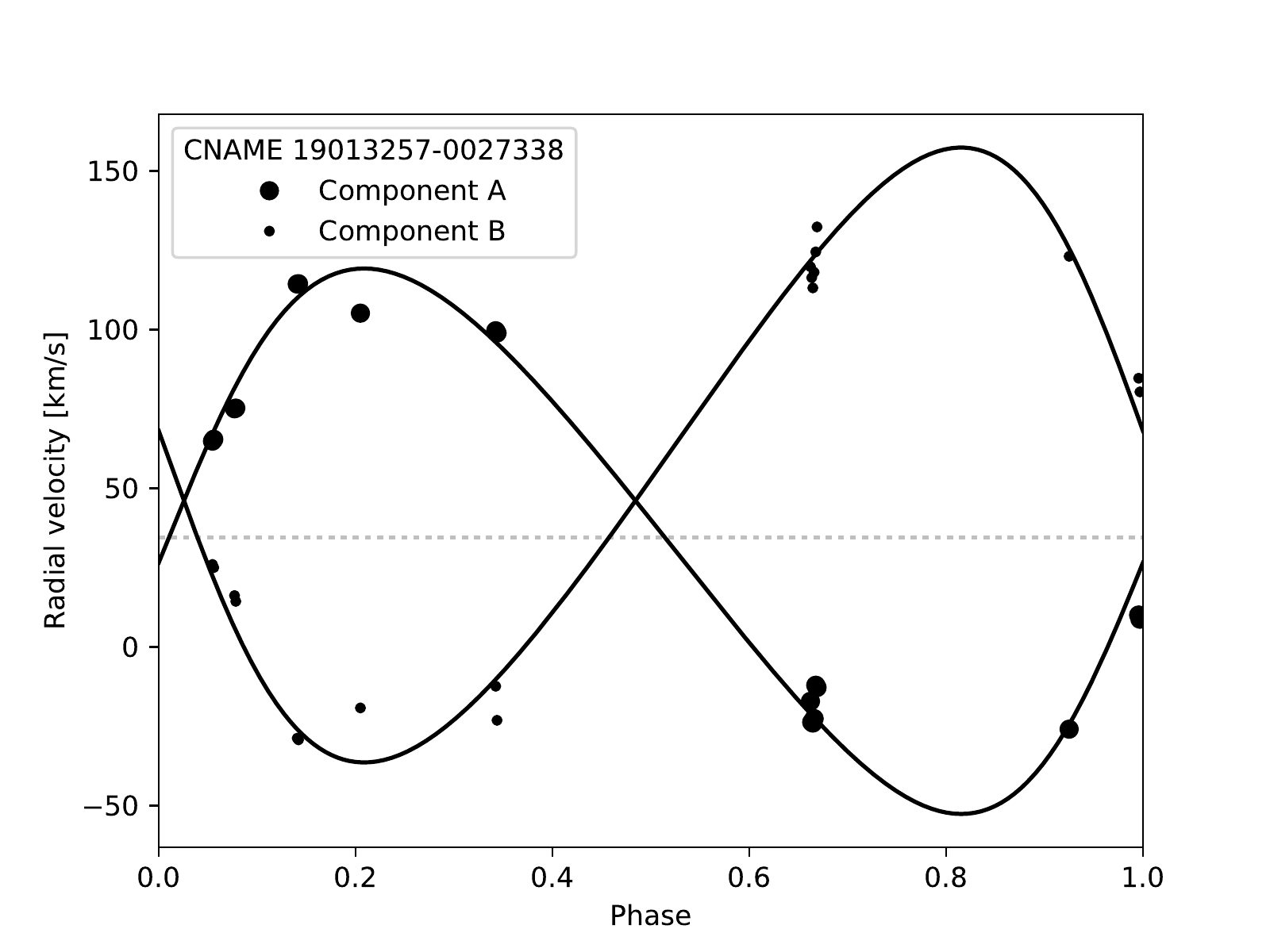}
\caption[]{\label{fig:Br81_orbit}
The SB2 orbit of 19013257-0027338 in Berkeley~81. Component A is represented by large circles and component B by small circles. The error on radial velocities amounts to $\pm0.25$~\kms. Horizontal dotted line is $V_0$.}
\end{figure}
\noindent{\it NGC 2451.} 
The CCF of 07470917-3859003 exhibits three clear peaks (the CCF is classified as 2030A), at 25.0, 96.1, and 136.6~\kms. The first velocity is compatible with membership in NGC~2451A. The DSS\footnote{Digitized Sky Survey: \url{https://archive.stsci.edu/cgi-bin/dss_form}} image reveals the presence of a slightly fainter star about 12'' south (a larger distance than the 1.2'' size of the fibre, so no contamination is possible). Given the fact that the two fainter peaks are not located symmetrically with respect to the cluster velocity, it is doubtful that the system could be an SB3 system in case of membership to NGC 2451.
\medskip\\
\noindent{\it NGC~2516.} NGC~2516 45 (system 07575737-6044162) is a star classified as A2~V \citep{hartoog1976} with $V=9.9$. The iDR4 recommended parameters (\Teff$=8500$~K, $\log{g}=4.1$ and solar metallicity) suggest that it could be a $\delta$~Scu star. Its CCF is most likely associated with a fast rotator with a superimposed sharper central peak. The SB3 nature of this candidate is therefore doubtful and a follow-up of this source should be performed before drawin any firm conclusion.
\medskip\\
\noindent{\it NGC 6705.}
In total, the \doe\ routine finds 52 SB3 candidates in NGC~6705,  one of the largest number of SB3 among all the targeted clusters (Table~\ref{tab:sb_field}). After a first-pass analysis we discarded all of them but one.  NGC 6705 1147 (system 18510286-0615250). The velocities corresponding to the three peaks observed in the CCF are listed in Table~\ref{tab:NGC_6705_1147}. They exhibit clear temporal variations.
The cluster velocity is 29.5 \kms\ \citep{2014A&A...569A..17C}.
 This velocity is close to that of the middle (C, \emph{i.e.}, faintest) peak in the CCF. That central peak is not varying as much as the most extreme peaks, and moreover, the shape of the C peak is not as sharp as are the A and B peaks. Considering the fact that the cluster NGC 6705 is a dense one, we believe that this third peak is from background contamination.  We therefore conclude that the detection of  NGC 6705 1147 as SB3 is spurious, and should be downgraded to SB2. The SB2 analysis is presented in Table~\ref{tab:NGC_6705_1147} where we computed the mass ratio, adopting 34~\kms\ (Table~\ref{tab:sb_field}) as the centre-of-mass (cluster) velocity. The observed velocity variations are consistent at all times with a mass ratio of the order of 1.32. 
\medskip\\
\noindent{\it NGC 6005.} The CCF of 15553867-5724434 (classified as 2030B) shows three peaks, at $-81.6$, $-14.4$ and $32.7$~\kms, to be compared with $-25.2$~\kms\ for the cluster velocity \citep{2014AJ....147..138C}. The spectra are at the minimum required $S/N$. These data are compatible with 15553867-5724434 being an SB3, member of NGC~6005.
\medskip\\
\noindent {\it NGC 6802.} The CCF of 19302315+2013406 (classified as 2030C) shows three distinct peaks, at $-22.4$, $22.0$ and $65.5$~\kms, to be compared with 12.4~\kms\ for the cluster velocity \citep{2014AJ....147...69H}. These data are compatible with 19302315+2013406 being an SB3, member of NGC~6802.
\medskip\\
\noindent {\it Trumpler 20.} The CCF of 12391904-6035311 (classified as 2030C) shows three distinct peaks, at $-85.78$, $-44.4$ and 14.8~\kms, to be compared with $-40.8$~\kms\ for the cluster velocity \citep{2005A&A...438.1163K}. These data are compatible with 12391904-6035311 being an SB3, member of Trumpler 20. An extended analysis of the GES data for this cluster may be found in \citet{2014A&A...561A..94D}.

\begin{table*}
\caption[]{\label{tab:NGC_6705_1147}
Velocities of the three peaks (A, B, C) in the CCF of NGC~6705 1147. 
The columns labeled $\Delta$ list the differential velocity with respect to the centre-of-mass (\emph{i.e.}, cluster) velocity, adopted as 34~\kms.
}
\center
\begin{tabular}{rlrrrrrcccc}
\hline
\hline
JD - 2\;456\;000 & Setup & $v_r(A)$ & $v_r(B)$ & $v_r(C)$ &  $\Delta v_r(A)$ & $\Delta v_r(B)$ & $M_A/M_B$  \\
\hline
77.409  & HR3  &$ 79.62$ & $-24.70$ & 33.83 & 45.62 & 58.70 & 1.29 \\
99.268  & HR3  &$ 95.35$ & $-47.33$ & 29.87 & 61.35 & 81.33 & 1.33 \\
99.280  & HR5A &$ 95.38$ & $-45.73$ & 23.68 & 61.38 & 79.73 & 1.30 \\
99.295  & HR6  &$ 93.65$ & $-44.84$ & 35.92 & 59.65 & 78.84 & 1.32 \\
99.298  & HR9B &$ 94.83$ & $-46.62$ & 40.28 & 60.84 & 80.62 & 1.33 \\
103.110 & HR10 &$-26.78$ & $106.91$ & 39.14 & 60.78 & 72.91 & 1.20 \\
442.394 & HR10 &$ 75.38$ & $-18.75$ & 40.61 & 41.38 & 52.75 & 1.27 \\
442.400 & HR10 &$ 72.20$ & $-20.23$ & 26.72 & 38.20 & 54.23 & 1.42 \\
442.406 & HR10 &$ 75.41$ & $-22.23$ & 33.44 & 41.41 & 56.23 & 1.36 \\
\hline
\end{tabular}
\end{table*}

\begin{figure}
 \includegraphics[width=9cm]{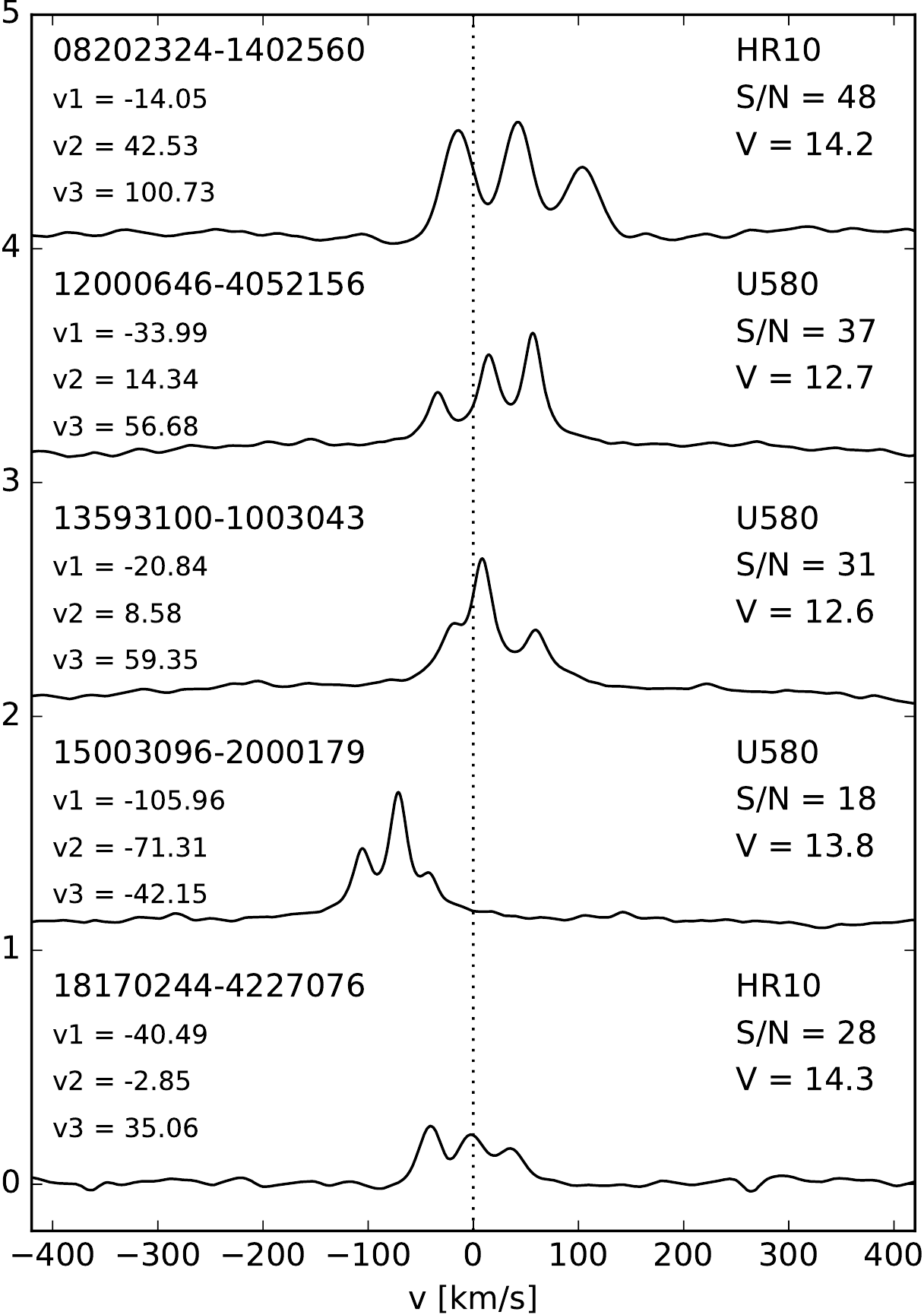}
 \caption{The CCFs of the five SB3 candidates (flagged 2030~A) in the field. Velocities of the components are given in \kms.}
  \label{fig:sb3_mw_ccf}
\end{figure}

\begin{figure}
 \includegraphics[width=9cm]{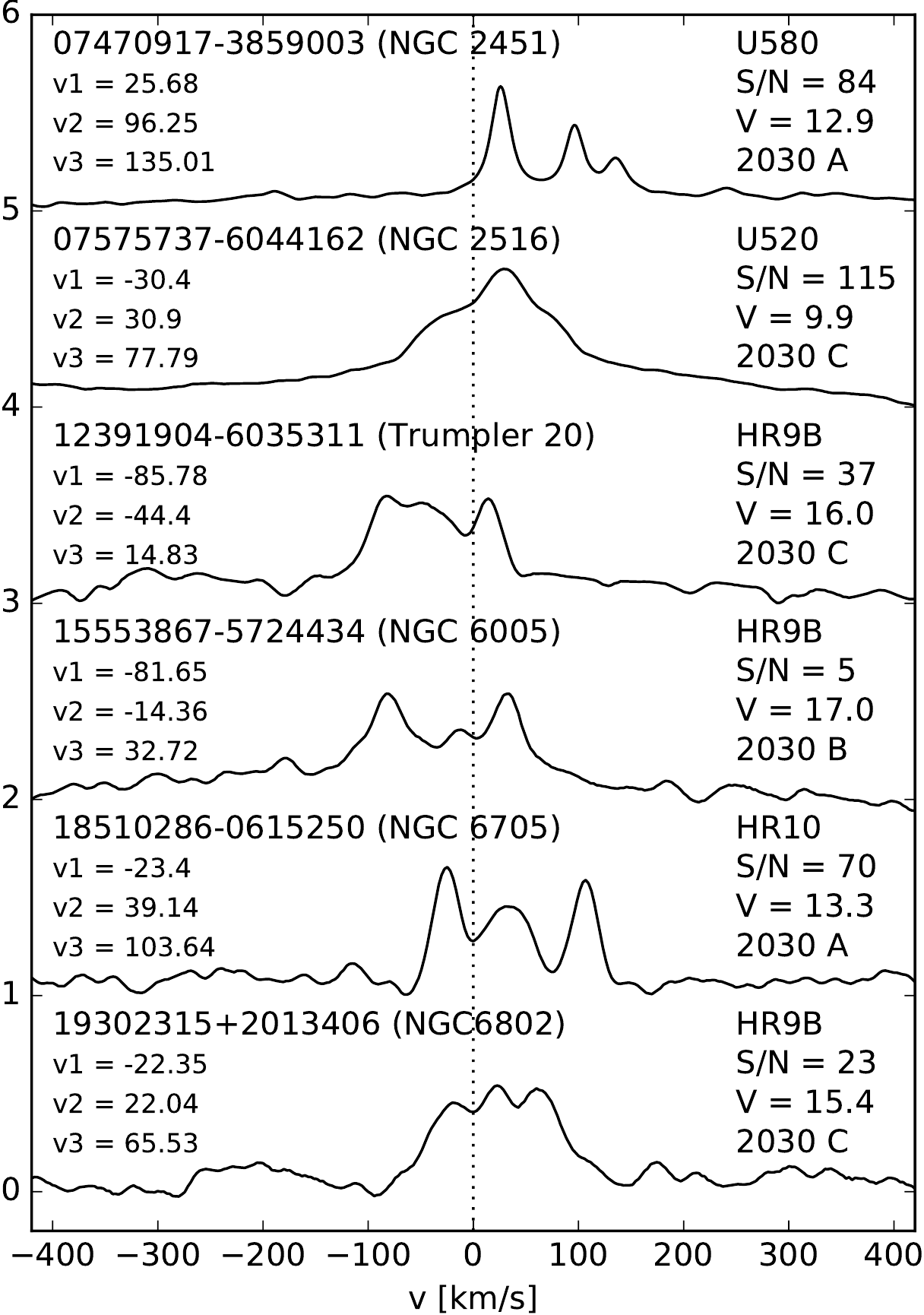}
 \caption{The CCFs of the six SB3 candidates in the stellar clusters. Velocities of the components are given in \kms. The vertical scale of the CCFs has been magnified for clarity.}
 \label{fig:sb3_oc_ccf}
\end{figure}

\subsection{The unique SB4 candidate HD~74438}
\label{Sect:SB4}

\begin{figure}[t]
 \includegraphics[width=1.\linewidth]{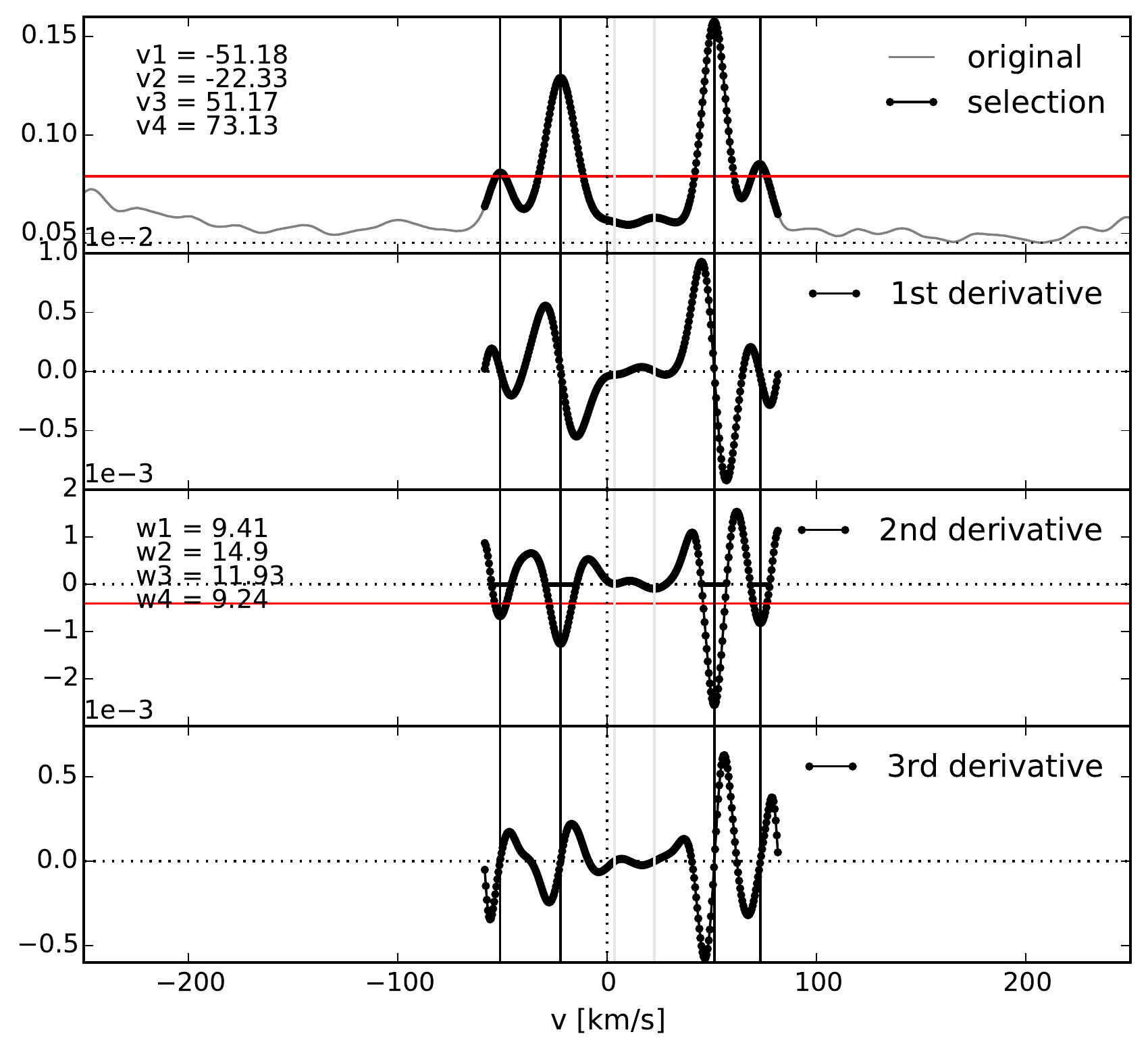}
 \caption{The CCF of the A2V star HD 74438, obtained on JD 2456707.102 with the setup U580. The four peaks are clearly visible.
\label{fig:HD74438}}
\end{figure}

\begin{table*}
\caption[]{\label{tab:HD74438}
Velocities of the four peaks (A, B, C, D) in the CCF of HD 74438 over the night of February 18, 2014 obtained with the U580 setup. The columns labeled $\Delta$ list the differential velocity with respect to the centre-of-mass (\emph{i.e.}, cluster) velocity.
}
\centering
\begin{tabular}{lllllllllll}
\hline \hline
JD - 2\;456\;707 & $v_r(A)$ & $v_r(B)$ & $v_r(C)$ & $v_r(D)$ & $\Delta v_r(A)$ & $\Delta v_r(B)$ & $\Delta v_r(C)$ & $\Delta v_r(D)$ & $M_A/M_B$ & $M_D/M_C$ \\
\hline
0.028 & 50.61 & $-21.40$ & $-44.25$  & 67.92 & 35.81 & 36.20 & 59.05 & 53.12 & 1.01 & 1.11\\
0.030 & 50.67 & $-21.14$ & $-44.53$   & 68.18 & 35.87 & 35.94 & 59.33 & 53.38 & 1.00 & 1.11\\
0.113 & 51.18 & $-22.18$ & $-52.08$ & 74.07 & 36.38 & 36.98 & 66.88 & 59.27 & 1.02 & 1.13\\
0.120 & 51.08 & $-22.40$ & $-52.31$ & 74.55 & 36.28 & 37.20 & 67.11 & 59.75 & 1.02 & 1.12\\
\hline
\end{tabular}
\end{table*}

We have detected one SB4 candidate: the A2V star HD~74438 (CNAME 08414659-5303449, with $V=7.58$) belonging to the open cluster IC 2391 \citep{Platais2007}.

The star has been observed 45 times within 2.5~h on February 18, 2014, with the U520 and U580 setups. Its peculiarity was already noticed by \citet{Platais2007}, since it lies 0.9~mag above the main sequence in a color-magnitude diagram, and therefore was already suspected to be a triple system  (since the maximum deviation for  a binary system with two components of equal brightness would amount to $2.5 \times \log 2 = 0.75$~mag). It is nevertheless considered a \emph{bona fide} member of the cluster by \citet{Platais2007}. Therefore, one may consider that the centre-of-mass velocity for the system is identical to the cluster velocity, namely $14.8\pm1$~\kms\ \citep{Platais2007}. A typical example of the CCF of HD~74438 is presented on Fig.~\ref{fig:HD74438}, with its four distinct CCF peaks clearly apparent.  The velocities of the peaks at different times over the night of February 18, 2014 are collected in Table~\ref{tab:HD74438}. In this Table, we first notice that the velocities of components A and B (which correspond to the highest peaks) vary slowly and oppositely to each other. Their amplitude of variations is similar. If we compute the velocity variations with respect to the cluster velocity (which should correspond to the center-of-mass velocity of the AB pair, neglecting the gravitational influence of components C and D -- columns $\Delta v_r(A)$ and $\Delta v_r(B)$ in Table~\ref{tab:HD74438}), we note that these variations obey a simple property: their ratio is almost constant. In a simple binary system, this property is expected since the ratio $\Delta v_r(B) / \Delta v_r(A)$ equals the mass ratio $M_A/M_B$. Here we find  $M_A/M_B  \sim 1.01$. Thus, the brightest components in the system, which correspond to the most prominent peaks A and B, are close to twins since their masses differ by 1\% only. We observe that the pair CD obeys the same property: $\Delta v_r(C) / \Delta v_r(D)$ is almost constant, even though the amplitude of variations is larger than that of the AB pair. Again, assuming no perturbations from the AB pair, we get  $M_D/M_C = \Delta v_r(C) / \Delta v_r(D) \sim 1.12$. It seems therefore that the  observed variations do make sense and give credit for a physical nature of the ABCD system as a double pair AB/CD. We could nevertheless expect some perturbations of one pair on the other, at least in the form of a trend of the center-of-mass velocities of each pair, if pair CD orbits around pair AB. 
The available observations do not span a time interval long enough to check that possibility.

Assuming that the ratio of the CCF amplitudes roughly scales with the luminosity ratio\footnote{If the spectral types of the components are very different, spectral mismatch may invalidate this hypothesis, but this is unlikely given the SB2 nature of the source which implies a luminosity ratio close to one and hence similar spectral types.}, and adopting a ratio of 3 between the peak amplitudes of A and D (see Fig.~\ref{fig:HD74438}), we get a magnitude difference between components A and D equal to $\Delta m = 2.5 \log 3 = 1.2$~mag. Consequently, the observed visual magnitude $m_V=7.58$ is mainly due to the pair AB. With the parallax of the system $\pi=5.716\pm0.298$~mas provided by Gaia DR1 \citep{Gaia2016}, the distance of this system is $175\pm 9$~pc. The absolute magnitude of AB pair is then $M_V(AB)=1.36$. Assuming similar masses, we have $M_V(A)=M_V(B)=2.12$. This corresponds to a spectral type A7 and to masses of $M_A=M_B=1.8$~M$_\odot$ if on the main sequence (luminosity class V). The absolute magnitudes of the components C and D are consequently $M_V(C)=M_V(D)=3.31$ corresponding to a spectral type F1 which correspond to a mass of about 1.5~M$_\odot.$ Inserting these values in the defining relation for the orbital velocity semi-amplitude (expressed in \kms):
\begin{equation}
\label{Eq:K}
K_i = 212.9 \left( \frac{M_i}{P({\rm d})}\right)^{1/3} \frac{q}{(1+q)^{2/3}} \frac{\sin i}{(1-e^2)^{1/2}},
\end{equation} 
it is possible to derive an upper limit to the orbital period. Indeed, for the AB pair, we adopt $e=0$, $q = 1$, $M_A = 1.8$~M$_\odot$, and $K_A > 36$~\kms (Table~\ref{tab:HD74438}), and obtain an upper limit on the orbital period of the AB pair, $P$~(d)~$ < 93 \sin^3{i}$. The same method applied on the CD pair (with $M_D = 1.5$~M$_\odot$, $e=0$, $q = 1.1$, and $K_D > 60$~\kms) yields $P$~(d)~$< 20\sin^3{i}$, in agreement with the fast variation observed in Table~\ref{tab:HD74438} for the C and D velocities. 

An even more constraining limitation on the orbital period may be derived from the fast variations exhibited by the D component over the 2.2~h time span covered by  the observations (Table~\ref{tab:HD74438}).  We first assume that 74.55~\kms\ corresponds to the maximum orbital velocity, from which we derive a semi-amplitude $K_D = 59.75$~\kms, corresponding to $\omega t_1 = \pi/2$. It is then possible to find $\omega = 2\pi / P$, and hence $P$, by assuming a sinusoidal velocity variation (in a circular orbit), reaching a velocity of 67.9~\kms\ at time $t_2 < t_1$, such that   $\omega t_2 = \arcsin \frac{67.9 - 14.8}{K_D} = 1.093$. From this, we derive  $\omega t_1 - \omega t_2 = \pi/2 - 1.093 = 0.477 = 2\pi /P \times (t_1 - t_2) = 2\pi/P \times 2.2$~h,  or $P = 29$~h as tentative period of the CD pair.

To conclude, we note that the above arguments allow us as well to estimate the deviation of HD~74438 in the color - magnitude diagram, for a system consisting of components with fluxes $F_A = F_B$, and  $F_D = F_C = 1/3 F_A$. The magnitude excess amounts to $2.5 \log (2 + 2/3) = 1.1$~mag, not far from the 0.9~mag reported by \citet{Platais2007}. The velocities of the components  would definitely be worth monitoring over a few hundred days.

\begin{figure}
\includegraphics[width=9cm]{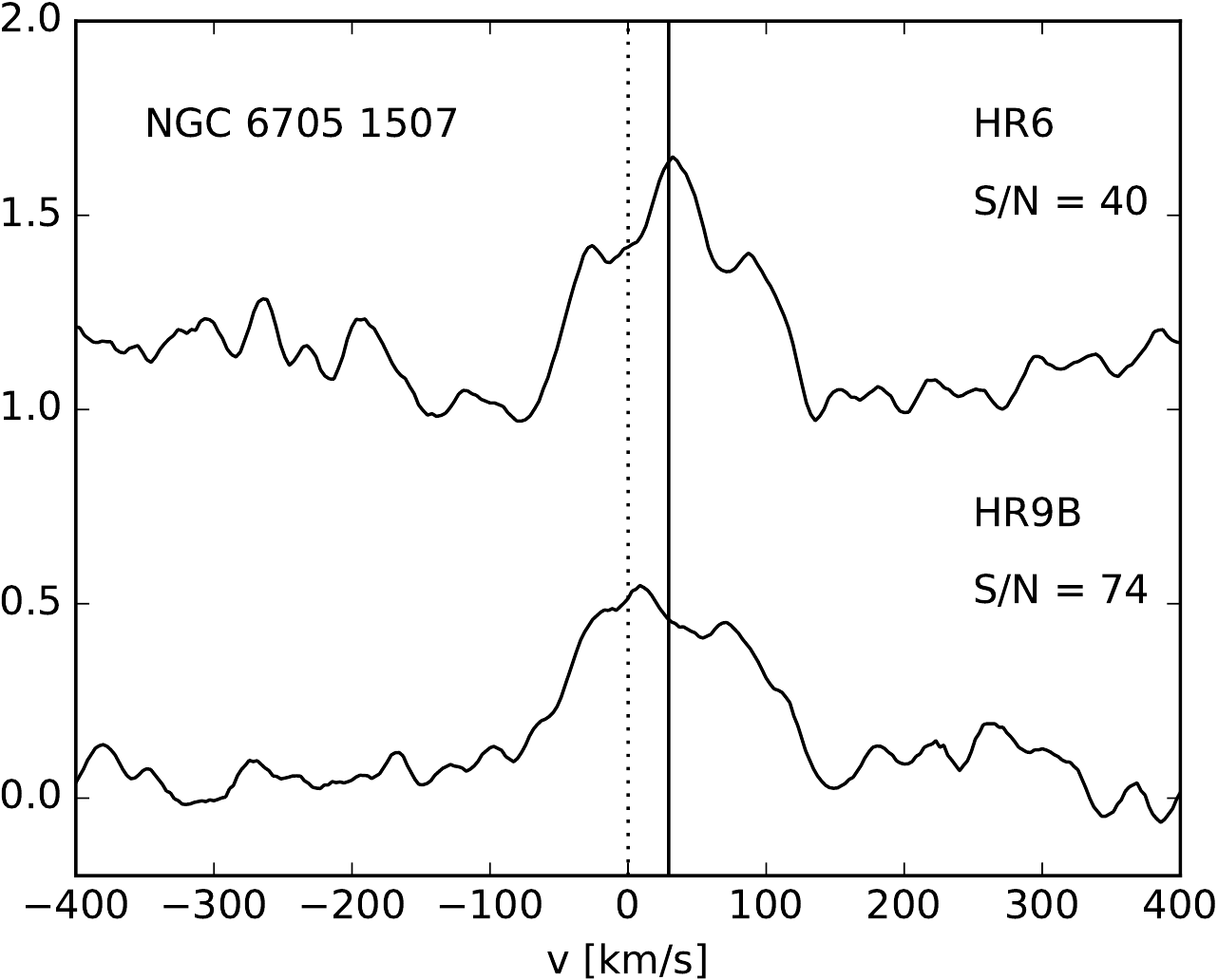}
\caption[]{\label{fig:NGC6705_1507}
The CCFs of star 1507 in the cluster NGC~6705, with its triple-peak CCF, most probably caused by pseudo-absorptions (caused by pulsation) superimposed on a rapid-rotator profile. The vertical plain line shows the cluster velocity.}
\end{figure}

\subsection{Multiplicity flagging by other GES working groups}
It is worth mentioning that different nodes within the GES WGs have identified/detected spectroscopic systems for restricted subsamples of iDR4 data. Because we wanted to rely on a homogeneous detection process, we did not include the SB$n$ detected by other WG in the present analysis. This detailed comparison will be performed for the next data release. 

WG~12, focusing on pre-main sequence stars in clusters, detected 176 SB2 (A: 168, B:2, C: 6), one SB3 and two SB4. The intersection with our list amounts to 66. In particular, the two SB4 detected by WG12 are classified as SB2 in our final list; we re-checked that only two peaks are visible on the CCFs computed from single exposures. WG~12 developed a specific method to remove nebular contamination by masking the nebular lines in HR15N spectra for the clusters NGC~2264, NGC~6530 and Tr14. Indeed, these nebular lines can produce a double-peaked CCF that can be misclassified as an SB2 candidate; see Klutsch et al. (in prep.) for more details. 

WG~13, dedicated to OBA-star spectrum analyses, identified about 30 SB2 in clusters (NGC~2547, NGC~3293, NGC~6705 and Tr 14). They detected one SB3 candidate (system 10344470-5805229 in NGC~3293) that we have rejected. Indeed, the three peaks were detected by our method only in two CCFs and only in the HR5A setup, whereas 10 CCFs of the same object displayed only one or two peaks in various other setups (HR3, HR6, HR9B and HR14A). This SB3 detection was therefore considered as not reliable enough considering our rejection criteria (discussed at the beginning of Sect.~\ref{sect:results}). However, we did not even select this object as an SB2 because the  velocity difference between the 2 peaks is too large ($> 290$~\kms), indicating possible spurious peak(s). 

In summary, the GES working groups, which are very focused, will inevitably reach higher detection rates for specific types of objects, but their methods do not apply to the whole GES survey. The method presented here, on the contrary, aims at providing homogeneous information for the whole survey, using all (GIRAFFE and UVES) individual spectra.

\begin{figure}
 \includegraphics[width=\linewidth]{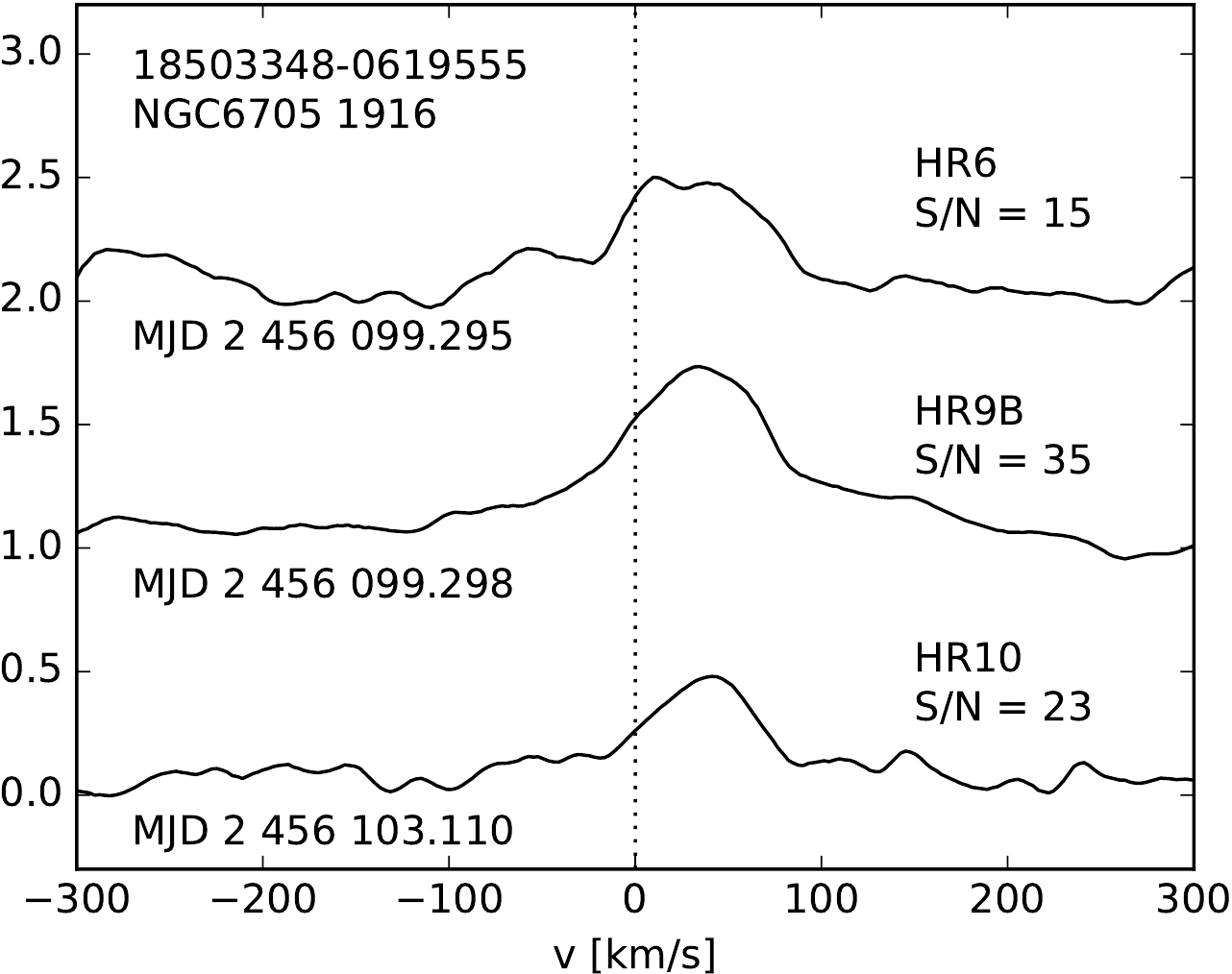}
 \caption{Example of CCFs of a $\delta$~Scu type star that can mimic an SB2 or even an SB3.}
  \label{fig:delta_scu}
\end{figure}
\begin{figure}
 \includegraphics[width=\linewidth]{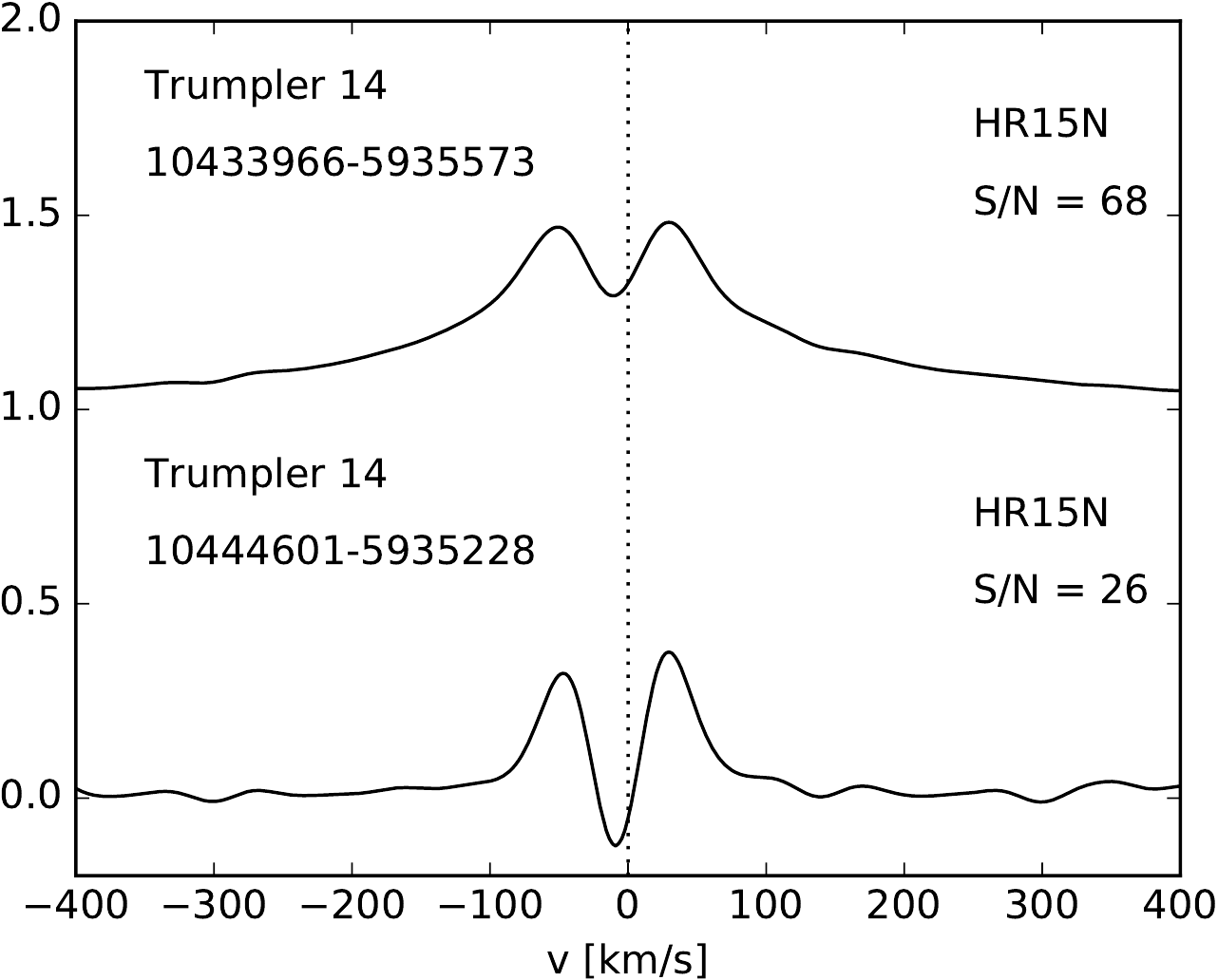}
 \caption{Example of CCFs in HR15N that mimic SB2 but are due to emission in \ha\ produced by nebular lines in the young cluster Trumpler~14.}
 \label{fig:red_issue}
\end{figure}

\subsection{Multiple peak CCFs unrelated to binarity}
\label{sect:ccf_degeneracy}
Double- and triple-component CCFs may sometimes be mimicked by physical processes unrelated to binarity. To clearly establish the binary nature of field stars, multiple observations covering a complete orbital cycle are mandatory in order to derive the orbital elements that fit best the radial velocities. In the case of stars belonging to associations and clusters containing hot and cold gas, the situation is worse: emission lines, which are not masked prior to the CCF computations, may produce troughs in the CCFs that could be interpreted as multiple peaks. Moreover, hot and pulsating stars like $\delta$ Scu stars, or young hot stars with discs, may also produce bumps in the CCFs. It is beyond the scope of the present paper to study the specific signatures of such processes on the CCFs, which also depend on the considered setup. However, we provide below some examples of multiple peak CCFs probably unrelated to binarity.
Besides, in order to remove some spectral signatures degrading the CCFs (emission lines, very strong lines, etc.), we plan to recompute consistently all GIRAFFE and UVES CCFs in a forthcoming paper.

\begin{figure}
 \includegraphics[width=\linewidth]{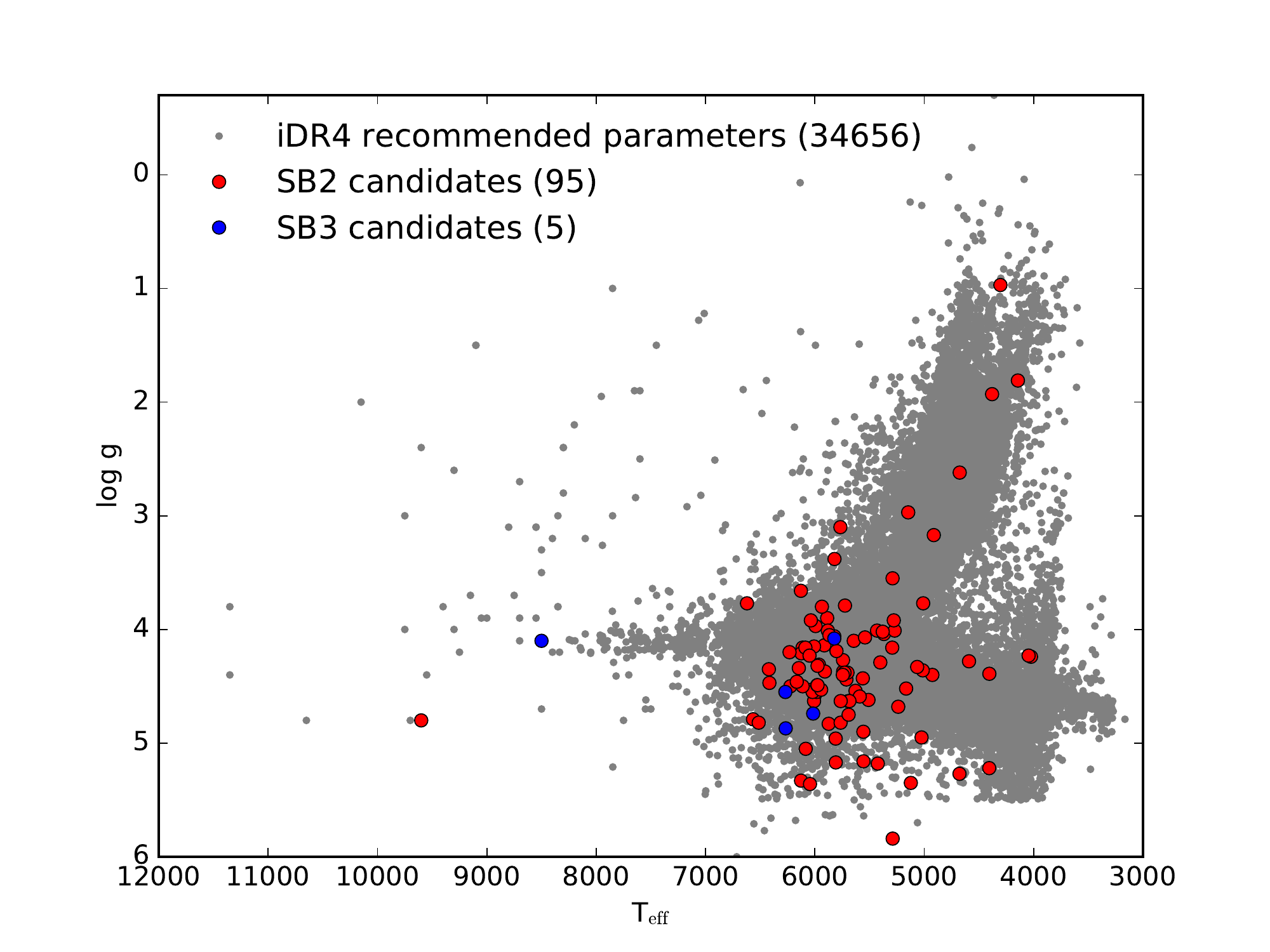}
  \caption{$\log{g}$ -- $T_\mathrm{eff}$ diagram of iDR4 stars with recommended atmospheric parameters. Among them rhe SB2 (red circles) and SB3 (blue circles) candidates are displayed.}
 \label{fig:logg_teff}
\end{figure}

For instance, NGC 6705 1507 (system 18505296-0617402) is classified as A0 \citep{2014A&A...569A..17C} and shows three peaks in its CCF (originally classified as 2030C; Fig.~\ref{fig:NGC6705_1507}) for the setting HR6, at $-25.1$, 33.8, and 86.9~\kms. The central, highest peak is close to the cluster velocity, and the other two are almost symmetrically located from the central peak, at $\pm 50$~\kms. The very edge of the CCF has a steep slope which is reminiscent of a fast rotator. Indeed, the full  base width of the CCF is about 180~\kms, a value typical for the rotation velocities of A stars.  Moreover, a spectrum in the HR9B setting, taken on the same night, confirms the above analysis, which makes us conclude that the triple-peak CCF of star 1507 in the cluster NGC 6705 is most probably caused by pseudo-absorptions superimposed on a rapid-rotator profile. 
A similar situation is encountered for the 2 other SB3 candidates 18510403-0616023 and 18511155-0606094. These three objects have been discarded from the final list.

An example of a star automatically classified as an SB2 with flag~C and very likely to be rather a $\delta$~Scu star, \emph{i.e.} a hot rapid rotator with pulsation and no emission in \ha, is 18503348-0619555 (NGC~6705~1916, $V=13.7$). This star has recommended parameters of \Teff $=7821$~K and $\log{g}=3.96$, compatible with a $\delta$~Scu-type star. The CCF in different setups at different epochs are shown on Fig.~\ref{fig:delta_scu}. The first CCF has 2 components (SB2), one broader than the other. The asymmetry of the second CCF could potentially lead the \doe\ code to identify 3 components (SB3). The last CCF is less ambiguous though it can be seen as an SB2 with  close radial velocities. This SB2 candidate has been removed from the final list of SB$n$ candidates.

\begin{figure}
\includegraphics[width=9cm]{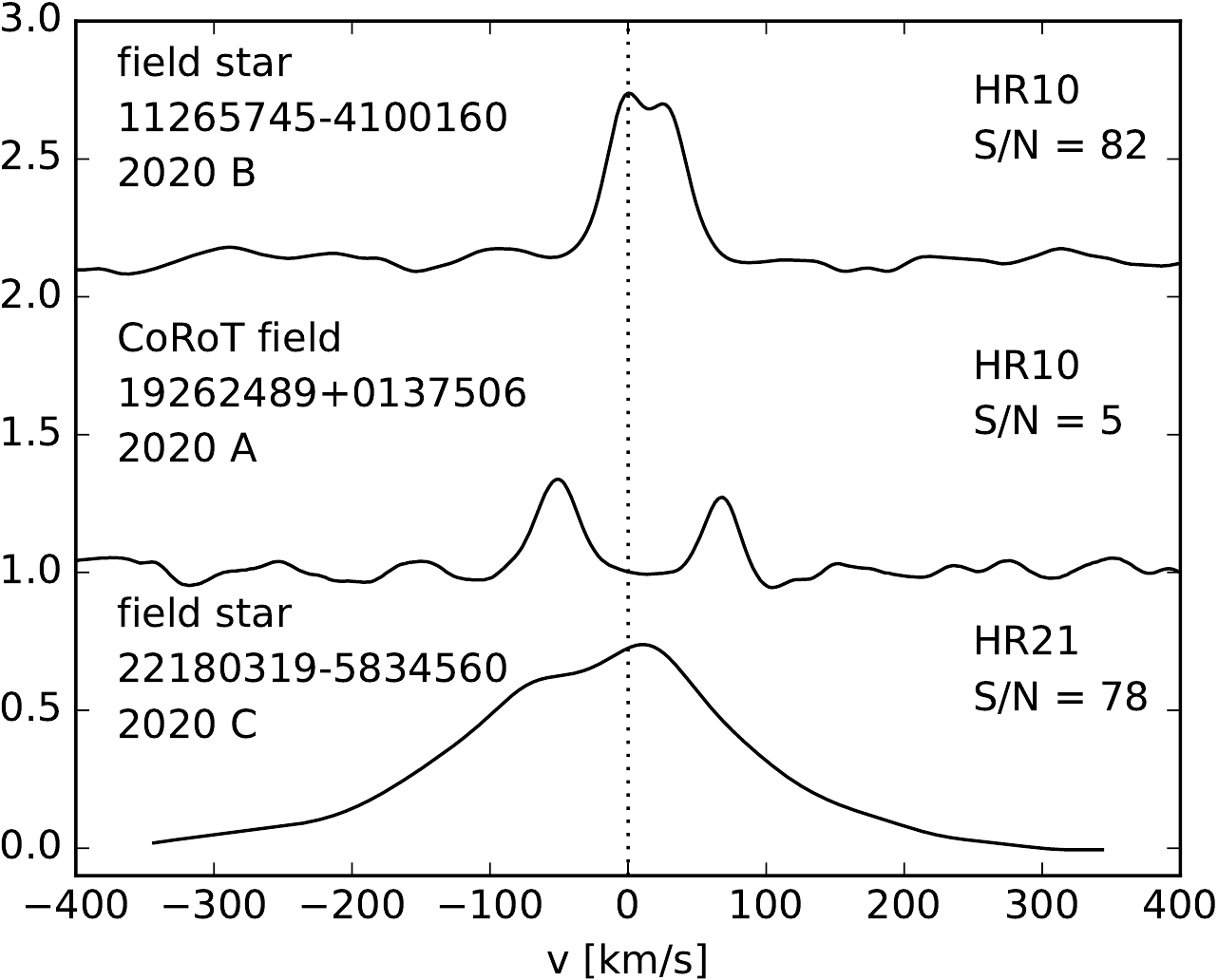}
\caption[]{\label{fig:sb2_giants}
Three giant SB2 candidates.}
\end{figure}

In Trumpler~14, spectra are strongly contaminated by nebular lines around H$\alpha$. This may result from a reduction issue (inadequate sky subtraction in a nebular background). The nebular lines in emission, located at the cluster velocity, superimpose on the absorption lines of the star, also at the cluster velocity. Because the nebular lines in emission are narrower than the stellar  ones in absorption, it results in a CCF with two clear peaks; sometimes the minimum between the two peaks gets even lower than the CCF continuum. Such false SB2 candidates could be unmasked (see Fig.~\ref{fig:red_issue} for the two examples 10433966-5935573 and 10444601-5935228) because in that cluster we found too many stars with radial velocities around $-40$ and 20~\kms, \emph{i.e.} symmetrical with respect to the cluster velocity ($\sim -10$~\kms). They can be explained by an emission at the cluster velocity obliterating the H$\alpha$ line resulting in a central absorption splitting the CCF (an emission line corresponds to absorption in the CCF.)

\begin{table*}
\caption{List of the nine known SB2 systems confirmed by GES.}
\label{tab:known_sb2}
 \begin{tabular}{lllrll}
 \hline\hline
 Name & GES field & CNAME & $V$ & Catalogue & Reference \\
 \hline\\
2MASS J06435849-0100515 &  CoRoT    &  06435847$-$0100516 & 13.05 &      &  \cite{loeillet2008}\\
CoRoT 102715243         &           &                   &  &                 &                     \\
CD-52 2472, IC~2391 56  &  IC~2391  &  08385566$-$5257516 & 10.06 & WEBDA      &  \cite{mermilliod2009} \\
NGC 2682 117            &  M~67     &  08511868+1147026 & 12.59 & SB9, WEBDA      &  \cite{mathieu1990} \\
NGC 2682 119            &  M~67     &  08511901+1150056 & 12.53 & SB9, WDS, WEBDA &  \cite{mathieu1990} \\
NGC 2682 ES 4004        &  M~67     &  08512291+1148493 & 12.69 & SB9, WDS, WEBDA &  \cite{mathieu1990} \\
NGC 2682 165            &  M~67     &  08512940+1154139 & 12.83 & WDS              &  \cite{gavras2010} \\
PU Car                  &  Cha I    &  11085326$-$7519374 & 12.17 & WDS             &  \cite{kohler2008}  \\    
2MASS J18505933-0622051 &  NGC~6705 &  18505933$-$0622051 & 17.06 &      &  \cite{koo2007}\\  
CoRoT 101129018         &  CoRoT    &  19263739+0152562 & 13.60 &      &  \cite{cabrera2009}\\ 
\\
\hline\\
 \end{tabular}

Note: 

 SB9: Ninth catalogue of spectroscopic binary orbits \citep{pourbaix2004};

 WDS: Washington visual Double Star catalogue \citep{mason2016};

 WEBDA: A site Devoted to Stellar Clusters in the Galaxy and the Magellanic Clouds: \url{http://webda.physics.muni.cz}

\end{table*}

\subsection{Distribution in the ($\log g, T_{\rm eff}$) plane}
\label{Sect:Teff-g}
The GES consortium provides recommended atmospheric parameters ($T_\mathrm{eff}$, $\log{g}$ and [Fe/H]) for 63\% of stars from the iDR4. They result from a delicate merging of atmospheric parameters obtained by different WGs using different methods, but all with the same model atmospheres and linelists. Among them, we identified a hundred of our confirmed SB$n$ candidates, representing 30\% of our detected SB$n$).

They are shown in the  $\log{g}$ -- $T_\mathrm{eff}$ plane (see Fig.~\ref{fig:logg_teff}). This figure reveals a sudden drop in the number of stars surveyed above 7000~K. This threshold corresponds to the transition between A and F stars, the latter being surveyed in a systematic way by the GES, the former being included only if they belong to specific clusters. For SB$n$, the atmospheric parameters provided by the GES pipeline are uncertain (or even wrong, as we show below) because (i) composite spectra cannot be fitted with single synthetic spectra, and (ii) spectra fitted by the automated pipelines are not the individual exposures but rather the stacked ones. Despite these shortcomings, the $\log{g}$ -- $T_\mathrm{eff}$ diagram nevertheless allows us to identify systems of interest.

The two SB2 and SB3 systems on the warm end of the $\log{g}$ -- $T_\mathrm{eff}$ diagram (with $T_\mathrm{eff} > 8000$~K) are worth discussing. Their CNAMEs are 
18280622+0642252 (NGC 6633 110, BD+06 3793, A3V),  classified as 2020A, and 
07575737-6044162 (NGC 2516 45, CD-60 1959, A2V), classified as 2030C. 

The system 18280622+0642252 shows two peaks of equal heights at $-70$~\kms\ and $38$~\kms. The former peak is particularly broad, and is probably associated with a rapidly rotating star. Since the cluster velocity ($-25.4$~\kms) lies in between the two peaks, and the double-peak CCF is very well-defined, we confirm the SB2  flag from the \doe\ routine.    

The system 07575737-6044162  exhibits a broad CCF most likely associated with a fast rotator. It has a sharp central peak. It may perhaps be an SB2, but certainly not an SB3 (see Fig.~\ref{fig:sb3_oc_ccf}).

The three giant SB2 candidates (19262489+0137506, 22180319-5834560 and 11265745-4100160) appearing in the $\log{g}$ -- $T_\mathrm{eff}$ diagram (with $\log{g}<2$) are surprising, since they 
should have a mass ratio very close to 1.
Their CCFs are displayed on Fig.~\ref{fig:sb2_giants}.
 To our knowledge, there are only few SB2 systems known so far involving two  giant stars: (i) HD~172481 (more precisely an F2Ia post-AGB star and an M giant; \mbox{\citep{{2001A&A...365..465R,2009A&A...498..489J}}}; (ii) HD~187669 \citep[a double-line eclipsing binary;][]{helminiak2015a}; (iii) TYC 6861-523-1 /  ASAS J182510-2435.5 \citep{ratajczak2013}; (iv) KIC 09246715 \citep[a double-lined spectroscopic and eclipsing binary;][]{helminiak2015b}.

The system 19262489+0137506 (a CoRoT target with $T_{\rm eff} = 4300$~K, $\log g = 1.0$), classified as 2020A, has indeed two peaks well separated by 117~\kms, of almost equal intensities, implying  a rather short period for a pair of giants (Fig.~\ref{fig:sb2_giants}, middle). Adopting $K = 117/2$~\kms, $q=1$, $\sin i = 1$, $e = 0$, and $M_1= 1$~M$_\odot$, Eq.~\ref{Eq:K} predicts a period of  the order of 7.5~d for the associated binary. This is rather short considering the giant nature of the two components. For instance, the minimum orbital period in the large sample of binaries with a K giant component in open clusters  \citep{2007A&A...473..829M} is just above 25~d. The situation is even worse for the sample of field M giants from \citet{2009A&A...498..489J} where the shortest orbital period is above 200~d. This trend of course reflects the increase of the stellar radius along the giant branch. 
Independendtly, the spectral type of the system was estimated to be M2III from broad-band photometry (Exo-Dat, \citealt{deleuil2009}).
In any case, this system is worth a follow-up investigation, especially looking for signs of mass-transfer activity (like possible  H$\alpha$ emission in its spectrum, but the 2 spectra available in HR15 are too noisy to see any such sign of activity).

The system 22180319-5834560, classified as 2020C (and $T_{\rm eff} = 4100$~K, $\log g = 1.8$), exhibits a very broad CCF coming from the strong \ion{Ca}{ii} triplet in the HR21 setup, with two bumps responsible for the SB2 classification (Fig.~\ref{fig:sb2_giants}, bottom). Observations in HR10 one day later does not show any sign of binarity. Inspection of the HR21 spectra  reveals that the bumps observed in the CCF may be due to emission in the \ion{Ca}{ii} triplet line cores, making the SB nature doubtful.

The system 11265745-4100160 ($V=13$), classified as 2020B (with $T_{\rm eff} = 4400$~K, $\log g = 1.9$, top CCF of Figure~\ref{fig:sb2_giants}), exhibits two close velocity components in HR10 (separated by about 32 \kms) but not visible in HR21. The validity of the atmospheric parameters may have been disturbed by the SB2 nature of the star.

\begin{figure}
 \includegraphics[width=\linewidth]{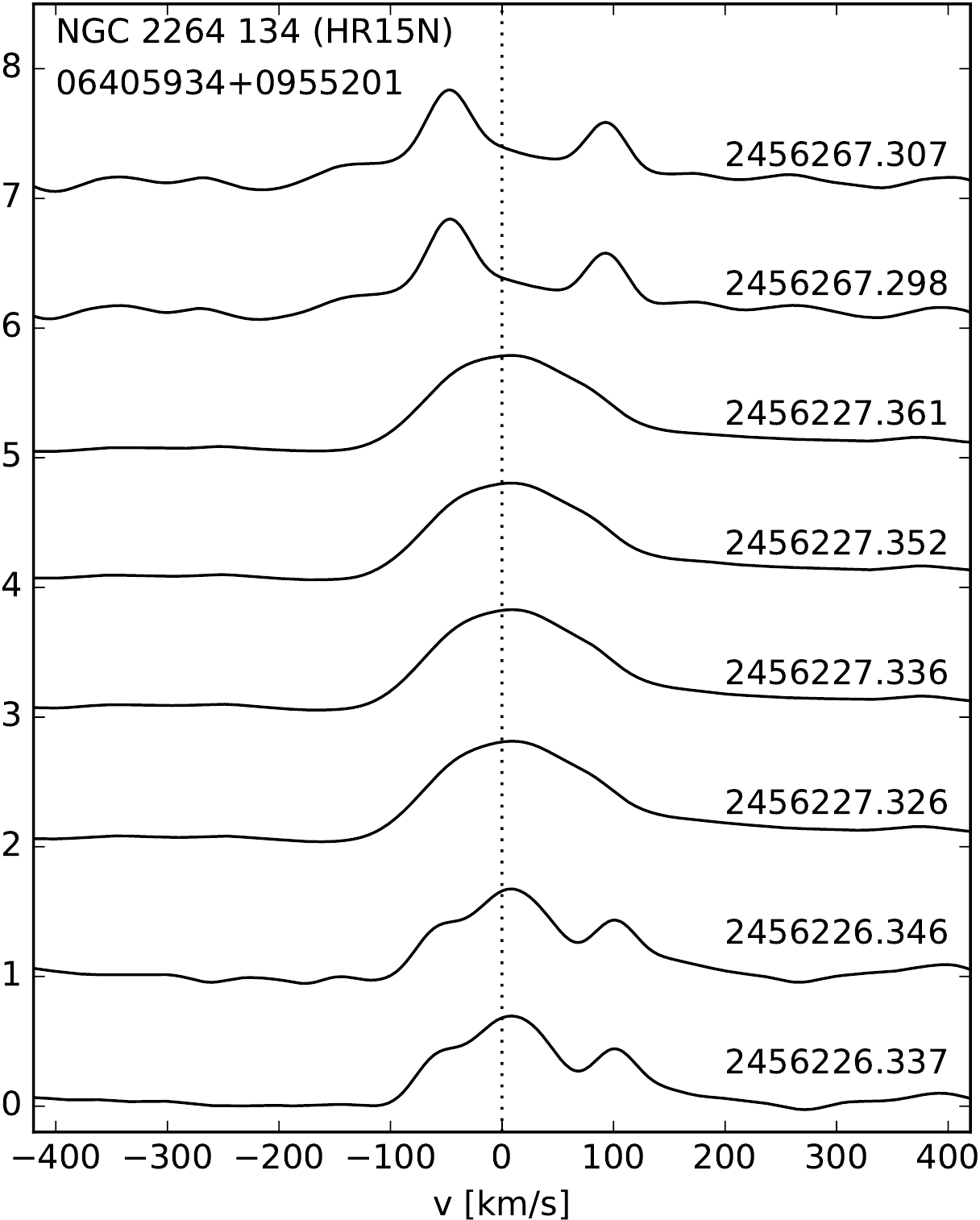}
 \caption{iDR4 CCFs of the pre-main sequence star NGC~2264~134 (06405934+0955201). Known as SB2 in WEBDA, this star shows clear evidence to be an SB3 candidate.}
 \label{fig:sb_missed_ccf}
\end{figure}

\subsection{Comparison with other catalogues}
To estimate the proportion of new SB$n$ candidates, we cross-checked our 352 distinct candidates with published online catalogues of stars. 
The intersection with the Simbad database \citep{wenger2000} provides 96 matches. Among them one is classified as double or multiple star (WDS J08513+1150, CNAME 08511901+1150056 belonging to M67), four as spectroscopic binary stars: 2MASS J06435849-0100515 (CNAME 06435847-0100516) in the Corot field, CD-52 2472 (CNAME 08385566-5257516) in the cluster IC2391, 2MASS J08512291+1148493 (CNAME 08512291+1148493) in M67 and NGC~2682 165 (CNAME 08512940+1154139) also in M67. Two are classified as eclipsing binary stars: 2MASS J18505933-0622051 (CNAME 18505933-0622051) in NGC~6705 and CoRoT 101129018 (CNAME 19263739+0152562). All these previously known binaries have been attributed by our \doe\ code a ``A'' confidence flag.

We cross-matched our detections with various other catalogues, using the X-Match and the Vizier Search online tools from the CDS\footnote{\url{http://cdsxmatch.u-strasbg.fr/xmatch}, \\ \url{http://vizier.u-strasbg.fr}} by uploading the J2000 coordinates built from the CNAME of our SB candidates. For each catalogue, we set the matching area within a radius of 3 arcsec. 

The comparison with the \emph{Ninth Catalogue of Spectroscopic Binary Orbits} \citep[SB9]{pourbaix2004} leads to three systems in common, namely 08511868+1147026, 08511901+1150056, and 08512291+1148493, which are members of the M67 cluster (NGC 2682) with a visual magnitude of about 12.5. 

The comparison with the \emph{Washington visual Double Star} catalogue (WDS, \citealt{mason2016})
leads to an intersection of two systems, namely WDS~J08513+1150 in M~67 and WDS~J11088-7519 in Cha~I (CNAME 08511901+1150056 and 11085326-7519374 respectively).

Cross-matches with the Geneva-Copenhagen Survey of the solar neighborhood III \citep{holmberg2009}, with the bibliographic catalogue of stellar radial velocities \citep{malaroda2006}, with the RAVE catalogue of SB2 candidates \citep{matijevic2010} and with  the Multiple Star Catalogue (MSC)  \citep{tokovinin1997} resulted in empty intersections. We stress that the limiting magnitudes of all these catalogues are much brighter than that of the GES ($V\sim 19$), therefore we expected a small intersection.

As far as the WEBDA cluster database is concerned, four of our SB$n$ candidates are known in WEBDA, with available orbital parameters (see Table~\ref{tab:known_sb2}). 
We also found that two SB2 known in WEBDA are observed in iDR4 but were discarded by the workflow. M67 111 has been observed (08511799+1145541) but the second peak is too low to be automatically detected. The same issue occurs with NGC~2264~134 (06405934+0955201), known to be an SB2 in WEBDA and known to be a pre-main sequence star. It has been observed eight times and seems to be an SB3 candidate because four CCFs have one peak, two CCFs have two peaks and two CCFs have three peaks (see Fig.~\ref{fig:sb_missed_ccf}).

For the sake of completeness, we also checked whether the \doe\ algorithm did retrieve the known SB$n$ candidates from the Geneva-Copenhagen Survey and from SB9. It turns out that only one SB2 (08511799+1145541 in M67) present in SB9 was not found by \doe. The reason thereof is the following: 10 observations in U580 were performed but the second peak is only visible and detected in two of them. Because only stars with more than 75\% of multiple peaks detection in a given setup were flagged as SB2 candidates, 08511799+1145541 was rejected. This shows that the 75\% criterion, chosen to be conservative, might be too restrictive in some cases, although it prevents many false positive detections.

Previously known SB2 systems flagged as such by the GES are listed in Table~\ref{tab:known_sb2}. We stress the fact that the analysis of the GES data provides a substantial number of new SB2 and SB3 candidates because SB detection was performed on a huge data sample ($\sim50\,000$ stars) characterized by a faint limiting magnitude with respect to previous surveys. The new SB2, SB3 and SB4 candidates clearly deserve more observations in order to derive their orbital elements.

\section{Conclusion}
We present a method aiming at identifying multiple-lined spectroscopic binaries (SB$n$, $n \ge 2$) based on the successive derivatives of the CCFs. A list of SB$n$ among the GES iDR4, both in the Galactic field and in the stellar clusters, is presented. In addition, orbital solutions for binary systems belonging to the open clusters NGC~2264 and Be~81 have been calculated.

The detection method has been tested on all the setups of the GIRAFFE and UVES spectrographs available within the GES. It turns out that UVES U580 and GIRAFFE HR10 are the most appropriate setups to detect multiplicity with velocity differences as low as 15 \kms\ and 23 \kms, respectively. 
Simulations show that the \doe\ algorithm reliably derives radial velocities (with a formal error of the order of 0.20~\kms\ at a typical $S/N$ of 10 for GIRAFFE and lower than 0.01~\kms\ at $S/N = 50$ for UVES setups; for multi-component CCFs, the formal error will be slightly increased and, in addition, the systematic error may reach a few~\kms\ at the detection limit).

The detection method leads to a number of false positive detections in stellar clusters. Using physical properties of the clusters and combining information from the spectra and CCFs of different setups, we discussed and discarded a fraction of candidates. A confusing SB2-like signature could be imprinted to the CCF by pulsations in $\delta$~Scuti variable stars, by \ha\ emission in circumstellar discs or interstellar absorption by cold clouds along the line of sight. In such cases, spurious peaks or bumps appear in the CCF.

We discovered 340 SB2, 11 SB3 and one SB4 out of 51\,000 stars with more than 205\,000 single exposures. The most confident binary candidates (`A' flag) most often show very clear composite spectra. Incidentally, we warn against the use of the GES recommended atmospheric parameters for these SB$n$ candidates. Indeed, one third of SB$n$ candidates do have GES recommended parameters, but the the presence of multiple components in spectral lines can potentially lead to erroneous parameters.

The frequency of SB$n$ ($n\ge2$) found by our method in the GES iDR4 sample is 0.7~\%.
Most of the SB$n$ candidates are new because they belong to a sample of stars much fainter than what was covered by previous catalogues.
If we extrapolate this percentage of 0.7\% SB$n$ binaries to the final GES pool of $10^5$ stars, we expect to reach about 1\,000 new SB$n$ systems in the upcoming data releases because the number of observed stars will increase by a factor of two and because we plan to further fine-tune our detection criteria. Indeed the aim of the present analysis was to detect binaries, minimizing the number of ``false positive'' detections (\emph{i.e.} stars wrongly classified as SB$n$). The method presented in this paper can be readily applied to the ESA \emph{Gaia} mission spectra. 

\begin{acknowledgement}
T.M., M.V.d.S. and S.v.E. are supported by a grant from the Fondation ULB.
This work has been partly funded by an {\it Action de recherche concertée} (ARC) from the {\it Direction g\'en\'erale de l'Enseignement non obligatoire et de la Recherche scientifique -- Direction de la recherche scientifique -- Communaut\'e fran\c{c}aise de Belgique.} 
T.M. is supported by the FNRS-F.R.S. as temporary post-doctoral researcher  under grant n$^\circ$ 2.4513.11. 
This work was supported by the Fonds de la Recherche Scientifique FNRS under Grant n$^\circ$ T.0198.13.
C.A. acknowledges to the Spanish grant AYA2015-63588-P within the European Founds for Regional Development (FEDER).
M.T.C acknowledge the financial support from the Spanish Ministerio de Economía y Competitividad, through grant AYA2013-40611-P.
R.S. acknowledges support from the Polish Ministry of Science and Higher Education.

This work was partly supported by the European Union FP7 programme through ERC grant number 320360 and by the Leverhulme Trust through grant RPG-2012-541. We acknowledge the support from INAF and Ministero dell' Istruzione, dell' Universit\`a' e della Ricerca (MIUR) in the form of the grant ``Premiale VLT 2012''. The results presented here benefit from discussions held during the \emph{Gaia}-ESO workshops and conferences supported by the ESF (European Science Foundation) through the GREAT Research Network Programme.

This research has made use of the Washington Double Star Catalogue maintained at the U.S. Naval Observatory.
This research has made use of the WEBDA database, operated at the Department of Theoretical Physics and Astrophysics of the Masaryk University.
This research has made use of the SIMBAD database, operated at CDS, Strasbourg, France.
This research has made use of Python, in particular the Python module pyfits.py which is a product of the Space Telescope Science Institute, which is operated by AURA for NASA.

This research has made used of the Digitized Sky Surveys which were produced at the Space Telescope Science Institute under U.S. Government grant NAG W-2166. The images of these surveys are based on photographic data obtained using the Oschin Schmidt Telescope on Palomar Mountain and the UK Schmidt Telescope. The plates were processed into the present compressed digital form with the permission of these institutions.

This work has made use of data from the European Space Agency (ESA)
mission {\it Gaia} (\url{https://www.cosmos.esa.int/gaia}), processed by
the {\it Gaia} Data Processing and Analysis Consortium (DPAC,
\url{https://www.cosmos.esa.int/web/gaia/dpac/consortium}). Funding
for the DPAC has been provided by national institutions, in particular
the institutions participating in the {\it Gaia} Multilateral Agreement.

The authors thank the referee for his comments which helped improving the manuscript.

\end{acknowledgement}

\bibliographystyle{aa}
\bibliography{biblio}

\appendix
\onecolumn

\section{SB2  and SB3 candidates in the field}
\begin{longtab}
 \centering
 \begin{longtable}{ccrrlcrrc}
 \caption{ List of SB2 candidates in the field ordered by right ascension.} \\
 \hline\hline
 CNAME & flag & \# exp. & \# sp. & setup & MJD & $v_r(1)$ & $v_r(2)$ & $V$ \\
 \hline\\
 \endfirsthead
 \caption{Continued.}\\
 \hline\hline
 CNAME & flag & \# exp. & \# sp. & setup & MJD & $v_r(1)$ & $v_r(2)$ & $V$ \\ 
 \hline\\
 \endhead
 \hline
 \endfoot
 00040663-0101512 & 2020B & 6 & 6 & HR21 & 56205.162 & $-65.45$ & $66.95$ & 16.10 \\
00195847-5423227 & 2020A & 4 & 4 & HR10 & 56532.287 & $93.29$ & $153.47$ & 14.20 \\
00202300-5436167 & 2020A & 4 & 4 & HR10 & 56532.308 & $285.74$ & $330.48$ & 15.30 \\
00301156-5001500 & 2020A & 2 & 4 & U580 & 56266.085 & $17.67$ & $47.52$ & 13.90 \\
00301724-0334401 & 2020C & 4 & 4 & HR21 & 56468.397 & $-48.62$ & $24.97$ & 15.40 \\
00324599-4354509 & 2020B & 4 & 8 & U580 & 56198.130 & $-5.84$ & $15.63$ & 12.90 \\
00503283-4955302 & 2020A & 4 & 4 & HR10 & 56268.134 & $-21.03$ & $32.27$ & 14.90 \\
00591557-0105576 & 2020B & 4 & 4 & HR21 & 56204.172 & $18.82$ & $130.23$ & 14.80 \\
01000070-0100143 & 2020C & 4 & 4 & HR21 & 56204.172 & $-97.34$ & $-33.48$ & 14.50 \\
01012693-5420463 & 2020C & 4 & 4 & HR21 & 56530.337 & $-92.63$ & $-23.25$ & 15.30 \\
01194076-0047374 & 2020C & 4 & 4 & HR21 & 56204.266 & $-22.60$ & $68.63$ & 15.20 \\
01200304-5435209 & 2020B & 4 & 4 & HR21 & 56552.310 & $-10.28$ & $80.61$ & 15.60 \\
01202092-0102102 & 2020C & 4 & 4 & HR21 & 56204.284 & $-56.79$ & $16.26$ & 15.80 \\
01300825-5009146 & 2020A & 4 & 4 & HR10 & 56580.192 & $-16.04$ & $74.55$ & 13.50 \\
01390790-5403014 & 2020A & 4 & 4 & HR10 & 56548.390 & $40.40$ & $83.81$ & 14.20 \\
01393831-4648457 & 2020B & 4 & 4 & HR10 & 56197.248 & $24.46$ & $60.55$ & 13.50 \\
01405323-5356575 & 2020B & 4 & 4 & HR10 & 56548.390 & $12.59$ & $51.88$ & 13.10 \\
01585747-5401493 & 2020B & 4 & 4 & HR21 & 56580.217 & $-65.75$ & $40.52$ & 14.00 \\
01592290-4658510 & 2020C & 4 & 8 & U580 & 56207.122 & $5.97$ & $26.76$ & 12.80 \\
02000945-5352567 & 2020A & 4 & 4 & HR10 & 56579.297 & $-7.69$ & $54.71$ & 14.00 \\
02002707-4655438 & 2020B & 4 & 4 & HR10 & 56207.144 & $36.62$ & $73.89$ & 14.10 \\
02003583-0053539 & 2020C & 4 & 4 & HR21 & 56224.275 & $-62.19$ & $111.35$ & 13.50 \\
02005449-0055403 & 2020A & 4 & 4 & HR10 & 56223.207 & $-23.45$ & $24.67$ & 15.00 \\
02105686-5012361 & 2020C & 4 & 4 & HR21 & 56531.288 & $-20.25$ & $86.83$ & 16.00 \\
02194365-0104381 & 2020C & 5 & 5 & HR21 & 56532.333 & $-23.84$ & $42.38$ & 15.00 \\
02290765-0318506 & 2020B & 4 & 4 & HR21 & 56226.222 & $6.73$ & $81.89$ & 15.70 \\
02290959-5004269 & 2020A & 4 & 4 & HR10 & 56578.216 & $13.53$ & $94.34$ & 14.50 \\
02302503-4956149 & 2020A & 4 & 4 & HR10 & 56578.216 & $-11.88$ & $72.23$ & 14.30 \\
02394731-0057248 & 2020A & 4 & 4 & HR10 HR21 & 56172.267 & $-58.70$ & $21.36$ & 13.80 \\
02503269-5010152 & 2020C & 4 & 4 & HR21 & 56576.204 & $-58.59$ & $15.40$ & 15.70 \\
03103980-5007403 & 2020B & 6 & 6 & HR10 & 56310.061 & $-15.11$ & $41.25$ & 15.90 \\
03175192-0034528 & 2020B & 4 & 4 & HR10 & 56225.186 & $37.94$ & $64.65$ & 14.70 \\
03175934-0024337 & 2020C & 4 & 4 & HR21 & 56226.132 & $-9.62$ & $86.60$ & 15.10 \\
03181102-0034546 & 2020A & 4 & 4 & HR10 & 56225.186 & $-113.51$ & $0.00$ & 14.00 \\
03200828-4656379 & 2020B & 4 & 4 & HR10 & 56197.296 & $-2.25$ & $40.37$ & 14.90 \\
03201610-5601321 & 2020B & 4 & 8 & U580 & 56580.261 & $-7.88$ & $21.61$ & 13.50 \\
03374095-2723284 & 2020A & 4 & 4 & HR10 & 56208.238 & $67.49$ & $115.44$ & 15.60 \\
03381845-2722333 & 2020A & 4 & 4 & HR10 HR21 & 56208.238 & $-46.86$ & $98.89$ & 12.90 \\
03394566-4710178 & 2020B & 4 & 4 & HR10 & 56207.312 & $288.73$ & $337.53$ & 16.30 \\
03401027+0002559 & 2020A & 4 & 8 & U580 & 56195.354 & $-35.88$ & $29.38$ & 13.30 \\
03592788-4650482 & 2020C & 4 & 4 & HR10 & 56194.274 & $39.19$ & $75.96$ & 15.10 \\
03595053-4701073 & 2020A & 4 & 4 & HR10 & 56194.274 & $-90.52$ & $27.62$ & 14.50 \\
04202910-0019338 & 2020A & 4 & 8 & U580 & 55998.026 & $-50.58$ & $100.14$ & 11.90 \\
04301327-5001191 & 2020A & 6 & 12 & U580 & 56264.244 & $118.87$ & $167.53$ & 13.10 \\
04404692-4609391 & 2020A & 4 & 4 & HR10 HR21 & 56577.238 & $59.05$ & $141.04$ & 15.30 \\
04410121-5004008 & 2020A & 4 & 4 & HR10 & 56223.304 & $-16.92$ & $51.75$ & 14.10 \\
04434718-0040232 & 2020B & 4 & 4 & HR10 HR21 & 56551.345 & $98.00$ & $136.04$ & 14.40 \\
05291006-6028494 & 2020B & 4 & 4 & HR21 & 56709.111 & $-21.19$ & $71.29$ & 13.20 \\
05294654-6025081 & 2020A & 4 & 4 & HR10 & 56709.019 & $32.19$ & $75.21$ & 15.10 \\
05313822-6021421 & 2020A & 4 & 4 & HR10 & 56709.019 & $57.18$ & $116.23$ & 16.00 \\
05402480-4726342 & 2020B & 4 & 8 & U580 & 56711.024 & $50.09$ & $71.43$ & 12.50 \\
05403344-4738199 & 2020B & 4 & 4 & HR10 & 56711.113 & $75.79$ & $118.51$ & 15.80 \\
05554481-6034418 & 2020C & 4 & 4 & HR21 & 56606.315 & $3.28$ & $110.72$ & 14.70 \\
05562593-6029184 & 2020A & 4 & 8 & U580 & 56606.315 & $-12.42$ & $34.22$ & 13.10 \\
07554475-0908077 & 2020A & 4 & 4 & HR10 & 56001.042 & $79.69$ & $125.36$ & 14.90 \\
07555317-0848462 & 2020C & 4 & 4 & HR21 & 56000.076 & $47.50$ & $119.55$ & 15.10 \\
07593692-0025252 & 2020A & 8 & 8 & HR10 HR21 & 55974.132 & $-12.32$ & $51.48$ & 14.40 \\
08191969-1412025 & 2020B & 4 & 4 & HR10 & 56758.012 & $47.82$ & $83.86$ & 14.00 \\
08194766-1411293 & 2020B & 4 & 4 & HR10 & 56758.012 & $15.96$ & $48.53$ & 16.00 \\
08231542-0535165 & 2020B & 4 & 8 & U580 & 56314.137 & $-5.39$ & $11.91$ & 12.90 \\
08231783-0523549 & 2020C & 4 & 4 & HR21 & 56341.085 & $-33.50$ & $31.74$ & 16.40 \\
08233762-0536506 & 2020A & 4 & 8 & U580 & 56314.137 & $12.10$ & $44.00$ & 13.10 \\
08395189-0756213 & 2020B & 4 & 4 & HR10 & 56378.103 & $-21.55$ & $15.86$ & 15.10 \\
08395720-0756505 & 2020C & 4 & 4 & HR10 & 56378.103 & $73.20$ & $104.50$ & 13.50 \\
08403017-1409445 & 2020A & 4 & 4 & HR10 & 56678.207 & $50.72$ & $91.12$ & 14.40 \\
08582336-1403021 & 2020B & 4 & 4 & HR10 & 56679.175 & $68.05$ & $98.08$ & 16.60 \\
09193694-1751496 & 2020B & 4 & 4 & HR10 & 56706.273 & $37.38$ & $82.60$ & 14.40 \\
09382162-1758544 & 2020C & 4 & 4 & HR21 & 56708.237 & $64.53$ & $133.81$ & 14.70 \\
09391804-1755456 & 2020B & 4 & 4 & HR21 & 56708.237 & $-26.58$ & $93.47$ & 16.60 \\
09393263-0505599 & 2020B & 5 & 5 & HR10 & 56793.997 & $-33.37$ & $20.06$ & 14.90 \\
09594300-4054056 & 2020B & 4 & 4 & HR10 & 55928.261 & $57.16$ & $84.96$ & 13.90 \\
09594650-4059014 & 2020A & 4 & 4 & HR10 HR21 & 55928.261 & $-43.94$ & $45.55$ & 14.20 \\
10004160-4053496 & 2020A & 4 & 4 & HR10 & 55928.282 & $-56.40$ & $0.72$ & 13.80 \\
10075849-0753079 & 2020C & 4 & 4 & HR21 & 56346.187 & $-3.44$ & $101.75$ & 16.90 \\
10090938-4121350 & 2020B & 4 & 4 & HR10 & 56343.190 & $9.68$ & $47.54$ & 17.13 \\
10091241-4132476 & 2020A & 4 & 4 & HR10 & 56343.190 & $41.90$ & $89.08$ & 16.89 \\
10092032-4138285 & 2020A & 4 & 4 & HR10 HR21 & 56343.190 & $-44.60$ & $55.97$ & 16.23 \\
10092718-4128583 & 2020A & 4 & 8 & U580 & 56343.190 & $5.54$ & $59.52$ & 13.80 \\
10224640-3541044 & 2020A & 4 & 8 & U580 & 56677.262 & $-14.88$ & $28.50$ & 13.60 \\
10232266-3541019 & 2020A & 4 & 8 & U580 & 56679.316 & $22.99$ & $55.67$ & 13.70 \\
10232300-3531571 & 2020C & 4 & 4 & HR21 & 56677.333 & $-33.71$ & $41.65$ & 14.40 \\
10394014-4108011 & 2020A & 4 & 4 & HR10 & 56376.050 & $-32.14$ & $16.59$ & 15.70 \\
10403618-4104492 & 2020A & 4 & 4 & HR10 HR21 & 56376.050 & $-56.89$ & $58.35$ & 16.00 \\
11001645-4102232 & 2020C & 5 & 5 & HR10 & 55972.231 & $7.28$ & $39.34$ & 14.90 \\
11010640-1322020 & 2020C & 4 & 4 & HR10 & 56343.284 & $1.09$ & $33.20$ & 18.60 \\
11035508-1800428 & 2020B & 4 & 4 & HR10 & 56816.953 & $16.00$ & $53.28$ & 14.70 \\
11230355-3455286 & 2020A & 4 & 4 & HR10 & 56798.975 & $-10.70$ & $69.68$ & 13.40 \\
11265745-4100160 & 2020B & 4 & 4 & HR10 & 56376.096 & $-3.01$ & $29.55$ & 13.00 \\
11315400-4359284 & 2020C & 4 & 4 & HR21 & 56378.058 & $-61.47$ & $80.43$ & 14.40 \\
11593504-4050266 & 2020C & 4 & 4 & HR21 & 55998.260 & $-18.04$ & $99.27$ & 16.70 \\
12000916-4101004 & 2020A & 4 & 8 & U580 & 55998.260 & $-47.51$ & $18.99$ & 12.30 \\
12001709-3711459 & 2020A & 4 & 4 & HR10 & 56798.028 & $11.64$ & $73.11$ & 16.40 \\
12005511-3711201 & 2020A & 4 & 8 & U580 & 56798.028 & $-0.76$ & $41.61$ & 13.80 \\
12111883-4109109 & 2020A & 4 & 4 & HR10 HR21 & 56099.020 & $21.79$ & $97.94$ & 14.20 \\
12113870-4103193 & 2020C & 4 & 4 & HR10 & 56099.020 & $-141.40$ & $-3.08$ & 14.30 \\
12121230-4104498 & 2020C & 4 & 4 & HR10 & 56099.020 & $-6.21$ & $33.32$ & 16.80 \\
12194390-3652280 & 2020A & 4 & 4 & HR10 & 56799.021 & $-17.48$ & $37.19$ & 16.50 \\
12270079-4054566 & 2020C & 4 & 4 & HR21 & 56026.160 & $-11.18$ & $70.95$ & 14.80 \\
12273877-4056402 & 2020C & 4 & 8 & U580 & 56026.160 & $-13.10$ & $5.17$ & 13.00 \\
12431359-1304540 & 2020B & 4 & 4 & HR10 & 56075.090 & $80.54$ & $117.97$ & 16.50 \\
12432209-4053149 & 2020A & 4 & 4 & HR10 & 56446.016 & $-43.16$ & $3.42$ & 14.70 \\
12435905-0553086 & 2020A & 4 & 4 & HR10 & 56445.971 & $11.28$ & $65.36$ & 15.20 \\
12562790-4516555 & 2020C & 6 & 6 & HR21 & 56468.068 & $-33.48$ & $29.84$ & 14.80 \\
13201190-0859503 & 2020A & 4 & 4 & HR10 HR21 & 56444.062 & $-63.29$ & $15.57$ & 15.90 \\
13203450-1302162 & 2020C & 4 & 4 & HR10 & 56444.108 & $18.37$ & $50.90$ & 14.30 \\
13272650-4059266 & 2020A & 4 & 4 & HR10 HR21 & 56074.137 & $-52.36$ & $45.90$ & 14.40 \\
13285153-4107423 & 2020A & 4 & 4 & HR10 & 56074.137 & $-122.48$ & $-77.71$ & 15.10 \\
14001419-4054092 & 2020B & 4 & 4 & HR10 & 56002.306 & $-101.41$ & $-61.30$ & 15.70 \\
14091400-3404548 & 2020A & 4 & 4 & HR10 HR21 & 56758.198 & $-12.18$ & $98.40$ & 15.70 \\
14194570-1451154 & 2020C & 4 & 4 & HR21 & 56756.274 & $-55.20$ & $28.19$ & 16.50 \\
14222902-4402086 & 2020A & 4 & 8 & U580 & 56469.067 & $-73.83$ & $-39.67$ & 13.00 \\
14271982-0854407 & 2020B & 4 & 4 & HR10 & 56443.065 & $-54.04$ & $-9.21$ & 14.50 \\
14402357-4009161 & 2020A & 4 & 4 & HR10 HR21 & 56471.007 & $-51.42$ & $6.26$ & 13.40 \\
14591899-2001019 & 2020C & 4 & 4 & HR21 & 56754.372 & $-71.26$ & $5.85$ & 16.60 \\
15001595-2001152 & 2020A & 4 & 4 & HR10 & 56754.264 & $-24.52$ & $13.59$ & 17.10 \\
15003201-1456355 & 2020A & 4 & 4 & HR10 HR21 & 56755.236 & $-124.58$ & $-53.53$ & 15.60 \\
15095102-1507425 & 2020A & 4 & 4 & HR10 & 56756.227 & $-91.84$ & $-34.22$ & 14.30 \\
15095773-2000080 & 2020B & 4 & 8 & U580 & 56757.241 & $-26.65$ & $3.27$ & 13.40 \\
15103048-1508193 & 2020A & 4 & 4 & HR10 & 56756.248 & $-6.32$ & $40.67$ & 14.50 \\
15104140-1502572 & 2020A & 4 & 4 & HR10 HR21 & 56756.227 & $-110.74$ & $-16.23$ & 14.00 \\
15104535-4054419 & 2020B & 4 & 4 & HR10 & 56445.093 & $-63.55$ & $-12.87$ & 14.70 \\
15105813-4048090 & 2020A & 4 & 4 & HR10 HR21 & 56445.093 & $-55.76$ & $58.10$ & 13.70 \\
15112349-4052387 & 2020A & 4 & 4 & HR10 & 56445.093 & $-131.32$ & $-48.91$ & 15.70 \\
15122047-4054438 & 2020B & 4 & 4 & HR21 & 56446.197 & $-50.78$ & $65.82$ & 14.80 \\
15161563-4125518 & 2020C & 4 & 4 & HR21 & 56444.196 & $2.50$ & $67.81$ & 15.00 \\
15164593-4122457 & 2020C & 4 & 4 & HR21 & 56444.196 & $-81.86$ & $-14.62$ & 14.20 \\
15291504-1953570 & 2020C & 4 & 4 & HR21 & 56817.237 & $20.76$ & $83.47$ & 16.80 \\
15300257-4303505 & 2020C & 4 & 4 & HR21 & 56375.273 & $-59.09$ & $12.07$ & 12.70 \\
15305329-1956301 & 2020C & 4 & 4 & HR21 & 56817.219 & $-82.63$ & $24.67$ & 14.20 \\
15305481-4130573 & 2020B & 2 & 2 & HR21 & 56854.987 & $-96.73$ & $92.52$ & 13.30 \\
15420717-4407146 & 2020A & 4 & 4 & HR10 & 56377.359 & $-17.37$ & $78.24$ & 14.80 \\
15490519-1359089 & 2020A & 4 & 4 & HR10 HR21 & 56798.207 & $-39.28$ & $67.37$ & 15.30 \\
15492053-0742483 & 2020A & 4 & 4 & HR10 HR21 & 56853.980 & $-16.69$ & $66.03$ & 14.40 \\
15495562-0724391 & 2020C & 4 & 4 & HR21 & 56853.148 & $-172.12$ & $-97.63$ & 16.50 \\
15502613-0740084 & 2020A & 4 & 4 & HR10 & 56854.001 & $-189.77$ & $-132.55$ & 15.40 \\
15504227-1937508 & 2020C & 4 & 4 & HR21 & 56852.040 & $44.81$ & $118.81$ & 14.60 \\
15545953-4106578 & 2020A & 4 & 4 & HR10 & 56024.218 & $-158.01$ & $21.97$ & 16.70 \\
16035830-4547485 & 2020C & 4 & 4 & HR21 & 56377.316 & $-92.54$ & $-6.58$ & 14.50 \\
17005619-0511542 & 2020C & 4 & 4 & HR21 & 56024.333 & $-63.21$ & $7.58$ & 14.70 \\
17334015-4253407 & 2020A & 7 & 7 & HR10 & 56024.378 & $-11.27$ & $72.68$ & 15.40 \\
17592273-4232176 & 2020C & 4 & 4 & HR21 & 56795.221 & $-21.63$ & $55.54$ & 17.40 \\
18103653-4455176 & 2020B & 4 & 8 & U580 & 56798.409 & $12.57$ & $34.07$ & 13.10 \\
18134362-4221083 & 2020C & 6 & 6 & HR21 & 56821.118 & $-102.95$ & $-15.45$ & 14.50 \\
18135851-4226346 & 2020B & 6 & 12 & U580 & 56856.988 & $-33.65$ & $-6.38$ & 12.90 \\
18162528-4239594 & 2020A & 2 & 2 & HR10 & 56821.258 & $-166.42$ & $61.92$ & 14.10 \\
18180629-4457294 & 2020B & 2 & 2 & HR21 & 56853.175 & $-99.34$ & $28.14$ & 14.10 \\
18201282-4708422 & 2020C & 4 & 4 & HR10 & 56446.173 & $-39.74$ & $32.80$ & 16.40 \\
18203927-4655397 & 2020A & 4 & 4 & HR10 HR21 & 56446.151 & $-59.02$ & $45.27$ & 15.30 \\
18402582-4709250 & 2020C & 4 & 4 & HR10 & 56498.087 & $-77.50$ & $-54.20$ & 17.00 \\
18410111-4238337 & 2020A & 4 & 4 & HR10 HR21 & 56854.225 & $-132.94$ & $96.51$ & 14.20 \\
18490733-3954253 & 2020A & 4 & 4 & HR10 HR21 & 56821.304 & $-52.98$ & $11.59$ & 14.10 \\
18590483-4711187 & 2020C & 2 & 2 & HR21 & 56852.228 & $-3.62$ & $78.82$ & 16.50 \\
18591414-4710472 & 2020C & 2 & 2 & HR21 & 56852.228 & $-126.16$ & $-6.03$ & 16.60 \\
19000942-4231227 & 2020A & 4 & 8 & U580 & 56796.289 & $64.26$ & $118.21$ & 13.20 \\
20183934-5400476 & 2020C & 4 & 4 & HR21 & 56795.348 & $-18.21$ & $53.56$ & 14.50 \\
20192137-4706271 & 2020B & 4 & 8 & U580 & 56169.233 & $-40.37$ & $-17.80$ & 12.80 \\
20194866-4651252 & 2020B & 4 & 4 & HR21 & 56173.176 & $-108.88$ & $42.10$ & 14.60 \\
20593297-4655410 & 2020A & 5 & 5 & HR10 HR21 & 56819.391 & $-89.93$ & $7.19$ & 16.10 \\
20594465-0044334 & 2020B & 4 & 4 & HR10 & 56855.317 & $-36.30$ & $-3.47$ & 15.00 \\
21100126-0156012 & 2020A & 2 & 2 & HR10 & 56075.346 & $-15.57$ & $57.46$ & 15.90 \\
21101784-0205349 & 2020A & 4 & 8 & U580 & 56075.346 & $-51.08$ & $14.05$ & 13.70 \\
21201559-4807298 & 2020C & 2 & 2 & HR21 & 56170.281 & $-207.13$ & $-126.88$ & 17.10 \\
21392385-5501257 & 2020A & 4 & 4 & HR10 & 56852.300 & $-113.35$ & $-54.44$ & 16.20 \\
21402535-0055041 & 2020B & 4 & 8 & U580 & 56855.364 & $-35.24$ & $-9.49$ & 12.70 \\
21523327-0321571 & 2020A & 4 & 4 & HR10 HR21 & 56101.381 & $-131.22$ & $-18.06$ & 12.70 \\
21523611-0327136 & 2020A & 4 & 4 & HR10 HR21 & 56101.381 & $-54.42$ & $27.90$ & 16.10 \\
21594936-4747133 & 2020A & 7 & 7 & HR10 HR21 & 56468.343 & $-80.96$ & $24.33$ & 14.80 \\
21595211-4745562 & 2020C & 7 & 7 & HR21 & 56103.390 & $-58.88$ & $9.37$ & 15.70 \\
22003339-4803527 & 2020A & 7 & 7 & HR10 & 56468.343 & $-40.29$ & $80.29$ & 12.90 \\
22180319-5834560 & 2020C & 4 & 4 & HR21 & 56853.375 & $-71.53$ & $15.23$ & 14.60 \\
22184292-5454411 & 2020C & 4 & 4 & HR21 & 56634.025 & $-66.04$ & $18.99$ & 15.10 \\
22184686-5506505 & 2020A & 4 & 4 & HR10 & 56607.047 & $60.95$ & $122.54$ & 14.20 \\
22291350-0507554 & 2020B & 4 & 4 & HR10 & 56502.314 & $-25.08$ & $26.66$ & 14.40 \\
22293255-5016362 & 2020C & 4 & 4 & HR21 & 56635.034 & $-41.36$ & $40.10$ & 14.60 \\
22494111-0506006 & 2020A & 4 & 4 & HR10 & 56548.228 & $-105.68$ & $-26.75$ & 15.80 \\
22495134-5544411 & 2020B & 4 & 4 & HR10 & 56576.109 & $-11.74$ & $32.94$ & 14.10 \\
22593725-0052333 & 2020A & 4 & 8 & U580 & 56501.304 & $-65.70$ & $-26.62$ & 13.90 \\
23291894-5018404 & 2020A & 4 & 4 & HR10 & 56503.371 & $31.27$ & $75.99$ & 16.10 \\
23303304-0504082 & 2020C & 4 & 4 & HR21 & 56225.047 & $-26.68$ & $51.21$ & 15.30 \\
23354061-4305405 & 2020A & 4 & 4 & HR10 & 56857.312 & $69.50$ & $112.65$ & 15.30 \\
23394097-0056031 & 2020C & 4 & 4 & HR21 & 56224.096 & $-32.62$ & $28.29$ & 15.90 \\
23481930-5617480 & 2020B & 4 & 4 & HR10 & 56547.261 & $42.89$ & $80.65$ & 15.10 \\
23501242-0503050 & 2020B & 4 & 4 & HR21 & 56267.025 & $-62.12$ & $58.52$ & 15.70 \\
23501961-5012563 & 2020A & 4 & 8 & U580 & 56602.084 & $-70.46$ & $70.20$ & 12.20 \\
23572607-4802051 & 2020C & 4 & 4 & HR21 & 56206.128 & $-27.59$ & $46.31$ & 15.30 \\
 \\
 \hline
  \label{tab:mw_sb2}
 \end{longtable}
 \tablefoot{The column `CNAME' is the GES name (constructed from the J2000 coordinates), `flag' is the final flag after eye inspection, `\# exp.' the number of exposures available for that star, `\# sp.' is the number of available spectra (larger than the number of exposures in the case of UVES data which provide two spectra per exposure), `setup' is the spectrograph setup,  `MJD' is the modified Julian Date of the unique observation listed, $v_r(1)$ and $v_r(2)$ are the velocities of the two components in \kms. The last column gives the visual magnitude of the source.}
\end{longtab}

\begin{table*}[h]
 \centering
 \renewcommand\thetable{A.2}

 \caption{List of SB3 candidates in the field ordered by right ascension.}
 \begin{tabular}{ccrrlcrrrc}
 \hline\hline
 CNAME & flag & \# exp. & \# sp. & setup & MJD & $v_r(1)$ & $v_r(2)$ & $v_r(3)$ & $V$ \\
 \hline\\
08202324-1402560 & 2030A & 4 & 4 & HR10 & 56758.012 & $-14.05$ & $42.53$ & $100.73$ & 14.20 \\
12000646-4052156 & 2030A & 4 & 8 & U580 & 55998.324 & $-33.99$ & $14.34$  & $56.68$ & 12.70 \\
13593100-1003043 & 2030A & 6 & 12 & U580 & 55999.277 & $-16.50$ & $11.43$ & $52.05$ & 12.60 \\
15003096-2000179 & 2030A & 4 & 8 & U580 & 56754.264 & $-105.96$ & $-71.31$ & $-42.15$ & 13.80 \\
18170244-4227076 & 2030A & 2 & 2 & HR10 & 56821.258 & $-39.69$ & $-1.83$ & $32.40$ & 14.30 \\
 \\
 \hline
 \end{tabular}
  \label{tab:mw_sb3}
\end{table*}

\newpage
\section{SB2 candidates in selected fields}

\begin{table*}[h]
 \centering
 \caption{\label{tab:other_sb} List of SB2 candidates in  selected fields\tablefootmark{a} ordered by right ascension.}
 \begin{tabular}{lcrrlcrrl}
 \hline\hline
 CNAME & flag & \# exp. & \# sp. & setup & MJD & $v_r(1)$ & $v_r(2)$ & $V$ \\
 \hline\\
\textbf{Bulge}\\
17542544-3750568 & 2020C & 2 & 2 & HR21 & 56819.206 & $-77.81$ & $9.78$ & 15.36 \\
17571482-4147030 & 2020B & 2 & 4 & U580 & 56173.006 & $-11.07$ & $30.13$ & 11.57 \\
17581333-3434348 & 2020B & 3 & 3 & HR21 & 56817.269 & $-32.77$ & $54.95$ & 15.46 \\
18041571-3000506 & 2020A & 2 & 2 & HR21 & 55724.240 & $-51.90$ & $145.50$ & 16.31 \\
18175005-3247501 & 2020C & 2 & 2 & HR21 & 56207.979 & $-17.67$ & $75.84$ & 15.10 \\
18380149-2820437 & 2020B & 2 & 2 & HR21 & 56758.359 & $-128.80$ & $-42.25$ & 14.18 \\
\\
\textbf{Cha~I} \\
11085326-7519374 & 2020B & 2 & 4 & U580 & 56047.093 & $-33.15$ & $51.72$ & 12.17 \\
11120384-7650542 & 2020B & 2 & 2 & HR15N & 56025.156 & $11.90$ & $60.79$ & 14.16 \\
\\
\textbf{CoRoT}\\
06435847-0100516 & 2020A & 18 & 18 & HR10 HR15N & 55999.997 & $-40.34$ & $54.97$ & 13.05 \\
19235724+0138241 & 2020A & 6 & 6 & HR10 HR15N HR21 & 56470.289 & $7.22$ & $103.61$ & 14.40 \\
19243943+0048136 & 2020C & 18 & 18 & HR15N & 56171.039 & $-52.08$ & $65.68$ & 15.11\tablefootmark{b} \\
19255064+0022240 & 2020C & 6 & 6 & HR15N & 56756.414 & $-47.86$ & $-1.75$ & 12.69 \\
19261871+0030211 & 2020A & 6 & 6 & HR10 HR21 & 56473.199 & $13.50$ & $90.26$ & 14.86 \\
19262489+0137506 & 2020A & 6 & 6 & HR10 & 56816.215 & $-50.50$ & $67.49$ & 14.90 \\
19263739+0152562 & 2020A & 6 & 6 & HR10 & 56473.166 & $-60.25$ & $221.28$ & 13.60 \\
\\
\textbf{$\gamma^2$~Vel}\\
08072516-4712522 & 2020A & 2 & 4 & U580 & 55972.105 & $-25.30$ & $20.17$ & 11.43 \\
08073722-4705053 & 2020C & 2 & 4 & U580 & 55929.251 & $54.85$ & $83.85$ & 11.83 \\
08074628-4700347 & 2020B & 2 & 2 & HR15N & 55929.251 & $-11.60$ & $56.77$ & 13.37 \\
08082580-4716381 & 2020C & 2 & 2 & HR15N & 55928.190 & $-0.20$ & $61.00$ & 16.21 \\
08091392-4715498 & 2020C & 2 & 2 & HR15N & 55928.146 & $15.65$ & $107.03$ & 16.93 \\
08091937-4719385 & 2020C & 2 & 2 & HR15N & 55972.080 & $-67.47$ & $92.02$ & 12.75 \\
08093154-4724289 & 2020C & 2 & 2 & HR15N & 55928.146 & $17.43$ & $107.48$ & 17.47 \\
08093589-4718525 & 2020B & 2 & 6 & HR15N U580 & 55972.080 & $-4.47$ & $40.48$ & 12.79 \\
08094221-4719527 & 2020C & 2 & 6 & HR15N & 55972.080 & $-29.17$ & $49.90$ & 12.40 \\
08094864-4702207 & 2020B & 2 & 2 & HR15N & 55927.156 & $-4.09$ & $52.59$ & 16.52 \\
08095076-4745311 & 2020C & 2 & 2 & HR15N & 55972.155 & $-44.06$ & $1.17$ & 12.90 \\
08095692-4717476 & 2020B & 2 & 2 & HR15N & 55972.080 & $-33.39$ & $48.57$ & 13.35 \\
08103996-4714428 & 2020B & 2 & 8 & HR15N U580 & 55972.056 & $-36.24$ & $60.15$ & 12.06 \\
08111009-4718006 & 2020B & 2 & 2 & HR15N & 55928.099 & $11.41$ & $65.79$ & 16.36 \\
08115305-4654115 & 2020A & 2 & 6 & U580 HR15N & 55927.111 & $0.52$ & $53.30$ & 12.93 \\
08115892-4715140 & 2020B & 2 & 2 & HR15N & 55928.099 & $-4.87$ & $60.44$ & 16.95 \\
\\
\textbf{$\rho$~Oph}\\
16244913-2447469 & 2020C & 2 & 2 & HR15N & 56103.158 & $-10.40$ & $47.36$ & 15.68 \\
 \\
 \hline
 \end{tabular}
 \tablefoot{\\
\tablefoottext{a}{See text for references about target selection and membership assessement in those fields.}\\
 \tablefoottext{b}{The visual magnitude of this star was wrongly assessed by CASU. The closest star resolved in Simbad is at a distance of 42.88 arcsec and corresponds to CoRoT 100791478.} 
 }
\end{table*}

\newpage

\section{SB2, SB3 and SB4 candidates in stellar clusters}
\begin{figure}
\center
\includegraphics[width=9cm]{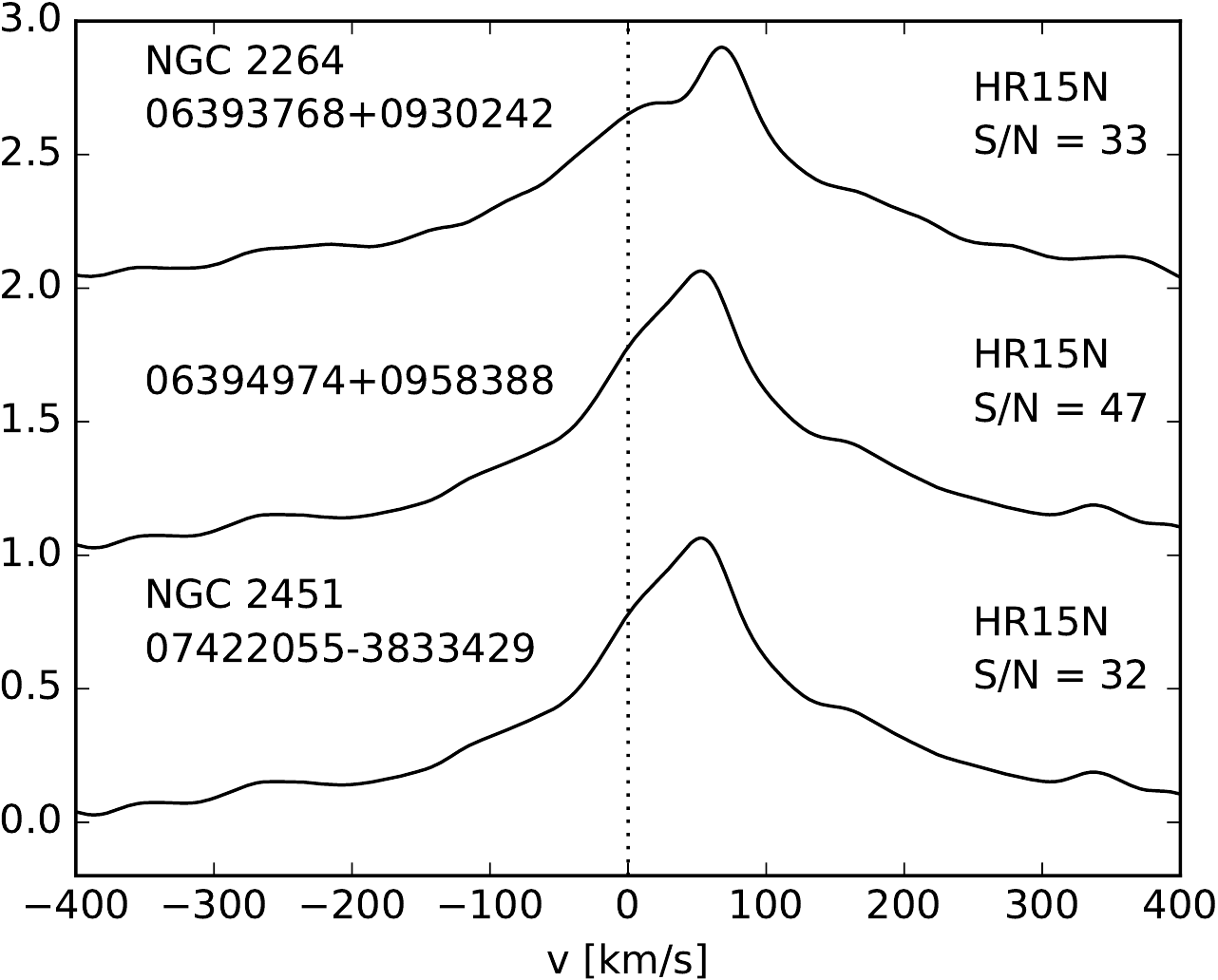}
\caption[]{\label{fig:NGC2264}
Two examples of stars in NGC~2264 and one in NGC~2451 flagged as SB2 by the \doe\ procedure but discarded from the final list.}
\end{figure}
\label{ap:sb_clusters}
\noindent {\it IC 2391.} 
This open cluster includes the unique SB4 candidate 08414659-5303449 in the current iDR4 GES data. At some epochs, the two weakest components are hardly visible, that is why we classified this source with both 2020A and 2040A flags (see Fig.~\ref{fig:HD74438}). This SB4 candidate is analyzed in detail in Sect.~\ref{Sect:SB4}. 
\medskip\\
\noindent {\it IC 2602.} All the three systems are consistent with cluster membership. System 10403116-6416249 seems to be a pair of rapidly rotating stars.
\medskip\\
\noindent {\it IC 4665.} System 17452506+0540233 has a broad CCF with a secondary bump in its tail, but the velocities are not centred on the cluster velocity. 
\medskip\\
\noindent {\it M67.} All four SB2s are confirmed through visual inspection, having composite spectra and having membership confirmed.
\medskip\\
\noindent {\it NGC 2243.} Only one clear SB2 candidate with a composite spectrum is retained. Two other candidates (06292559-3116070 and 06294409-3116276) are not retained, since the major peak of the CCF is at the cluster velocity, with  a secondary bump offset by $-60$~\kms\ and $-100$~\kms, respectively.
\medskip\\
\noindent {\it NGC 2264.}  The \doe\ procedure has flagged a lot of stars as SB2 in this (and in all other) young clusters. Many of these stars have broad CCFs with a secondary bump, as illustrated in Fig.~\ref{fig:NGC2264}. As the centre of this very broad CCF is close to the cluster velocity, these stars are thought to be both rapidly rotating and pulsating  ($\delta$~Scu variables), and this combination is responsible for the peculiar and specific CCFs observed in young clusters, whose turn-off is located higher up on the main sequence to allow the presence of $\delta$~Scu stars.
Not all of them are A stars though, and therefore we suggest the alternative hypothesis
that this peculiar CCF profile is related to the disk still surrounding these young stars. In that case, the CCFs offer an interesting diagnostic to study/detect these disks (see \citealt{rebull2002}).
06405650+0911389 (HD 261905) has its main peak at the cluster velocity, and a clearly defined, well-separated second peak at a velocity of 71.9~\kms. Although the field is not especially crowded, the DSS image\footnote{\url{http://archive.stsci.edu/dss}} reveals that the stellar image might be not perfectly round, and seems contaminated by a nearby source.
06421531+0942581 has a secondary peak close to the cluster velocity, but the main peak is totally offset (99~\kms). That peak might be due to a somewhat brighter star (NGC 2264 SBL 560) located about 4 arcsec west of the target (probably not a member, given its largely offset velocity).
\medskip\\
\noindent {\it NGC 2451.} 
The situation for this cluster is quite special since there are in fact two different clusters, located at different distances, superimposed at the same location on the sky \citep{dias2002}.
These authors report a velocity of +22.7~\kms\ for the nearest NGC 2451A cluster and 14.0~\kms\ for the farthest NGC 2451B cluster. 
07401559-3735416, a genuine SB2 system, cannot be a member of NGC~2451.
On JD 2456634, the CCF exhibits peaks  at  21 and 62~\kms, while at  JD 2456638 and JD 2456677, the peaks are located around $-2$ and 85~\kms, implying a centre-of-mass velocity of the order of 40~\kms, significantly offset with respect to the velocities of NGC~2451A and NGC~2451B.
07422055-3833429 bears similarities with the cases discussed in relation with NGC~2264, namely a very broad CCF (base width of about 400~\kms), a main peak at 105~\kms, well offset with respect to the cluster velocity, 
and another bump at 20~\kms, close to the cluster velocity. The spectrum seems to show \ha\ emission. This star has been discarded from the final list.
\medskip\\
\noindent{\it NGC 2516.}
07593671-6021483 is probably a genuine SB2, with the peaks (22 and 50~\kms) centred on the cluster velocity (23.6~\kms). 
07594121-6109251 has a broad (base width 100~\kms) CCF, with two bumps ($-23$ and $-5$~\kms) not centred on the cluster velocity and is maybe contaminated by nebular absorption lines.
\medskip\\
\noindent {\it NGC 2547.}
08081564-4908244 is a genuine SB2, but probably not a member of NGC~2547, since the component velocities (52 and 122~\kms) do not encompass the cluster velocity (15.7~\kms). 
\medskip\\
\noindent {\it NGC 3293.} 
10361099-5814310 classified as 2020C is probably a $\delta$ Scu star (the recommended parameters are  \Teff\,$=8985$~K, $\log{g}=4.01$) and shows emission in \ha. 
Rather than SB2 systems, 10353288-5813498 and 10353397-5813178 are rapidly rotating (and probably pulsating) stars (because their CCFs are distorted). They are pre-main sequence star candidates \citep{2007A&A...467.1397D}.
\medskip\\
\noindent {\it NGC 3532.} The source 11085927-5849560 is identified for the first time as an SB2 candidate.
\medskip\\
\noindent {\it NGC 6005.} Three SB2 candidates have only been observed with the HR9B setup (around the Mg~I~b triplet) where it is difficult to assess if the spectra are composite or not. 15555518-5725349 shows a broad CCF due to H$\alpha$ in HR15N with a main peak at $-69$~\kms, and a bump at $-27$~\kms, close to the cluster velocity. 
\medskip\\
\noindent {\it NGC 6530.}
Numerous spurious detections of SB candidates are due to the presence of nebular lines in emission in HR15.
Also, there are thin and deep absorption lines around 6678, 6715 and 6730 \AA.
Nebular lines are present around 6717 and 6730~\AA\ in HR15 and have led to some reduction issues since negative fluxes are observed at these wavelengths in some stars (18044420-2415380, 18045889-2415261, 18043887-2427164). Their associated CCFs are then not reliable and have been discarded from the final list. 
There is strong and deep  \ha\ absorption in several stars. 
Surprisingly, many discarded SB2 components have velocities close to $-60$~\kms. This raises the question of the presence of another possible velocity component for that cluster.
\medskip\\
\noindent {\it NGC 6633.} 
18280622+0642252 (NGC 6633 110, $V=10.1$, A3) is an interesting case of a fast rotator which could be a $\delta$~Scu-type star according to the iDR4 recommended parameters (\Teff\,$=9600$~K, $\log{g}=4.80$ and solar metallicity). Only the upper grating of U580 is available showing very thin and deep absorption lines superimposed on the less deep and rotationally broadened Na~I~D doublet probably caused by nebular line contamination. 
\medskip\\
\noindent {\it NGC 6705.}
The composite spectra and the associated CCFs of one of the five SB2 A candidates are presented in Fig.~\ref{fig:sb2_setups}. This is an illustration of a very favorable case because 18503230-0617112 (NGC~6705~1936) has been observed in eight setups and shows a two-component CCF in all of them.
18511434-0617090 has four observations with the HR15N (\ha) setup. In all cases, the main peak is around 33~\kms, close thus to the cluster velocity, whereas the CCF exhibits a secondary bump around $-50$~\kms. But the contrast  of that bump is variable, suggesting that its origin may be related to stellar variability \citep[but $B - V \sim 1$, suggesting that the star is a red giant, and \ha\ variability is not expected;][]{2014A&A...569A..17C}.
On the contrary, if the system were an SB2, its kinematics is not compatible with membership in  the cluster.
\medskip\\
\noindent{\it NGC 6752.} 
19105940-5957059 is the star A13 in \citet{2006AA...451..499M} which has not been detected as binary. The CCFs show clearly the presence of two peaks (flagged 2020~B). 23 observations covering more than 1500 days are available, but we unsuccessfully tried to fit an orbit. Indeed the radial velocities of the components stay constant within few \kms. Moreover the star is located in a very dense region of this globular cluster and we conclude that this is an  ``optical'' SB2.
\medskip\\
\noindent{\it NGC 6802.}
Two SB2 and one SB3 candidates have been found in this cluster. 
\medskip\\
\noindent {\it Trumpler 14.}
A large number of false SB2 detections were identified due to the presence of very strong nebular lines and reduction issues in HR15N where spectra have \ha\ with negative flux and core emission (see Sect.~\ref{sect:ccf_degeneracy} and figures within for discussion). Nebular emission in Trumpler~14 and more generally in the Carina nebula is investigated in details by \citet{damiani2016}.
\medskip\\
\noindent{\it Trumpler 20.}
12384378-6037077 has one component located at the cluster velocity.
The unique CCF of 12393764-6038190 could either be indicative of a rapidly rotating star, with some asymmetries in the line profile or of a cluster member ($-36$~\kms) blended with a non-member ($-77$~\kms). The same remark holds true for 12393362-6041446. The secondary peak of 12391767-6036083 is probably from a non-member. The main peak in the CCF of 12391992-6029552 at $-3.4$~\kms\ is probably from a non-member. 
\medskip\\
\noindent{\it Trumpler 23.} 
The radial velocity of this cluster has just been assessed to $-61.3\pm 1.9$ \kms\ within the GES consortium (Overbeek et al., submitted). Therefore the SB2 candidate 16004521-5332044 may be considered as a member of this cluster.

\begin{table*}[b]
\caption[]{\label{tab:SB2cluster}
SB2, SB3 and SB4 candidates in clusters ordered by increasing identifier. The column `CNAME' is the GES name (constructed from the J2000 coordinates), `flag' is the final flag after eye inspection, `\# exp.' the number of exposures available for that star, `\# sp.' is the number of available spectra (larger than the number of exposures in the case of UVES data which provide two spectra per exposure), `setup' is the spectrograph setup,  `MJD' is the modified Julian Date of the unique observation listed, $v_r$ is the cluster velocity, $v_r(1)$ and $v_r(2)$ are the velocities of the two components. The `Member' column states whether the SB candidate belongs to the cluster or not (see Sect.~\ref{sec:sb2_candidates}). The last column `Remark' contains additional information after detailed inspection of their spectra and CCFs: CS: composite spectrum, 1RRC/2RRC: one or two rapidly-rotating component(s), PULS: pulsating star, $\delta$~Scu: probable $\delta$~Scu type star, \ha e: \ha\ with emission, NaDe: Na~I~D with emission, NLC: nebular line contamination, ILC: interstellar line contamination, XR: X-ray source, ORB: orbit calculated, ST: see text for additional information. The ``?'' character is indicative of some uncertainty in the preceeding characterisation.}
\setlength{\tabcolsep}{3pt}
{\footnotesize
\begin{tabular}{llllllrrccl}
\hline
\textbf{Cluster} & log age &   &  &       &     &  \multicolumn{2}{c}{$v_r$ (\kms)}\\
CNAME & flag & \# exp. & \# sp. & setup & MJD & $v_r(1)$  & $v_r(2)$  & SB2/3/4 & Member & Remark \\
      &                   &       &       &      &        &             (\kms)  &  (\kms) \\ 
\hline\\
\textbf{IC 2391}    &  7.74 & & &         &     &   \multicolumn{2}{c}{$  14.49 \pm 0.14$}\\
 08385566-5257516 & 2020B &  4  &  8  & U580 & 56705.032  &  $-14.84$ &   44.91 & SB2 & y & CS \\
 08393881-5310071 & 2020A &  6  & 12  & U580 & 56705.032  &  $-23.25$ &   39.12 & SB2 & y & CS, 1RRC\\
 08414659-5303449 & 2020A & 45  & 90  & U520 & 56707.028  &  $-21.64$ &   50.75 & SB2 & y & CS \\
 08414659-5303449 & 2040A & 45  & 90  & U520 U580 & 56707.028 &  $-21.64$ &   50.75 & SB4 & y & CS, ST
 \medskip\\
\textbf{IC 2602}    &  7.48 & & &        &     & \multicolumn{2}{c}{$  18.12 \pm 0.30$} \\
10403116-6416249  & 2020C &  1  &  1  & HR15N & 53827.129 &  $-67.55$ &   98.63 & SB2 & y & CS \\
10450829-6422416  & 2020A &  1  &  1  & HR15N & 53839.031 &  $-69.80$ &   66.01 & SB2 & y & CS \\
10460575-6420184  & 2020B &  2  &  4  & U580  & 56711.229 &  $-18.14$ &   22.87 & SB2 & y & CS, 1RRC 
\medskip\\
\textbf{IC 4665}    &  7.60 & & &         &     & \multicolumn{2}{c}{$ -15.95 \pm 1.13$}  \\
17450496+0541287  & 2020A &  6  &  6  & HR15N & 56469.201 & $-109.77$ &   46.17 & SB2 & y & CS, \ha e \\
17452506+0540233  & 2020C &  2  &  2  & HR15N & 56471.099 & $   7.72$ &   87.39 & SB2 & n & CS? \\
17453692+0542424  & 2020A &  2  &  4  & U580  & 56471.099 & $ -49.75$ &   18.17 & SB2 & y & CS \\
17455717+0601224  & 2020B &  2  &  2  & HR15N & 56471.233 & $ -43.74$ &   53.34 & SB2 & y & CS \\
17472992+0607069  & 2020B &  2  &  2  & HR15N & 56473.072 & $ -64.42$ &    8.52 & SB2 & y & CS 
\medskip\\
\textbf{M 67}       &  9.60 & & &        &     & \multicolumn{2}{c}{$  33.8 \pm 0.5$} \\
 08511868+1147026 & 2020A &  3  &  6  & U580l & 54866.304 &  $-12.47$ &   88.80 & SB2 & y & CS, XR \\
 08511901+1150056 & 2020A &  4  &  8  & U580l & 54866.221 &  $-28.00$ &   97.99 & SB2 & y & CS \\
 08512291+1148493 & 2020A &  4  &  8  & U580l & 54866.221 &  $ 15.78$ &   53.47 & SB2 & y & CS \\
 08512940+1154139 & 2020A &  5  & 10  & U580l & 54853.182 &  $ 15.99$ &   49.90 & SB2 & y & CS 
 \medskip\\
\textbf{NGC 2243}   &  9.60 & & &       &     & \multicolumn{2}{c}{$  59.5 \pm 0.8$}  \\
 06290412-3114343 & 2020B &  4  &  4  & HR15N & 56603.226 &   19.78 &  118.51 & SB2 & y & CS \\
 \medskip\\
\textbf{NGC 2264}   &  6.48 & & &        &     & \multicolumn{2}{c}{$  24.69 \pm 0.98$} \\
 06404608+0949173 & 2020A & 22  & 24  & HR15N U580  & 55915.177 &  $-88.21$ &  102.35 & SB2 & y & CS, 2RRC, ORB, XR, ST \\
 06413150+0954548 & 2020B & 22  & 24  & HR15N U580 & 55915.177  &  $-24.07$ &   58.82 & SB2 & y & CS, NLC? \\
 06413207+1001049 & 2020B &  4  &  6  & U580  & 56267.205 &   77.74 &  133.75 & SB2 & n & CS, 1RRC \\
 06414775+0952023 & 2020A &  8  & 10  & HR15N U580  & 56268.205 &  $-52.95$ &   84.71 & SB2 & y & CS, NaDe \\
\medskip\\
 \textbf{NGC 2451}    &  7.8 (A)  & & &        &     & \multicolumn{2}{c}{22.70 (A)}  \\ 
                        &  8.9 (B)  & & &        &     & \multicolumn{2}{c}{14.00 (B)}\\
07371334-3831467  & 2020B &  4  &  4  & HR15N & 56634.212 &  $-16.05$ &   46.59 & SB2& y & CS \\
07382664-3839208  & 2020B &  4  &  4  & HR15N & 56634.212 &  $  0.71$ &   57.57 & SB2& y & \\
07384076-3743189  & 2020C &  4  &  4  & HR15N & 56677.217 &  $ -6.51$ &   61.72 & SB2?& y & \\
07401559-3735416  & 2020A &  6  & 12  & U580  & 56634.259 &  $ 21.18$ &   61.77 & SB2& n & CS, ST\\
07405697-3721458  & 2020A &  2  &  4  & U580l & 56677.309 &  $101.60$ &  146.85 & SB2& n & CS \\
07413421-3719442  & 2020C &  2  &  2  & HR15N & 56635.182 &  $ 49.04$ &  118.40 & SB2?& n & \\
07431451-3810155  & 2020B &  2  &  2  & HR15N & 56635.226 &  $-11.94$ &   47.02 & SB2& y & \\
07454636-3809168  & 2020C &  2  &  2  & HR15N & 56679.221 &  $-34.99$ &   46.81 & SB2& y & \\
07455390-3812406  & 2020B &  4  &  8  & U580  & 56637.222 &  $ -2.21$ &   31.73 & SB2& y & CS \\
07455995-3854469  & 2020B &  2  &  2  & HR15N & 56679.290 &  $-23.86$ &   68.51 & SB2& y & CS \\
07463487-3905202  & 2020A &  2  &  2  & HR15N & 56679.290 &  $-10.76$ &   75.22 & SB2& y & CS \\
07470917-3859003  & 2030A &  2  &  4  & U580  & 56637.287 &  $ 25.04$ &    96.07,136.62 & SB3& y & CS 
\medskip\\
 \hline
 \end{tabular}
 }
 \end{table*}
 
 \addtocounter{table}{-1}
 
 \begin{table*}
 \caption[]{Continued.}
{\footnotesize
\setlength{\tabcolsep}{3pt}
\begin{tabular}{llllllrrccl} 
\hline\\
\textbf{Cluster} & log age &   &  &       &     &  \multicolumn{2}{c}{$v_r$ (\kms)}\\
CNAME & flag & \# exp. & \# sp. & set-up & MJD & $v_r(1)$  & $v_r(2)$  & SB2/3/4 & Member & Remark \\
      &                   &       &       &      &        &             (\kms)  &  (\kms) \\ 
\hline\\
\textbf{NGC 2516}   &  8.20 & & &         &     & \multicolumn{2}{c}{$  23.6 \pm 1.0$}  \\
07540665-6043081  & 2020A &  2  &  2  & HR15N & 56342.032 &  $-40.52$ &   63.01 & SB2& y & CS, \ha e\\
07551150-6028375  & 2020C &  2  &  2  & HR15N & 56374.017 &  $ -4.99$ &   54.22 & SB2& y & CS\\
07563381-6046027  & 2020B &  2  &  2  & HR15N & 56375.037 &  $ -3.54$ &   50.50 & SB2& y & CS \\
07575737-6044162  & 2030C &  2  &  4  & U520  & 56375.037 &  $-30.40$ & 30.90, 77.79  & SB3& y & ST \\
07593411-6042583  & 2020B &  4  &  4  & HR15N & 56375.037 &  $-39.31$ &   61.19 & SB2& y & CS \\
07593671-6021483  & 2020B &  2  &  4  & U580  & 56376.004 &  $ 21.84$ &   49.71 & SB2& y & CS \\
07594121-6109251  & 2020C &  3  &  4  & U580  & 56374.128 &  $-23.26$ &   $-5.11$ & SB2 & n & CS?, NLC? \\
07594744-6049228  & 2020B &  4  &  4  & HR15N & 56375.011 &  $ -1.23$ &   48.64 & SB2& y & CS \\
07595659-6049283  & 2020C &  2  &  2  & HR15N & 56375.078 &  $ -9.56$ &   49.97 & SB2& y & CS
\medskip\\
\textbf{NGC 2547}   &  7.54 & & &         &     & \multicolumn{2}{c}{$  15.65 \pm 1.26$}  \\
08081564-4908244  & 2020A &  4  &  6  & U580 HR15N  & 56310.201 &   51.01 &  119.87 & SB2 & n & CS, ST\\
\medskip\\
\textbf{NGC 3293}   &  7.00 & & &         &     & \multicolumn{2}{c}{$ -12.00 \pm 4.00$}  \\
10343408-5814431  & 2020A & 12  & 12  & HR3 HR5A HR9B HR14A & 55972.322 &   $-9.65  $&   51.10  & SB2 & n & CS?, \ha e\\
10345341-5812222  & 2020C & 12  & 12  & HR9B  & 56024.110 &  $-21.54$ &   39.34 & SB2 & y &  \ha e? \\
10350728-5810574  & 2020B &  9  &  9  & HR6   & 56024.034 &  $-18.58$ &  151.33 & SB2 & y & 2RRC?\\
10361099-5814310  & 2020C &  9  &  9  & HR6 HR14A & 56000.121 &  $-78.31$ &   41.59 & SB2? & y & $\delta$ Scu? \ha e ST\\
10361385-5819052  & 2020B & 12  & 12  & HR5A HR14A  & 55972.322 &  $-70.47$ &   34.69 & SB2 & y & CS?, 1RRC\\
10361494-5814170  & 2020B &  7  &  7  & HR6 HR14A   & 55999.147 &  $-54.32$ &   20.01 & SB2 & y & \ha e \\
10361791-5814296  & 2020C & 12  & 12  & HR14A & 55972.322 &  -46.06 &   57.26 & SB2 & y & \\
10362294-5825333  & 2020B &  7  &  7  & HR3   & 55998.113 &  $-40.41$ &   39.06 & SB2 & y & CS? \\
10362842-5805112  & 2020B &  7  &  7  & HR3 HR5A HR6  & 55998.218 &  $-40.13$ &   29.55 & SB2 & y & CS\\
\medskip\\
\textbf{NGC 3532}   &  8.48 & & &          &     & \multicolumn{2}{c}{$  -4.8 \pm 1.4$} \\
11085927-5849560  & 2020B &  9  & 18  & U580  & 56440.953 &  $-11.12$ &   28.73 & SB2 & y & CS
\medskip\\
\textbf{NGC 4815}   &  8.75 & & &         &     & \multicolumn{2}{c}{$   -29.4 \pm 4$}  \\
12573865-6454061  & 2020B & 12  & 12  & HR9B  & 56025.203 & $-103.71$ &   37.33 & SB2 & y & noisy \\
12572682-6456300  & 2020C & 10  & 10  & HR15N & 56028.203 &  $-89.49$ &   $-2.06$ & SB2 & y & \ha e?\\
\medskip\\
\textbf{NGC 6005}   &  9.08 & & &         &     & \multicolumn{2}{c}{$ -24.1 \pm 1.3$}  \\
15553867-5724434  & 2030B &  4  &  4  & HR9B  & 56795.265 &  $-81.6 $ & $-14.4$, 32.7&SB3?& y? & noisy, ST\\
15554550-5728087  & 2020B &  4  &  4  & HR9B  & 56795.265 &  $-63.56$ & $   3.69$ & SB2 & y & CS? \\
15554669-5725386  & 2020A &  2  &  2  & HR9B  & 56794.295 &  $-50.27$ & $   0.40$ & SB2 & y & CS? \\
15555518-5725349  & 2020C &  6  &  6  & HR15N & 56816.147 &  $-68.85$ & $ -26.73$ & SB2?& n?& noisy, ST \\
15561896-5725399  & 2020A &  2  &  2  & HR9B  & 56794.295 & $-104.43$ & $ -28.40$ & SB2 & n & CS? \\
\medskip\\
\textbf{NGC 6530}   &  6.30 & & &        &     & \multicolumn{2}{c}{$ -4.21 \pm 6.35$}  \\
18040734-2422217  & 2020C &  1  &  1  & HR15  & 52787.320 & $ -40.$0 9  &    5.40 & SB2 & y \\
18040988-2425323  & 2020A &  1  &  1  & HR15  & 52787.390 & $ -62.$7 6  &   50.74 & SB2 & y & CS, NLC\\
18045495-2423096  & 2020C &  3  &  3  & HR15  & 52787.390 & $ -61.$2 8  &    0.50 & SB2? & y?\\
18045528-2412512  & 2020A &  2  &  2  & HR15N & 56173.078 & $-105.$5 2  &    6.67 & SB2 & y & CS \\
18052912-2428104  & 2020C &  2  &  2  & HR15N & 56502.262 & $ -14.$1 3  &   56.55 & SB2 & n & \ha e\\
\medskip\\
\textbf{NGC6633}    &  8.78 & & &        &     & \multicolumn{2}{c}{$ -28.8 \pm 1.5$} \\
18263193+0637329  & 2020B &  2  &  2  & HR15N & 56444.288 &  $-72.17$ &   78.21 & SB2 & y & \ha e\\
18263896+0630410  & 2020B &  2  &  2  & HR15N & 56445.184 & $ -74.59$ &   $-1.45$ & SB2 & y & NLC? \\
18264081+0632435  & 2020B &  2  &  2  & HR15N & 56445.184 & $   4.85$ &   58.63 & SB2 & n & CS \\
18265864+0640458  & 2020B &  2  &  2  & HR15N & 56444.288 & $ -89.10$ &   $-0.78$ & SB2 & y & CS \\
18270724+0638394  & 2020A &  2  &  2  & HR15N & 56445.184 &  $-49.28$ &   42.61 & SB2 & y & CS \\
18271075+0627061  & 2020C &  4  &  4  & HR15N & 56442.297 & $   2.20$ &   58.00 & SB2 & n & \\
18272122+0637268  & 2020A &  5  &  5  & HR9B HR15N  & 56444.273 &  $-45.42$ &   59.67 & SB2 & y  & CS \\
18272783+0644321  & 2020C &  2  &  2  & HR15N & 56444.309 & $ -34.67$ &   37.55 & SB2 & y & CS\\
18274341+0641115  & 2020C &  5  &  5  & HR9B  & 54279.258 & $ -80.28$ &  $-31.25$ & SB2 & n \\
18280622+0642252  & 2020A &  2  &  4  & U520u & 56444.342 & $ -55.62$ &   39.20 & SB2 & y & 1RRC, PULS, ILC \\
18280970+0638061  & 2020B &  3  &  3  & HR15N & 56444.333 & $ -66.95$ &   28.70 & SB2 & y & CS \\
18281038+0647407  & 2020B &  2  &  4  & U580  & 56854.131 & $ -38.61$ &  $-13.44$ & SB2 & y & CS \\
18282150+0645278  & 2020C &  2  &  2  & HR15N & 56446.241 & $  -8.46$ &  103.81 & SB2 & n & \ha e?\\
18282354+0646402  & 2020B &  3  &  3  & HR15N & 56444.333 & $-104.96$ &   30.15 & SB2 & y & CS \\
18283303+0645562  & 2020C &  4  &  6  & U520  & 56444.333 & $  -7.29$ &   10.97 & SB2 & n & CS, ILC\\ 
\hline
 \end{tabular}
 }
 \end{table*}
 
 \addtocounter{table}{-1}
 
 \begin{table*}
 \caption[]{Continued.}
{\footnotesize
\setlength{\tabcolsep}{3pt}
\begin{tabular}{llllllrrccl} 
\hline\\
\textbf{Cluster} & log age &   &  &       &     &  \multicolumn{2}{c}{$v_r$ (\kms)}\\
CNAME & flag & \# exp. & \# sp. & set-up & MJD & $v_r(1)$  & $v_r(2)$  & SB2/3/4 & Member & Remark \\
      &                   &       &       &      &        &             (\kms)  &  (\kms) \\ 
\hline\\
\textbf{NGC 6705}   &  8.47 & & &         &     & \multicolumn{2}{c}{$  34.9 \pm 1.6$}  \\
18503230-0617112  & 2020A & 12  & 12  & HR6 HR9B HR10 HR21  & 56103.110 &  $-38.24$ &  117.09 & SB2 & y & CS \\
18503690-0621100  & 2020C &  2  &  2  & HR15N & 56102.120 &  $-29.51$ &   83.84 & SB2 & y & CS\\
18503840-0617048  & 2020C &  7  &  7  & HR10 HR15N HR21 & 56443.382 &  $-18.97$ &   85.47 & SB2 & y & CS\\
18504649-0611443  & 2020C &  2  &  2  & HR15N & 56101.333 &  $-42.77$ &   37.24 & SB2 & y?  & CS\\
18505726-0609408  & 2020C &  2  &  2  & HR15N & 56102.120 &  $-14.52$ &  104.31 & SB2 & y \\
18505561-0614552  & 2020B &  2  &  2  & HR15N & 56101.243 &  $-10.39$ &   80.95 & SB2 & y & CS \\
18505933-0622051  & 2020C &  4  &  4  & HR15N & 56077.363 &  $ -2.32$ &  144.85 & SB2 & y &\ha e?\\
18510072-0609118  & 2020C &  2  &  2  & HR15N & 56101.333 &  $-12.01$ &   60.80 & SB2 & y &CS\\
18510223-0614547  & 2020A & 10  & 10  & HR3 HR6 HR9B HR10  & 56099.365  &   $-8.13$ &   69.14 & SB2 & y & CS, 1RRC \\
                    &       &       &       & HR14A \\
18510286-0615250  & 2020A & 12  & 12  & HR3 HR6 HR9B HR10 & 56442.400 &  $-19.65$ & 71.55   &SB2 & y & CS \\
                    &       &       &       & HR21 \\ 
18510286-0615250  & 2030A & 12  & 12  & HR3 HR6 HR9B HR10 & 56099.311 &  $-44.37$ & 40.28, $93.39$& SB3 & y & CS, ST\\
                    &       &       &       & HR21 \\ 
18510401-0615387  & 2020C &  2  &  2  & HR15N & 56075.275 &  $ -7.77$ &   62.68 & SB2 & y &noisy\\
18510405-0617156  & 2020C &  2  &  2  & HR15N & 56075.255 &  $-59.09$ &   42.95 & SB2 & y &\ha e?\\
18510456-0617121  & 2020A & 12  & 12  & HR3 HR6 HR9B HR10  & 56103.110  &   $-7.76$ &   81.62 & SB2 & y & CS \\
                    &       &       &       & HR14A  HR15N  HR21 \\ 
18510462-0616124  & 2020B & 10  & 10  & HR3 HR6 HR9B HR14A  & 56099.365 &   $-3.64$ &   91.28 & SB2 & y &  \\
18511134-0616106  & 2020A & 12  & 12  & HR3 HR6 HR9B HR10  & 56103.110  &    2.11 &   71.07 & SB2 & y & CS\\
                    &       &       &       & HR14A HR21 \\
18511220-0617467  & 2020B &  2  &  2  & HR15N & 56101.288 &  $ -3.56$ &   78.60 & SB2 & y & CS\\
18511434-0617090  & 2020C &  4  &  4  & HR15N & 56075.300 &  $-51.22$ &   33.12 & ?   & y & \ha e?, ST\\
18512166-0624074  & 2020C &  4  &  4  & HR15N & 56075.300 &    7.39 &   53.69 & SB2 & y \\
18513193-0612518  & 2020C &  2  &  2  & HR15N & 56077.363 &   $-2.33  $&   96.69  & SB2 & y &\ha e?\\
\medskip\\
\textbf{NGC 6752}   &  10.13  & & &        &     & \multicolumn{2}{c}{$ -24.5 \pm 1.9$} \\
19105940-5957059  & 2020B & 54  & 108 & U580  & 54624.335 &  $-39.76$ &  $-18.25$ & SB2 & y  & 
\medskip\\
\textbf{NGC 6802}   &  8.95 & & &         &     & \multicolumn{2}{c}{$  11.9 \pm 0.9$}  \\
19302315+2013406  & 2030C &  4  &  4  & HR9B  & 56794.388 &  $-22.35$ &22.04, 65.53& SB3 & y \\
19303540+2016178  & 2020A &  4  &  4  & HR9B  & 56794.388 &  $-10.52$ &   34.61 & SB2 & y & CS \\
19304355+2016530  & 2020A &  6  &  6  & HR9B  & 56797.303 &  $-60.71$ &   86.93 & SB2 & y \\
\medskip\\
\textbf{Trumpler 14}  &  6.67 & & &         &     & \multicolumn{2}{c}{$         -15.0$}  \\
10434299-5953132  & 2020B & 15  & 15  & HR6   & 56445.026 & $-118.89$ &  113.84 & SB2 & y  \\
10442462-5930359  & 2020A & 19  & 19  & HR5A HR6 HR14A & 56442.090  & $-139.47$ &   94.25 & SB2 & y  & CS, 1RRC, \ha e\\
10443037-5937267  & 2020A & 19  & 19  & HR6 HR14A & 56442.094 & $-126.45$ &  161.46 & SB2 & y  & CS, 2RRC, \ha e \\
\medskip\\
\textbf{Trumpler 20}&  9.20 & & &        &     & \multicolumn{2}{c}{$ -40.2 \pm 1.3$} \\
12382369-6041067  & 2020C &  1  &  1  & HR9B  & 54962.018 & $ -58.36$ &   $-8.07$ & SB2 & y & noisy\\
12382945-6036007  & 2020A &  1  &  1  & HR9B  & 54960.122 &  $-72.68  $&   13.06  & SB2 & y &\\
12383365-6031092  & 2020B &  1  &  1  & HR9B  & 54929.059 & $ -53.14$ &   36.44 & SB2 & y & CS? \\
12384378-6037077  & 2020B &  1  &  1  & HR9B  & 54959.979 & $ -40.24$ &   24.74 & SB2 & n & CS \\
12384744-6036400  & 2020B &  1  &  1  & HR9B  & 54960.028 &  $-43.73  $&    9.52  & SB2 & n &\\
12385726-6038597    & 2020C &  1    &  1    & HR9B  & 54960.075 &  $-63.54 $  & $  -4.50$   & SB2?& y \\
12390677-6042208  & 2020B &  1  &  1  & HR9B  & 54959.979 &  $-21.45  $&   19.41  & SB2 & n \\
12390898-6037473  & 2020B &  7  &  7  & HR9B HR15N  & 56002.184 & $-124.91$ &   41.80 & SB2 & y & CS?\\
12391247-6037429  & 2020C &  1  &  1  & HR9B  & 54962.068 &  $-67.29  $&  $ -3.30 $& SB2 & y &\\ 
12391904-6035311  & 2020B &  1  &  1  & HR9B  & 54960.075 &  $-85.78  $&14.83&SB2  & y\\ 
12391904-6035311  & 2030C &  1  &  1  & HR9B  & 54960.075 &  $-85.78  $&$-44.40$, 14.83&SB3  & y\\ 
12393449-6039575  & 2020B &  1  &  1  & HR9B  & 54962.068 &  $-60.30  $&  $ -7.33 $& SB2 & y \\
12393764-6038190  & 2020C &  1  &  1  & HR9B  & 54960.075 &  $-77.26  $&  $-36.00 $& SB2  & y & 1RRC?, ST\\
12391767-6036083  & 2020C &  1  &  1  & HR9B  & 54962.068 &  $-42.12  $&   $-4.61 $& SB2 & y & ST\\
12391992-6029552  & 2020C &  1  &  1  & HR9B  & 54962.068 &  $-49.20  $&   $-3.38 $& SB2 & y & ST\\
12394909-6040513  & 2020A &  7  &  7  & HR9B HR15N  & 56002.184 &  $-69.73$ &  $-12.26$ & SB2 & y \\
12401228-6034325  & 2020C & 12  & 12  & HR15N & 56377.231 &  $-65.37  $&   $-2.61$  & SB2 & y \\
12402686-6036013  & 2020C &  1  &  1  & HR9B  & 54962.018 &  $-67.57  $&  $-11.38 $& SB2 & y \\
12403561-6044331  & 2020C &  1  &  1  & HR9B  & 54962.018 &  $-60.11  $&  $-26.56$  & SB2 & y & noisy\\
12404299-6046290  & 2020A &  1  &  1  & HR9B  & 54929.107 & $-119.68$ &   17.91 & SB2 & y & CS? \\
\hline
 \end{tabular}
 }
 \end{table*}
 
 \addtocounter{table}{-1}
 
 \begin{table*}
 \caption[]{Continued.}
{\footnotesize
\setlength{\tabcolsep}{3pt}
\begin{tabular}{llllllrrccl} 
\hline\\
\textbf{Cluster} & log age &   &  &       &     &  \multicolumn{2}{c}{$v_r$ (\kms)}\\
CNAME & flag & \# exp. & \# sp. & set-up & MJD & $v_r(1)$  & $v_r(2)$  & SB2/3/4 & Member & Remark \\
      &                   &       &       &      &        &             (\kms)  &  (\kms) \\ 
\hline\\
\textbf{Trumpler 23}  &  8.90 & &               &         &     & \multicolumn{2}{c}{$-61.3\pm1.9$} \\
16004521-5332044  & 2020A &  4  &  4  & HR9B  & 56551.985 & $-137.87$ &   26.29 & SB2  & y & 2RRC
\medskip\\
\textbf{Berkeley 25}  &  9.70 & & &           &     & \multicolumn{2}{c}{$ 134.3\pm0.2$}  \\
06413639-1628236  & 2020B & 20  & 20  & HR9B  & 56576.317 &   $-6.62$ &   27.00 & SB2  & n & noisy \\
06414138-1624323  & 2020B & 20  & 20  & HR9B  & 56576.297 &  $-10.52$ &   28.45 & SB2  & n & noisy
\medskip\\
\textbf{Berkeley 81}  &  8.93 & &&           &     & \multicolumn{2}{c}{$  48.3\pm0.6  $}   \\
19013140-0028066  & 2020C &  8  &  8  & HR15N & 56170.005 &  $-11.28$ &   76.17 & SB2? & y &\\
19013257-0027338  & 2020A & 24  & 24  & HR9B HR15N  & 56170.005 &  $-18.49$ &  118.26 & SB2  & y & ORB \\
\hline
\end{tabular}
}
\end{table*}

\end{document}